\documentclass{egpubl}
 
\JournalPaper         %

\CGFccby

\usepackage{fix-cm}
\usepackage[T1]{fontenc}
\usepackage{dfadobe}  

\usepackage{cite}  %
\BibtexOrBiblatex

\electronicVersion
\PrintedOrElectronic
\ifpdf \usepackage[pdftex]{graphicx} \pdfcompresslevel=9
\else \usepackage[dvips]{graphicx} \fi

\usepackage{egweblnk} 

\usepackage{amsmath} 	%
\usepackage{amsfonts} 	%
\usepackage{dsfont} 	%
\usepackage[ruled]{algorithm2e} %
\usepackage{xfrac} %

\usepackage{soul}       %
\usepackage{microtype} 	%
\usepackage{lastpage}
\usepackage{units} 	    %
\usepackage{textcomp}
\usepackage{gensymb} 	%
\usepackage{pifont} 	%

\usepackage{array} 	    %
\usepackage{booktabs}
\usepackage{multirow} 	%
\usepackage{makecell} 	%
\usepackage{arydshln}
\usepackage{wrapfig}  	%
\usepackage{float} 	    %
\usepackage[percent]{overpic}

\usepackage[export]{adjustbox}

\makeatletter
\let\egpubl@makecaption\@makecaption
\makeatother

\usepackage{subcaption}

\makeatletter
\let\@makecaption\egpubl@makecaption
\makeatother

\usepackage{enumitem}   %

\usepackage[x11names]{xcolor} 	%
\usepackage{tcolorbox}    %
\usepackage{fancybox}
\usepackage{dashbox}
\usepackage{transparent}

\newcommand{\cmark}{\ding{51}}%
\newcommand{\xmark}{\ding{55}}%

\newlength{\tmpintextsep}
\newlength{\tmpcolumnsep}
\tmpintextsep=\intextsep
\tmpcolumnsep=\columnsep

\newcommand{\figref}[1]{Fig.~\ref{#1}}
\newcommand{\secref}[1]{Sec.~\ref{#1}}
\newcommand{\tabref}[1]{Table~\ref{#1}}

\newcommand{\appref}[1]{App.~\ref{#1}}

\newsavebox{\largestimage}

\title[Single-line drawing generation via semantics-driven optimization]%
      {Single-Line Drawing Generation via Semantics-Driven Optimization}

\author[T.\ Magne et al.]
{\parbox{\textwidth}{\centering Tanguy Magne\orcid{0009-0001-0231-026X}, Alexandre Binninger\orcid{0000-0002-9833-4126}, Ruben Wiersma\orcid{0000-0001-7900-7253} and Olga Sorkine-Hornung\orcid{0000-0002-8089-3974}}
        \\
{\parbox{\textwidth}{
    \centering ETH Zurich, Zurich, Switzerland\\
    tanguy.magne@inf.ethz.ch, alexandre.binninger@inf.ethz.ch, ruben.wiersma@inf.ethz.ch, olga.sorkine@inf.ethz.ch
}}}

\begin{document}

\teaser{
    \vspace{-2ex}
    \centering
    \begin{subfigure}[t]{0.24\textwidth}
        \vspace{0pt}%
        \begin{overpic}[width=\textwidth, height=\textwidth]{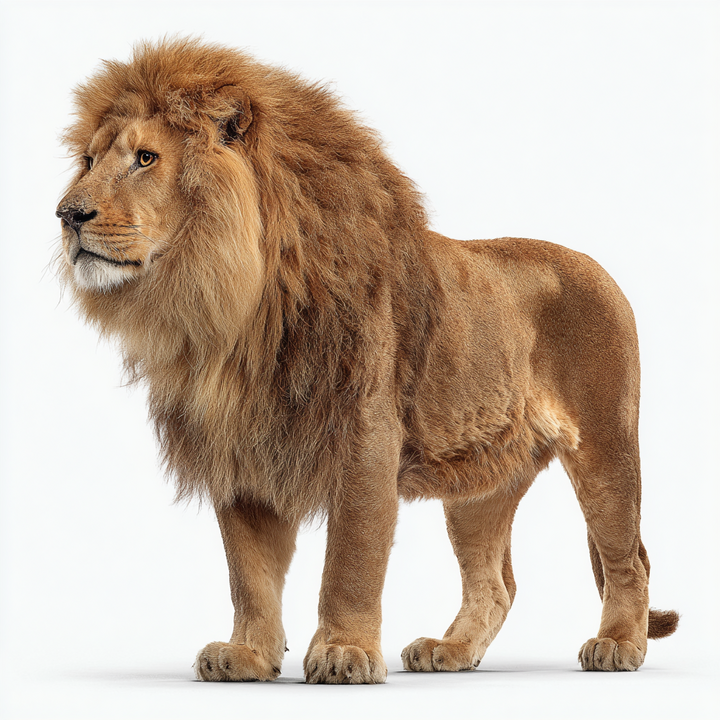}%
         \put (0,6) {\setlength{\fboxsep}{1pt}\transparent{0.6}\colorbox[RGB]{255,255,255}{\transparent{1.0}\parbox{\dimexpr\linewidth-2\fboxsep}{\footnotesize{\textit{Single line drawing of a lion, continuous black line on a white background.}}}}}
        \end{overpic}
    \end{subfigure}
    \hspace{0.01\linewidth}%
    \begin{subfigure}[t]{0.24\linewidth}
        \vspace{0pt}%
        \begin{overpic}[width=0.49\textwidth, height=0.49\textwidth]{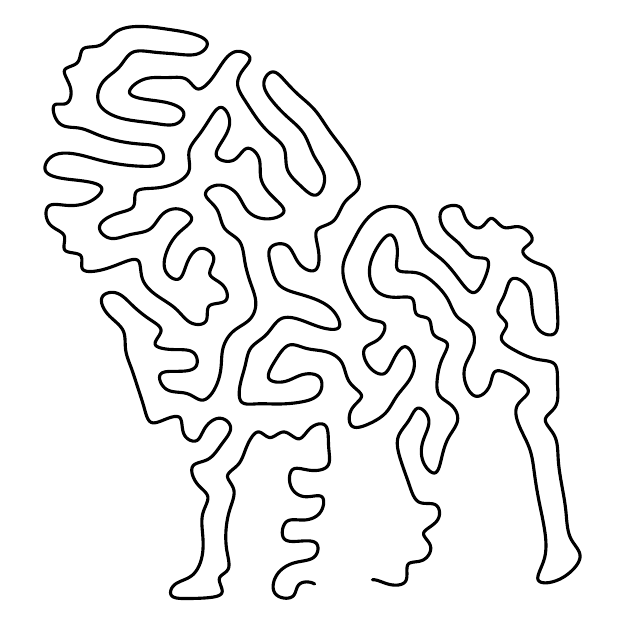}%
         \put (60,90) {\scriptsize 0 iter}%
        \end{overpic}%
        \begin{overpic}[width=0.49\textwidth, height=0.49\textwidth]{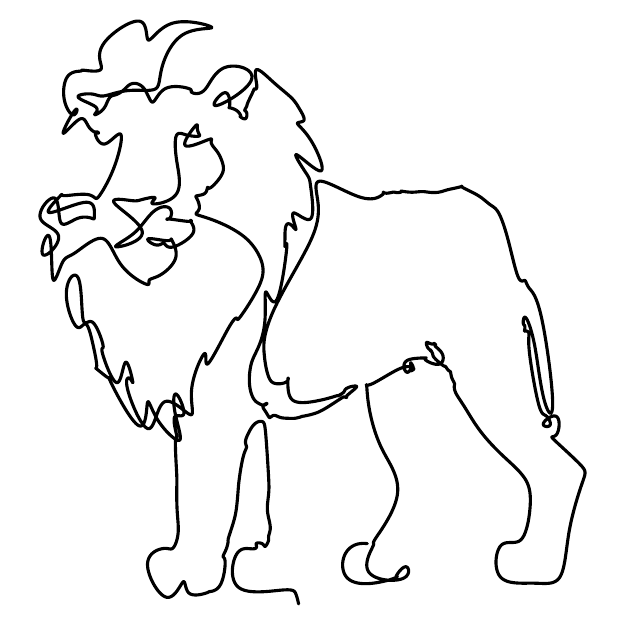}%
         \put (60,90) {\scriptsize 1000 iter}%
        \end{overpic}\\[0.01\textwidth]%
        \begin{overpic}[width=0.49\textwidth, height=0.49\textwidth]{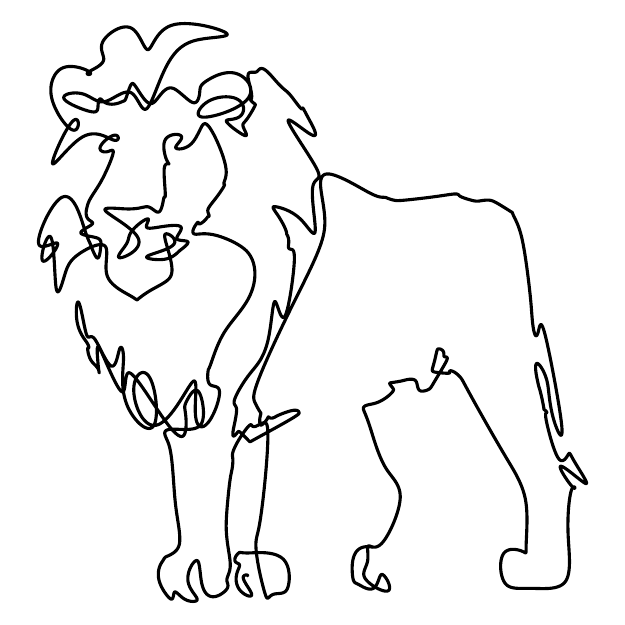}%
         \put (60,90) {\scriptsize 2000 iter}%
        \end{overpic}%
        \begin{overpic}[width=0.49\textwidth, height=0.49\textwidth]{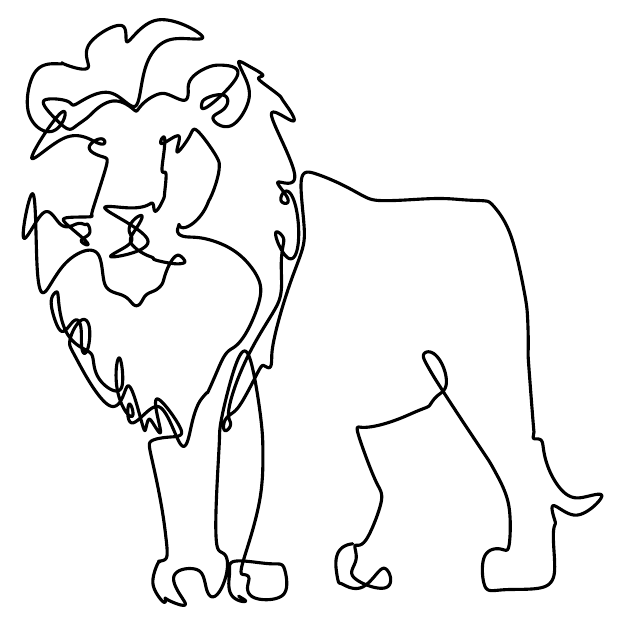}%
         \put (60,90) {\scriptsize 3000 iter}%
        \end{overpic}%
    \end{subfigure}%
    \hspace{0.01\linewidth}%
    \begin{subfigure}[t]{0.24\linewidth}
        \vspace{0pt}%
        \adjincludegraphics[width=\textwidth,trim={{.1\width} {.1\height} {.1\width} {.1\height}},clip]{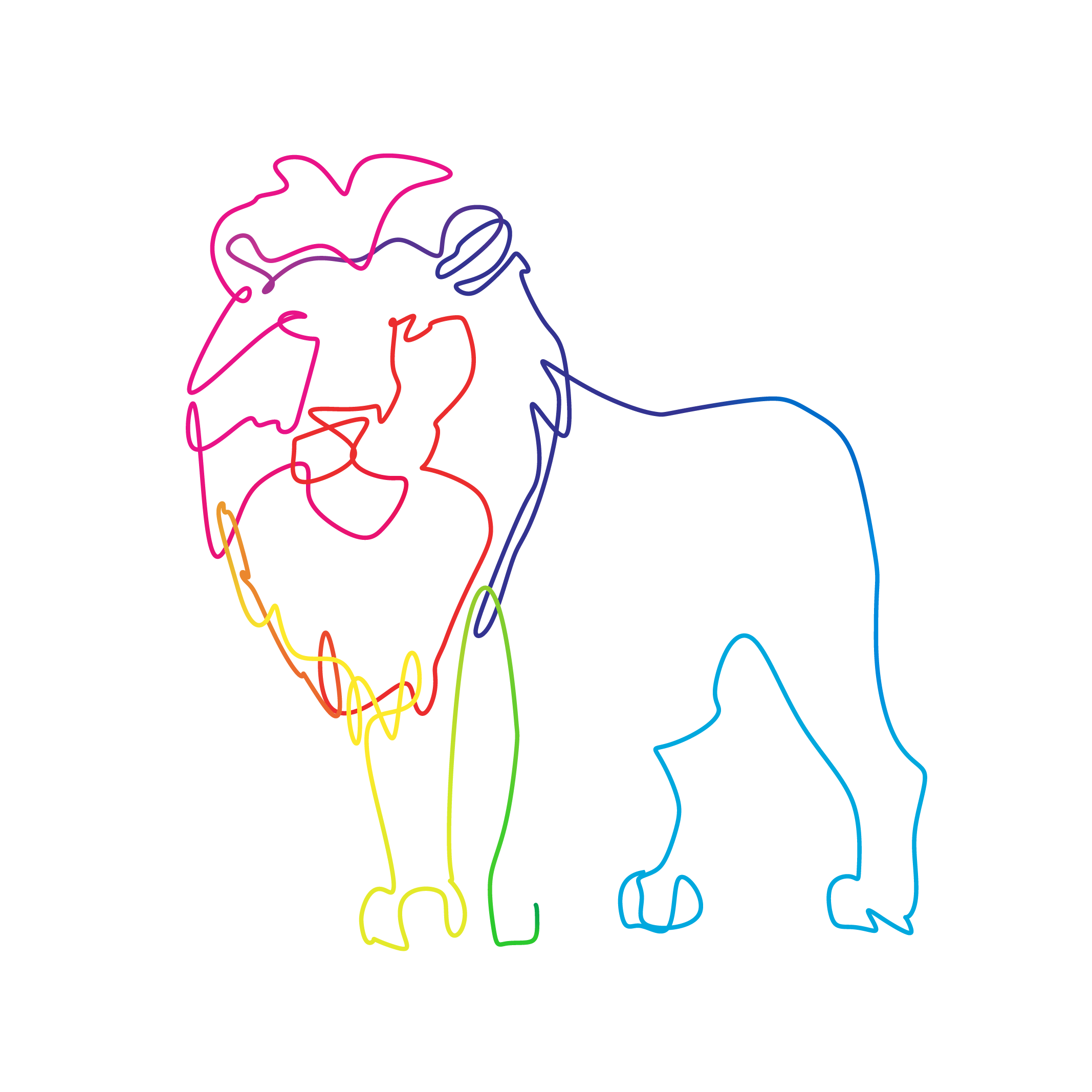}%
    \end{subfigure}%
    \hspace{0.01\linewidth}%
    \begin{subfigure}[t]{0.24\linewidth}
        \vspace{0pt}%
        \adjincludegraphics[width=\textwidth, height=\textwidth, trim={{0.03\width} {0.03\width} {0.03\width} {0.03\width}},clip]{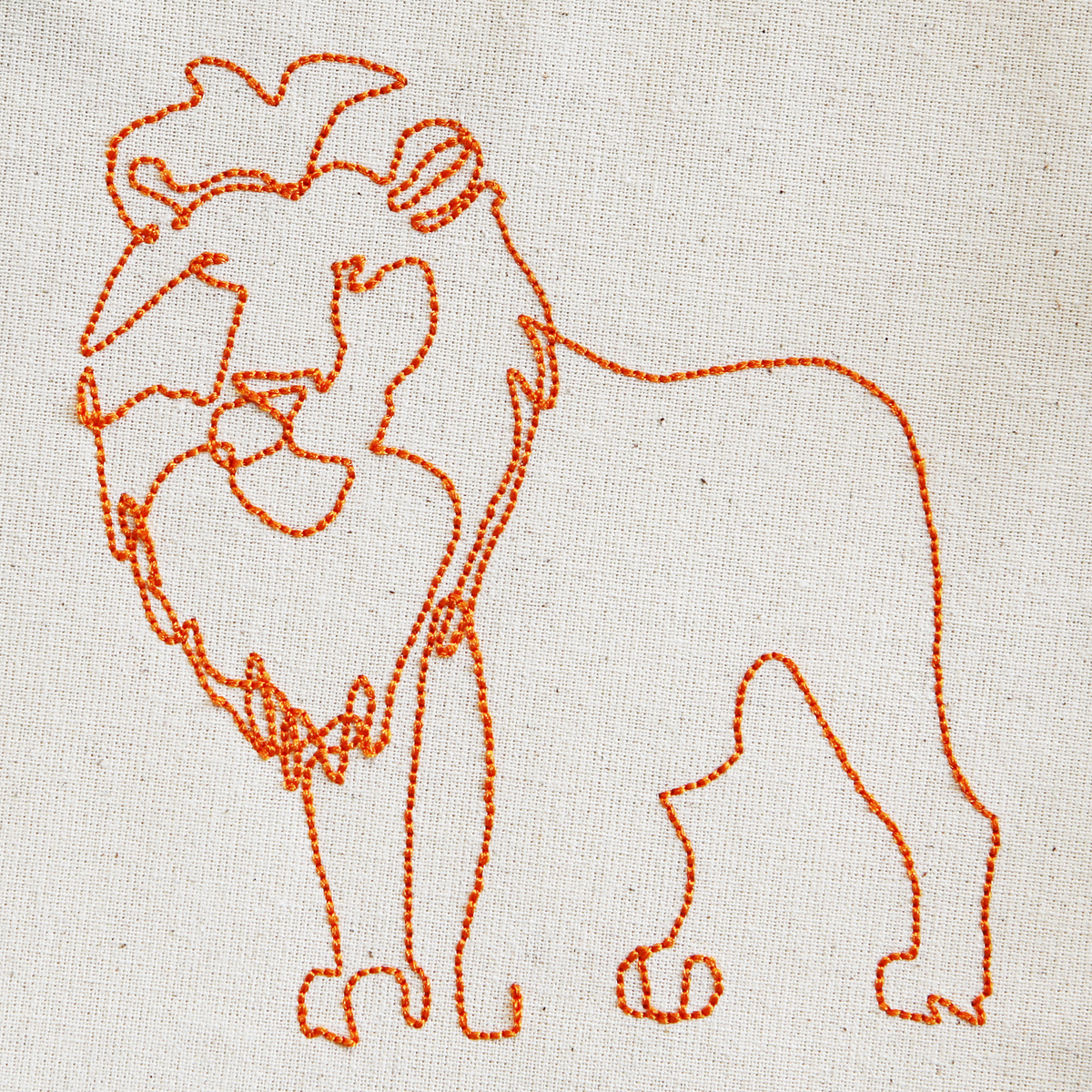}
    \end{subfigure}%
    \hspace{0.03\linewidth}%
    \caption{Single-line drawings, where lifting the pen from the paper is not allowed, lead to minimalist creations, often exhibiting surprising and expressive abstractions. Our method generates single-line drawings in vector format based on text and/or image input. From left to right: Input image and prompt, the single-line drawing at various stages of the optimization, the final result (the parameterization of the curve is represented by a color gradient) and an example of embroidery fabrication created from the generated single-line drawing.}\label{fig:teaser}
    \vspace{2ex}
}

\maketitle

\begin{abstract}
Line drawings are a highly expressive art form that requires the artist to abstract and distill the essence of their subject. We present the first semantics-driven method for automatically generating single-line drawings in vector format, guided either by a text prompt describing the concept or an input image depicting it. Our approach leverages score distillation sampling to optimize the parameters of a uniform rational B-spline (URBS) curve, ensuring that the drawing consists of a single continuous stroke by design. This representation provides fine-grained control over the level of detail, while additional loss terms allow us to steer the final artistic style.
We demonstrate that our method outperforms state-of-the-art text-to-image models and optimization pipelines for this task, producing results that are both more aesthetically pleasing and more faithful to the style of continuous line drawing artists. Furthermore, because our method generates a vectorized curve, it directly supports downstream fabrication processes such as embroidery, laser engraving and wire bending. Our code and results are available at https://github.com/tanguymagne/SLDgen.


\end{abstract}

\section{Introduction}
\label{sec:intro}

\begin{figure}[t]
    \centering
	\small
	\setlength{\tabcolsep}{1pt}
    \setlength{\fboxsep}{0pt}
    \begin{tabular}{cc}
    \adjincludegraphics[width=0.49\linewidth,trim={{0.0\width} {.15\height} {0.0\width} {0.0\height}},clip]{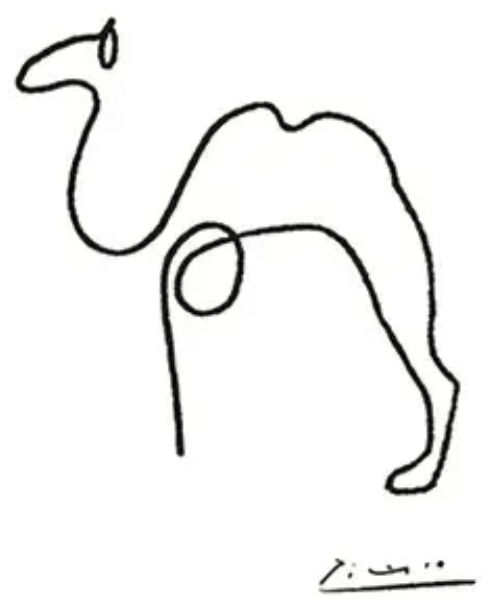}%
     & \adjincludegraphics[width=0.49\linewidth,trim={{0.0\width} {0.03\height} {0.0\width} {0.05\height}},clip]{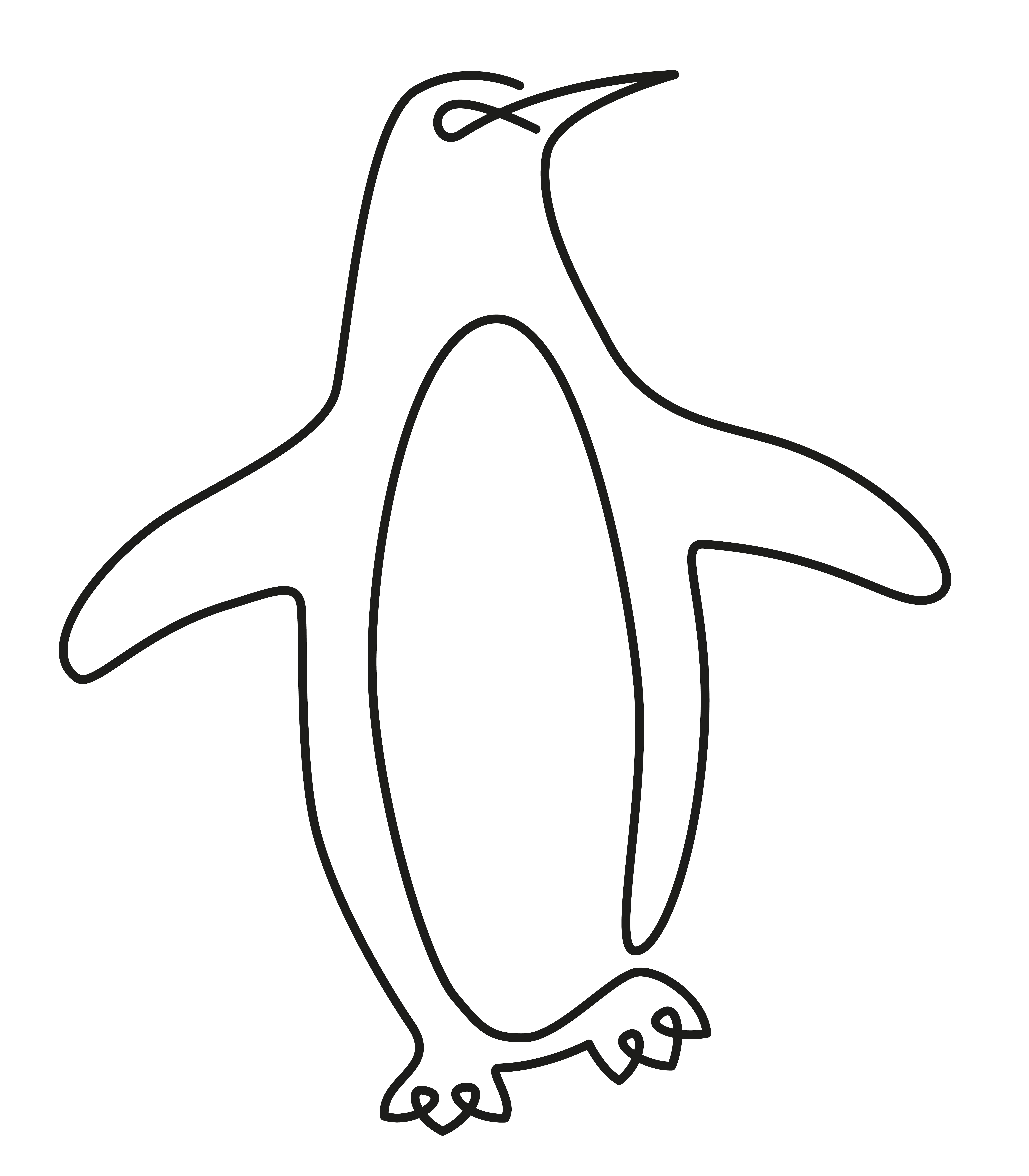} \\
    \adjincludegraphics[width=0.49\linewidth,trim={{0.075\width} {0.0\height} {0.025\width} {0.1\height}},clip]{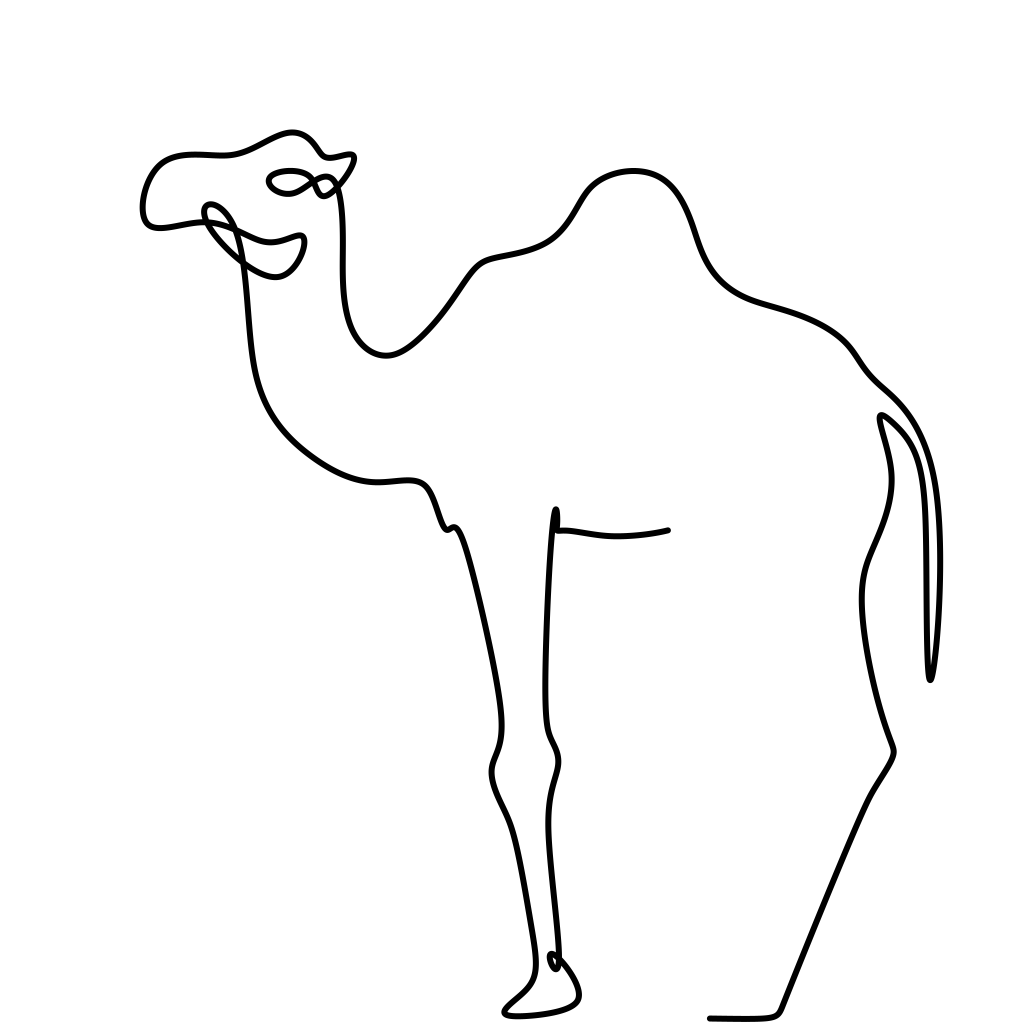}%
     & \adjincludegraphics[width=0.49\linewidth, height=0.49\linewidth,trim={{0.1\width} {0.08\width} {0.1\width} {0.12\width}},clip]{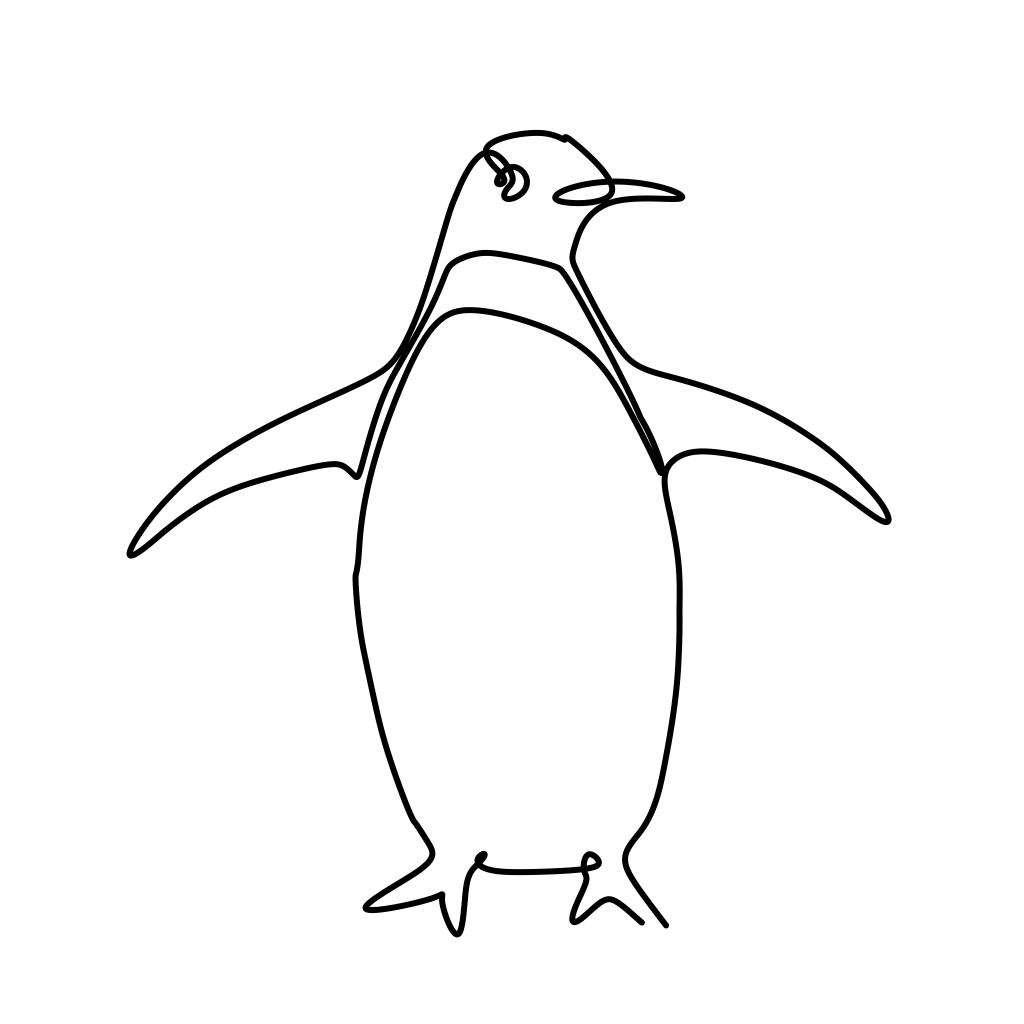}%
    \end{tabular}
    \caption{Top: examples of single-line drawings from Picasso (left) and Felix Hornung (right). Bottom: examples of single-line drawings generated by our method.}\label{fig:picasso}
\end{figure}

Line drawings are a fundamental form of artistic expression. They allow artists to capture the essence of a subject through a small set of strokes, abstracting away details while retaining semantic meaning and aesthetic quality. As such, they are widely used both as preparatory sketches and as finished artworks in their own right. Single-line drawings are a distinctive form of line art in which an entire object or scene is represented with a single continuous stroke, like drawing an image without lifting the pen from the paper. This minimalist style, famously explored by Picasso (see \figref{fig:picasso}), requires both abstraction and aesthetic sensitivity: a single curve must simultaneously convey structural correctness, semantic interpretability and artistic expressiveness.

\begin{figure}[t]
    \centering
	\small
	\setlength{\tabcolsep}{1pt}
    \setlength{\fboxsep}{0pt}
    \begin{tabular}{cccc}
    \includegraphics[width=0.32\linewidth]{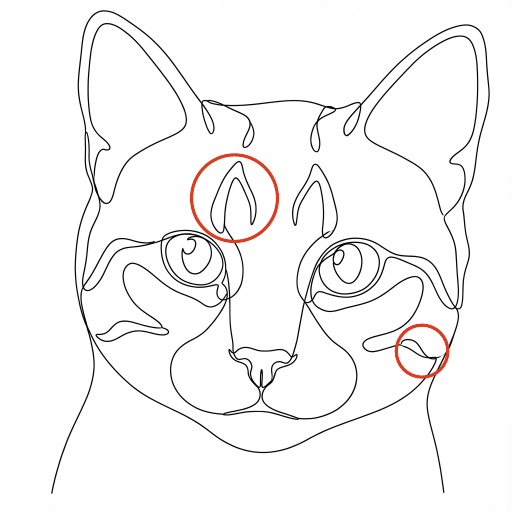}
    & \includegraphics[width=0.32\linewidth]{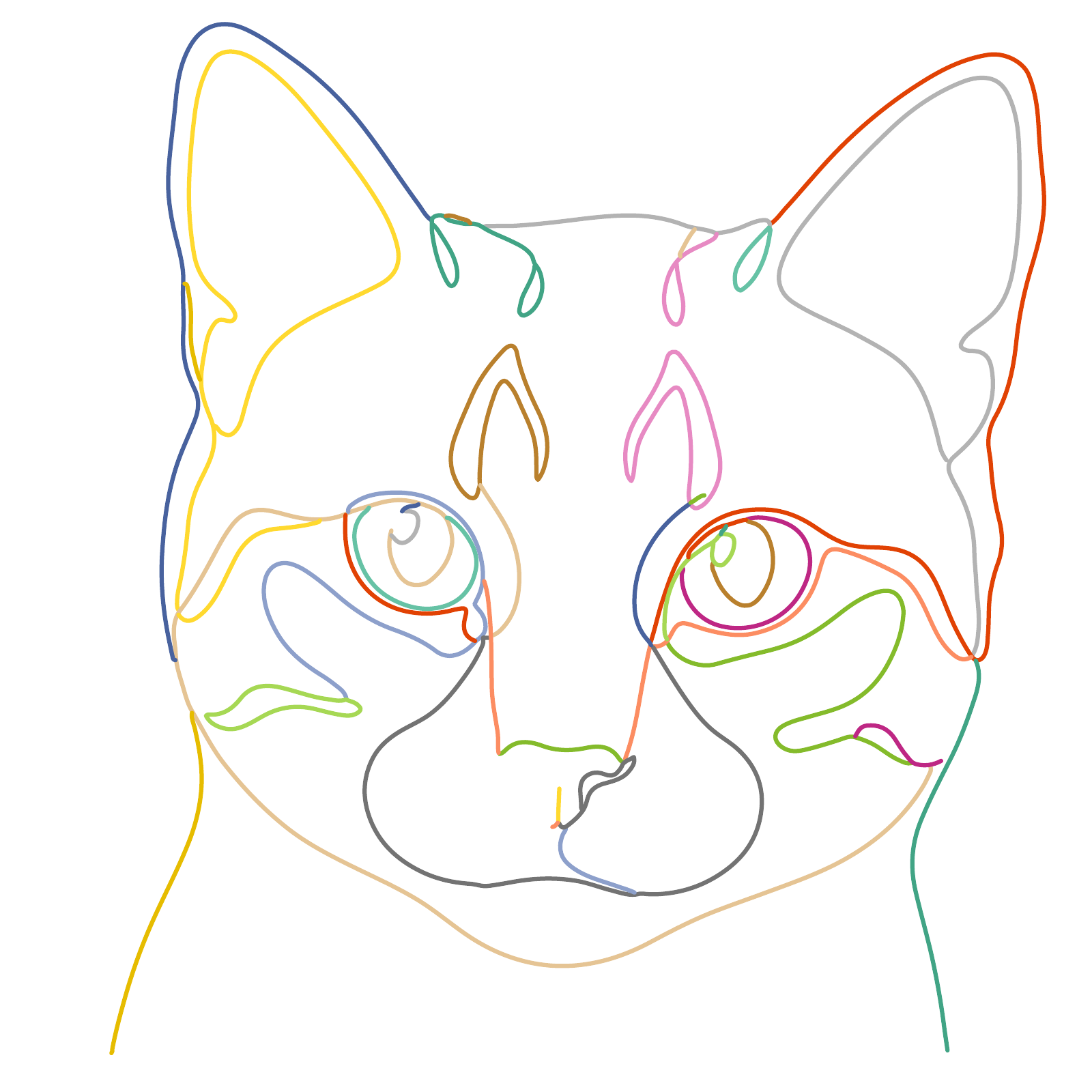}
    & \includegraphics[width=0.32\linewidth]{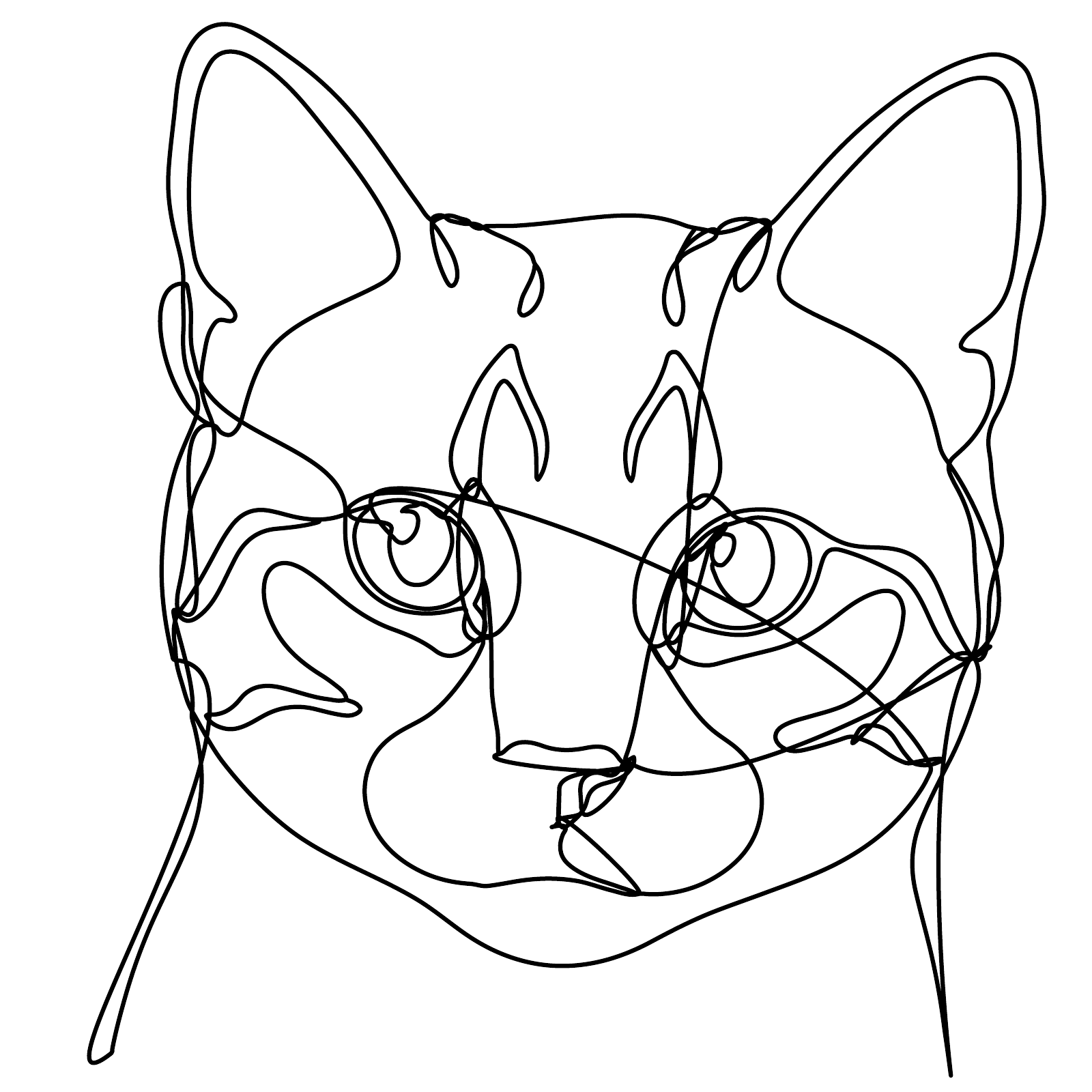}
    \end{tabular}
    \caption{Left: Gemini generates drawings with multiple curves (highlighted by red circles) in raster format. Middle: Vectorizing the raster output with Adobe Illustrator produces many curves (illustrated with different colors). Right: Connecting these curves gives a messy single-line drawing, with unnecessary strokes.}
    \label{fig:chatgpt_connectedlines}
\end{figure}

With the recent advances in large pretrained vision–language models, several algorithmic approaches have been proposed that can generate stylized line drawings. Using text-to-image models with the prompt ``a single line drawing of...'' already yields results that look like single-line drawings. However, these images are not single-line drawings: every image we generated with the latest models contains multiple strokes. Adding more guidelines to the prompt (e.g., ``continuous line'') or more data does not yield a single-line: we finetuned Stable Diffusion 3.5 using LoRA~\cite{hu.etal2021} on a dataset of single lines and still obtained multiple strokes (see \secref{subsec:lora}). This is due to the limitations of diffusion models operating in pixel space. Enforcing a constraint as rigid as continuity within a single stroke is fundamentally difficult because their sampling process is stochastic and conditioning-oriented, but not constraint-preserving. As we show in our applications, this flaw is a deal-breaker; this is not just a semantic matter, but has practical consequences. 
A seemingly straightforward strategy for producing single-line drawings would be to first use a pretrained text-to-image model to synthesize a raster image in the desired style, vectorize it and connect the obtained curves by solving a traveling salesperson problem on a graph where each node represents one of the curves. In practice, however, this pipeline rarely succeeds (see \figref{fig:chatgpt_connectedlines}). The reason is that single-line drawings are not collections of connected curves, but a single continuous curve where every part of the line matters and should be carefully drawn.

\begin{figure}[t]
    \centering
	\small
	\setlength{\tabcolsep}{1pt}
    \setlength{\fboxsep}{0pt}
    \begin{tabular}{cccc}
    \includegraphics[width=0.24\linewidth]{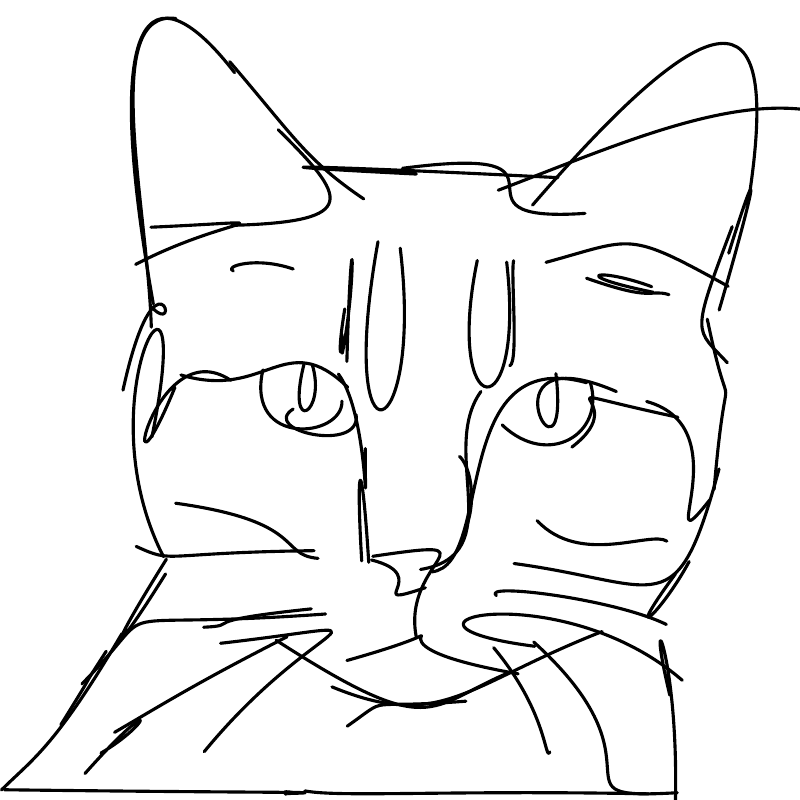}
    & \includegraphics[width=0.24\linewidth]{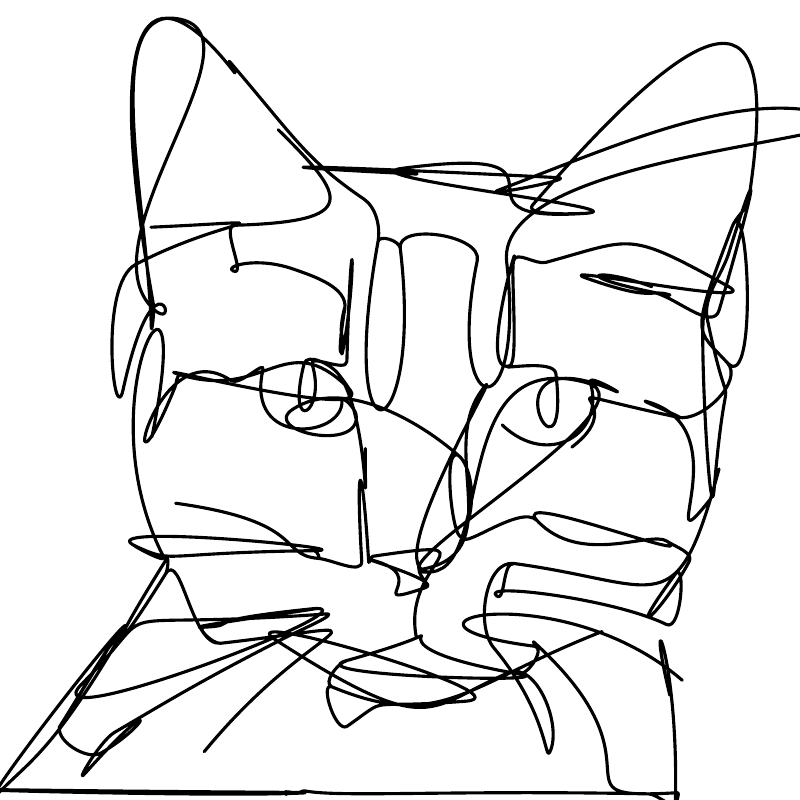}
    & \includegraphics[width=0.24\linewidth]{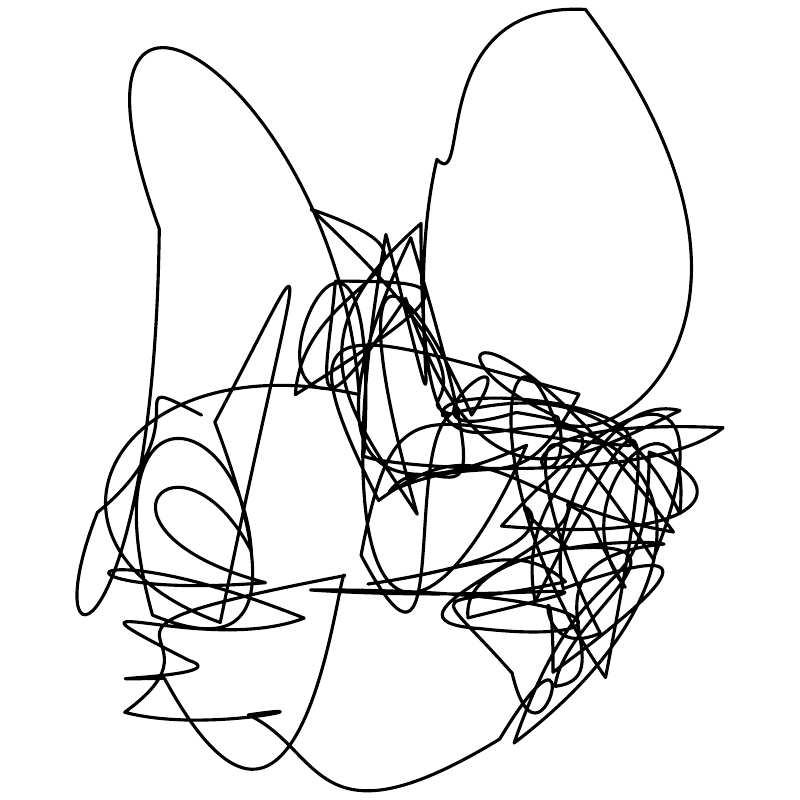}
    & \includegraphics[width=0.24\linewidth]{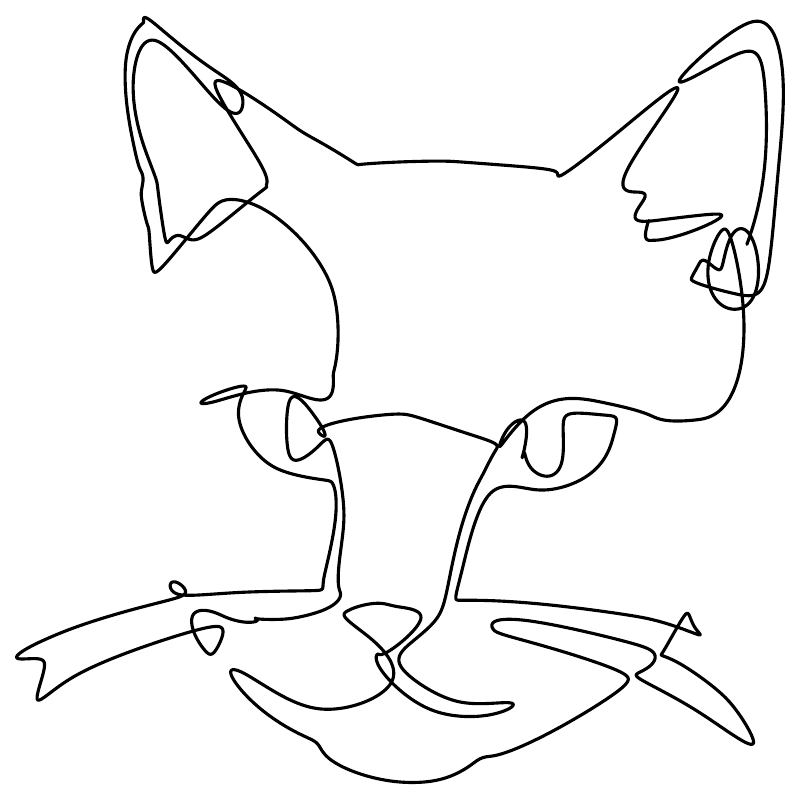}
    \end{tabular}
    \caption{Left: ControlSketch \cite{arar.etal2025} generates drawings with multiple curves. Middle left: Connecting these curves gives a messy single-line drawing, full of overdrawn and unnecessary strokes. Middle right: Directly generating single-line drawing with ControlSketch produces poor results. Right: Our result.}
    \label{fig:controlsketch_connectedlines}
    \vspace{-1ex}
\end{figure}

Existing sketch-based generation methods are designed to create drawing containing multiple strokes. The most advanced ones such as ControlSketch \cite{arar.etal2025} are based on score distillation sampling (SDS)~\cite{poole2023dreamfusion} to combine the generation quality of diffusion models with the controllability of classic parameterizations. However using ControlSketch for single-line drawing generation by simply parameterizing the image as a single B\'ezier curve (\figref{fig:controlsketch_connectedlines}, Middle right) or connecting multiple B\'ezier curves after they were generated (\figref{fig:controlsketch_connectedlines}, left and middle left) does not yield the desired style. The single B\'ezier curve does not fit single-line drawings well and connecting multiple B\'ezier strokes simply looks like a regular sketch, not a single-line drawing which is characterized by sparse use of smooth, simple curves.

In this work, we propose the first semantics-driven method for generating vector drawings that are, by construction, single-line drawings. Our approach builds on SDS \cite{poole2023dreamfusion}, which transfers the guidance from large pretrained diffusion models to specialized, structured representations. Rather than optimizing pixels in a raster image, we directly optimize the parameters of a single uniform rational B-spline (URBS) curve, guaranteeing by construction that the result is one continuous stroke. This representation is particularly well suited to our task: cubic B-splines provide smooth curves, each control point influences only a local region of the curve, and rational weights allow adjusting their importance during optimization. By optimizing both control point positions and weights, the curve can automatically adapt its level of detail while remaining compact.
To integrate SDS in this setting, we rely on differentiable rasterization. The curve is sampled and rasterized using DiffVG \cite{li.etal2020}, a differential rasterizer, which allows the gradients of the diffusion model to update the curve parameters. To favor the emergence of single-line drawing characteristics during optimization, we adapt the diffusion model with a LoRA model \cite{hu.etal2021} trained on a small set of single-line drawings. However, the resulting single-line images do not yet conform to the desired artistic properties. Therefore, we complement the SDS loss with additional regularizers that encourage the clean style characteristic of single-line drawings and prevent unnecessary details and redundant control points, resulting in a simpler curve where possible. Together, these regularizers help guide the optimization not only toward semantic alignment with the input, but also toward an aesthetically pleasing single-line drawing.

Our approach can operate from either a text description or an input image, producing as output a vectorized single-line drawing suitable for downstream use (see \figref{fig:teaser}). In our experiments, we show that the proposed method consistently outperforms existing approaches for single-line drawing generation, both in terms of semantic alignment and visual quality. Qualitative comparisons highlight that our outputs better reflect the style of continuous line artists, while quantitative evaluations confirm improvements across text–image similarity, image–image similarity and aesthetic metrics. Through ablation studies, we analyze the role of each component in our optimization framework, and we demonstrate how varying certain losses enables stylistic control over the drawings. Finally, we illustrate the practical utility of our vectorized single-line outputs by fabricating them with embroidery and laser engraving.

\section{Related work}
\label{sec:related}

To situate our contribution, we review existing approaches to sketch generation and single-line drawing vectorization. Additional background on diffusion models and their integration into optimization pipelines using score distillation sampling is provided in \appref{app:background}.

\subsection{Direct data-driven sketch generation}
\label{subsec:datadriven}

The task of generating sketch-like representations from images has long been studied in the context of non-photorealistic rendering, with classic techniques such as Canny edge detection \cite{Canny1986Edge}, suggestive contours \cite{Decarlo2003Suggestive}, apparent ridges \cite{ApparentRidges:2007} and extended difference of Gaussians (XDoG) filters \cite{Winnemoeller:2012:XED} producing stylized line renderings. More recently, data-driven approaches have greatly expanded the stylistic diversity of such methods and improved their semantic alignment with the depicted content \cite{Chan_2022_CVPR}. Some classical methods also enable the recovery of the drawing order of existing vector line drawings \cite{fu.etal2011}. 

Early generative models sought to directly synthesize sketches from training data. Supervised image-to-sketch translation approaches have been proposed \cite{Kampelmuhler_2020_WACV}, but they require paired photo–sketch datasets, which are difficult to collect at scale \cite{sangkloy2020sketchy}. Style transfer methods \cite{Huang_2025_Survey} can offer an alternative by converting input images into sketch-like renderings. Still, these remain limited to rasterized outputs and cannot enforce structural constraints inherent to vector drawings.
A large body of work has instead focused on class-based sketch generation in vector space. Architectures such as Sketch-RNN \cite{ha2017neuralrepresentationsketchdrawings} (RNN-based), Sketchformer \cite{Ribeiro_2020_CVPR} (transformer-based), SketchBERT \cite{Lin_2020_CVPR}, and methods incorporating CNNs with cycle-consistency \cite{Song_2018_CVPR} all learn to generate sketches conditioned on object categories. Mo et al. \cite{mo.etal2021} proposed a method based on recurrent neural networks to vectorize images and generate vector drawings. However, its generation capabilities are limited to photographs of faces, and it produces very simple results that only capture the head pose. More recent work leverages diffusion models, either directly in vector parameter space, as in SketchKnitter \cite{wang2023sketchknitter}, or in implicit representations, such as the unsigned distance field used in CoProSketch \cite{zhan2025coprosketchcontrollableprogressivesketch}.
Other approaches adopt adversarial training. GANs have been used to generate pixel-based contour drawings \cite{li2019photosketchinginferringcontourdrawings}, while SketchGANs \cite{Varshaneya2021SketchGANs} operate in vector space for more flexible downstream editing.
These feed-forward methods rely on large-scale sketch datasets, most notably those collected from amateur users across many categories, such as QuickDraw \cite{jongejan2016quickdraw}, or Sketchy \cite{Eitz2012Sketch}. Other datasets instead consist of expert-drawn sketches for a specific class of objects, such as ProSketch \cite{Yue_Prosketch_2021} or OpenSketch \cite{gryaditskaya.etal2019}. All these datasets are limited by their drawing quality \cite{jongejan2016quickdraw, Eitz2012Sketch} or by their size and variability. OpenSketch \cite{gryaditskaya.etal2019} is limited to household items and the dataset of Berger et al. \cite{berger.etal2013} is composed entirely of portraits. 
In addition, since these method are trained on a specific dataset they do not generalize well to subjects or styles out of the training dataset distribution and, in particular, to single-line drawings. Collecting a sufficiently large dataset of single-line drawings in vector format for training is infeasible due to the low availability of such data.

\begin{figure*}[t]
    \centering
    \includegraphics[width=\textwidth]{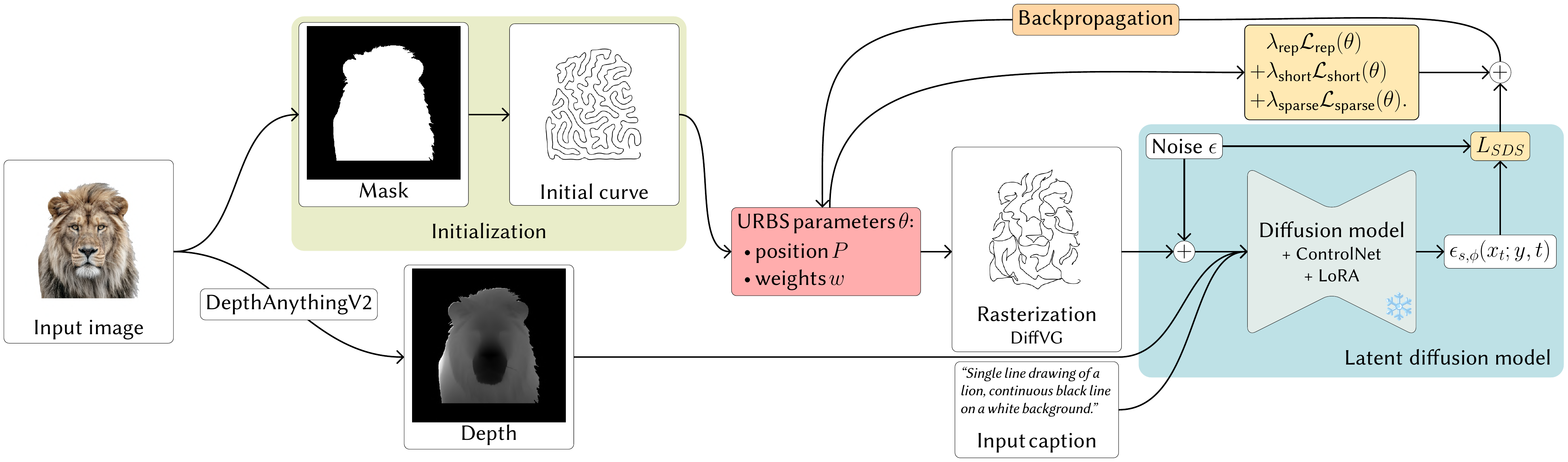}
    \caption{Overview of our generation pipeline. Points are sampled uniformly within the segmented subject region and traversed by solving a TSP problem to initialize the control points of our URBS curve. The parameters are then optimized via gradient descent on the SDS loss to capture the semantics of the subject and various regularization losses to achieve the aesthetic quality of single-line drawings.}
    \label{fig:pipeline}
\end{figure*}

\subsection{VLM for vector sketches}
\label{subsec:vlm}

Beyond specialized data-driven strategies, another line of work leverages pretrained vision–language models (VLMs) to generate sketches without requiring dedicated datasets. A prominent example is contrastive language–image pre-training (CLIP) \cite{radford2021learningtransferablevisualmodels}, which jointly embeds text and images into a shared latent space where similarity can be measured, typically via cosine distance. Several works use the CLIP similarity score as a loss to guide the optimization of vector primitives toward alignment with a target text prompt or reference image using a differentiable renderer such as DiffVG \cite{li.etal2020}. CLIPDraw \cite{frans.etal2022} generates vectorized sketches by optimizing the parameters of colored Bézier curves. CLIPasso \cite{vinker.etal2022} focuses on optimizing a small number of strokes to produce abstracted drawings by leveraging CLIP activations at specific layers. CLIPascene \cite{vinker.etal2023} trains an MLP under CLIP supervision to balance simplicity and fidelity.

Our method builds on a related, but distinct paradigm: score distillation sampling (SDS) \cite{poole2023dreamfusion}, which distills guidance from pretrained diffusion models into specialized representations (see \appref{app:background}). SDS has been successfully applied to generating vector graphics, most notably in VectorFusion \cite{jain.etal2023}, where parameters of vector primitives are optimized under diffusion model guidance. By constraining primitives to particular geometric shapes, VectorFusion can naturally enforce sketch-like outputs. This approach has since been extended and refined in works such as SVGDreamer, SVGDreamer++ \cite{xing.etal2024, xing.etal2024a} and DiffSketcher \cite{xing.etal2023}, which specifically targets sketch generation. Some approaches, such as SwiftSketch \cite{arar.etal2025}, use VLM-guided optimization to create training data for custom generative models. ControlSketch, the method used by Arar et al. \cite{arar.etal2025} to generate training data, uses a framework similar to ours. However, they optimize multiple cubic Bézier curves with a different initialization and without regularization losses.
These methods highlight the potential of optimization frameworks to impose strong structural constraints. In a similar spirit, our approach leverages SDS not only to generate sketches but to explicitly enforce the strict single-line drawing constraint, which cannot be achieved by raster-based models.

\subsection{Single-line drawing}
\label{subsec:single-line}

The use of single-line drawings has long attracted interest, particularly within procedural art. A common approach consists of distributing nodes according to the density of a target grayscale image and solving a traveling salesperson problem (TSP) to produce a single continuous path that matches this density \cite{bosch.herman2004}. 
This strategy was later generalized through Voronoi diagrams and dithering, enabling line drawings with richer variations in local density \cite{kaplan.bosch2005}. An alternative to TSP-based formulations is to construct a branching tree via path finding over pixels selected by a cost function that balances tonal fidelity and structural saliency, from which a continuous line drawing can be derived \cite{li.mould2014,li2012cld}.
Lebrat et al. \cite{lebrat.etal2019} introduced an optimal transport framework for approximating a target density with structured measures, including curves, thereby enabling single-line renderings of images for applications such as wood engraving.
All these methods generate non-intersecting curves. Their style differs from our target single-line drawing style and are closer to space-filling curves \cite{Noma2024SurfaceFilling}, which have been studied extensively for 3D printing \cite{chermain.etal2023} and CNC milling \cite{zhao.etal2018}.
Closer to our target application, Liu et al. \cite{liu.etal2017b} proposed extracting edges from labeled images to generate continuous paths for whole-cloth quilting. In contrast to our approach, these prior methods generally yield much denser results, as their primary objective is to reproduce the tonal density of the input image rather than to abstract it.
A series of works by Wong and Takahashi \cite{wong.takahashi2011,wong.takahashi2013,wong.takahashi2013a} investigate continuous line illustrations from images, aiming to abstract image content while retaining salient features, a goal similar to ours. Contrary to our approach, these works rely primarily on user input or edge extraction rather than semantics-driven optimization.
Recently, Berio et al.\ \cite{berio.etal2025} proposed a method that leverages CLIP-based optimization to model human arm trajectories for automated drawing generation. Concurrent with our work, another method leveraging score-distillation sampling for single-line drawing generation has been proposed in \cite{berio.etal2025a}. Neither method proposes a specific loss for constant-width single-line drawings, as targeted by us, and the latter work \cite{berio.etal2025a} only shows varying-width line drawings. Our method can also generate such varying-width single-line drawings (see \secref{subsec:varying_width}), and our curve representation enables the automatic adjustment of the curve complexity.
Beyond image-based drawings, the problem of designing a single curve has been extended to surface representations: mesh surfaces can be converted into closed 3D curves through surface-filling algorithms \cite{AKLEMAN2013316,Noma2024SurfaceFilling}, while multi-view semantic optimization has been explored for generating 3D wire art intended for fabrication \cite{tojo.etal2024}. Compared to the method of Tojo et al.\ \cite{tojo.etal2024}, our method employs a different curve representation, initialization strategy and regularization losses. Other methods produce multiple curves in 3D space to abstract images or 3D objects \cite{hsiao.etal2018, choi.etal2024, liu.etal2025a}.

\section{Method}
\label{sec:method}

Our goal is to generate aesthetically pleasing single-line drawings. Our main challenges are to define the right parameterization (\secref{subsec:curve}) and to set up the initialization (\secref{subsec:init}), regularization (\secref{subsec:loss}), and optimization (\secref{subsec:lora}) to ensure the final output remains a true single-line drawing and adheres to the aesthetic of such drawings. An overview of our method is presented in \figref{fig:pipeline}.

\subsection{Curve representation}
\label{subsec:curve}

Line drawing generation methods commonly use B\'ezier curves~\cite{vinker.etal2022, arar.etal2025}. Instead, we propose to represent the line with a uniform rational B-spline (URBS) curve.
A uniform rational B-spline is a weighted (rational) B-spline with uniformly spaced control points. Let $P_i$ be the position of control point $i$ and $w_i$ the weight controlling its influence. The URBS curve $c(t)$ is then given by
\begin{equation*}
    c(t)=\frac{\sum_i w_i B_{i,k}(t) P_i}{\sum_i w_iB_{i,k}(t)},
\end{equation*}
where $B_{i,k}$ are the B-spline basis functions of degree k. These functions are defined recursively on the knot vector $\mathbf{t} = (t_0, \dots, t_N)$, with $N = n + k + 1$, where $n$ is the number of control points. 

The basis function is given as
\begin{align*}
B_{i,0}(t) &=
\begin{cases}
1 & \text{if } t_i \leq t < t_{i+1}, \\
0 & \text{otherwise}.
\end{cases} \\
B_{i,k}(t) &= \frac{t - t_i}{t_{i+k} - t_i} B_{i,k-1}(t) + \frac{t_{i+k+1} - t}{t_{i+k+1} - t_{i+1}} B_{i+1,k-1}(t).
\end{align*}
In our case, we use cubic ($k=3$) polynomials to allow for sufficient smoothness, while limiting the spatial influence and computational cost of each control point. In addition, we use a uniform knot vector, meaning that $\mathbf{t} = (0, \sfrac{1}{N},\sfrac{2}{N},\dots,1)$. For more information on uniform rational B-splines, please refer to Farin's work on curves and surfaces for CAGD \cite{farin2001}.

We choose URBS as our curve representation for the following reasons. First, cubic B-splines are smooth, which is a key property of the single-line drawing aesthetic. Second, each control point has local influence on the curve, allowing for easier optimization. Finally, we use the weights in rational B-splines to adapt the complexity of the curve where necessary. This is key, as it allows the optimization to adapt the complexity of the curve through the differentiable weights. This adaptivity would not be possible for a B\'ezier curve, because the choice of including a control point is binary (either included or not).

\begin{figure}[t]
    \centering
    \begin{subfigure}[t]{0.49\linewidth}
        \centering
        \adjincludegraphics[width=\textwidth,trim={{0.0\width} {.0\height} {0.0\width} {.0\height}},clip]{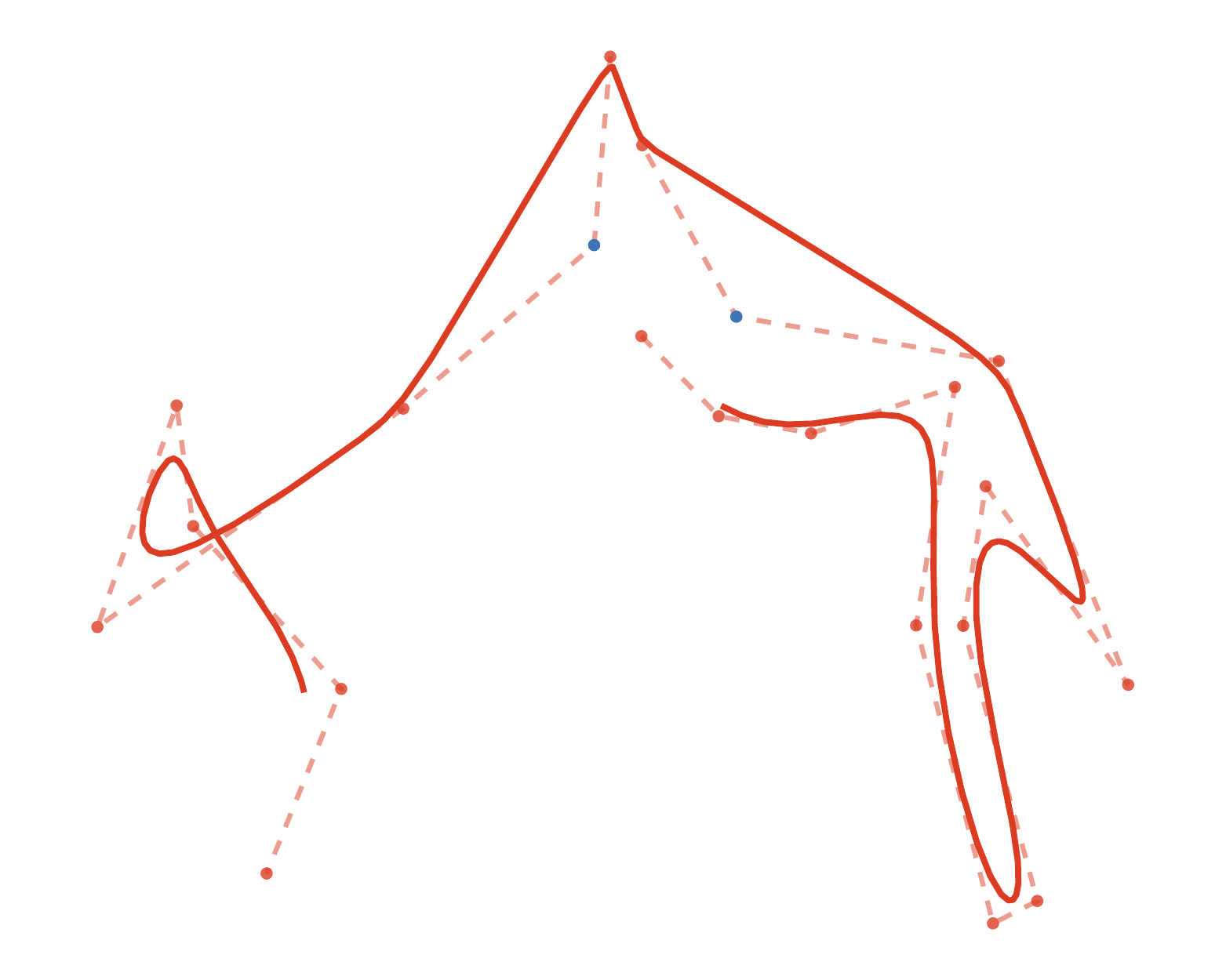}
    \end{subfigure}
    \begin{subfigure}[t]{0.49\linewidth}
        \centering
        \adjincludegraphics[width=\textwidth,trim={{0.0\width} {.0\height} {0.0\width} {.0\height}},clip]{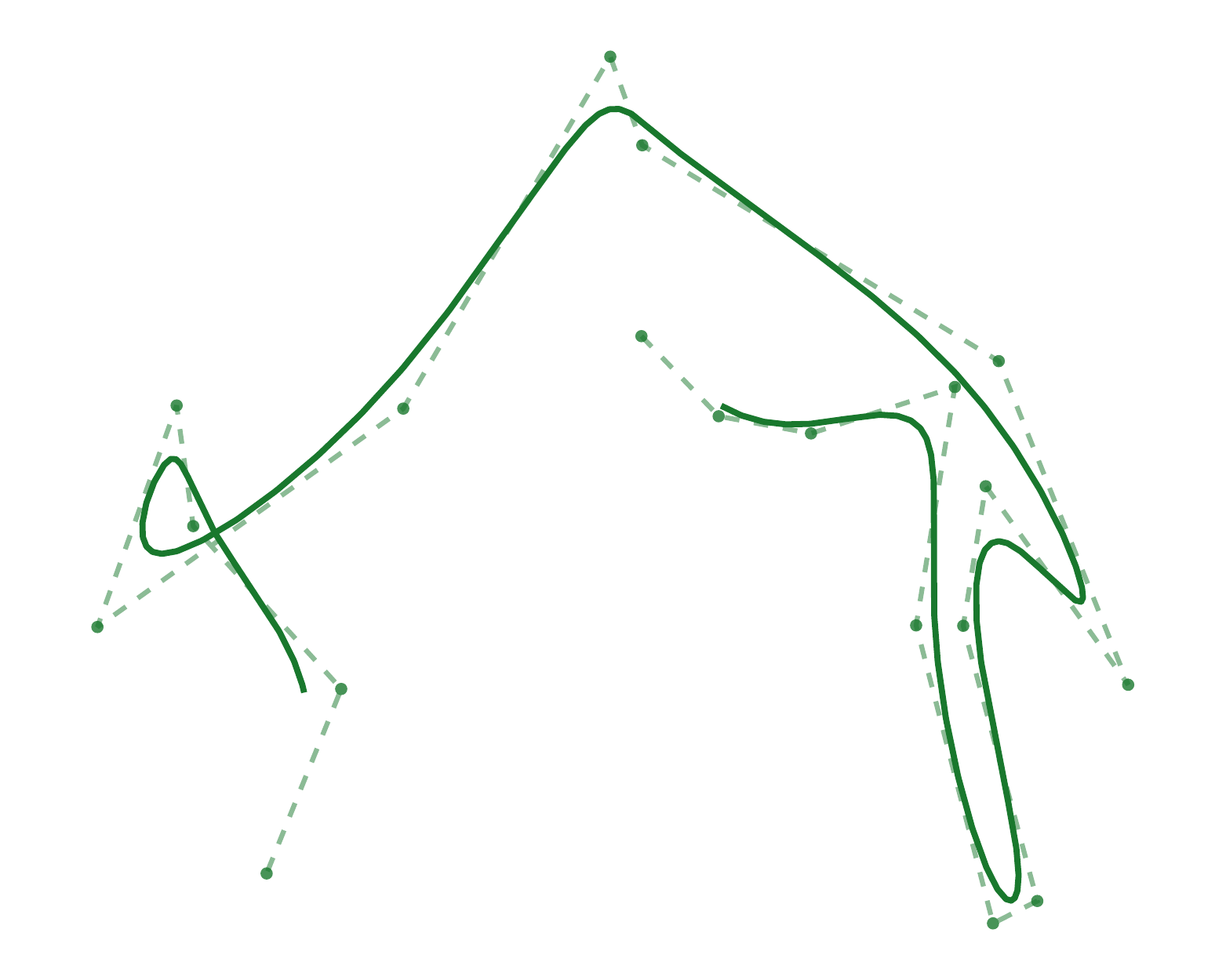}
    \end{subfigure}
    \caption{Two uniform rational B-spline curves. Left: the control points in blue have very low associated weights, leading to straight lines and sharp corners near them. Right: removing these low-weight control points results in a smoother-looking curve.}\label{fig:cp_prune}
    \vspace{-3ex}
\end{figure}

We use this adaptivity for a second core contribution of our algorithm: we sparsify, or prune, control points during the optimization. All weights are initialized to 1 and clamped between 0.001 and 1 during optimization. Once a weight becomes too small ($w_i < 0.002$), it is removed from the representation. This mechanism allows for adaptive density without requiring specialized initialization procedures \cite{arar.etal2025, vinker.etal2022} or manual finetuning of parameters, such as the number of segments. In addition, pruning keeps the curve visually smooth. As visible in \figref{fig:cp_prune}, if the weight associated with a control point becomes too small, the segment connected to it becomes nearly straight. Removing the control point from the representation makes the curve smoother, while preserving its shape.

\subsection{Initialization}
\label{subsec:init}

Because our sparsification approach lets the optimization decide where to place detail, we only require the initialization to roughly cover the shape. For this purpose, we solve a traveling salesperson problem (TSP) on a uniformly distributed sampling of the shape (\figref{fig:pipeline}, initialization). This is similar to the initialization proposed by the robotic-arm drawing approach by Berio et al.\ \cite{berio.etal2025}, which follows the TSP art~\cite{kaplan.bosch2005} method. 
However, we do not sample according to a saliency map of the input image, since our goal is simply to cover the shape.

We first segment the subject in the input image using the BRIA Background Removal v1.4 model \cite{briaai2024rmbg14} (\figref{fig:pipeline}, mask).  If an input image is not provided, a text-to-image diffusion model is used to generate an image representing the input subject, and the shape used to sample the points is extracted from it.
To distribute points uniformly within the obtained mask, we apply the weighted Voronoi stippling method \cite{secord2002}, which iteratively centers the points within their respective Voronoi regions. These points are then connected by solving a TSP with the Concorde solver \cite{concorde}. Because this curve is closed, we remove the longest edge from the cycle for open-ended drawings. As a result, the curve spans the entire object (\figref{fig:pipeline}, initial curve), which lets the optimizer quickly relocate control points to the important regions in a few iterations.

\subsection{Loss function}
\label{subsec:loss}
Now that we can guarantee a single curve, we aim to improve the aesthetic similarity to single-line drawings. We achieve this with specialized loss functions, described in this section (\figref{fig:pipeline}, blue and yellow blocks).

Let $\theta= (\mathbf{P}, \mathbf{w})$ be the set of parameters that we wish to optimize, where $\mathbf{P} \in \mathds{R}^{n \times 2}$ and $\mathbf{w} \in \mathds{R}^n$ are the stacked control point positions and weights, respectively. The complete loss is defined as
\begin{equation*}
\begin{split}
    \mathcal{L}(\theta) = & \mathcal{L}_\text{SDS}(\theta, \mathcal{I}_\text{txt}, \mathcal{I}_\text{img})\\
    &+ \lambda_\text{rep}\mathcal{L}_\text{rep}(\theta)\\
    &+ \lambda_\text{short}\mathcal{L}_\text{short}(\theta)\\
    &+ \lambda_\text{sparse}\mathcal{L}_\text{sparse}(\theta).
\end{split}
\end{equation*}
We define each term below.

The score distillation sampling loss $\mathcal{L}_\text{SDS}$ guides the curve optimization so that the curve represents the input image or text prompt. Computing this loss involves rasterizing the curve, which is done with DiffVG \cite{li.etal2020}. As DiffVG does not support differentiable rasterization of URBS out of the box, we densely sample the curve and rasterize each segment between two points. The diffusion model used for SDS takes a text prompt $\mathcal{I}_\text{txt}$ and an input image $\mathcal{I}_\text{img}$ as conditioning. If no text description is provided, a prompt is created using a captioning model. To condition the diffusion model on the input image, we first predict a depth map using Depth Anything V2 \cite{yang.etal2024} and feed the depth map to the diffusion model through a ControlNet \cite{zhang.etal2023}. While other types of conditioning exist (e.g., Canny edge conditioning), we used a depth-conditioning because it yields strong results, requires no parameter tuning and has already proven effective for sketch generation \cite{arar.etal2025}.

\begin{figure*}[t]
    \newcommand{\cwidth}{0.162\linewidth}
    \newcommand{\cspace}{\hspace{0.005\linewidth}}
    \newcommand{\lrspace}{1pt}
    \setlength{\fboxsep}{-0.1pt}
    \setlength{\fboxrule}{0.1pt}
    \centering
    \setlength{\tabcolsep}{1pt}
    \begin{tabular}{cccccc}
    \begin{subfigure}[t]{\cwidth}
        \begin{subfigure}{\textwidth}
            \begin{overpic}[width=\textwidth, height=\textwidth]{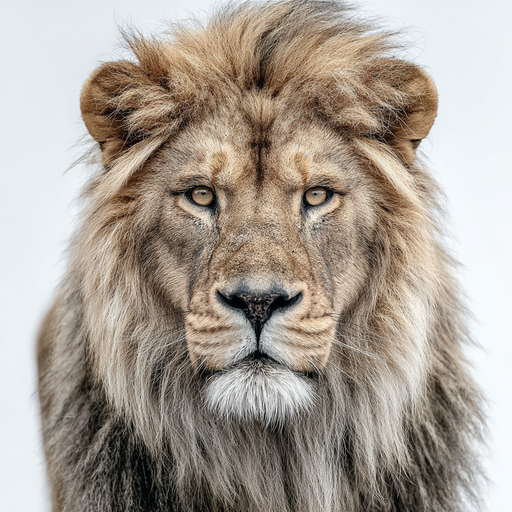}%
             \put (0,6) {\setlength{\fboxsep}{1pt}\transparent{0.5}\colorbox[RGB]{255,255,255}{\transparent{1.0}\parbox{\dimexpr\linewidth-2\fboxsep}{\scriptsize{\textit{Single line drawing of a lion head, continuous black line ...}}}}}
            \end{overpic}
        \end{subfigure}        
        \\[\lrspace]
        \begin{subfigure}{\textwidth}
            \begin{overpic}[width=\textwidth, height=\textwidth]{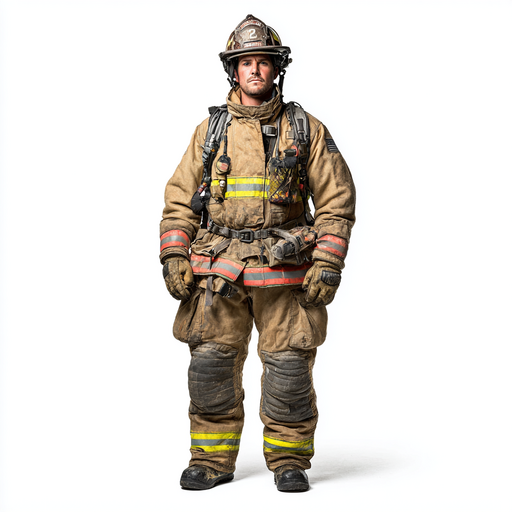}%
             \put (0,6) {\setlength{\fboxsep}{1pt}\transparent{0.5}\colorbox[RGB]{255,255,255}{\transparent{1.0}\parbox{\dimexpr\linewidth-2\fboxsep}{\scriptsize{\textit{Single line drawing of a firefighter, continuous black line ...}}}}}
            \end{overpic}
        \end{subfigure}
        \vspace{-15pt}
        \caption*{\scriptsize Input image and prompt}
    \end{subfigure}%
    & \begin{subfigure}[t]{\cwidth}
        \begin{subfigure}{\textwidth}
            \fbox{\includegraphics[width=\textwidth, height=\textwidth]{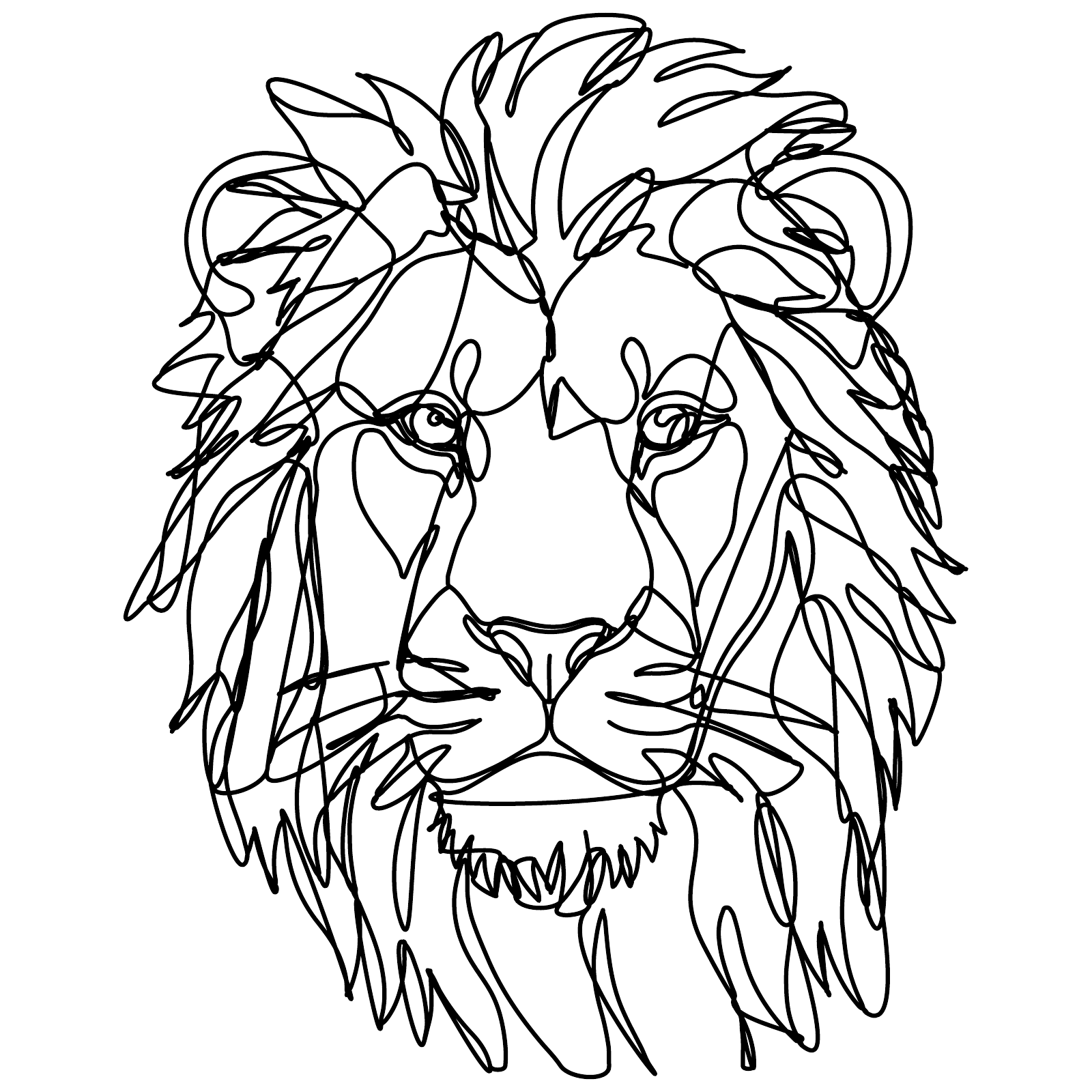}}%
        \end{subfigure}    
        \\[\lrspace]
        \begin{subfigure}{\textwidth}
            \fbox{\includegraphics[width=\textwidth, height=\textwidth]{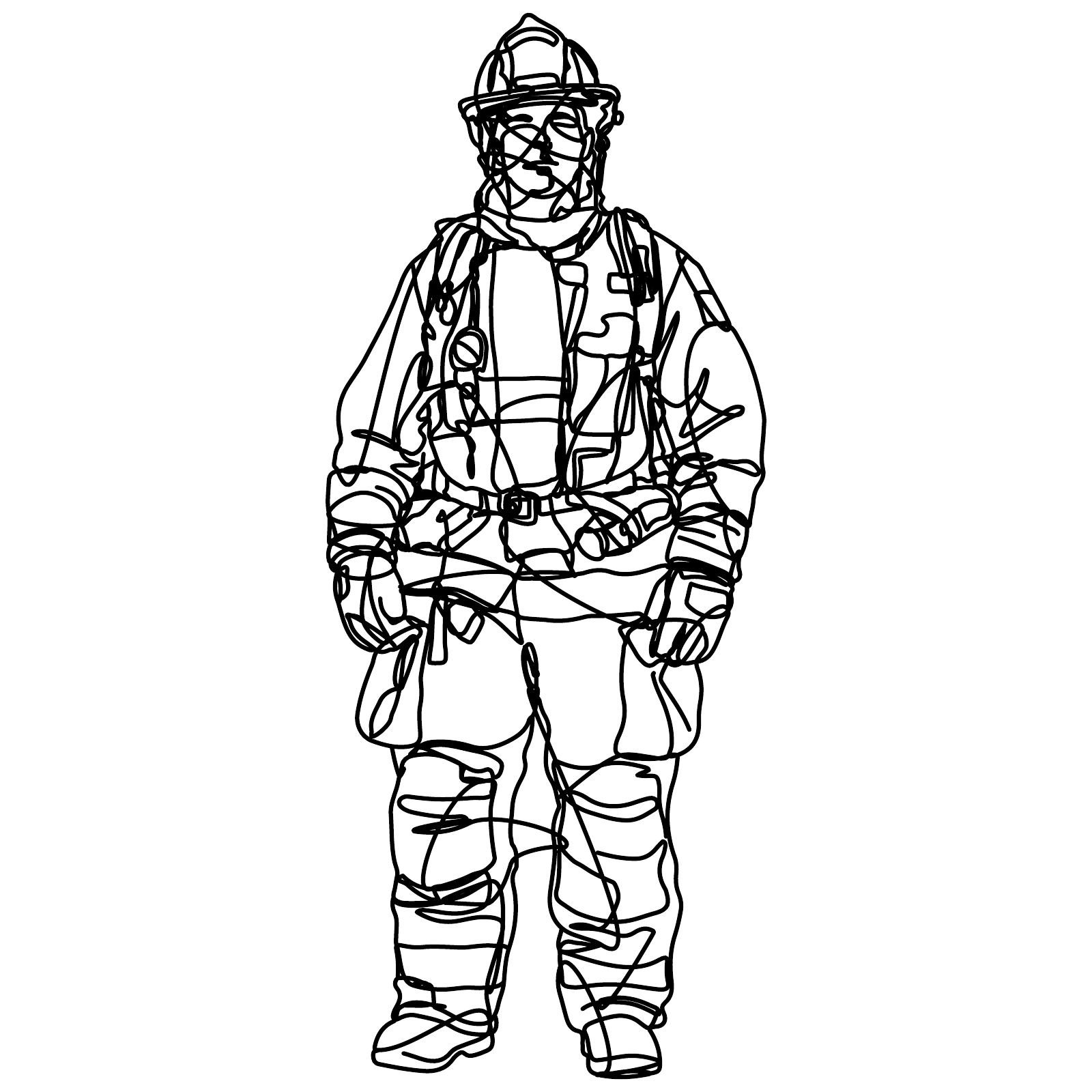}}%
        \end{subfigure}
        \vspace{-15pt}
        \caption*{\scriptsize Gemini (connected)}
    \end{subfigure}%
    & \begin{subfigure}[t]{\cwidth}
        \begin{subfigure}{\textwidth}
            \fbox{\includegraphics[width=\textwidth, height=\textwidth]{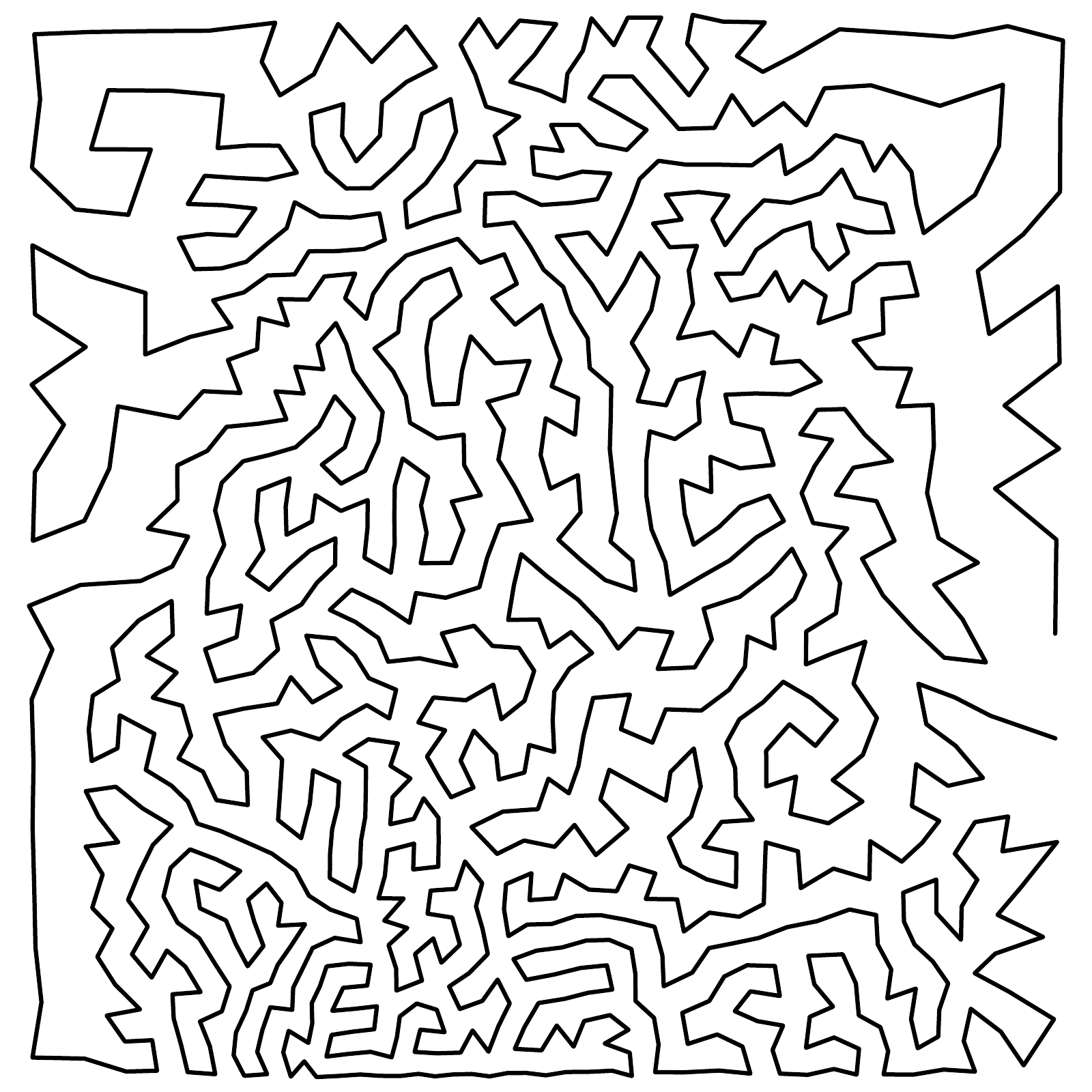}}%
        \end{subfigure}    
        \\[\lrspace]
        \begin{subfigure}{\textwidth}
            \fbox{\includegraphics[width=\textwidth, height=\textwidth]{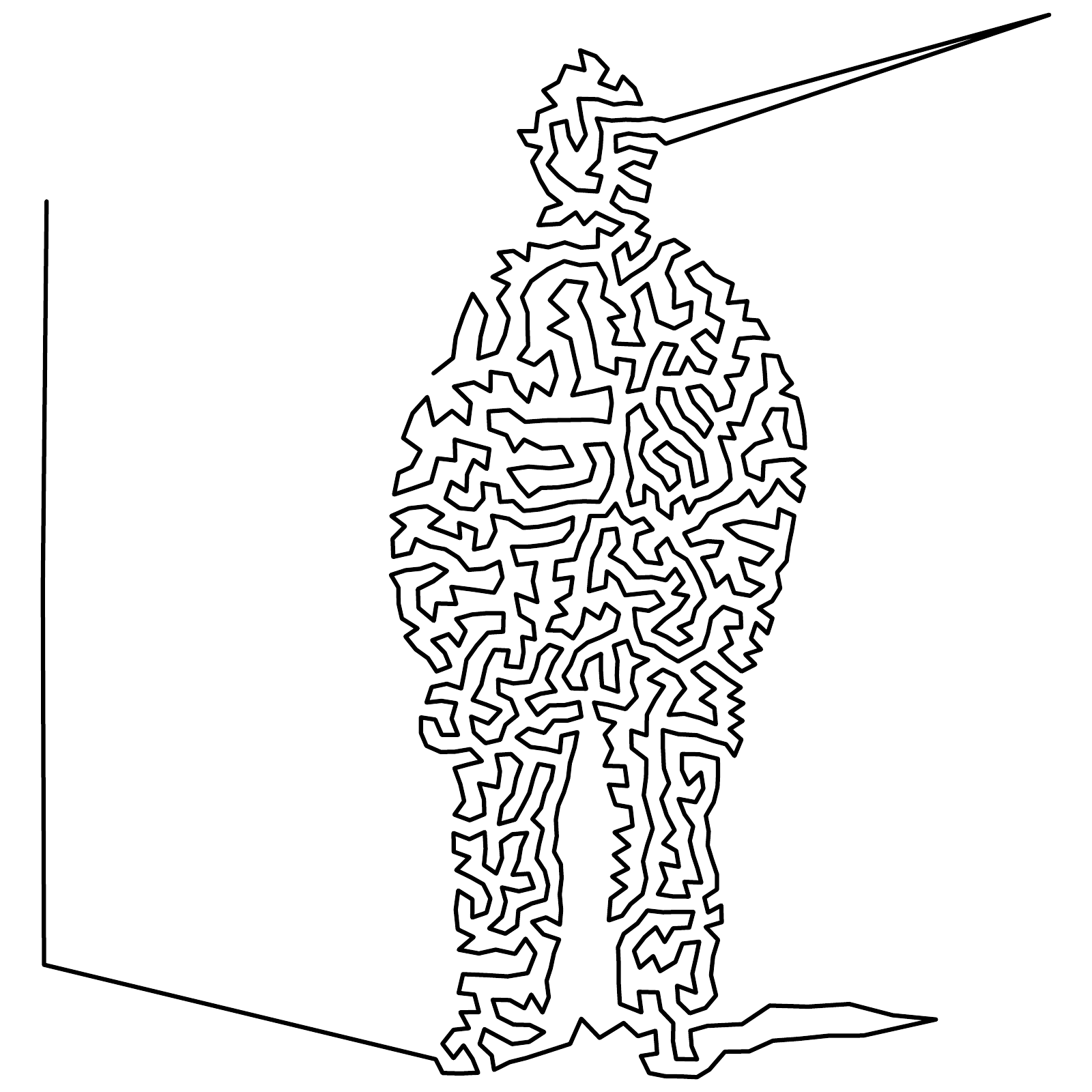}}%
        \end{subfigure}
        \vspace{-15pt}
        \caption*{\scriptsize TSP art \cite{kaplan.bosch2005}}
    \end{subfigure}%
    & \begin{subfigure}[t]{\cwidth}
        \begin{subfigure}{\textwidth}
            \fbox{\includegraphics[width=\textwidth]{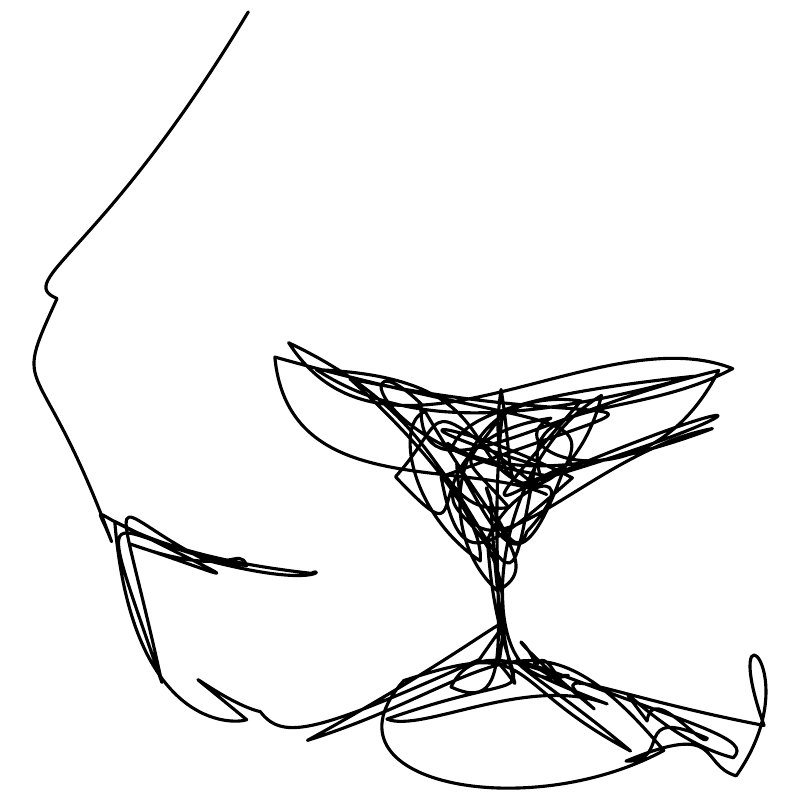}}%
        \end{subfigure}    
        \\[\lrspace]
        \begin{subfigure}{\textwidth}
            \fbox{\includegraphics[width=\textwidth]{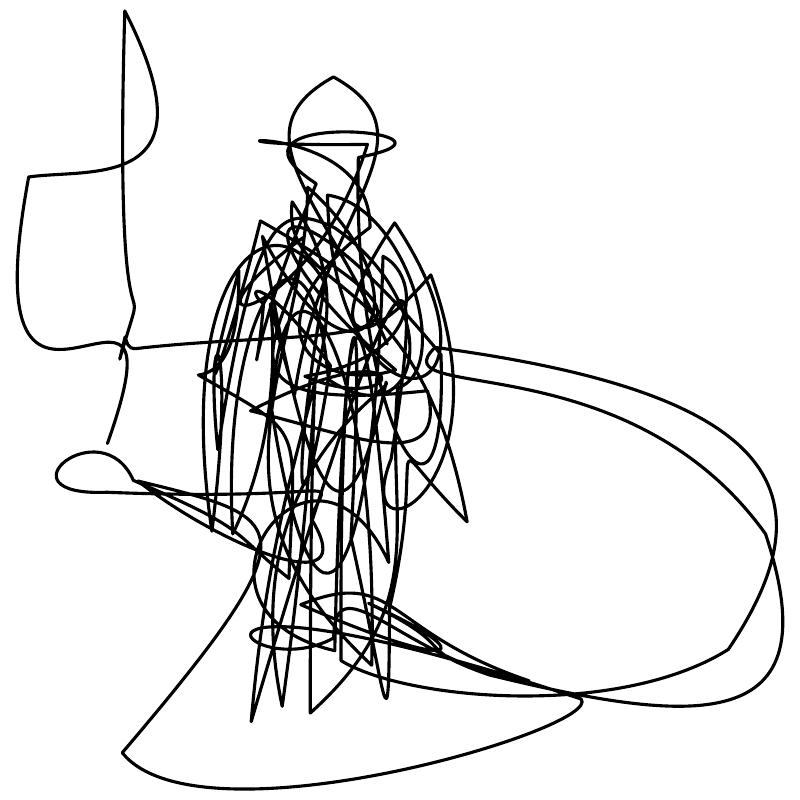}}%
        \end{subfigure}
        \vspace{-15pt}
        \caption*{\scriptsize ControlSketch \cite{arar.etal2025}}
    \end{subfigure}
    & \begin{subfigure}[t]{\cwidth}
        \begin{subfigure}{\textwidth}
            \fbox{\includegraphics[width=\textwidth]{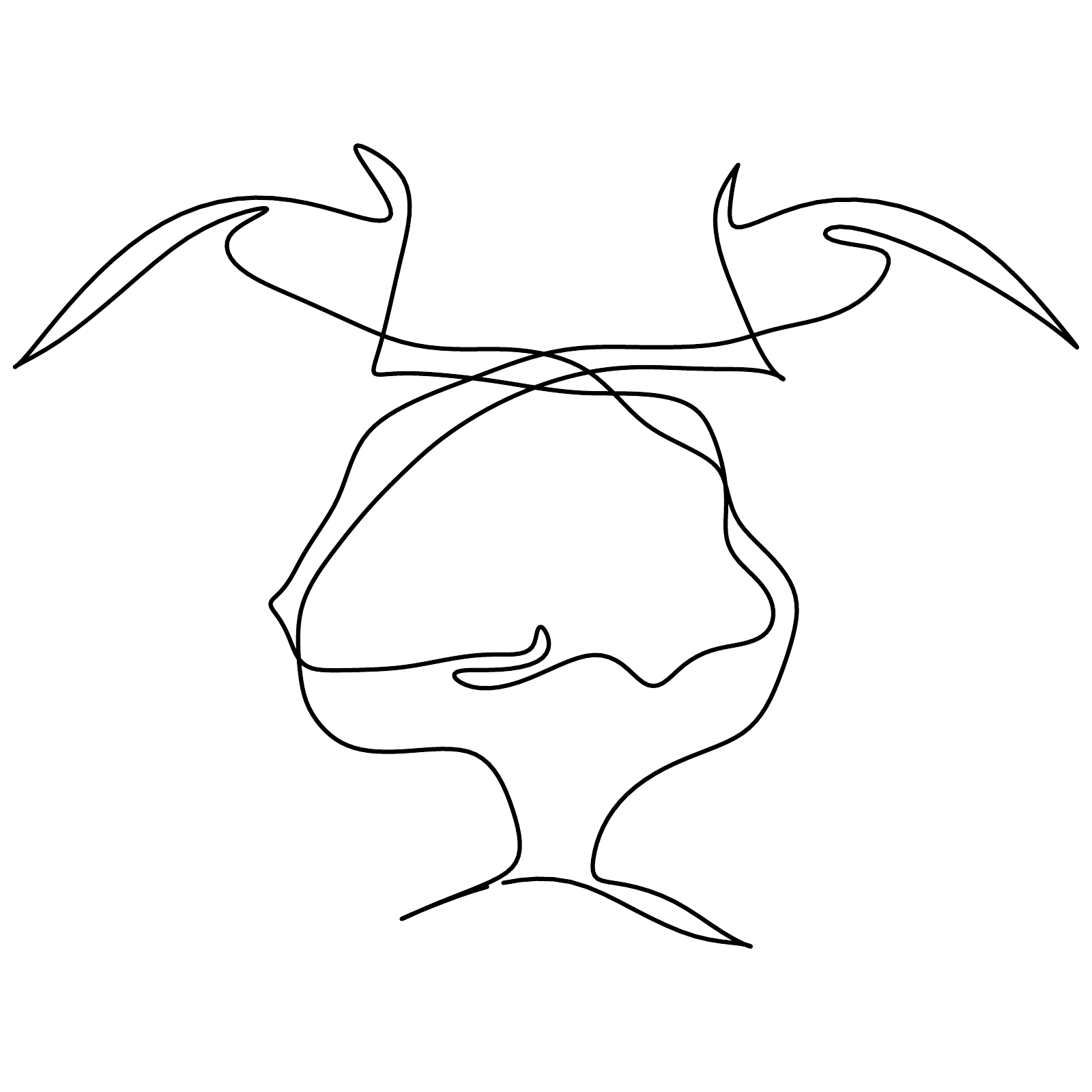}}%
        \end{subfigure}    
        \\[\lrspace]
        \begin{subfigure}{\textwidth}
            \fbox{\includegraphics[width=\textwidth]{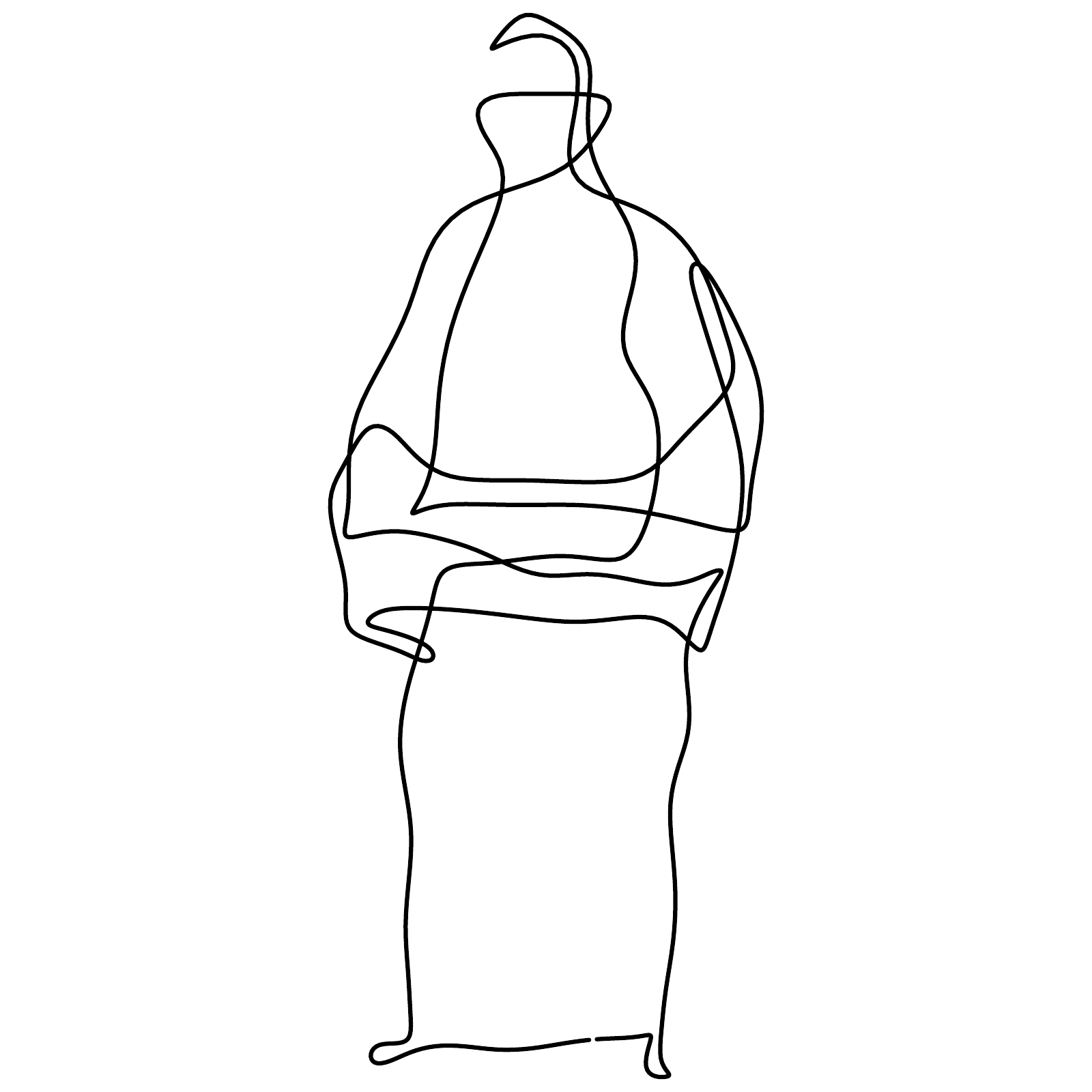}}%
        \end{subfigure}
        \vspace{-15pt}
        \caption*{\scriptsize 3D Wire Art \cite{tojo.etal2024}}
    \end{subfigure}
    & \begin{subfigure}[t]{\cwidth}
        \begin{subfigure}{\textwidth}
            \fbox{\includegraphics[width=\textwidth]{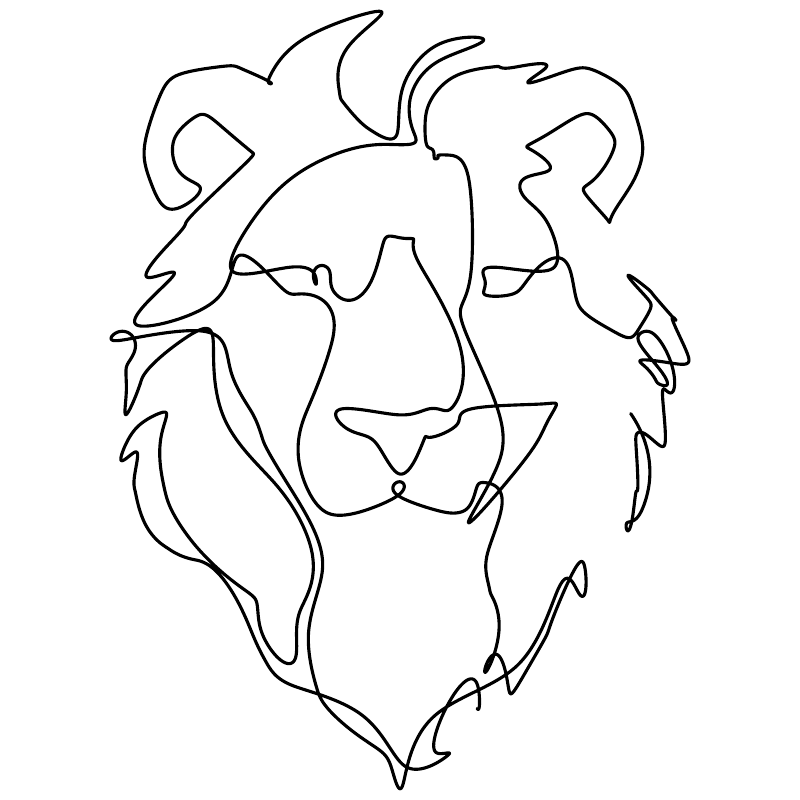}}%
        \end{subfigure}    
        \\[\lrspace]
        \begin{subfigure}{\textwidth}
            \fbox{\includegraphics[width=\textwidth]{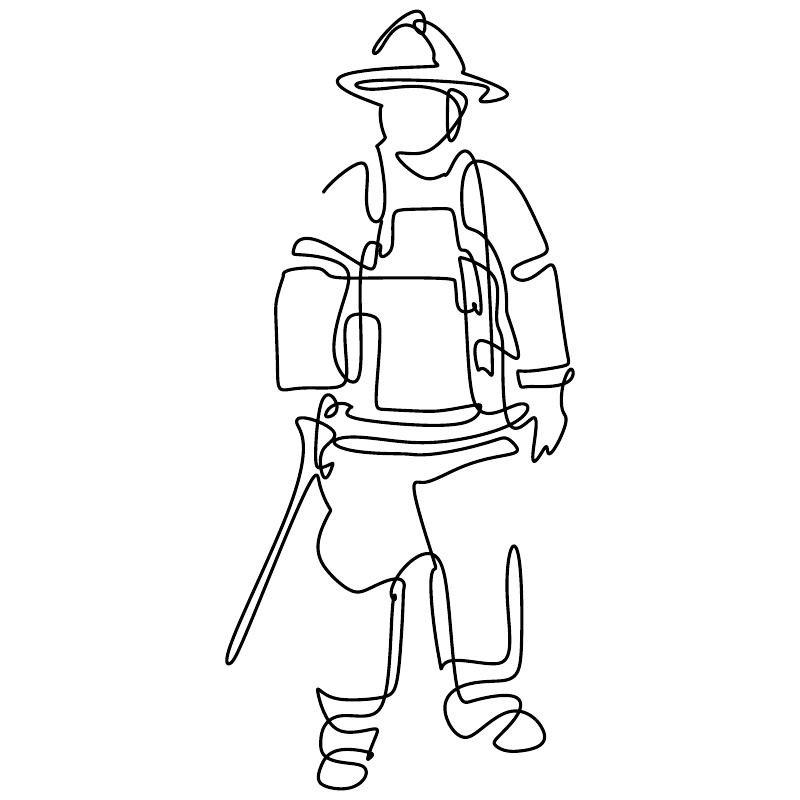}}%
        \end{subfigure}
        \vspace{-15pt}
        \caption*{\scriptsize Our method}
    \end{subfigure}%
    \end{tabular}
    \caption{Qualitative comparison of various line drawing generation methods. Text-to-image models (Gemini in this case) produce raster images and contain multiple lines. We present here the results obtained after applying the single-line conversion strategy (see \secref{sec:results}).}\label{fig:comparison} 
\end{figure*}

A key characteristic of single-line drawings is that their curve does not overdraw itself. On the other hand, crossings are allowed. This characteristic is enforced in our approach by the repulsion loss $\mathcal{L}_\text{rep}$, based on repulsive curves \cite{yu2021repulsive} and the implementation by \cite{tojo.etal2024}. This implementation encourages self-clearance while maintaining computational efficiency by summing a finite-support kernel over points perpendicular to the curve's tangent. While Tojo et al. \cite{tojo.etal2024} used this repulsion to enforce the fabricability of their 3D curve, we use this loss for stylistic purposes. The repulsion loss counteracts the tendency for curve segments to be overdrawn and positioned close to each other to fill regions of the image, as observed in ControlSketch (see \figref{fig:comparison}). The weight $\lambda_\text{rep}$ allows the user to control the style of the output drawing (see \secref{subsec:stylization}). 

We add the shortening loss $\mathcal{L}_\text{short}$ to keep the curve compact, avoiding unnecessary details. It is computed as an approximation of the spline curve's length: $\mathcal{L}_\text{short} = \sum_{i=0}^{m-2} \|p_{i+1} - p_i\|$, where $(p_i)_{i=0}^{m-1}$ are the $m$ sampled points along the curve. This loss works in tandem with weight pruning and the sparsity loss to get a simple curve with detail where necessary.

$\mathcal{L}_\text{sparse}$ is a sparsity loss, computed as the $L^1$ norm over the weight vector $\mathbf{w}$: $\mathcal{L}_\text{sparse}(\theta) = \|\mathbf{w}\|_1$. The $L^1$ norm pushes weights to vanish, while allowing some weights to be non-zero, depending on the other terms in the loss function. The weight of this loss, $\lambda_\text{sparse}$, is initially set to zero and linearly increases after each iteration to reach its final value. This gives more flexibility in the initial optimization steps, while removing redundant control points later on.

\begin{figure}
    \centering
    \setlength{\fboxsep}{0pt}
    \begin{subfigure}[t]{0.49\linewidth}
        \centering
        \adjincludegraphics[width=\textwidth,trim={{0.075\width} {.18\height} {0.075\width} {0.3\height}},clip]{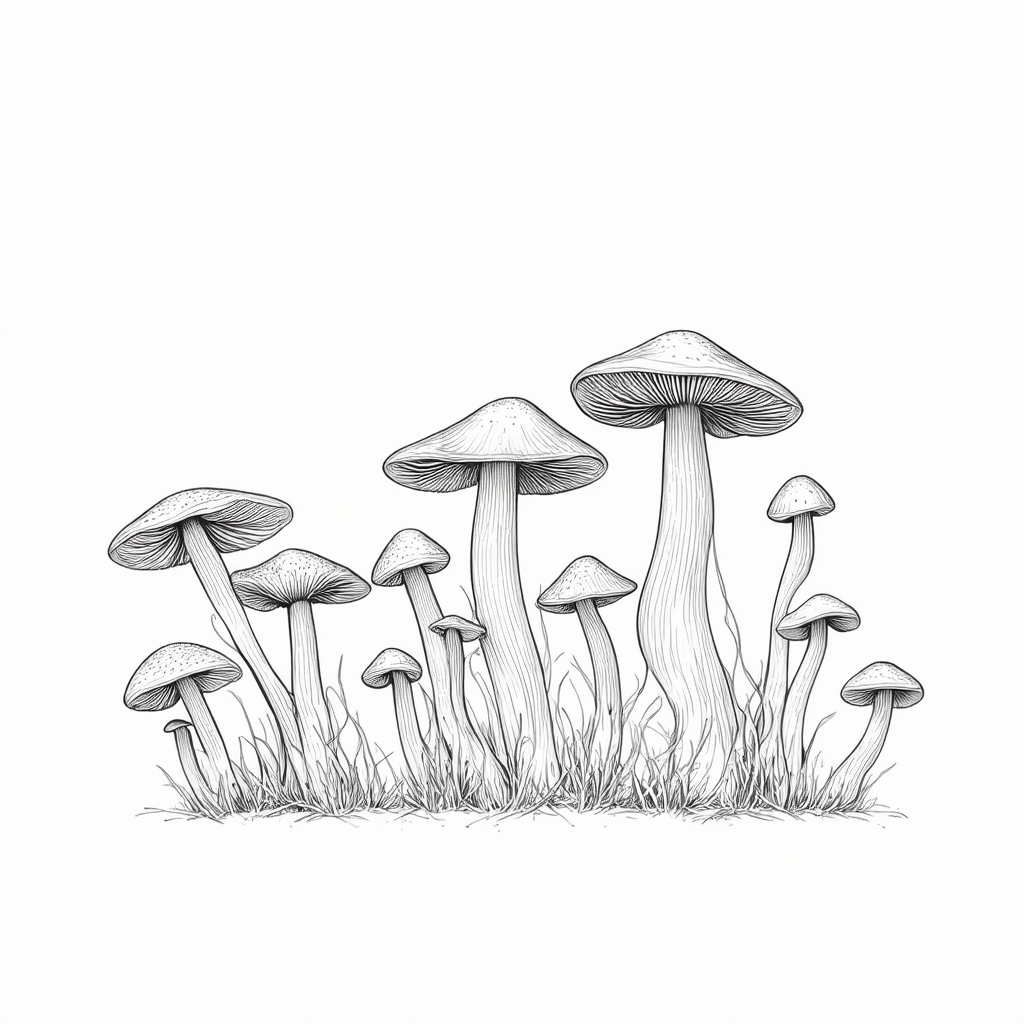}
    \end{subfigure}
    \begin{subfigure}[t]{0.49\linewidth}
        \centering
        \adjincludegraphics[width=\textwidth,trim={{0.01\width} {0.2\height} {0.01\width} {0.2\height}},clip]{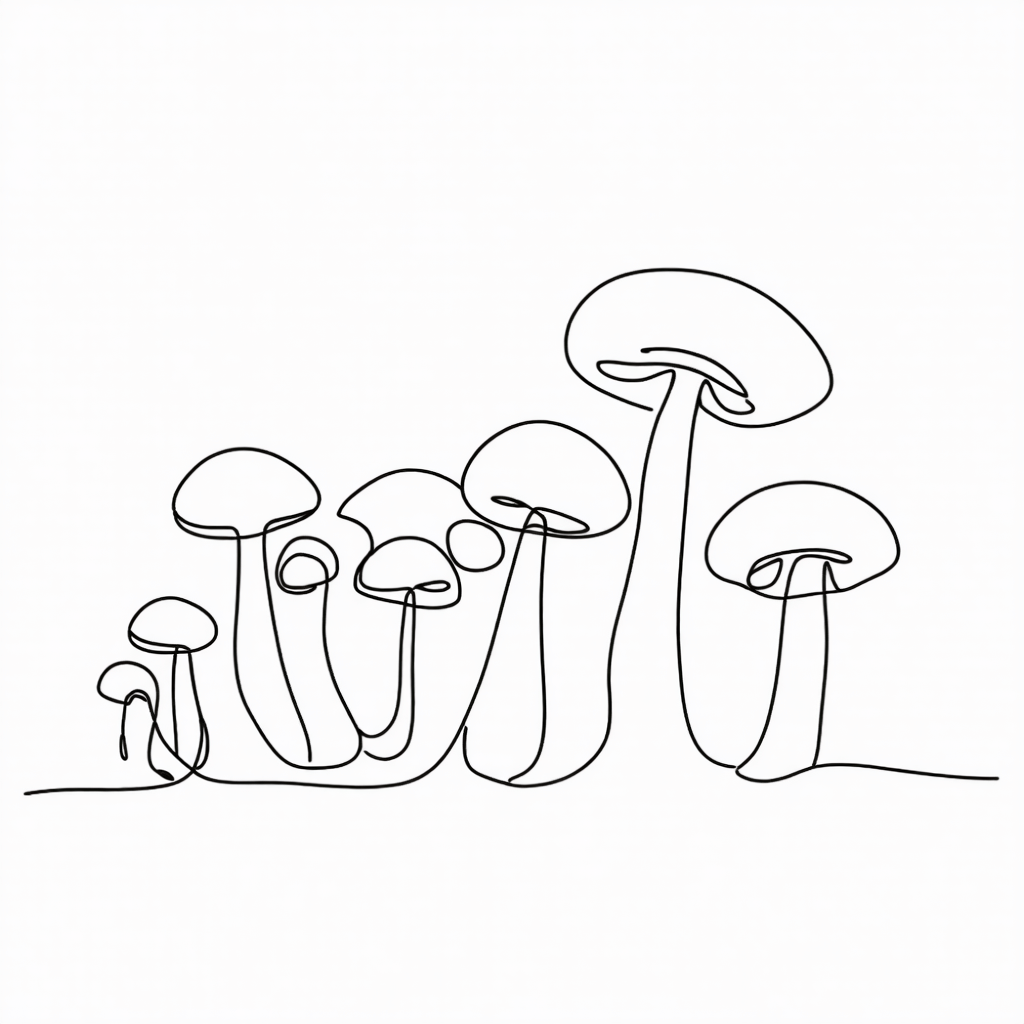}
    \end{subfigure}
    \caption{Left: an image generated with the original weights of the diffusion model. Right: a raster image generated using the LoRA weights trained on a dataset of single-line drawings. Both were generated with the same prompt: ``A single line drawing of a group of mushrooms of different sizes growing together, depicted with one flowing line.''}\label{fig:lora}
\end{figure}

\subsection{Diffusion model finetuning}
\label{subsec:lora}

As a final step, we train a custom low rank adaptation (LoRA) model \cite{hu.etal2021, stable-diffusion-3-5} to steer the generated drawings more toward the single-line drawing style through SDS. LoRA models efficiently finetune diffusion models by optimizing a small set of weights on top of the frozen, pre-trained weights of the original model. 
We base the LoRA on Stable Diffusion 3.5 and collected a dataset of 220 single-line drawings from the internet (see \appref{app:lora}, \figref{fig:lora_training}). For the descriptions used to train the model, we generate captions for each drawing using Llama 3.2 11B Vision instruction-tuned model \cite{llama-hf}. 
\figref{fig:lora} presents an example of an image generated using the original diffusion model and an image generated using the diffusion model with the LoRA weights for the same prompt. While the result produced by the LoRA model is not a single-line drawings, it is much closer to the style of such drawings. As demonstrated in \secref{subsec:ablation}, this helps our method produce results that more closely resemble the single-line drawings in the dataset.

\definecolor{gold}{HTML}{ffd700}
\definecolor{silver}{HTML}{c0c0c0}
\definecolor{bronze}{HTML}{CD7F32}
\definecolor{other}{HTML}{FFFFFF}
\let\originalfbox\fbox
\renewcommand\fbox{\fcolorbox{white}{white}}
\newcommand{\best}[1]{\setlength{\fboxsep}{0pt}\fbox{\setlength{\fboxsep}{2pt}\colorbox{gold!50}{#1}}}
\newcommand{\second}[1]{\setlength{\fboxsep}{0pt}\fbox{\setlength{\fboxsep}{2pt}\colorbox{silver!70}{#1}}}
\newcommand{\third}[1]{\setlength{\fboxsep}{0pt}\fbox{\setlength{\fboxsep}{2pt}\colorbox{bronze!50}{#1}}}
\newcommand{\other}[1]{\setlength{\fboxsep}{0pt}\fbox{\setlength{\fboxsep}{2pt}\colorbox{other}{#1}}}

\newcommand{\dashedother}[1]{\setlength{\fboxsep}{0pt}\dbox{\setlength{\fboxsep}{2pt}\colorbox{other}{#1}}}

\begin{table*}[t]
\centering
\caption[Quantitative comparison table]{Quantitative comparisons of single-line drawing generation methods. The metrics are given as averages across all samples on the benchmark presented in \secref{subsec:quantitative}, except for the FID which directly measures the distribution difference between the generated drawings and a reference dataset of single-line drawings. \best{Gold} indicates best score among vector single-line drawing methods that do not use the connecting strategy. \dashedother{Dashed} indicates higher score for methods that do not directly produce single-line drawings in vector format, or require vectorization and component connection to obtain a single-line drawing}.
    \begin{tabular}{l r r r r r r r r r}
    \toprule
    & \multicolumn{2}{c}{Output} & \multicolumn{1}{c}{Text-Image} & \multicolumn{2}{c}{Image-Image} & \multicolumn{2}{c}{Visual} \\
    \cmidrule[0.1pt](lr){2-3}\cmidrule[0.1pt](lr){4-4}\cmidrule[0.1pt](lr){5-6}\cmidrule[0.1pt](lr){7-8}
    Method & Vector  & Single Line  & CLIP $\uparrow$ & CLIP $\uparrow$ & DinoV2 $\uparrow$ & Aesthetic $\uparrow$ & FID $\downarrow$ \\
    \midrule
        ChatGPT (gpt-image-1.5) & \xmark & \xmark & \other{32.5} & \dashedother{83.2} & \dashedother{0.859} & \other{5.52} & \other{157.4}\\
        Nano Banana (gemini-3-pro-image-preview) & \xmark & \xmark & \other{32.1} & \dashedother{83.5} & \dashedother{0.871} & \other{5.50} & \other{169.9}\\
        Flux.2 [max] (flux-2-max) & \xmark & \xmark & \other{32.5} & \dashedother{79.6} & \other{0.765} & \other{5.82} & \other{133.0}\\\hdashline[0.5pt/1pt]
        Vinker et al. (multi-stroke) \cite{vinker.etal2022} & \cmark & \xmark & \other{32.3} & \dashedother{80.0} & \dashedother{0.824} & \other{4.51} & \other{266.3}\\
        Arar et al. (multi-stroke) \cite{arar.etal2025} & \cmark & \xmark & \other{31.7} & \dashedother{79.6} & \dashedother{0.817} & \other{4.52} & \other{209.4}\\\hdashline[0.5pt/1pt]
        ChatGPT connected & \cmark & \cmark & \other{33.1} & \dashedother{79.3} & \dashedother{0.827} & \other{5.78} & \other{173.9}\\
        Nano Banana connected & \cmark & \cmark & \other{32.8} & \dashedother{80.7} & \dashedother{0.830} & \other{5.56} & \other{186.3}\\
        Flux.2 [max] connected & \cmark & \cmark & \other{33.1} & \other{76.8} & \other{0.759} & \dashedother{6.13} & \other{144.0}\\
        Vinker et al. (multi-stroke) \cite{vinker.etal2022} connected & \cmark & \cmark & \other{32.1} & \other{73.7} & \other{0.755} & \other{4.76} & \other{223.6}\\
        Arar et al. (multi-stroke) \cite{arar.etal2025} connected & \cmark & \cmark & \other{32.2} & \other{76.2} & \other{0.767} & \other{4.84} & \other{219.3}\\\hdashline[0.5pt/1pt]
        Kaplan \& Bosch \cite{kaplan.bosch2005} - 1000 points & \cmark & \cmark & \other{24.1} & \other{58.1} & \other{0.632} & \other{4.37} & \other{346.8}\\
        Vinker et al. (single-stroke) \cite{vinker.etal2022} & \cmark & \cmark & \other{29.6} & \other{66.6} & \other{0.675} & \other{3.96} & \other{290.1}\\
        Arar et al. (single-stroke) \cite{arar.etal2025} & \cmark & \cmark & \other{28.2} & \other{64.7} & \other{0.651} & \other{4.09} & \other{273.5}\\
        Tojo et al. \cite{tojo.etal2024} - Image & \cmark & \cmark & \other{28.9} & \other{66.7} & \other{0.682} & \other{5.05} & \other{161.7}\\
        Tojo et al. \cite{tojo.etal2024} - Text & \cmark & \cmark & \other{29.7} & \other{65.0} & \other{0.646} & \other{4.92} & \other{160.2}\\
        \midrule
        Ours & \cmark & \cmark & \best{33.1} & \best{78.0} & \best{0.785} & \best{5.82} & \best{120.8}\\
        \bottomrule
    \end{tabular}
\label{tab:comparisontable}
\end{table*}
\let\fbox\originalfbox

\section{Results and discussion}
\label{sec:results}

Our results are both qualitatively and quantitatively compared to a wide range of alternatives: TSP methods \cite{kaplan.bosch2005}, general raster-based text-to-image models (ChatGPT, Gemini, Flux) and VLM-based differentiable sketching methods (CLIPasso~\cite{vinker.etal2022}, ControlSketch \cite{arar.etal2025} and Fabricable 3D Wire Art \cite{tojo.etal2024}). Only the TSP method does not require any adjustments to produce 2D single-line drawings. General text-to-image models output raster images instead of vector graphics, and it is impossible to ensure that the output is a single-line drawing. We modify CLIPasso and ControlSketch to output a single line, represented by a series of cubic Bézier curves. We use the Fabricable 3D Wire Art approach, originally designed for a multi-view setup, in a single-view context. For the VLM-based sketching methods~\cite{vinker.etal2022, arar.etal2025, tojo.etal2024}, we set the number of control points to the average number of control points used by our approach, which translates to 64 B\'ezier curves for CLIPasso and ControlSketch and 193 control points for Fabricable 3D Wire Art. While the code for SVGDreamer \cite{xing.etal2024} is available, we do not compare to it as the dependencies were outdated and conflicting, making it impossible to run. SVGDreamer is similar to ControlSketch \cite{arar.etal2025}, as it uses score distillation sampling and creates drawings with multiple strokes. Therefore, we expect the results to be comparable. Implementation details for our method are provided in \appref{app:implementation_details}.

\textit{Connecting-the-line strategy.\ } In addition, for raster-based text-to-image models, we propose a simple procedure to convert their outputs to vector single-line drawings. The input is first vectorized using Adobe Illustrator. We then connect the obtained strokes following the procedure described in the introduction. To achieve this, we solve a traveling salesperson problem (TSP) on a graph where each node represents one stroke. The graph is fully connected. The weight of an edge between two strokes is the minimum distance between their endpoints. The output of the TSP is an ordered list of strokes, which are then connected continuously using cubic Bézier curves. This strategy is also applied to the results of the original CLIPasso \cite{vinker.etal2022} and ControlSketch \cite{arar.etal2025} methods.

\subsection{Qualitative comparison}
\label{subsec:qualitative}

Visual comparisons to results from Gemini, TSP art, ControlSketch and Fabricable 3D Wire Art are shown in \figref{fig:comparison}. Additional comparisons are available in \figref{fig:extra_comparison}, and more methods are included in the supplemental material. Gemini's results are presented after applying the single-line drawing conversion strategy. These results do not capture the style of single-line drawings. They contain too many details of the input image and do not simplify and abstract it.
The TSP art \cite{kaplan.bosch2005} method produces results that follow the intensity of the input image, but the results do not present self-intersections and the method lacks abstraction capabilities. In addition, for images with relatively uniform intensity, the results are hard to recognize, especially with only 1000 sampled points. 
ControlSketch \cite{arar.etal2025} often fails to depict the intended subject and the aesthetic is not very pleasing, partly due to the way the method is initialized: strokes are sampled according to an attention map. While this works properly for multiple strokes, it does not translate well to a single-stroke scenario.
Fabricable 3D Wire Art \cite{tojo.etal2024} sometimes produces reasonable results for simple input images, as can be seen in \figref{fig:extra_comparison} (e.g., the candle or bird). While the outline correctly reflects the depicted subject, the internal details of the results often lack interpretability. In addition, the method often fails to correctly depict. more complex subjects (e.g., tree, cat, cow).

In contrast, our method produces vector, single-line results whose aesthetic matches that of single-line drawings created by artists. The curve represents the entire subject, and our optimization method produces meaningful details inside the shape. This is due to the curve representation and control-point pruning strategy, which allows the level of detail to be adapted to the input. For instance, in \figref{fig:extra_comparison}, the candle (83 active control points) consists of a much simpler curve than the firefighter (210 active control points). 

\renewcommand\fbox{\fcolorbox{white}{white}}
\begin{table}[t]
\centering
    \caption{Results of the perceptual study evaluation. Each column represents the percentage of samples that participants ranked in 1st, 2nd, 3rd or 4th position based on the closeness to single-line drawing style criterion. Our method was ranked first in most cases.}
    \resizebox{\linewidth}{!}{%
    \setlength{\tabcolsep}{1mm}
    \begin{tabular}{l r r r r}
        \toprule
        Rank & \multicolumn{1}{c}{1} & \multicolumn{1}{c}{2} & \multicolumn{1}{c}{3} & \multicolumn{1}{c}{4}  \\\midrule
        ControlSketch (single stroke) \cite{arar.etal2025} & \other{3.7} & \other{10.9} & \other{15.2} & \other{70.2} \\
        Flux.2 [max] connected & \other{16.3} & \other{55.9} & \other{23.4} & \other{4.4} \\
        Nano Banana connected & \other{3.7} & \other{18.5} & \other{52.8} & \other{25.0} \\
        Ours & \best{76.3} & \other{14.7} & \other{8.6} & \other{0.4} \\
        \bottomrule
    \end{tabular}
    }
\label{tab:userstudytable}
\end{table}
\let\fbox\originalfbox

\subsection{Quantitative evaluation}
\label{subsec:quantitative}

We use the following metrics for quantitative comparison. First, we are interested in how close the results are to the given text prompt. For that purpose, we define text-image similarity based on CLIP via cosine similarity (normalized dot-product) between the CLIP embedding of the input prompt and the output image.

To measure the similarity between the input and output images, we cannot rely on traditional image-based similarity measures, such as SSIM or PSNR, since the compared images are in different stylistic domains. Thus, we use CLIP~\cite{radford2021learningtransferablevisualmodels} and DinoV2~\cite{oquab2024dinov2} embeddings to encode the semantic content of each image and compute the cosine similarity between the respective embeddings.

Finally, we are interested in the aesthetic quality of the single-line drawings. In other words: are the results visually pleasing and do they look like single-line drawings? The `Aesthetic' score is computed using a network trained to mimic human preferences in terms of visual aesthetics \cite{schuhmann2022laion, discus2024aesthetic}. The Fréchet inception distance (FID) \cite{Heusel_2017_FID} is a measure of distance between two distributions of embeddings. An inception network is used to compute the embeddings of all output images in a generated dataset, as well as the embeddings of a real dataset of single-line drawings created by artists. Then, the statistics (mean and standard deviation) of these embeddings are compared. This metric tells us how close our resulting images are to the desired aesthetic style, i.e., distribution of single-line drawings.

\textit{Benchmark.\ } We compute the set of metrics on a benchmark of 50 prompt-image pairs, representing various subjects. The input images present the subject on a white background. The associated prompts follow this structure: ``\textit{Single line drawing of a [\dots], continuous black line on a white background.}'' where [\dots] is replaced with a short description of the subject (e.g., `lion', `ballerina').

\textit{Quantitative results.\ } We present our results in \tabref{tab:comparisontable}. Our approach outperforms all methods that directly produce actual single-line drawings on all metrics. Some text-to-image models, which are not constrained to be single-line drawings, perform better than our method in some metrics, especially in terms of image-image similarity. This is to be expected, since unconstrained text-to-image methods have far more freedom to represent the input, but they do not adhere to the single-line constraint. In some cases, the output resembles an edge map of the input image, rather than a single-line drawing. Observing the visual quality metrics, our method achieves similar or better results than these text-to-image models, even while producing a single curve. After applying the single-line drawing connection strategy to the output of these text-to-image models, their image-image similarity decreases a bit and their visual scores increase a little, which is expected. However the visual quality metrics still remain similar or worse than ours.

\renewcommand\fbox{\fcolorbox{white}{white}}
\begin{table}[t]
\centering
    \caption{Ablation study. We removed each component separately from our pipeline and run our method. The metrics are computed on the same benchmark as in \secref{subsec:quantitative}.}
    \resizebox{\linewidth}{!}{%
    \setlength{\tabcolsep}{1mm}
    \begin{tabular}{ c c c c c c c c}
        \toprule
        \multicolumn{3}{c}{Parameters} & \multicolumn{1}{c}{Text-Image} & \multicolumn{2}{c}{Image-Image} & \multicolumn{2}{c}{Visual} \\
        \cmidrule[0.1pt](lr){1-3}\cmidrule[0.1pt](lr){4-4}\cmidrule[0.1pt](lr){5-6}\cmidrule[0.1pt](lr){7-8}
        LoRA  & $\mathcal{L}_\text{sparse}$  & $\mathcal{L}_\text{short}$  & CLIP $\uparrow$ & CLIP $\uparrow$ & DinoV2 $\uparrow$ & Aesthetic $\uparrow$ & FID $\downarrow$ \\
        \midrule
        \xmark & \cmark & \cmark & \other{32.0} & \other{75.5} & \other{0.741} & \other{5.57} & \other{141.1}\\
        \cmark & \xmark & \cmark & \other{33.0} & \other{77.1} & \best{0.789} & \other{5.54} & \other{149.7}\\
        \cmark & \cmark & \xmark & \other{33.0} & \other{75.5} & \other{0.769} & \other{5.37} & \other{200.7}\\\hdashline[0.5pt/1pt]
        \cmark & \cmark & \cmark & \best{33.1} & \best{78.0} & \other{0.785} & \best{5.82} & \best{120.8}\\
        \bottomrule
    \end{tabular}
    }
\label{tab:ablationtable}
\end{table}

\begin{figure}[t]
    \centering
    \begin{subfigure}{0.24\linewidth}
        \includegraphics[width=\textwidth]{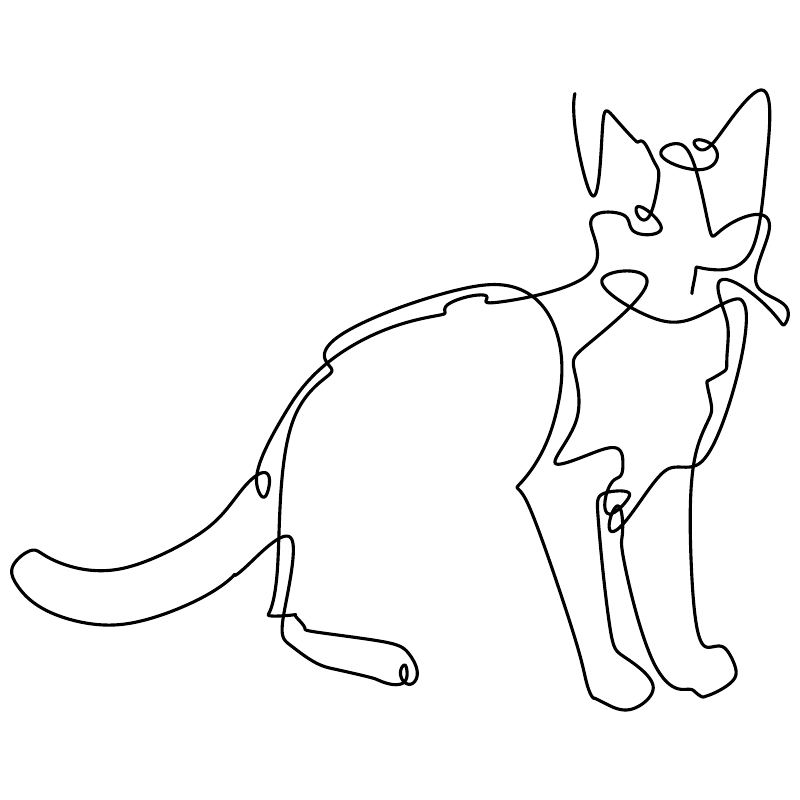}%
        \vspace{-6pt}%
        \caption*{\scriptsize no LoRA}%
    \end{subfigure}
    \begin{subfigure}{0.24\linewidth}
        \includegraphics[width=\textwidth]{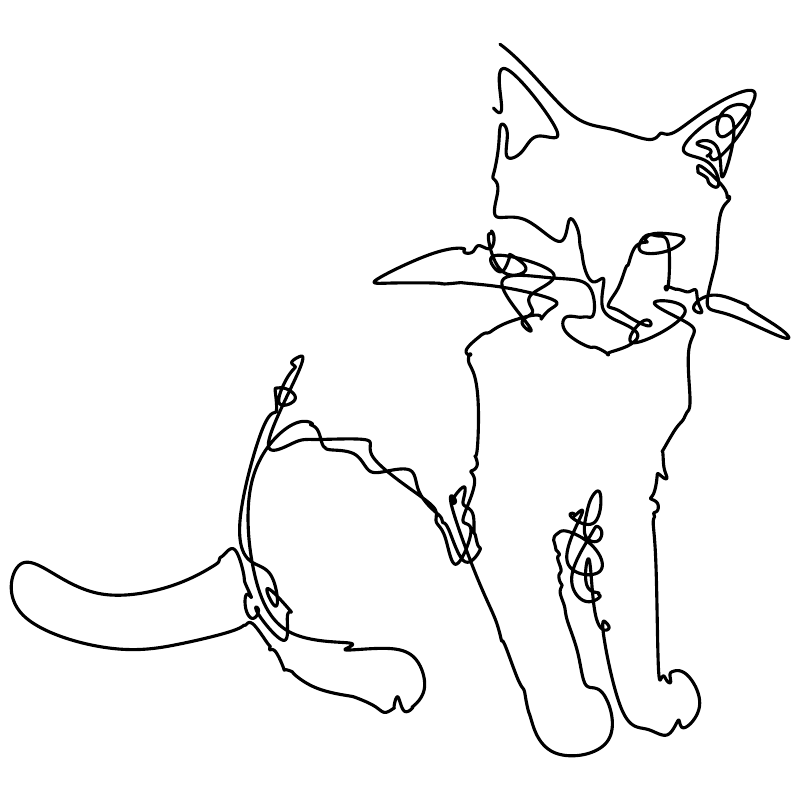}%
        \vspace{-6pt}%
        \caption*{\scriptsize no sparsity loss}%
    \end{subfigure}
    \begin{subfigure}{0.24\linewidth}
        \includegraphics[width=\textwidth]{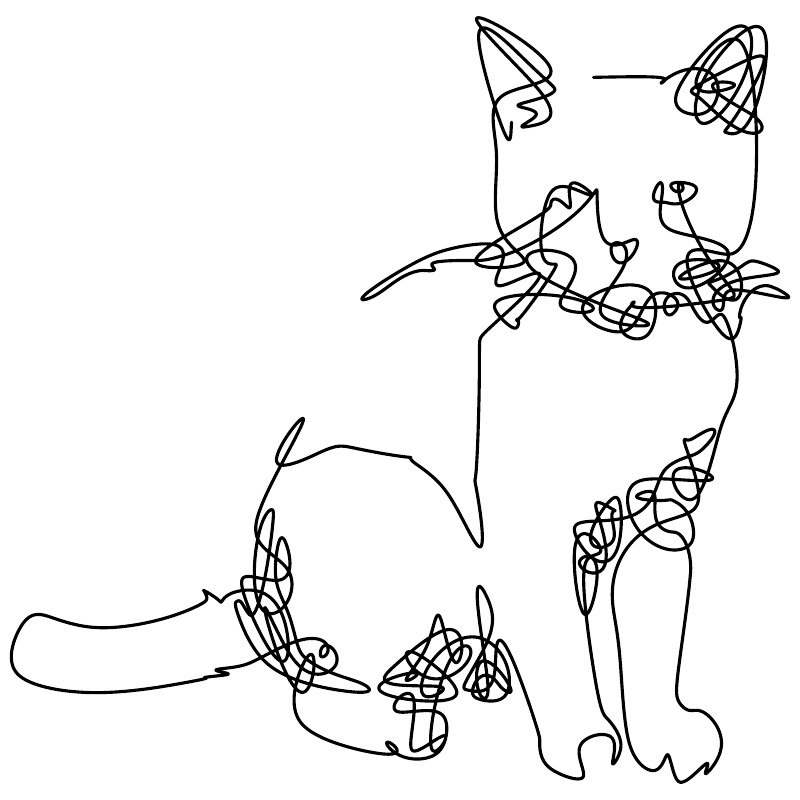}%
        \vspace{-6pt}%
        \caption*{\scriptsize no shortening loss}%
    \end{subfigure}
    \begin{subfigure}{0.24\linewidth}
        \includegraphics[width=\textwidth]{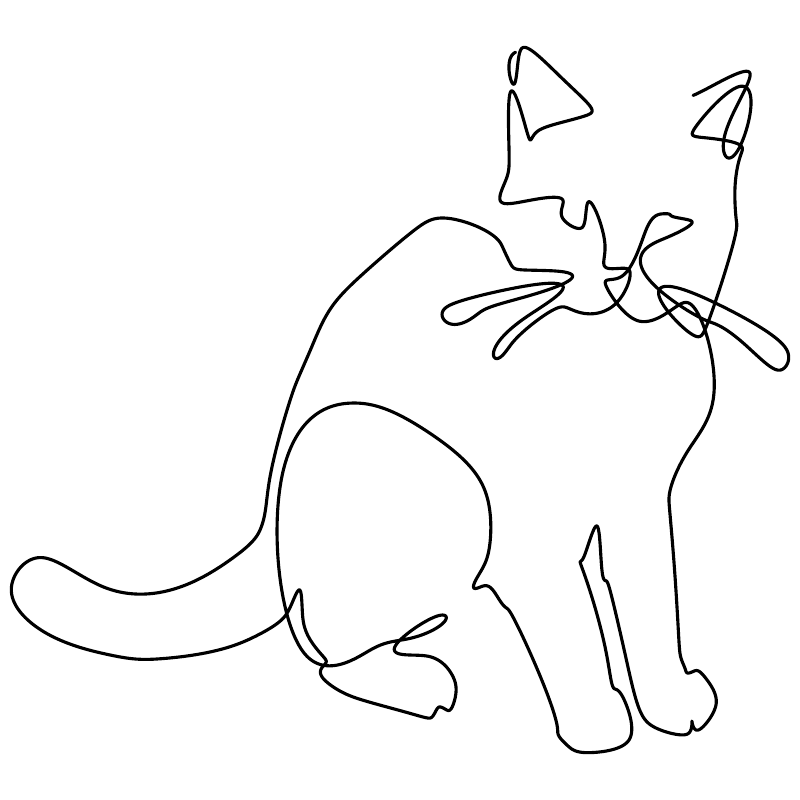}%
        \vspace{-6pt}%
        \caption*{\scriptsize our full loss}%
    \end{subfigure}
    \caption{Visual ablation study. Removing any component from our pipeline leads to worse-looking curves and too many details.}\label{fig:ablation}
\end{figure}

\renewcommand\fbox{\fcolorbox{white}{white}}
\begin{table}[t]
\centering
    \caption{Ablation study: curve representations. The metrics are computed on the same benchmark as in \secref{subsec:quantitative}.}
    \resizebox{\linewidth}{!}{%
    \setlength{\tabcolsep}{1mm}
    \begin{tabular}{l c c c c c}
        \toprule
        & \multicolumn{1}{c}{Text-Image} & \multicolumn{2}{c}{Image-Image} & \multicolumn{2}{c}{Visual} \\
        \cmidrule[0.1pt](lr){2-2}\cmidrule[0.1pt](lr){3-4}\cmidrule[0.1pt](lr){5-6}
        Curve  & CLIP $\uparrow$ & CLIP $\uparrow$ & DinoV2 $\uparrow$ & Aesthetic $\uparrow$ & FID $\downarrow$ \\
        \midrule
        Bézier & \other{32.7} & \other{77.3} & \other{0.781} & \other{5.52} & \other{121.0}\\
        B-Spline & \other{32.9} & \other{76.9} & \other{0.778} & \other{5.71} & \other{126.9}\\\hdashline[0.5pt/1pt]
        URBS (Ours) & \best{33.1} & \best{78.0} & \best{0.785} & \best{5.82} & \best{120.8}\\
        \bottomrule
    \end{tabular}
    }
\label{tab:ablationtable_curve}
\end{table}
\let\fbox\originalfbox

\begin{figure}[t]
    \setlength{\fboxsep}{0pt}
	\setlength{\tabcolsep}{1pt}
    \centering
    \begin{tabular}{c c c}
        \adjincludegraphics[width=0.32\linewidth,trim={{0.14\width} {0.0\width} {0.14\width} {0.0\width}},clip]{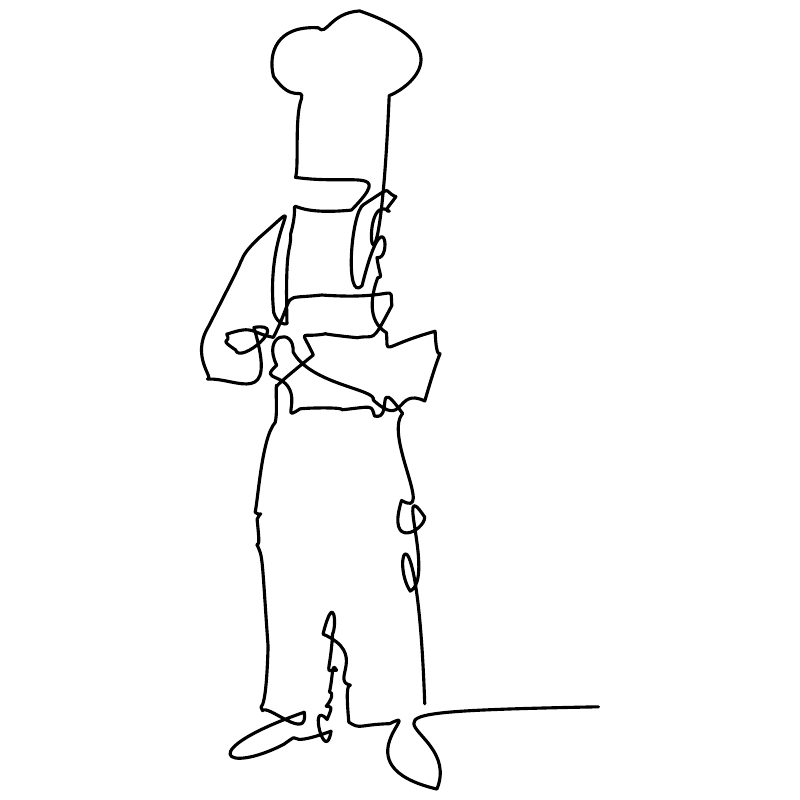} &%
        \adjincludegraphics[width=0.32\linewidth,trim={{0.14\width} {0.0\width} {0.14\width} {0.0\width}},clip]{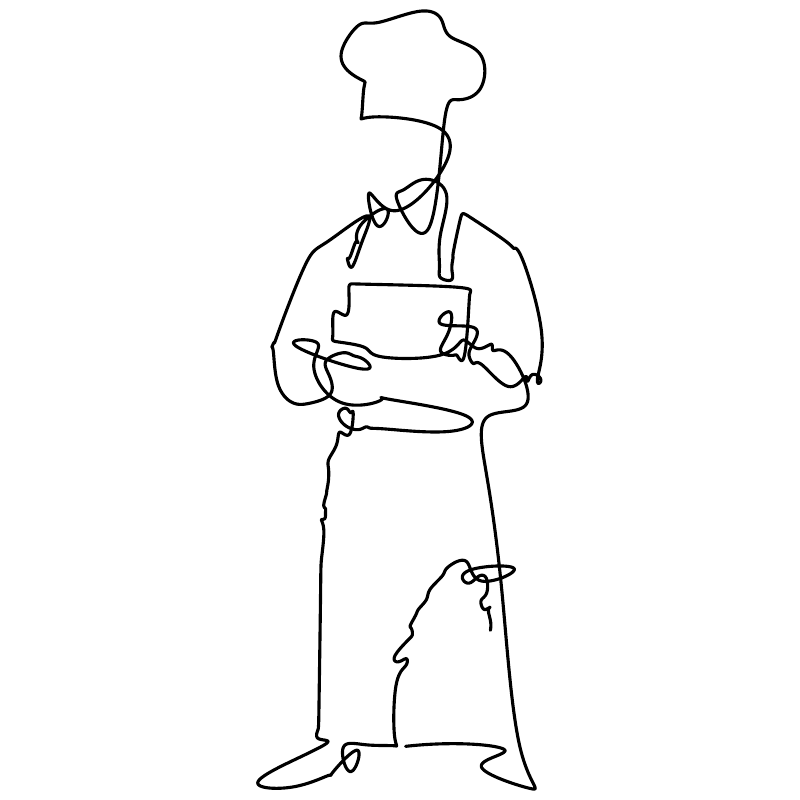} &%
        \adjincludegraphics[width=0.32\linewidth,trim={{0.14\width} {0.0\width} {0.14\width} {0.0\width}},clip]{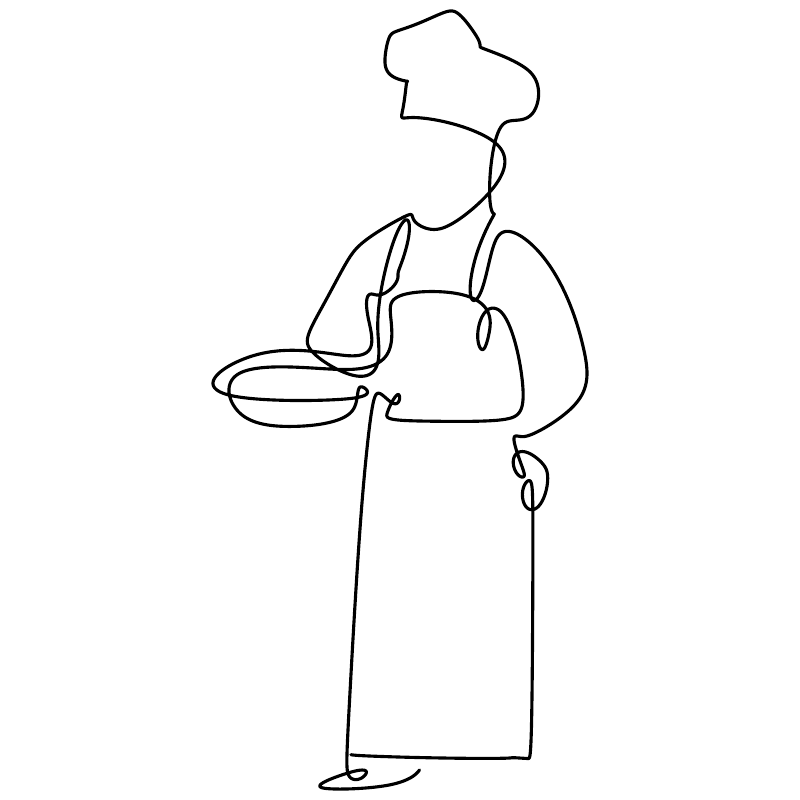}\\[1pt]%
        \adjincludegraphics[width=0.32\linewidth,trim={{0.14\width} {0.0\width} {0.14\width} {0.0\width}},clip]{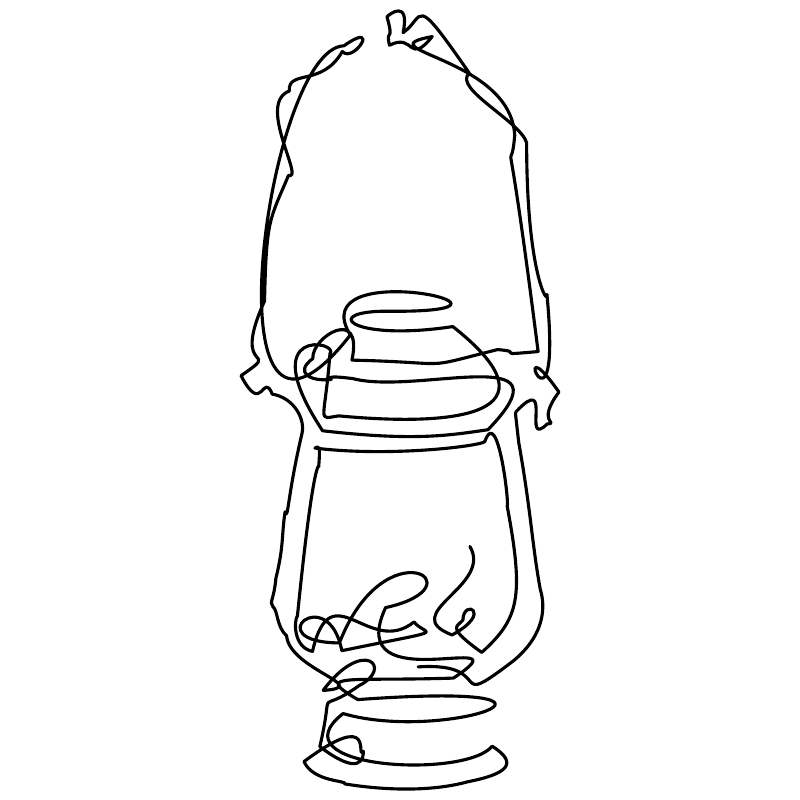} &%
        \adjincludegraphics[width=0.32\linewidth,trim={{0.14\width} {0.0\width} {0.14\width} {0.0\width}},clip]{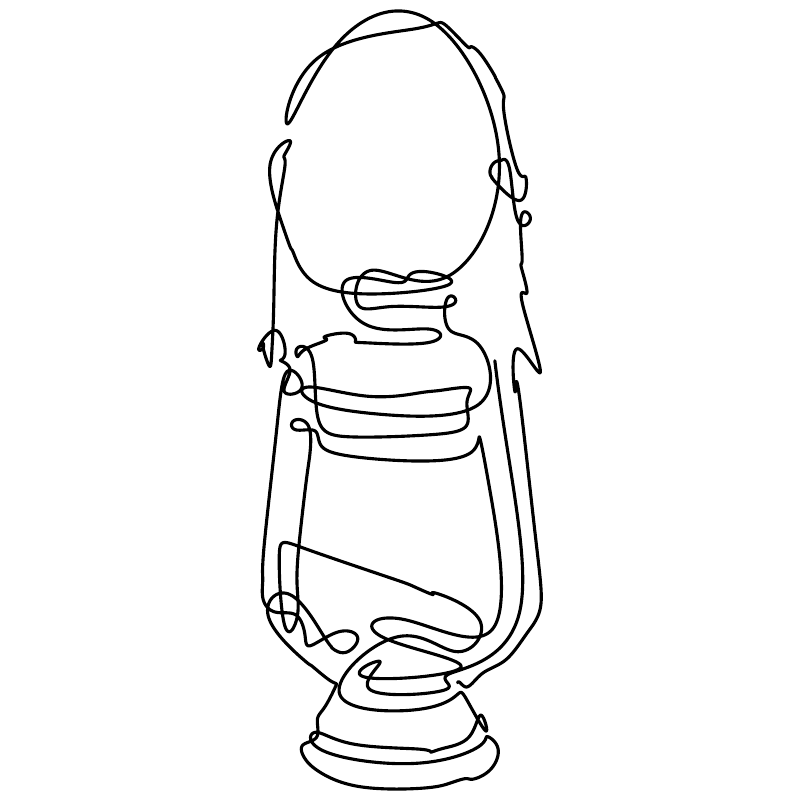} &%
        \adjincludegraphics[width=0.32\linewidth,trim={{0.14\width} {0.0\width} {0.14\width} {0.0\width}},clip]{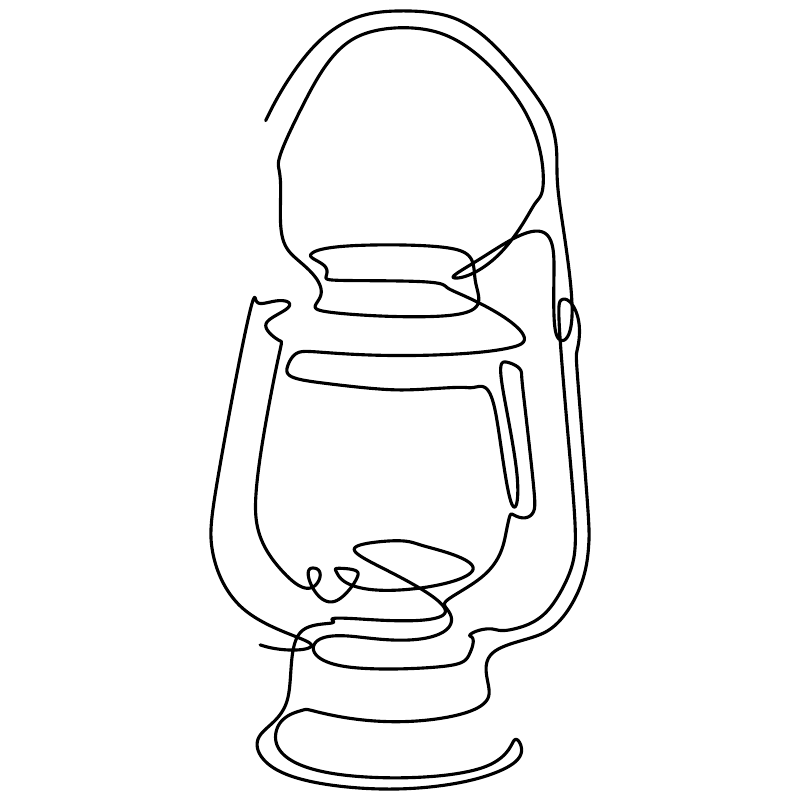} \\%
        Cubic Bézier spline & B-spline & URBS (ours)
    \end{tabular}
    \caption{Different curve representations.} 
    \label{fig:curve_representation}
\end{figure}

\begin{figure}[t]
    \setlength{\fboxsep}{0pt}
    \centering
    \begin{overpic}[width=0.25\linewidth,trim=62pt 46pt 54pt 46pt,clip]{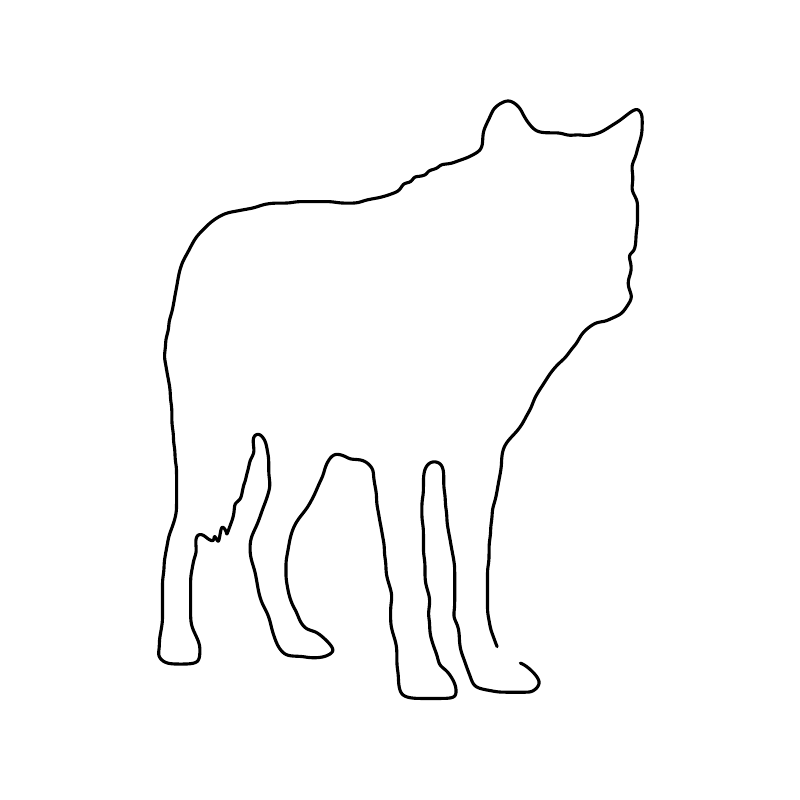}%
         \put (5,90) {\small init.}%
    \end{overpic}%
    \adjincludegraphics[width=0.25\linewidth,trim={{0.15\width} {0.12\width} {0.15\width} {0.12\width}},clip]{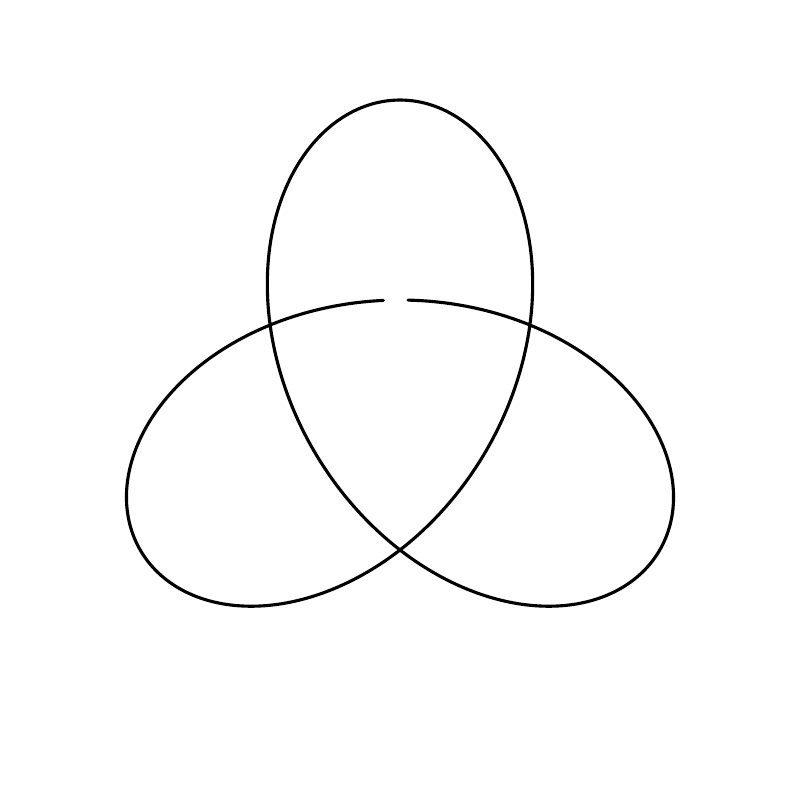}%
    \adjincludegraphics[width=0.25\linewidth,trim={{0.15\width} {0.12\width} {0.15\width} {0.12\width}},clip]{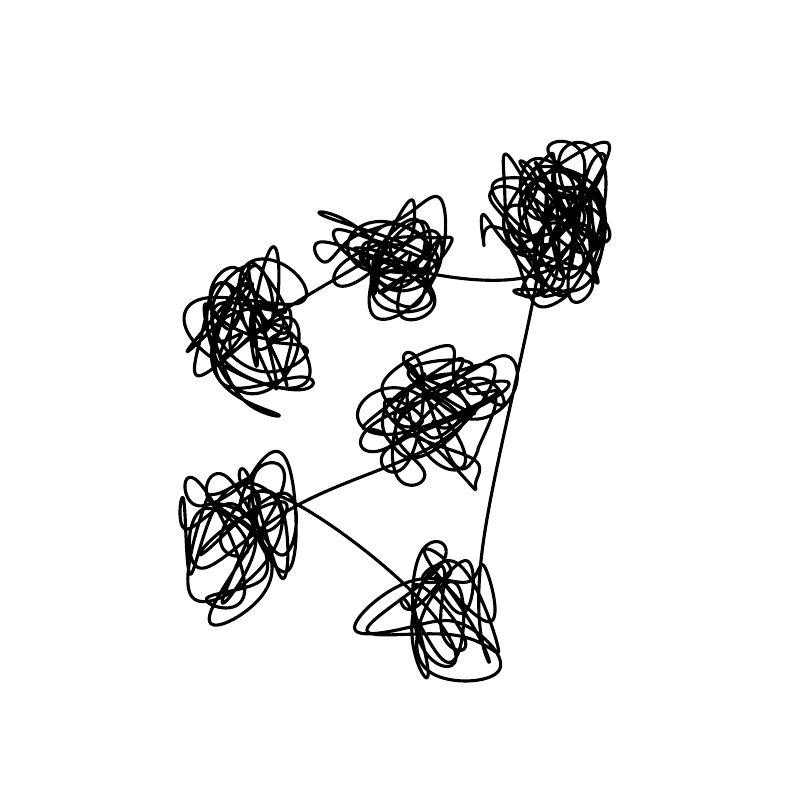}%
    \adjincludegraphics[width=0.25\linewidth,trim={{0.15\width} {0.12\width} {0.15\width} {0.12\width}},clip]{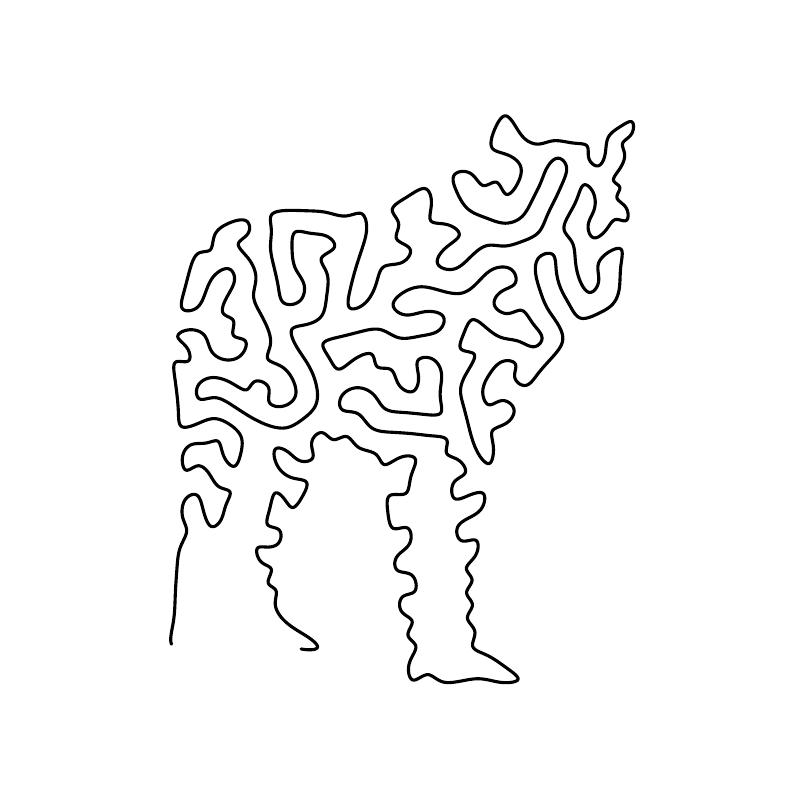}\\%
    \begin{overpic}[width=0.25\linewidth,trim=62pt 34pt 54pt 46pt,clip]{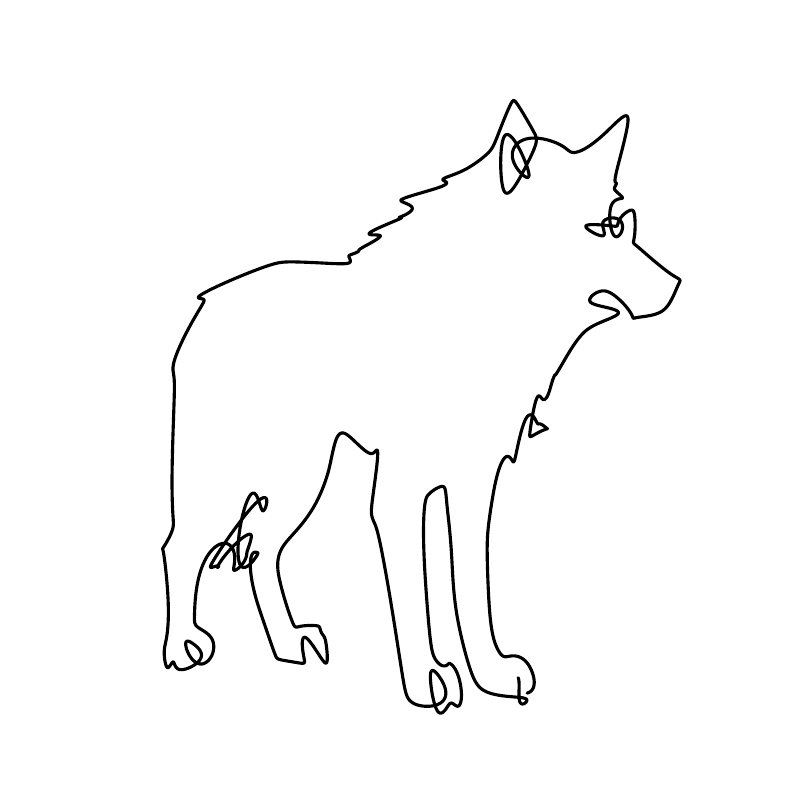}%
         \put (5,85) {\small result}%
    \end{overpic}%
    \adjincludegraphics[width=0.25\linewidth,trim={{0.15\width} {0.09\width} {0.15\width} {0.12\width}},clip]{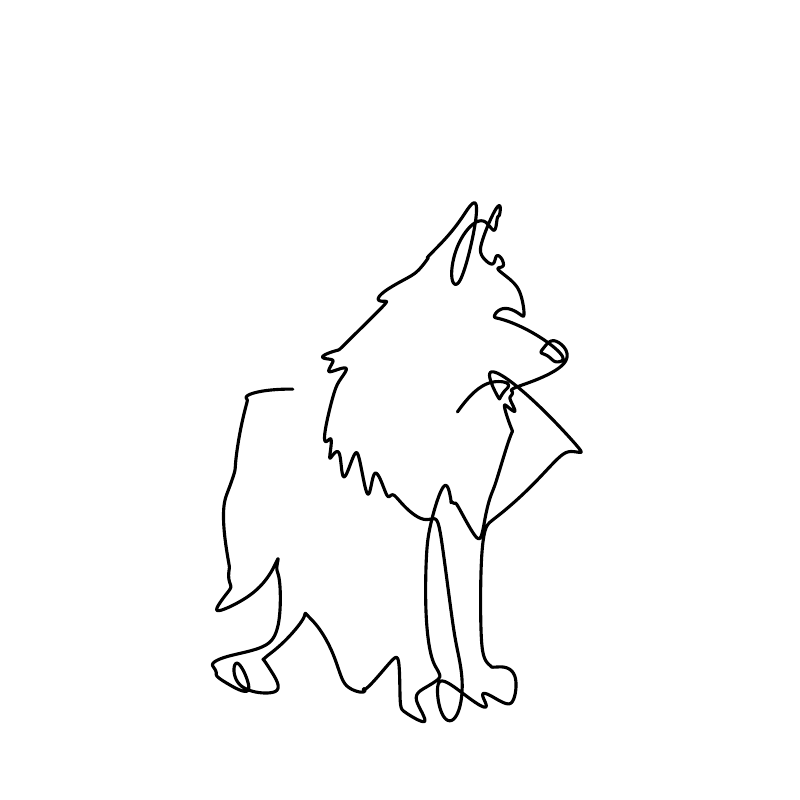}%
    \adjincludegraphics[width=0.25\linewidth,trim={{0.15\width} {0.09\width} {0.15\width} {0.12\width}},clip]{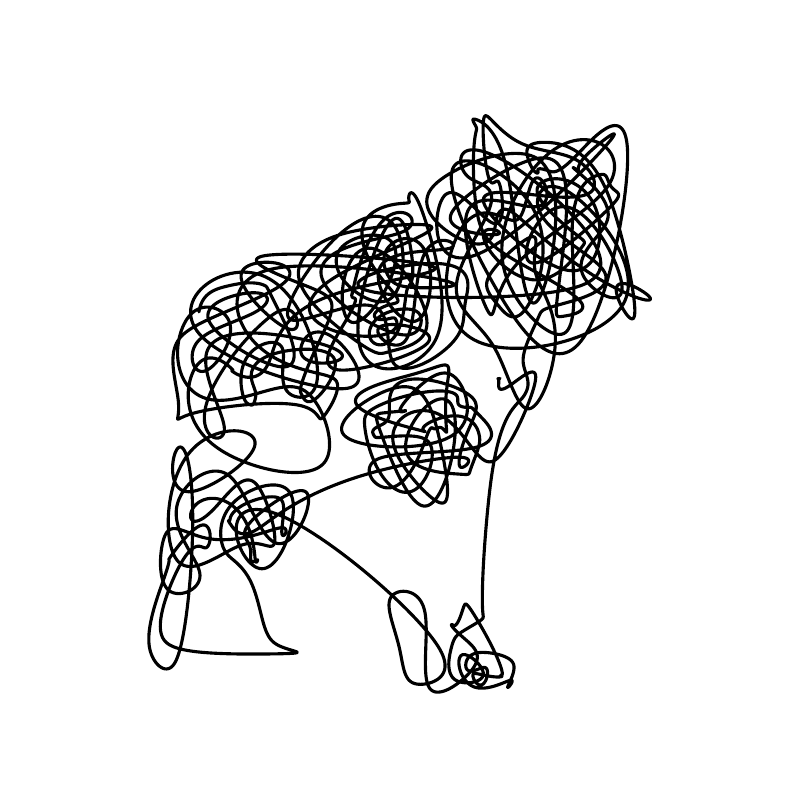}%
    \adjincludegraphics[width=0.25\linewidth,trim={{0.15\width} {0.09\width} {0.15\width} {0.12\width}},clip]{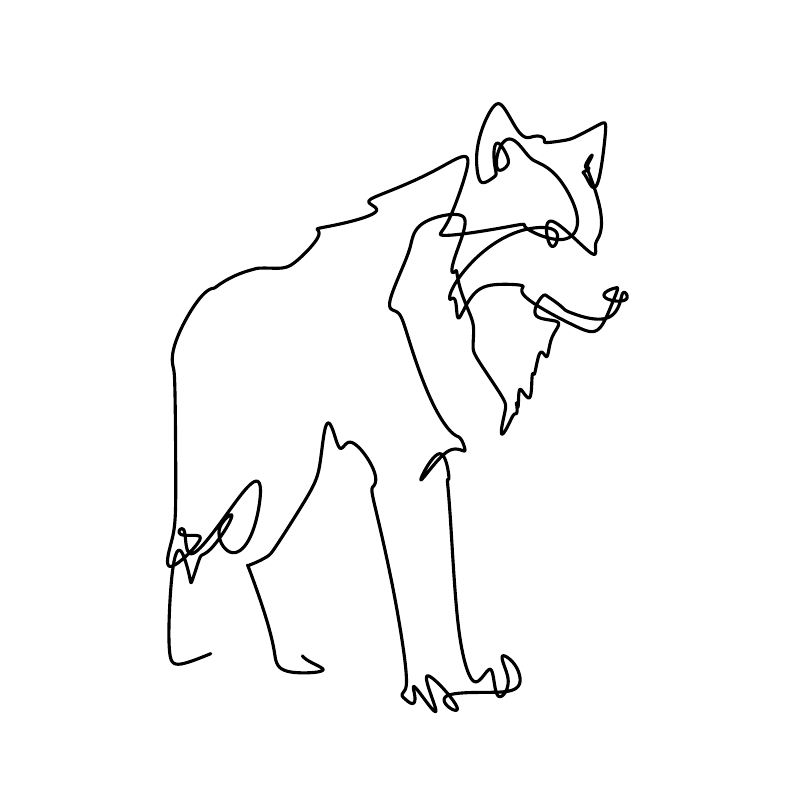}%
    \caption{Different initialization strategies lead to very different results. Top row: initial curve. Bottom row: final optimized curve. 
    }\label{fig:initialization}
\end{figure}

\begin{figure}[t]
    \centering
	\setlength{\tabcolsep}{0.005\textwidth}
    \begin{tabular}{c c c}
        \includegraphics[width=0.32\linewidth]{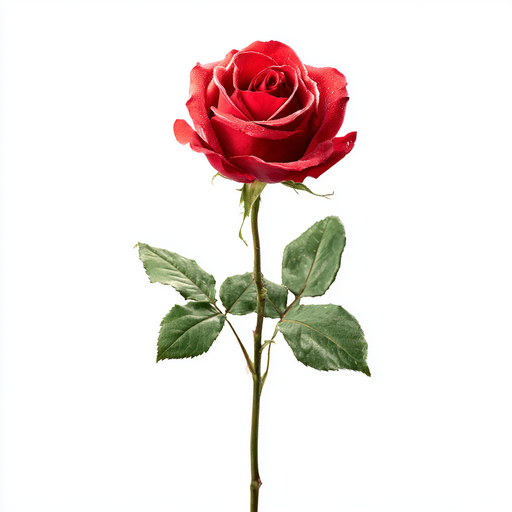} &%
        \includegraphics[width=0.32\linewidth]{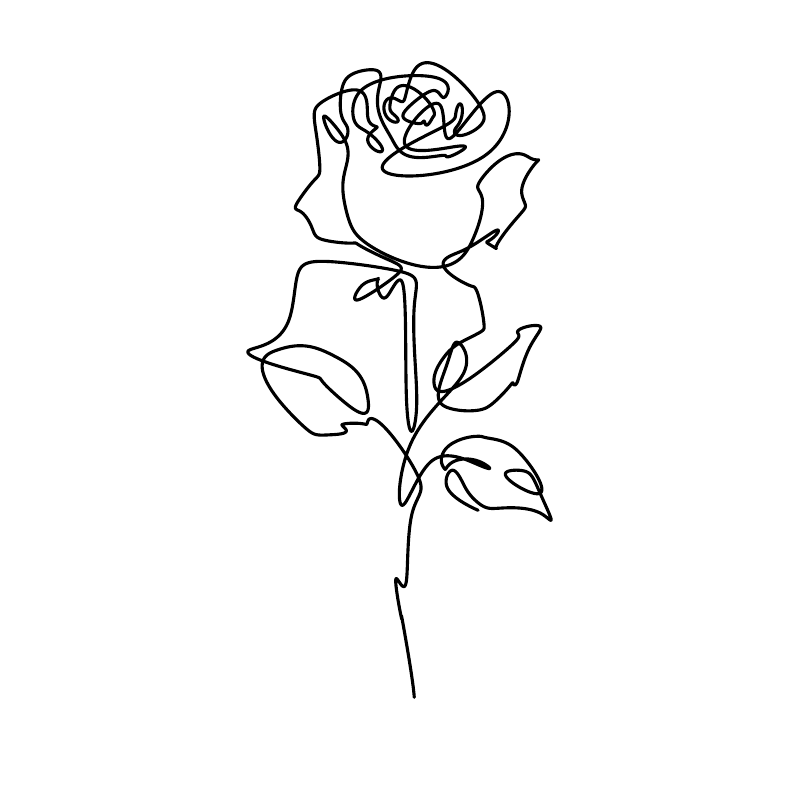} &%
        \includegraphics[width=0.32\linewidth]{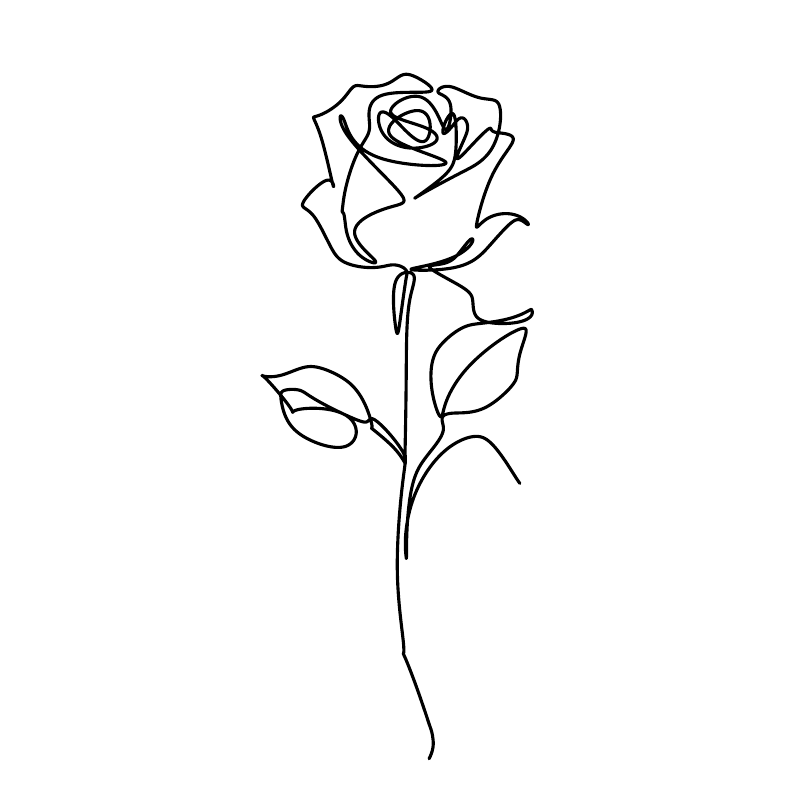}\\%
        \includegraphics[width=0.32\linewidth]{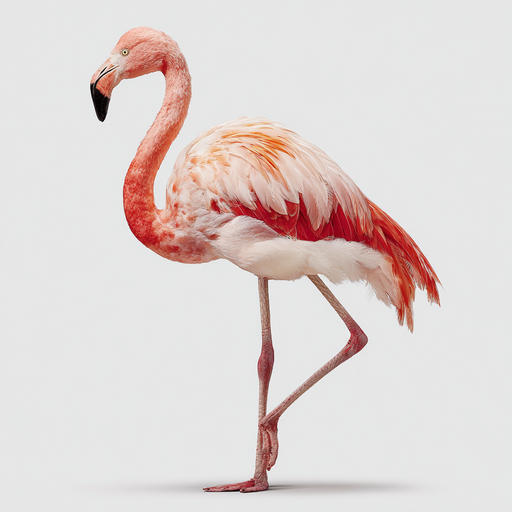} &%
        \includegraphics[width=0.32\linewidth]{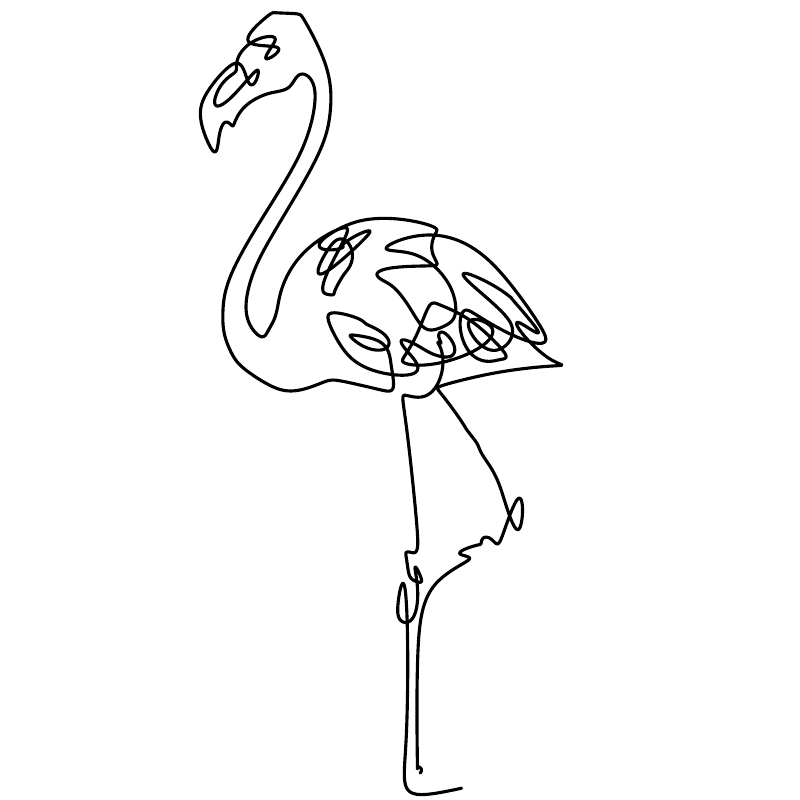} &%
        \includegraphics[width=0.32\linewidth]{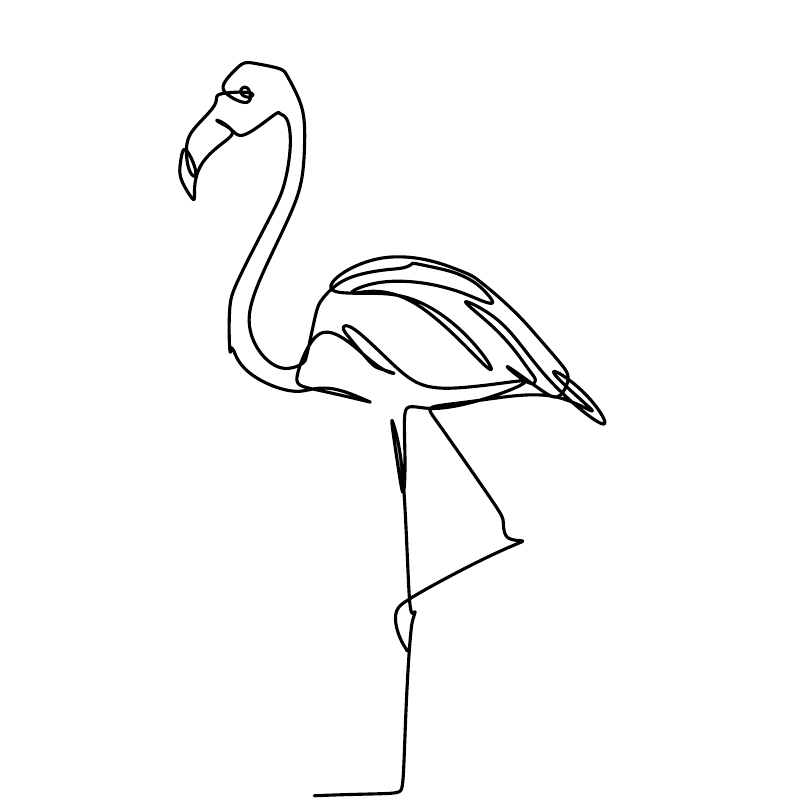}\\%
        input & ours & w/o repulsion loss
    \end{tabular}
    \caption{Removal of repulsion loss. From left to right: Input image, our default results, our results without the repulsion loss.
    }\label{fig:style_1}
\end{figure}

\begin{figure}[t]
    \centering
	\setlength{\tabcolsep}{1pt}
    \begin{tabular}{c c c}
        \includegraphics[width=0.32\linewidth]{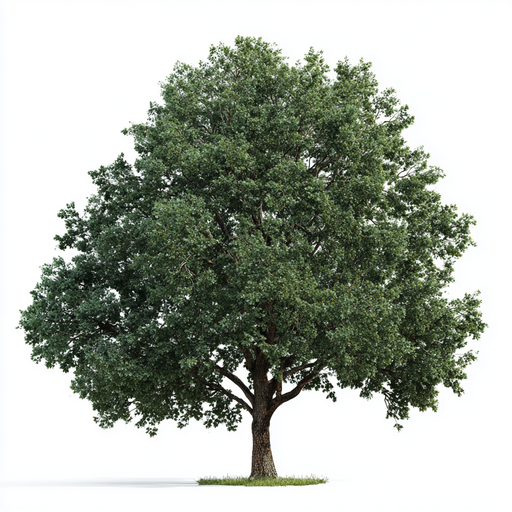} &%
        \adjincludegraphics[width=0.32\linewidth,trim={{0.025\width} {0.0\width} {0.025\width} {0.05\width}},clip]{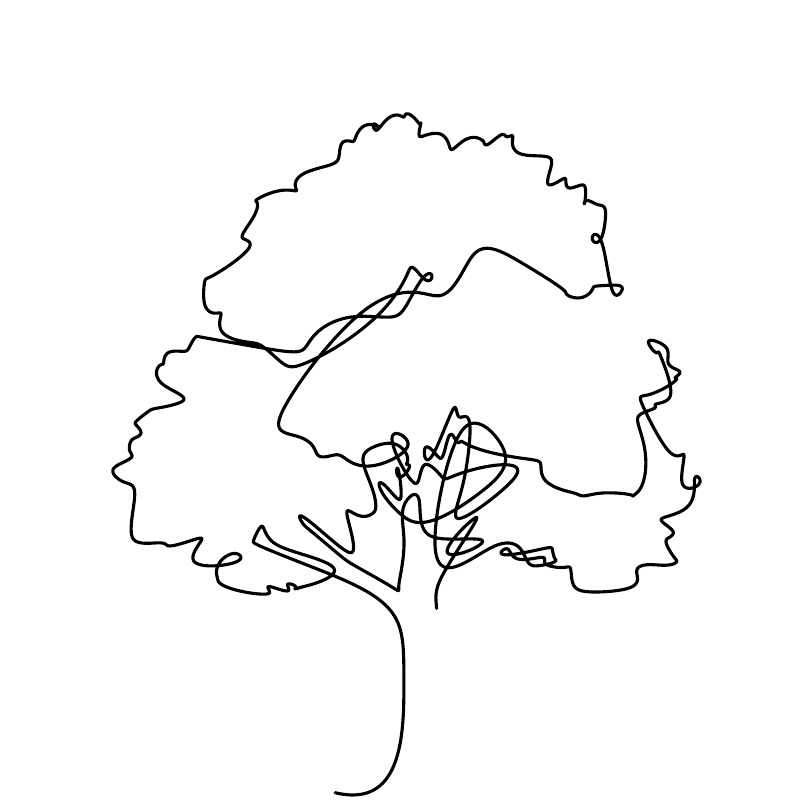} &%
        \adjincludegraphics[width=0.32\linewidth,trim={{0.025\width} {0.0\width} {0.025\width} {0.05\width}},clip]{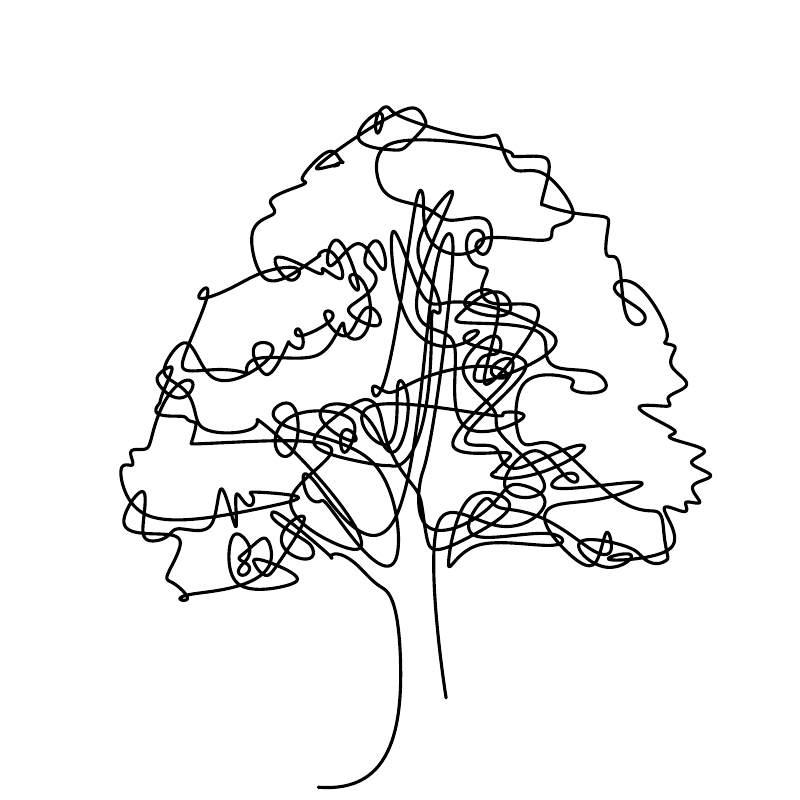}\\[3pt]%
        \adjincludegraphics[width=0.32\linewidth,trim={{0.02\width} {0.0\width} {0.02\width} {0.04\width}},clip]{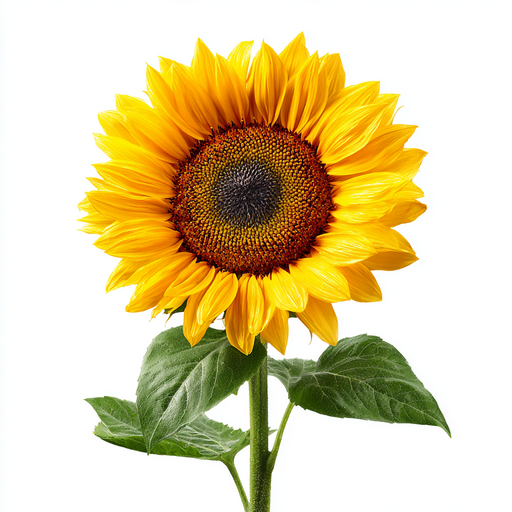} &%
        \adjincludegraphics[width=0.32\linewidth,trim={{0.05\width} {0.02\width} {0.05\width} {0.08\width}},clip]{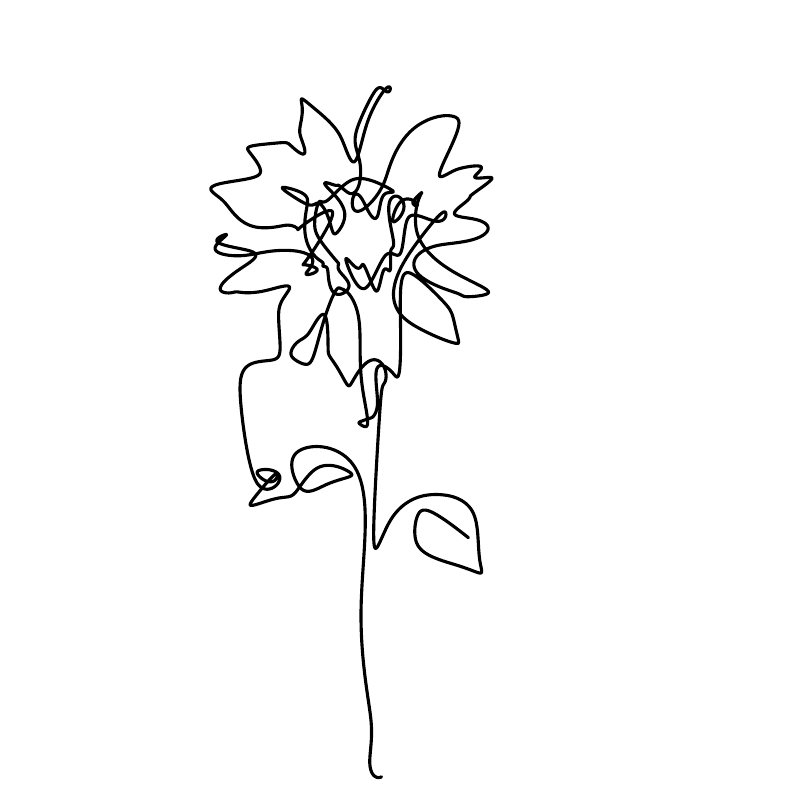} &%
        \adjincludegraphics[width=0.32\linewidth,trim={{0.0\width} {0.0\width} {0.0\width} {0.0\width}},clip]{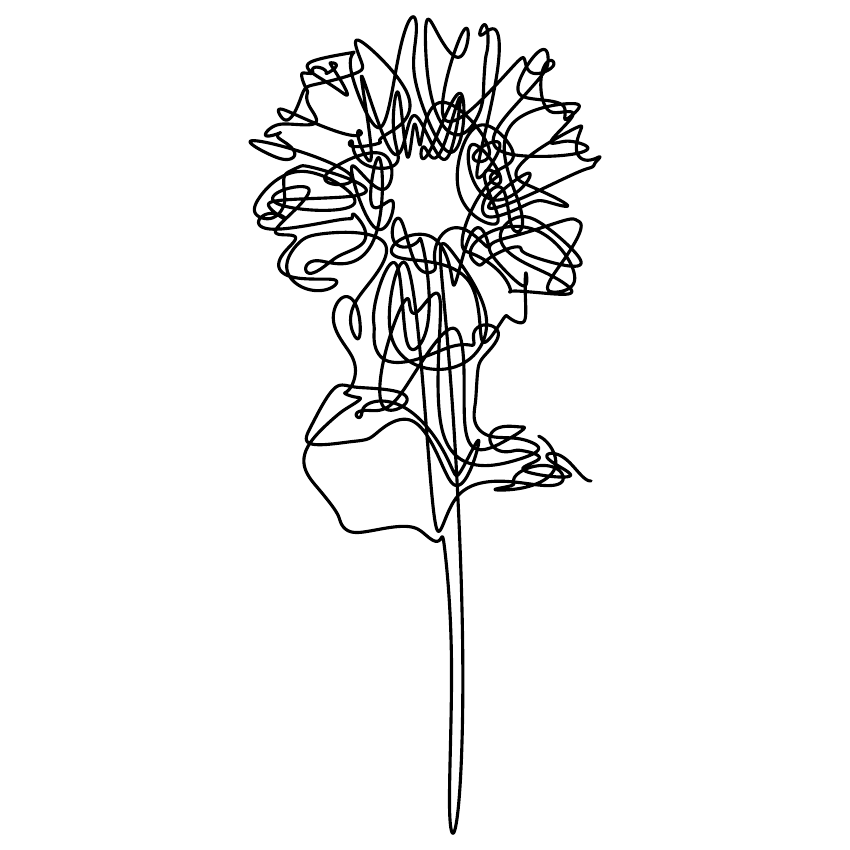}\\%
        input & ours & w/o shortening loss
    \end{tabular}
    \caption{Amplification of repulsion loss effect by removing length shortening loss. From left to right: Input image, our default results, our results without the length shortening loss.
    }\label{fig:style_2}
\end{figure}

\begin{figure}[t]
    \centering
	\small
    \setlength{\fboxsep}{0pt}
    \includegraphics[width=\linewidth]{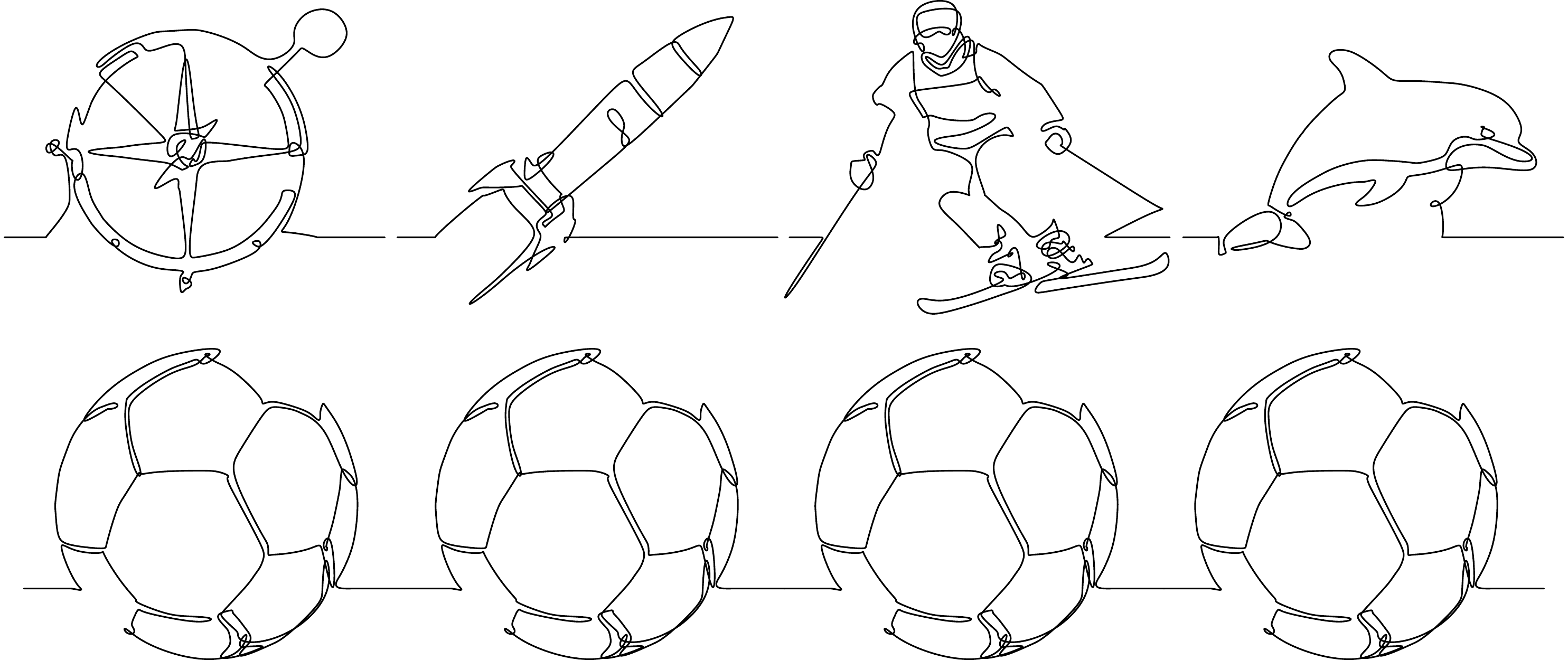}
    \caption{Our method can generate single-line drawings with horizontal start and end (top), allowing to connect them easily or to repeat a motif (bottom).}
    \label{fig:horizontal}
\end{figure}

\subsection{Perceptual study}
\label{subsec:user_study}

To verify the quality of our results further, we conduct a perceptual study. We compare our method with Gemini and Flux, both after being converted to single-line drawings with the method described in \secref{sec:results}. We choose these methods as Gemini has the best quantitative metrics, and Flux is the text-to-image model with the best visual scores. We also compare with ControlSketch \cite{arar.etal2025} results, as ControlSketch is one of the few methods that generate single-line drawings directly. We ask 49 participants to evaluate 15 images randomly sampled from the benchmark dataset regarding their adherence to the single-line drawing aesthetic (``Which drawing looks most like a single-line drawing?''). At the beginning of the questionnaire, users are asked to identify the single-line drawing in each of seven pairs of single-line/non-single-line drawings. This enables more precise interpretation of the results by indicating whether the users understand the task. The results of the study are presented in \tabref{tab:userstudytable}. Respondents consistently ranked our method as looking most like single-line drawings created by artists. Details about the user study can be found in \appref{app:user_study}.

\subsection{Ablation study}
\label{subsec:ablation}

To assess the influence of the different aspects of our method, we perform ablation studies, where we disable various components of our technique. The results of these experiments are presented in \tabref{tab:ablationtable}. Disabling any component of our method leads to worse scores. The length shortening loss affects the visual metrics the most and leads to results that are different from those produced by artists, as the FID becomes quite low. We also present visual results for these ablations in \figref{fig:ablation}. We observe that disabling the single-line drawing LoRA leads to a drawing that resembles the shape, but contains less meaningful details. Disabling the sparsity loss means that more control points are kept after pruning, which results in more redundant details. Disabling the length shortening loss leads to a long, spaghetti-like curve. Our final results contain the right amount of well-placed details, with nice looking curves. 

\textit{Curve representation.\ } We evaluate the impact of the underlying curve representation by comparing the URBS representation against regular B-splines and cubic Bézier splines (a series of connected cubic Bézier segments). While the results quality is an important factor in choosing URBS, our main reason for choosing this representation is that it supports optimizing the number of controls points, whereas for the alternative representations, the number of control points must be fixed a priori. We choose to use 193 control points for the B-splines and 64 cubic Bézier segments for the cubic Bézier splines to match the average number of control points produced by our optimization with URBS curves. As reported in \tabref{tab:ablationtable_curve}, our URBS representation outperforms alternative representations across all quantitative metrics. \figref{fig:curve_representation} provides a visual comparison demonstrating that our representation enables the generation of smoother curves without unwanted details and jitter.

\textit{Initialization.\ } We compare different initialization strategies in \figref{fig:initialization}. Simply initializing from the contour of the segmented subject yields interesting results, but the optimization struggles to add details inside the shape (\figref{fig:initialization}, left). The trefoil knot initialization proposed by Tojo et al. \cite{tojo.etal2024} struggles to capture the overall shape, as the initial geometry is highly decorrelated from the input image (\figref{fig:initialization}, middle left). Using the approach proposed by Arar et al. \cite{arar.etal2025} to sample points and connect them leads to a ``spaghetti'' curve that the optimizer cannot untangle (\figref{fig:initialization}, middle right). Our initialization method (\secref{subsec:init}) covers the entire target subject exclusively, providing sufficient structure while retaining the freedom to move the curve where needed (\figref{fig:initialization}, right).

\subsection{Stylization}
\label{subsec:stylization}

We do not include the repulsion loss in the ablation study because this loss can be seen as a stylization parameter. While we argue that including the repulsion loss produces the best results in general, tuning its weight or even removing it entirely can lead to interesting and aesthetically pleasing outcomes. In some cases, it can produce better results than our default weights.
\figref{fig:style_1} shows two examples where disabling the repulsion loss changes the style of the generated single-line drawing. Without the repulsion loss, we observe more cusps and fewer loops. This gives the petal of the rose and the feather of the flamingo a more appealing and realistic look.
\figref{fig:style_2}, on the contrary shows two examples where the repulsion loss is given more weight by disabling the length shortening loss. This allows the method to create more details and loops. This is a feature that is particularly well suited for organic- and plant-like subjects, such as the tree and the sunflower.

In addition, the user may specify and fix the endpoints of the curve. To accommodate this when solving the traveling salesperson problem, we introduce an auxiliary node connected to the chosen endpoints with zero cost, which guarantees that they serve as the start and end of the path. Intermediate points can likewise be fixed during optimization. This allows the creation of single-line drawings with horizontal lines at the beginning and end of the curve, as presented in the top row of \figref{fig:horizontal}. Such horizontal terminations are common in single-line drawings because they make it easy to link multiple drawings or to create repeating patterns (\figref{fig:horizontal}, bottom row).

\subsection{Varying width}
\label{subsec:varying_width}

\begin{figure}[t]
    \setlength{\fboxsep}{0pt}
    \centering
    \begin{subfigure}{0.32\linewidth}
        \adjincludegraphics[width=\linewidth,trim={{0.225\width} {0.1\height} {0.225\width} {0.1\height}},clip]{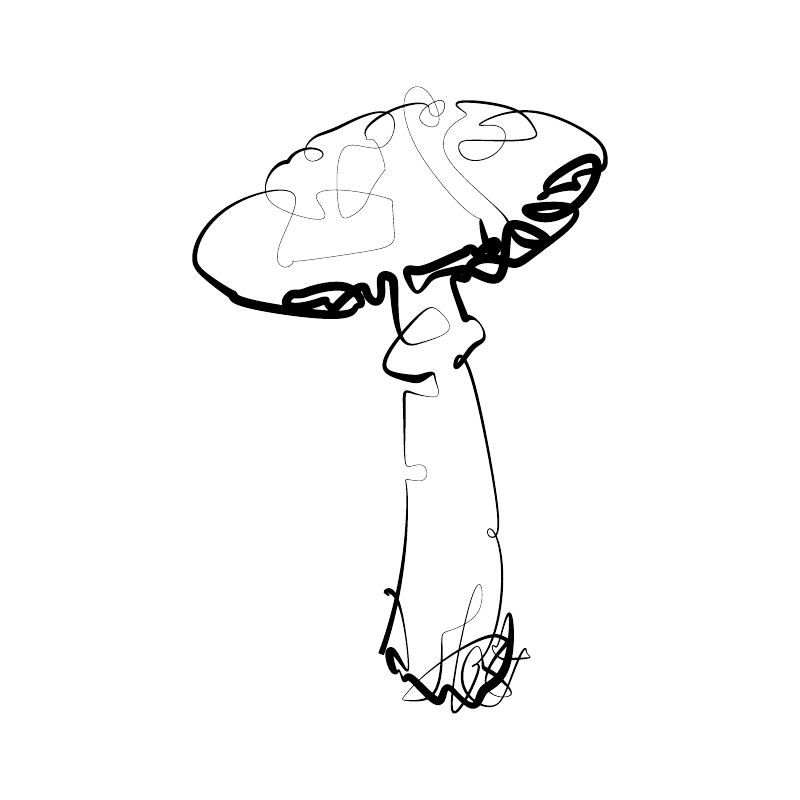}
    \end{subfigure}
    \begin{subfigure}{0.32\linewidth}
        \adjincludegraphics[width=\linewidth,trim={{0.24\width} {0.12\height} {0.24\width} {0.12\height}},clip]{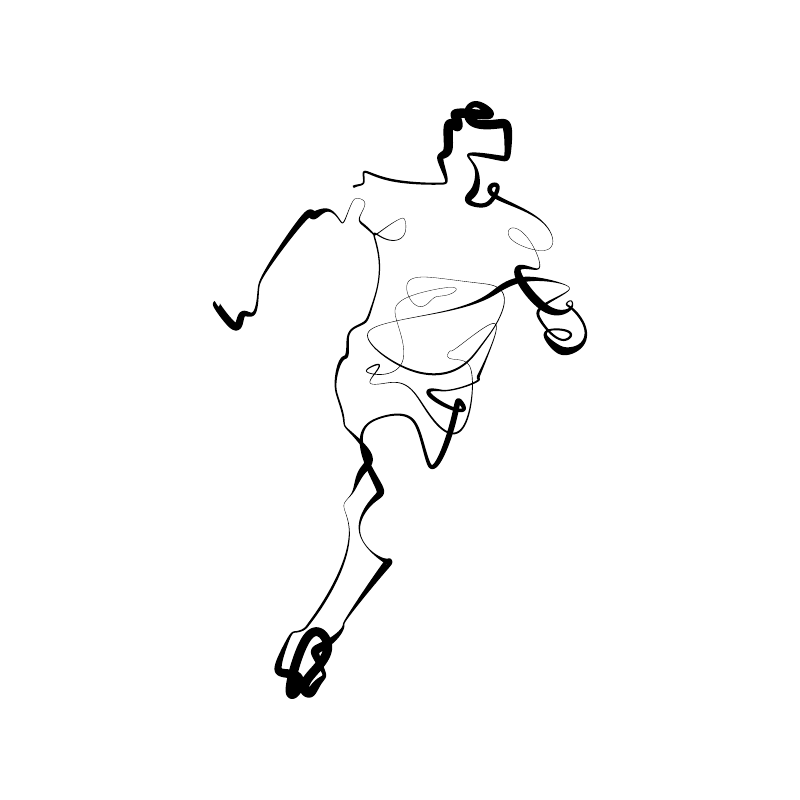}
    \end{subfigure}
    \begin{subfigure}{0.32\linewidth}
        \adjincludegraphics[width=\linewidth,trim={{0.2\width} {0.06\height} {0.2\width} {0.06\height}},clip]{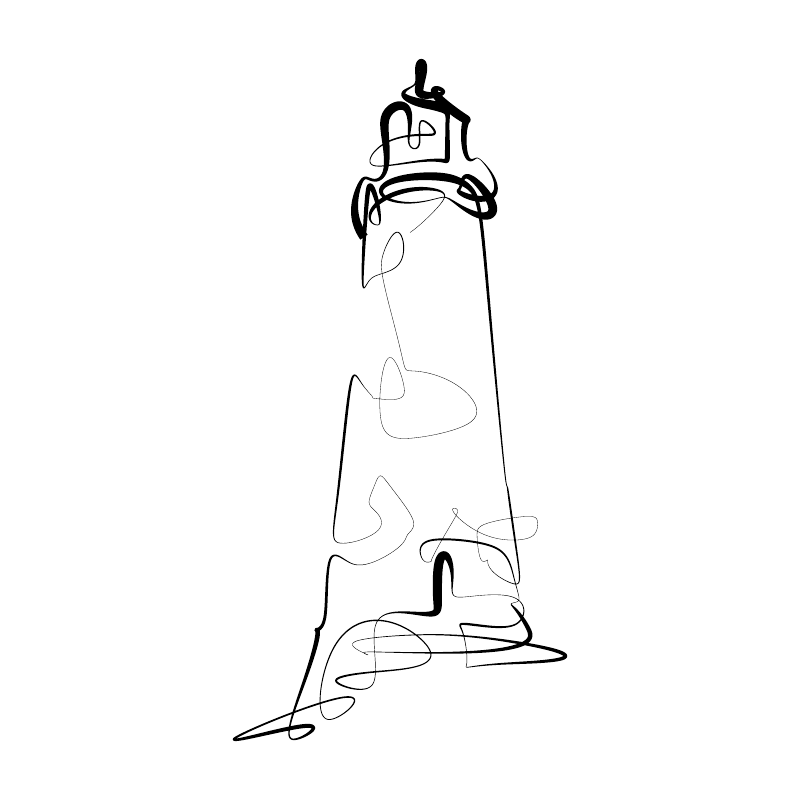}
    \end{subfigure}
    \caption{Our method can be modified to obtain curves with varying width along the line.}
    \label{fig:varying_width}
\end{figure}

Since DiffVG supports the rasterization of curves with varying widths, our method can be extended to support this type of curve stylization. To that end, a width parameter is associated with each control point and can be optimized alongside the control point positions. These width values are initialized to 1 and clipped between 0.1 and 3. The learning rate for these variables is set to 3/512. We present some results of width variation in \figref{fig:varying_width}.
Having a varying line width along the curve allows a distinct aesthetic, similar to that of Berio et al. \cite{berio.etal2025}. While this type of curve may appear smoother and more natural, we also argue that it makes the problem significantly easier. Specifically, the optimizer can almost hide segments of the curve by minimizing their width, rather than accurately localizing them to fit the target.

\subsection{Calligraphy}
\label{subsec:calligraphy}

\begin{figure}[t]
    \centering
    \begin{subfigure}{0.45\linewidth}
        \adjincludegraphics[width=\linewidth,trim={{0.0\width} {0.0\height} {0.0\width} {0.0\height}},clip]{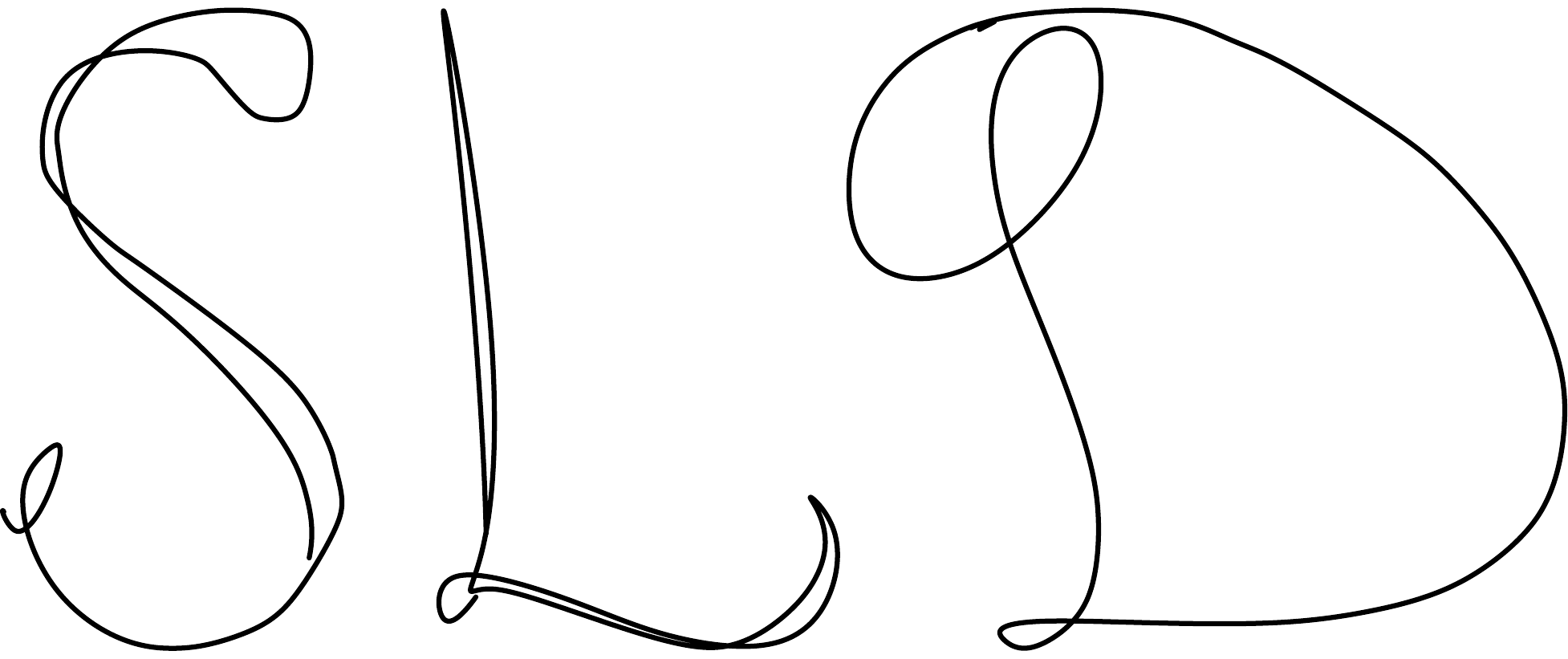}
    \end{subfigure}
    \hspace{0.08\linewidth}
    \begin{subfigure}{0.45\linewidth}
        \adjincludegraphics[width=\linewidth,trim={{0.0\width} {0.0\height} {0.0\width} {0.0\height}},clip]{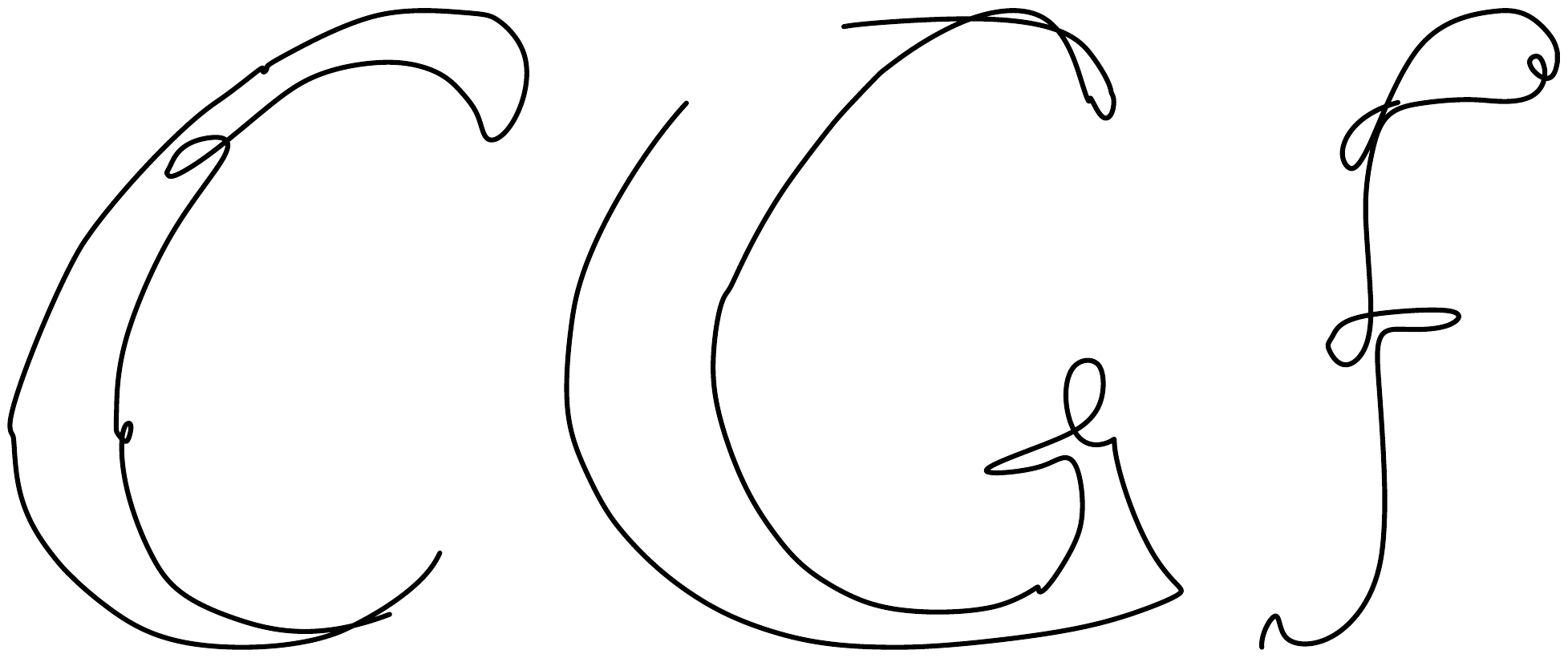}
    \end{subfigure}
    \caption{Our method can generate handwritten style letters with a single stroke.}
    \label{fig:calligraphy}
\end{figure}

While not originally designed for calligraphy, our method successfully produces handwritten-style letters. To that end, the parameters of our system need to be tuned slightly. The number of initial control points is reduced to 97 or 193. In case only 97 initial control points are used, the sparsity loss weight $\lambda_\text{sparse}$ is also reduced to 1000. The input image is simply a rasterized version of the letter in an arbitrary font, and the input prompt is \textit{``a cursive style handwritten letter [...]''}, where [...] is replaced with the letter to write. More importantly, we replace the depth ControlNet with a Canny edge ControlNet, since the depth is not meaningful for a letter. \figref{fig:calligraphy} shows 6 letters in this setting and an embroidery of the letter \textit{A} can be found in \figref{fig:embroidery_fabrication}. 

\begin{figure}[t]
    \centering
	\small
	\setlength{\tabcolsep}{2pt}
    \begin{tabular}{cc}
    \adjincludegraphics[width=0.44\linewidth,trim={{0.1\width} {.0\height} {0.1\width} {.0\height}},clip]{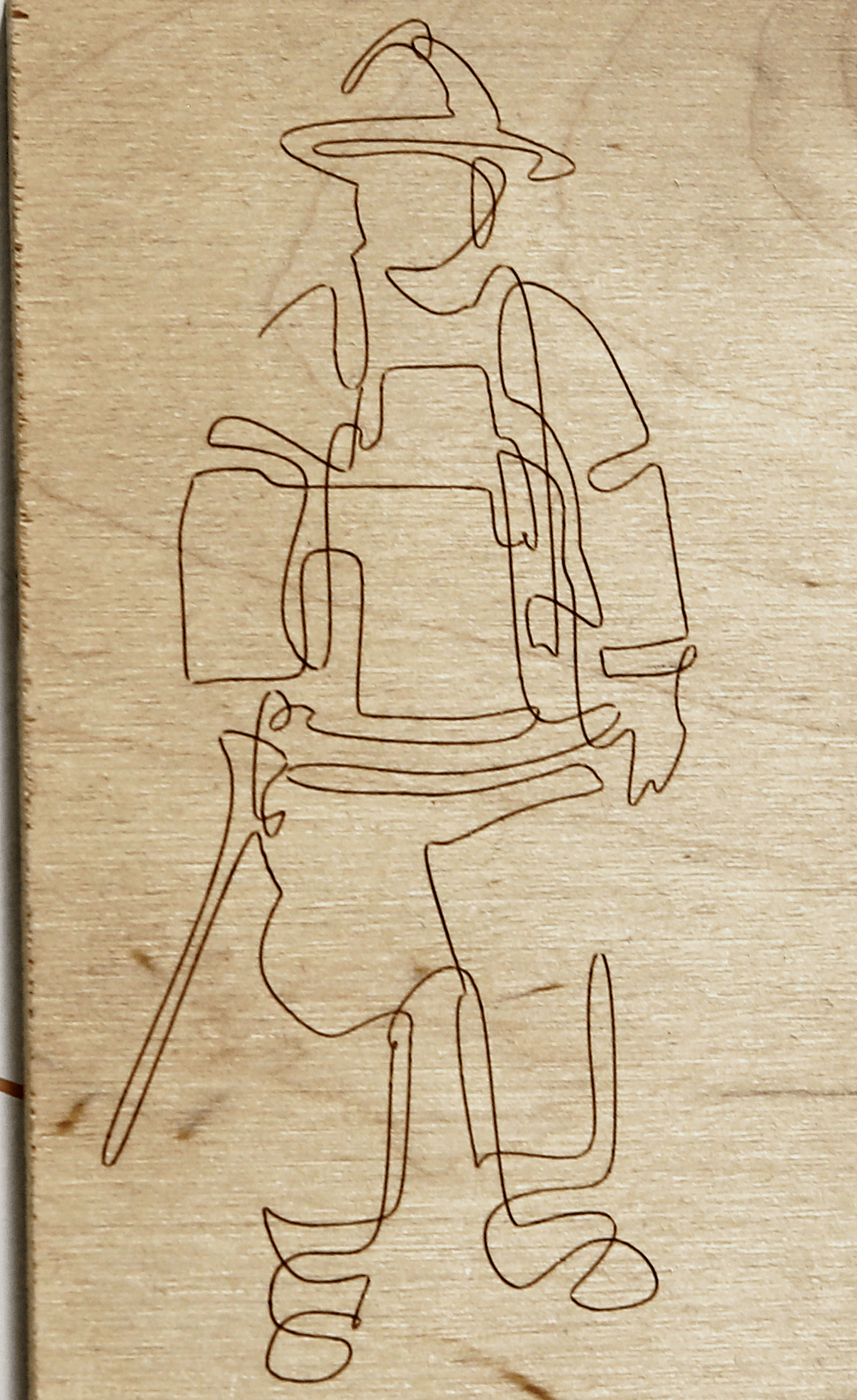}
     & \adjincludegraphics[width=0.44\linewidth,trim={{0.06\width} {.0\height} {0.14\width} {.0\height}},clip]{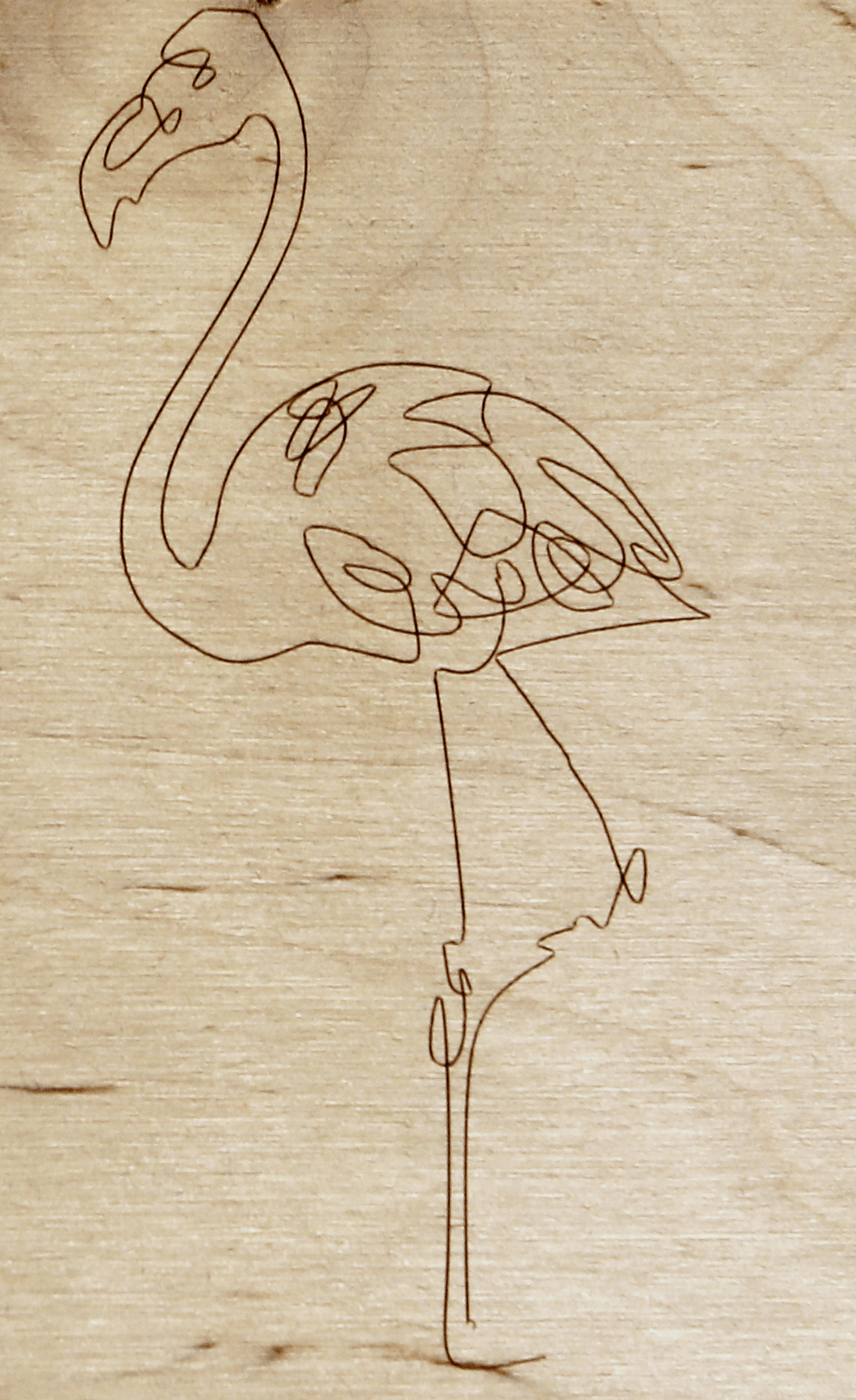} \\
     \includegraphics[width=0.44\linewidth]{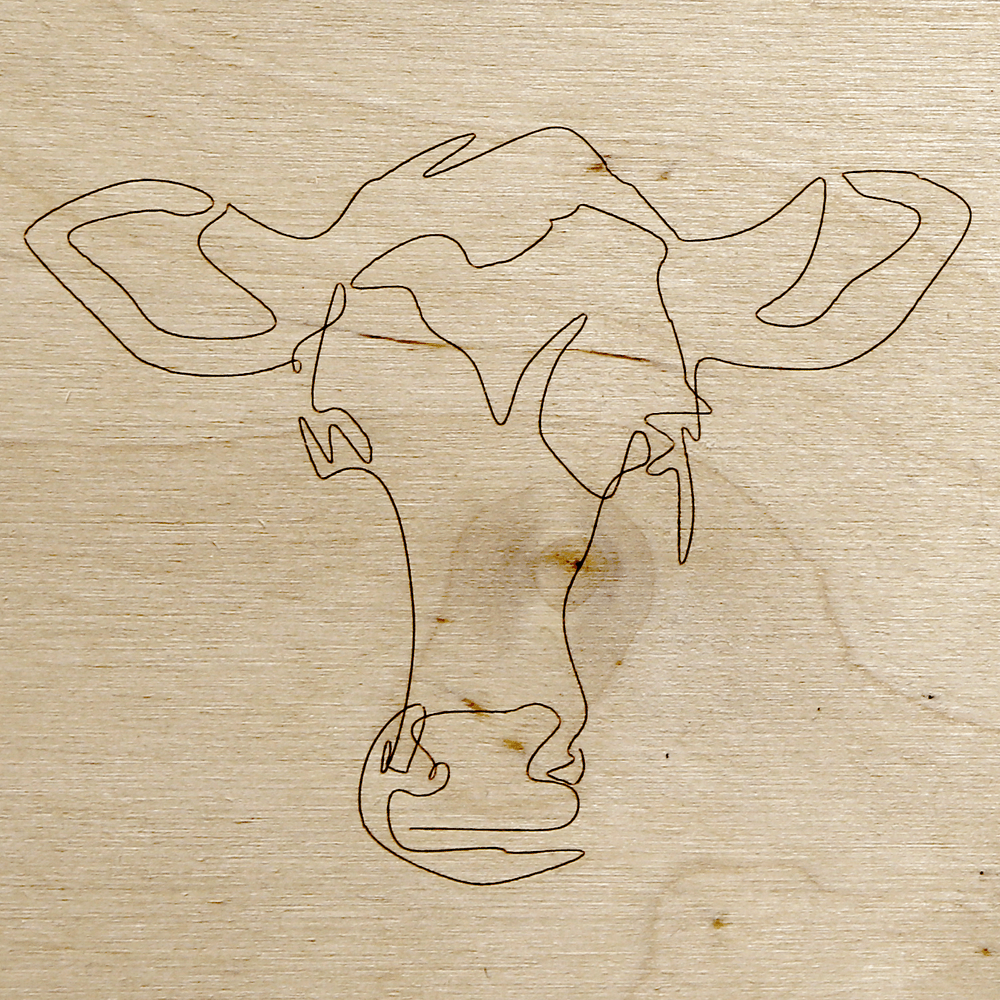} 
     & \includegraphics[width=0.44\linewidth]{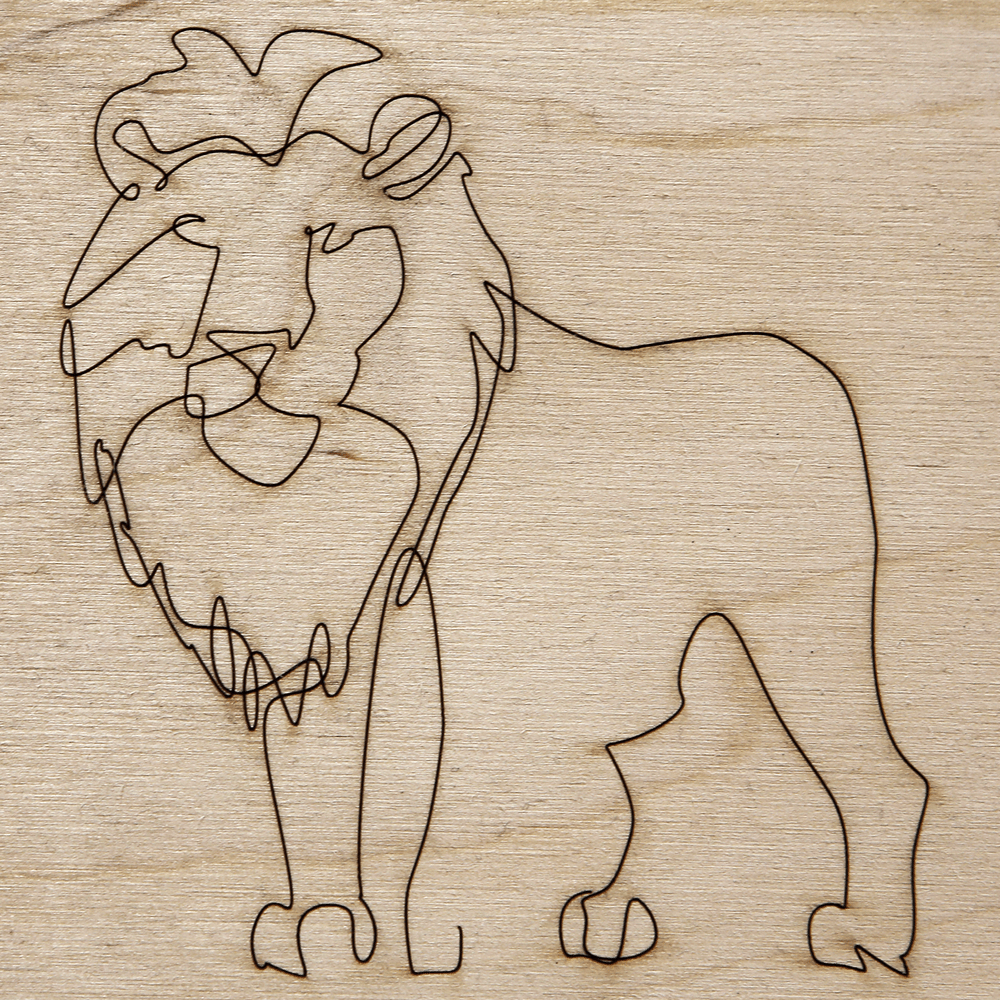}
    \end{tabular}
    \caption{Wood laser engraving results obtained from our generated single-line drawings. The continuous curves ensure fast fabrication while faithfully capturing the target shapes.}
    \label{fig:wood_laser_engraving}
\end{figure}

\subsection{Fabrication}
\label{subsec:fabrication}

\begin{figure}[t]
    \centering
	\small
	\setlength{\tabcolsep}{1pt}
    \begin{tabular}{cccc}        
        \begin{subfigure}{0.495\linewidth}
            \centering
            \includegraphics[width=\textwidth]{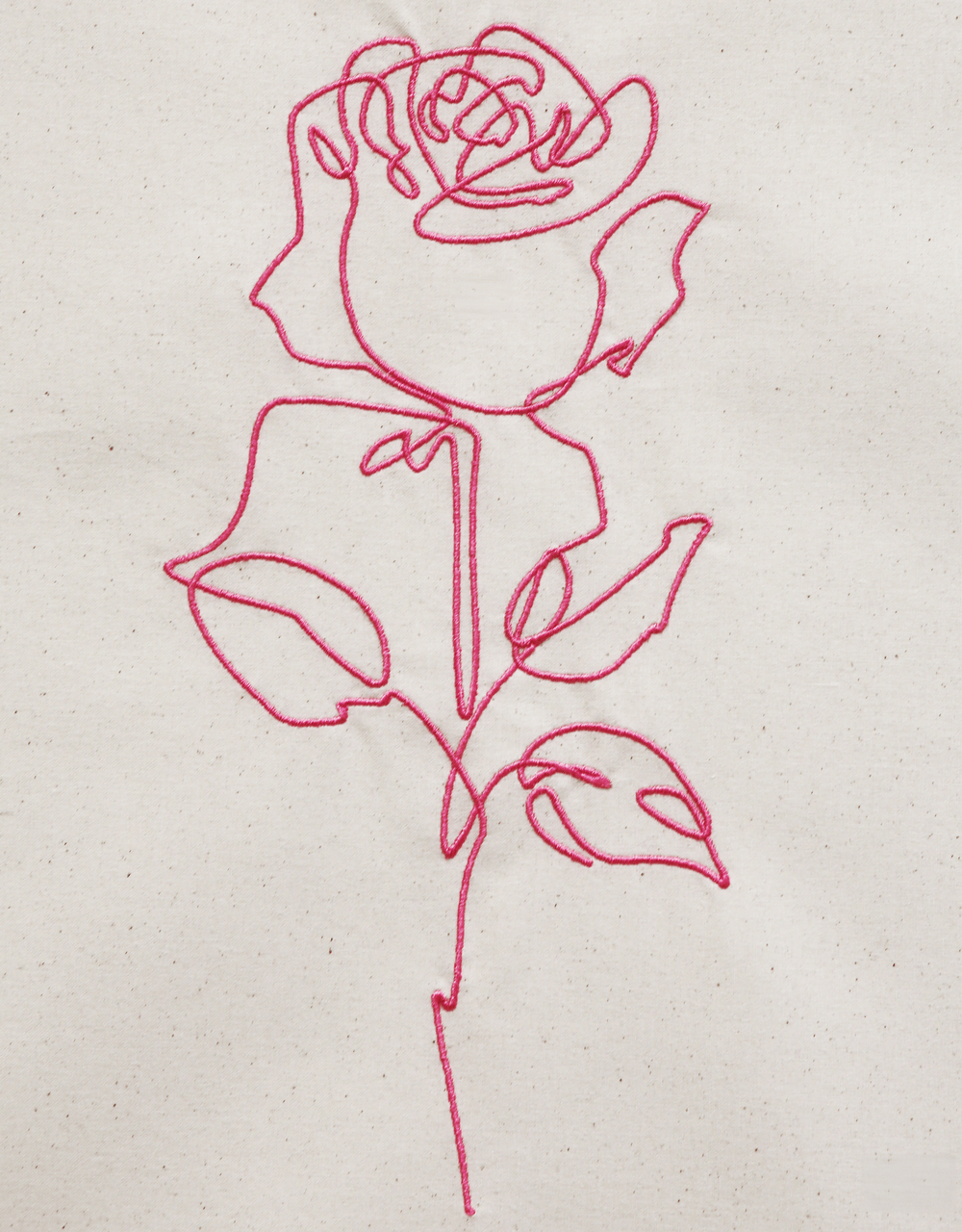}%
            \vspace{-5pt}%
            \caption*{Satin stitch}%
        \end{subfigure}& 
        \begin{subfigure}{0.495\linewidth}
            \centering
            \includegraphics[width=\textwidth]{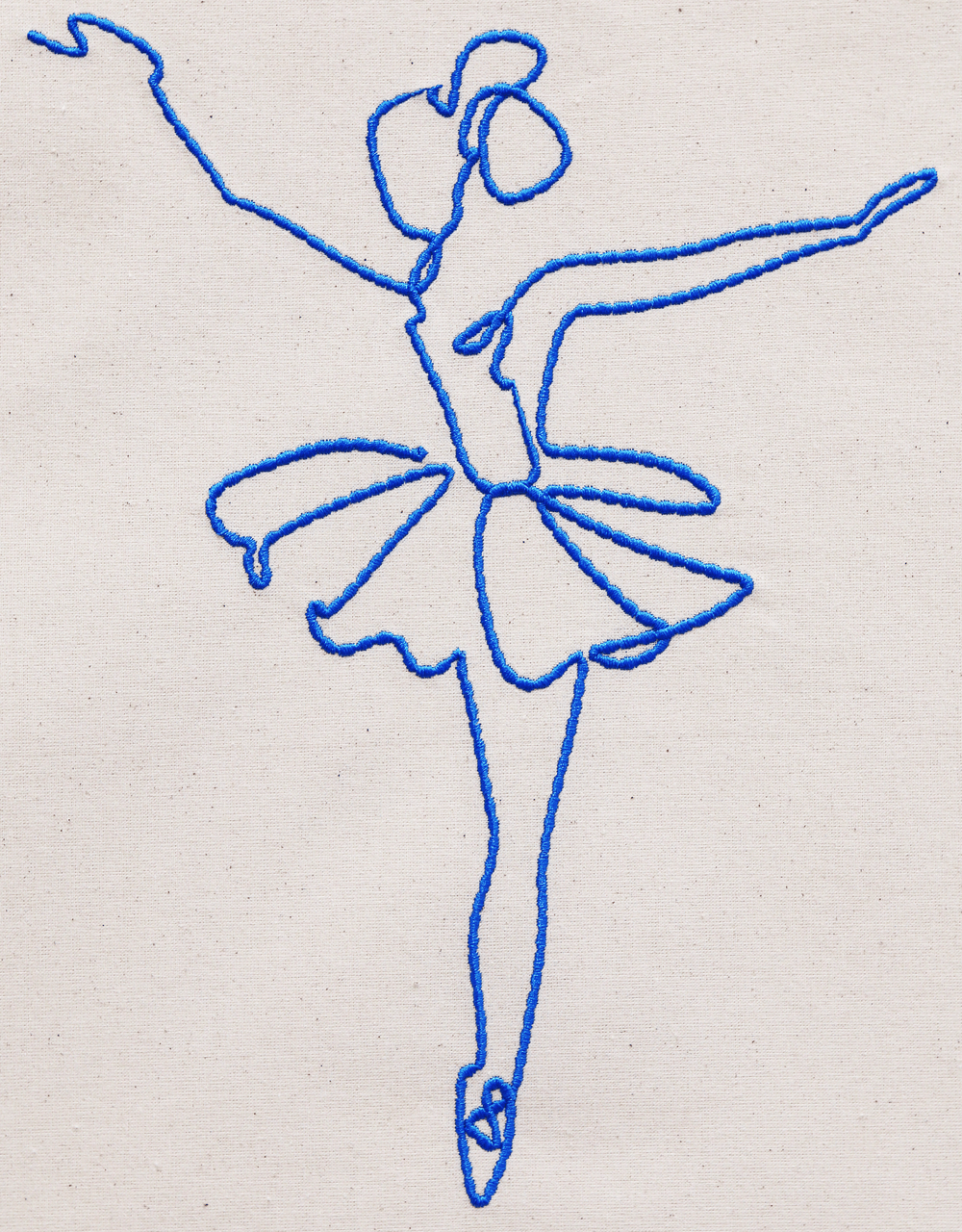}%
            \vspace{-5pt}%
            \caption*{Pearl satin stitch}%
        \end{subfigure}\\
        \begin{subfigure}{0.495\linewidth}
            \centering
            \includegraphics[width=\textwidth]{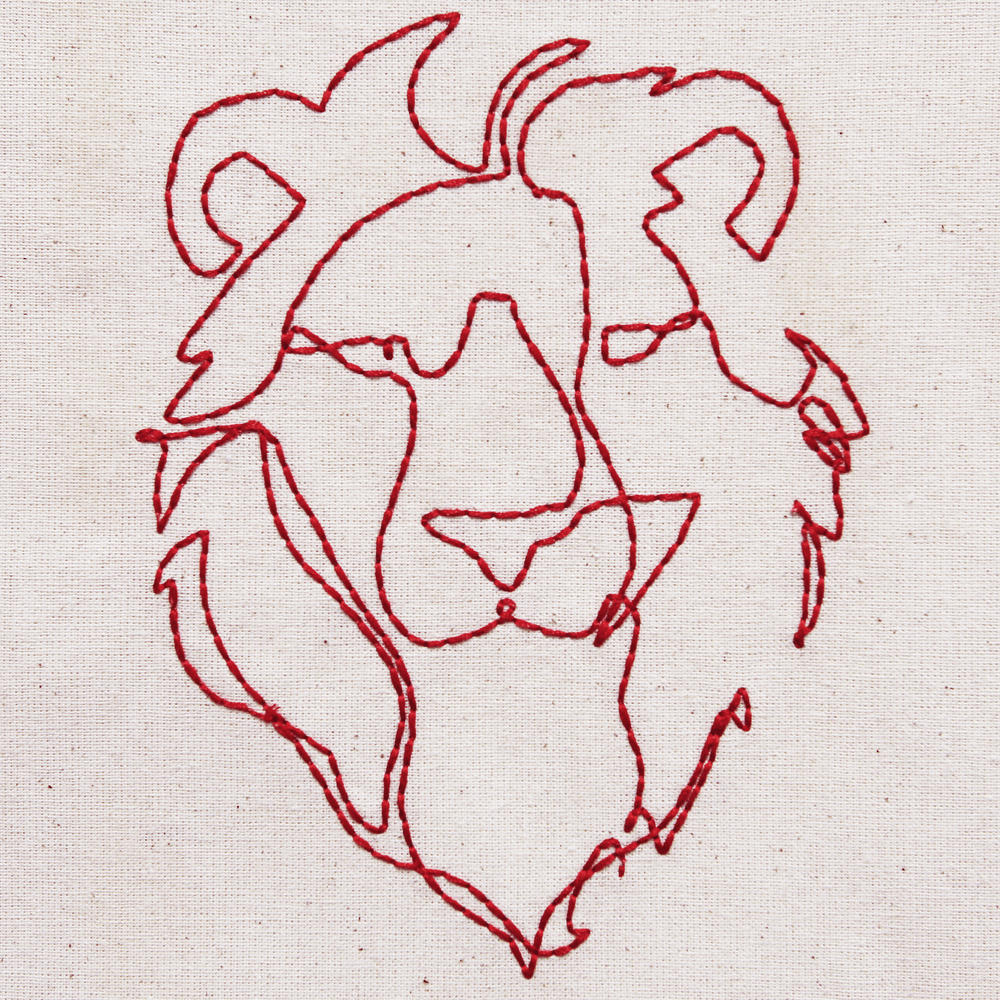}%
            \vspace{-5pt}%
            \caption*{Bean stitch}%
        \end{subfigure}& 
        \begin{subfigure}{0.495\linewidth}
            \centering
            \includegraphics[width=\textwidth]{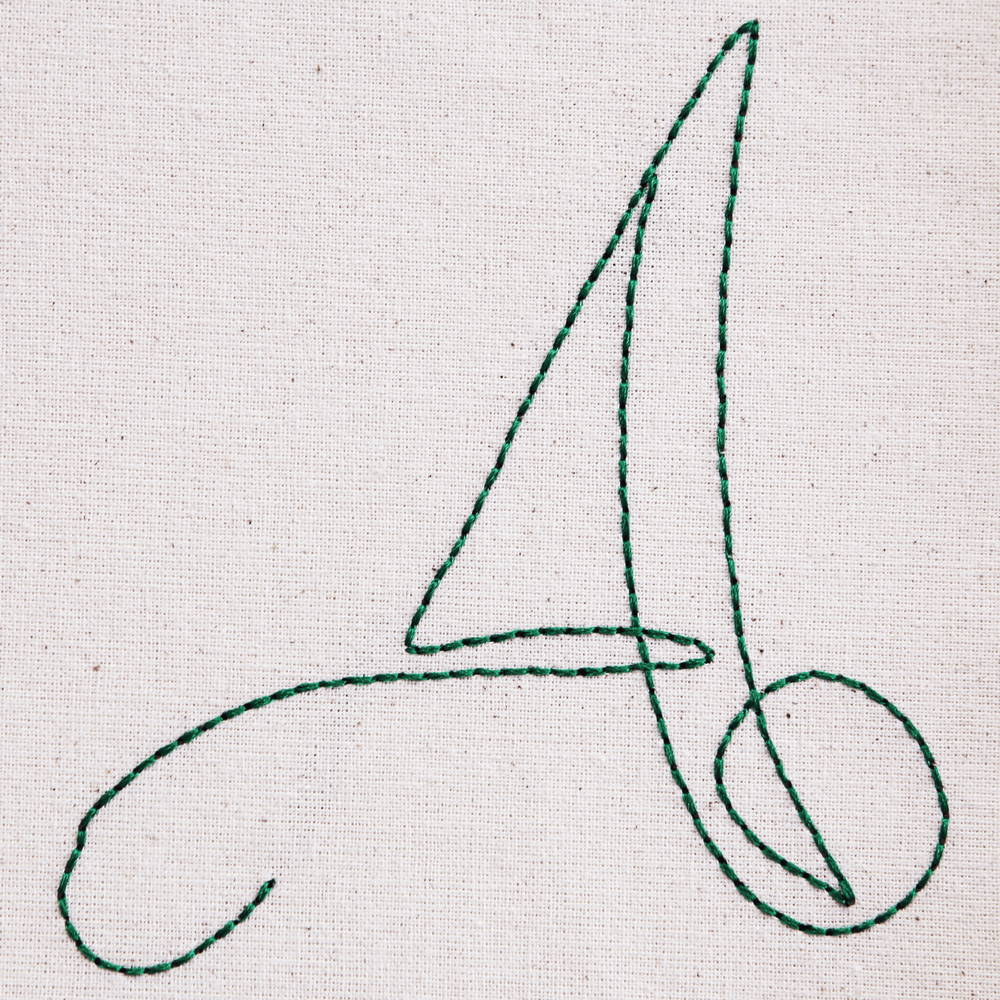}%
            \vspace{-5pt}%
            \caption*{Bean stitch}%
        \end{subfigure}
    \end{tabular}
    \caption{Embroidery fabrication examples produced from our generated single-line drawings. The designs demonstrate how varying stitch styles can translate vectorized line art into textile patterns.}
    \label{fig:embroidery_fabrication}
\end{figure}

Beyond their distinctive aesthetic appeal, the strict constraints of single-line drawings also make them especially well-suited for fabrication.
In embroidery, a continuous line representation naturally corresponds to a single uninterrupted stitch path, which avoids the need for thread cutting or repositioning. As shown in \figref{fig:teaser} and \figref{fig:embroidery_fabrication}, our method produces vectorized curves that can be directly interpreted by digital embroidery machines. Furthermore, the flexibility of our representation allows different stitching styles (e.g., satin or bean stitches) to be applied along the same path, resulting in a variety of visual textures while preserving the one-line constraint.
The same benefit applies to wood laser engraving, where maintaining a continuous trajectory prevents excessive acceleration and repositioning of the laser head. In this setting, our generated curves can be directly used as engraving paths, leading to smoother and faster fabrication than raster-based alternatives. For example, engraving the result presented at the bottom right of \figref{fig:wood_laser_engraving} takes only 49 seconds when using our vector input, compared to more than 8 minutes with a raster image. A video demonstrating both fabrication processes is included in the supplementary material. Examples of such engravings are presented in \figref{fig:wood_laser_engraving} to demonstrate how our approach yields clean, fabrication-ready vector paths suitable for precise physical rendering.

\begin{figure}[t]
    \centering
	\small
	\setlength{\tabcolsep}{1pt}
    \setlength{\fboxsep}{0pt}
    \begin{tabular}{ccc}
    \adjincludegraphics[width=0.32\linewidth,trim={{0.0625\width} {0.0\height} {0.0625\width} {0.0\height}},clip]{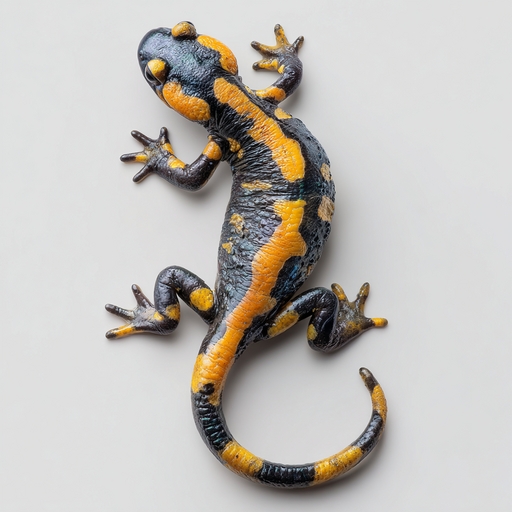}
    & \adjincludegraphics[width=0.32\linewidth,trim={{0.22\width} {0.04\height} {0.0\width} {0.09\height}},clip]{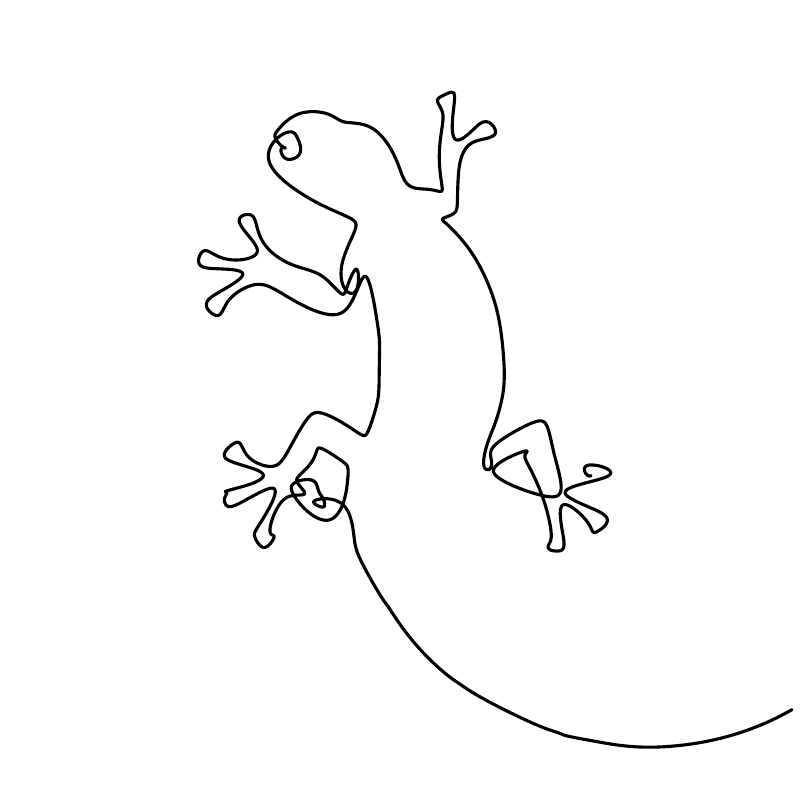}
    & \adjincludegraphics[width=0.32\linewidth,trim={{0.0\width} {0.0\height} {0.0\width} {0.\height}},clip]{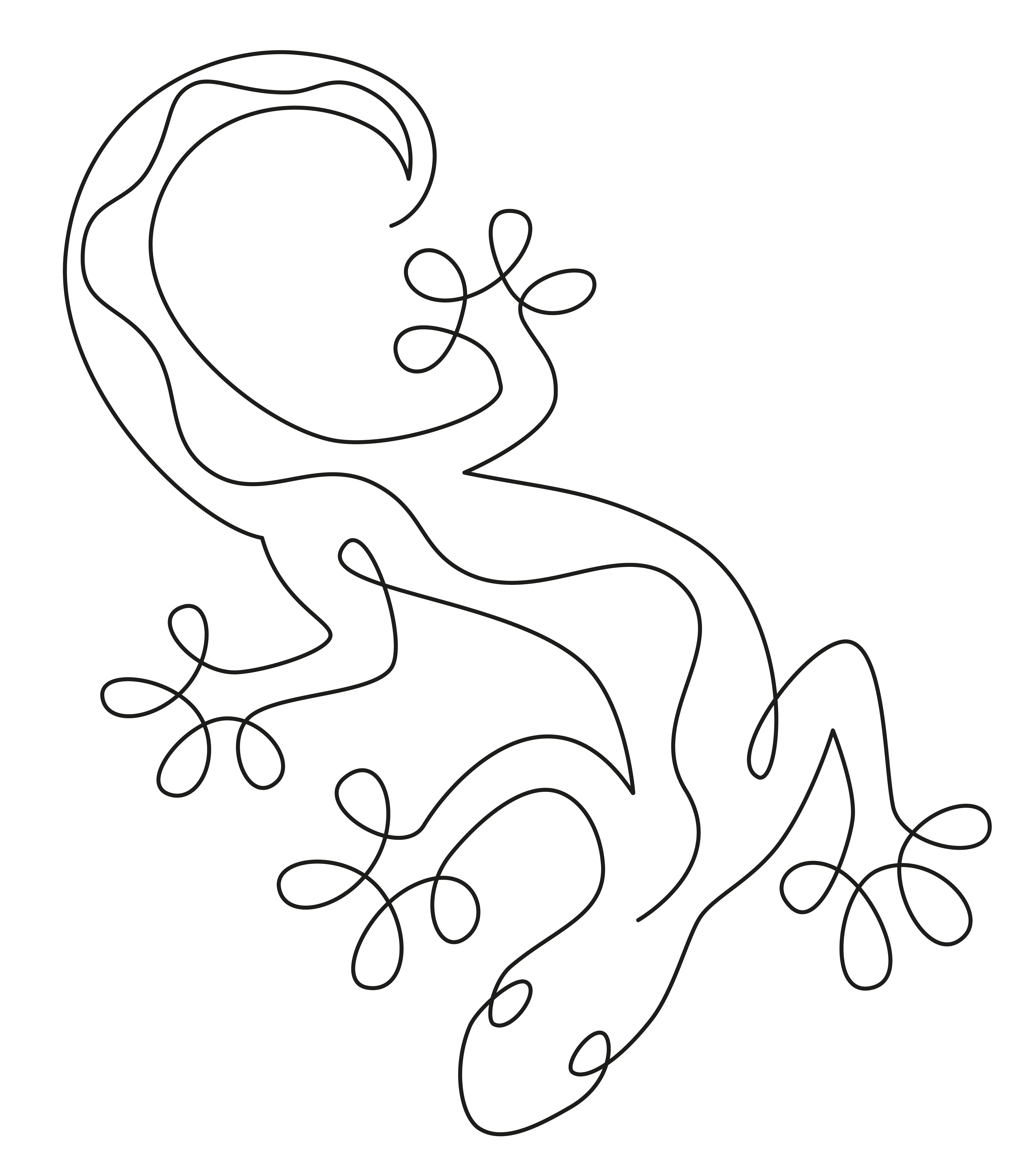}
    \end{tabular}
    \caption{For niche subjects, such as this salamander, our method inherits the flaws of the diffusion model and struggles to create a good-looking result with meaningful inside details. For reference, a single drawing created by an artist is shown on the right.}
    \label{fig:limitation}
\end{figure}

\subsection{Limitations and future work}
\label{subsec:limitation}

Our method is able to generate single-line drawings that follow the aesthetic of single-line drawings created by artists and represent the input subject well in general. However, it does inherit the flaws of the diffusion model that is used to compute the SDS loss and might struggle to produce good results for more niche subjects. For instance, the result for the salamander presented in \figref{fig:limitation} lacks some details. Our method can benefit from the continuous development of more powerful diffusion models, which could improve the quality of our results without requiring any change to the method.

Our method takes about 15 minutes to run. This is relatively long for a user interaction, but similar to most score distillation sampling methods. This issue can be mitigated by using a smaller diffusion model or running fewer iterations. As shown in the teaser, 1000 steps already produce a good-looking result and would reduce the running time by a factor of 4. We use 4000 steps to ensure the best results. In addition, building a differentiable rasterizer specifically tailored to URBS curves, similar to Tojo et al.'s implementation for B-splines \cite{tojo.etal2024}, should reduce the running time. Alternatively, a method like SwiftSketch~\cite{arar.etal2025} can be trained on the output of our model for faster inference.

Finally, our method only produces static images. Generation of animated sketches has been explored recently \cite{gal.etal2024a}, and we believe that application of similar methods to single-line drawings could produce very interesting results, as demonstrated by the Italian series ``La Linea'' \cite{cavandoli1971lalinea}.

\section{Conclusion}
\label{sec:conclusion}

In this paper, we presented a method for generating continuous single-line drawings from an input prompt and/or a reference image. Our method uses uniform rational B-splines to represent the curve, which allows our approach to control the amount of detail required to represent a subject. We use the score distillation sampling loss to encourage resemblance to the target concept and several regularizers, including a repulsive loss, to control the style of the drawings. Changing the influence of the repulsive loss allows the user to control the output style of the drawing. We showed qualitatively and quantitatively that our method produces results that more closely resemble the input concept and are more aesthetically pleasing than existing methods. Because our method outputs a vector representation of the drawing, it enables further fabrication tasks, such as embroidery or laser engraving.
We believe this work represents a meaningful step in the automatic generation of aesthetically pleasing line drawings. To foster further research in this field, we made our code publicly available at https://github.com/tanguymagne/SLDgen.

\section*{Acknowledgements}

We are grateful to the anonymous reviewers for their constructive suggestions. We also thank Danielle Luterbacher for her assistance with the embroidery process.
This work was supported in part by the European Research Council (ERC) under the European Union’s Horizon 2020 Research and Innovation Programme (ERC Consolidator Grant, agreement No. 101003104, MYCLOTH).
Open access publishing facilitated by ETH Zurich, as part of the Wiley- ETH Zurich agreement via the Consortium Of Swiss Academic Libraries.

\section*{Ethics Statement}
Our method builds upon publicly available pretrained diffusion models and vision-language models, which are widely used in the research community. These models may inherit biases from their training data. Our perceptual study involved human participants who evaluated and compared generated drawings. This study was conducted with appropriate informed consent, and no personal identifying information was collected or retained beyond the study period.
 
\begin{figure*}[t]
    \newcommand{\cwidth}{0.162\linewidth}
    \newcommand{\cspace}{\hspace{0.005\linewidth}}
    \newcommand{\lrspace}{1pt}
    \setlength{\fboxsep}{-0.1pt}
    \setlength{\fboxrule}{0.1pt}
    \centering
    \setlength{\tabcolsep}{1pt}
    \begin{tabular}{cccccc}
    \begin{subfigure}[t]{\cwidth}
        \begin{subfigure}{\textwidth}
            \begin{overpic}[width=\textwidth, height=\textwidth]{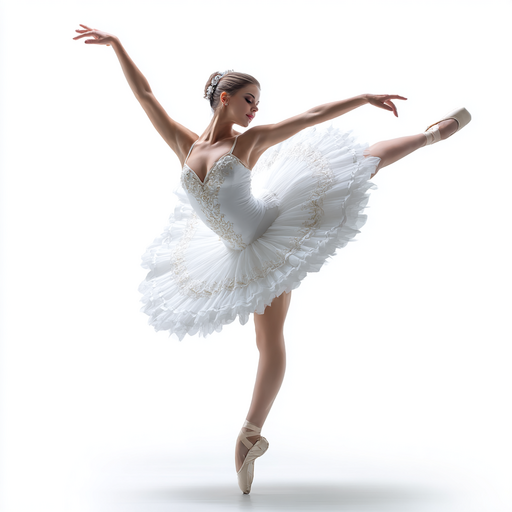}%
             \put (0,6) {\setlength{\fboxsep}{1pt}\transparent{0.5}\colorbox[RGB]{255,255,255}{\transparent{1.0}\parbox{\dimexpr\linewidth-2\fboxsep}{\scriptsize{\textit{Single line drawing of a ballerina, continuous black line ...}}}}}
            \end{overpic}
        \end{subfigure}
        \\[\lrspace]
        \begin{subfigure}{\textwidth}
            \begin{overpic}[width=\textwidth, height=\textwidth]{figures/results/input_image/firefighter.png}%
             \put (0,6) {\setlength{\fboxsep}{1pt}\transparent{0.5}\colorbox[RGB]{255,255,255}{\transparent{1.0}\parbox{\dimexpr\linewidth-2\fboxsep}{\scriptsize{\textit{Single line drawing of a book, continuous black line ...}}}}}
            \end{overpic}
        \end{subfigure}
        \\[\lrspace]
        \begin{subfigure}{\textwidth}
            \begin{overpic}[width=\textwidth, height=\textwidth]{figures/results/input_image/tree.png}%
             \put (0,6) {\setlength{\fboxsep}{1pt}\transparent{0.5}\colorbox[RGB]{255,255,255}{\transparent{1.0}\parbox{\dimexpr\linewidth-2\fboxsep}{\scriptsize{\textit{Single line drawing of a tree, continuous black line ...}}}}}
            \end{overpic}
        \end{subfigure}
        \\[\lrspace]
        \begin{subfigure}{\textwidth}
            \begin{overpic}[width=\textwidth, height=\textwidth]{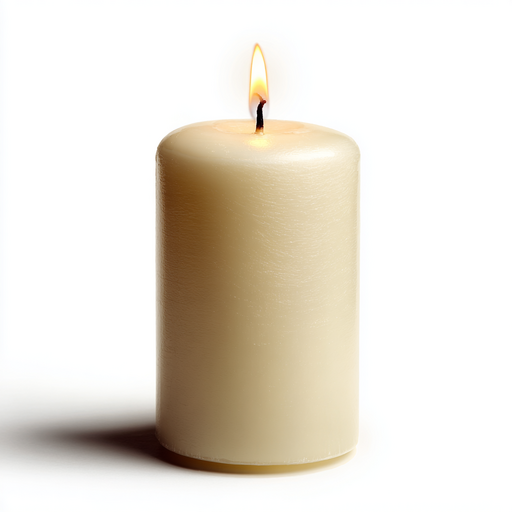}%
             \put (0,6) {\setlength{\fboxsep}{1pt}\transparent{0.5}\colorbox[RGB]{255,255,255}{\transparent{1.0}\parbox{\dimexpr\linewidth-2\fboxsep}{\scriptsize{\textit{Single line drawing of a candle, continuous black line ...}}}}}
            \end{overpic}
        \end{subfigure}
        \\[\lrspace]%
        \begin{subfigure}{\textwidth}%
            \begin{overpic}[width=\textwidth, height=\textwidth]{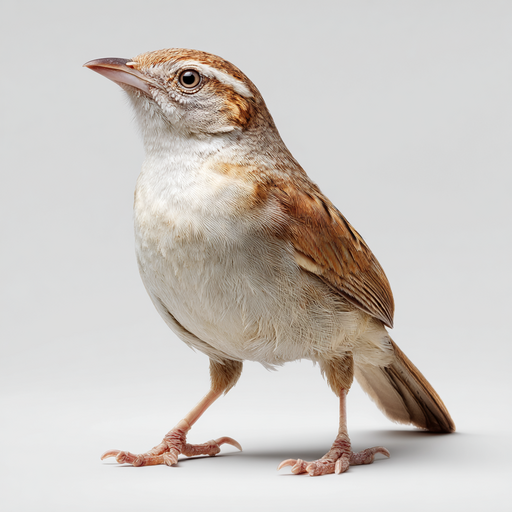}%
            \put (0,6) {\setlength{\fboxsep}{1pt}\transparent{0.5}\colorbox[RGB]{255,255,255}{\transparent{1.0}\parbox{\dimexpr\linewidth-2\fboxsep}{\noindent\scriptsize{\textit{Single line drawing of a bird, continuous black line ...}}}}}%
            \end{overpic}%
        \end{subfigure}%
        \\[\lrspace]%
        \begin{subfigure}{\textwidth}
            \begin{overpic}[width=\textwidth, height=\textwidth]{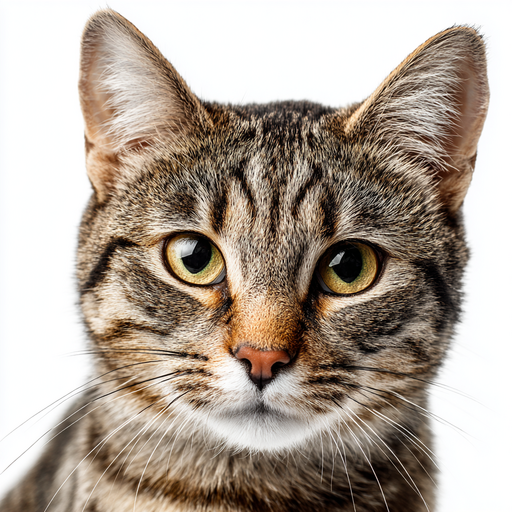}%
             \put (0,6) {\setlength{\fboxsep}{1pt}\transparent{0.5}\colorbox[RGB]{255,255,255}{\transparent{1.0}\parbox{\dimexpr\linewidth-2\fboxsep}{\scriptsize{\textit{Single line drawing of a cat head, continuous black line ...}}}}}
            \end{overpic}
        \end{subfigure}
        \\[\lrspace]
        \begin{subfigure}{\textwidth}
            \begin{overpic}[width=\textwidth, height=\textwidth]{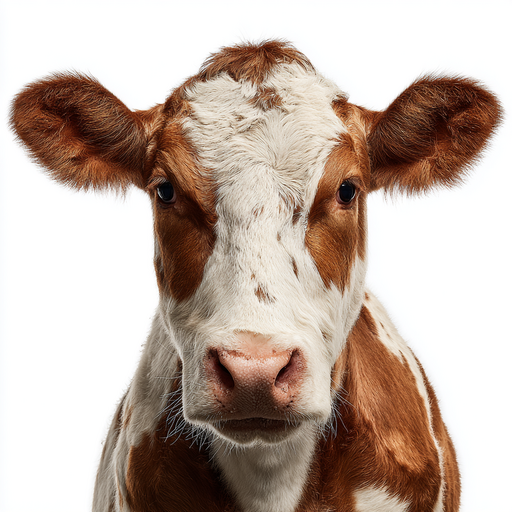}%
             \put (0,6) {\setlength{\fboxsep}{1pt}\transparent{0.5}\colorbox[RGB]{255,255,255}{\transparent{1.0}\parbox{\dimexpr\linewidth-2\fboxsep}{\scriptsize{\textit{Single line drawing of a cow head, continuous black line ...}}}}}
            \end{overpic}
        \end{subfigure}
        \vspace{-15pt}
        \caption*{\scriptsize Input image and prompt}
    \end{subfigure}%
    & \begin{subfigure}[t]{\cwidth}
        \begin{subfigure}{\textwidth}
            \fbox{\includegraphics[width=\textwidth, height=\textwidth]{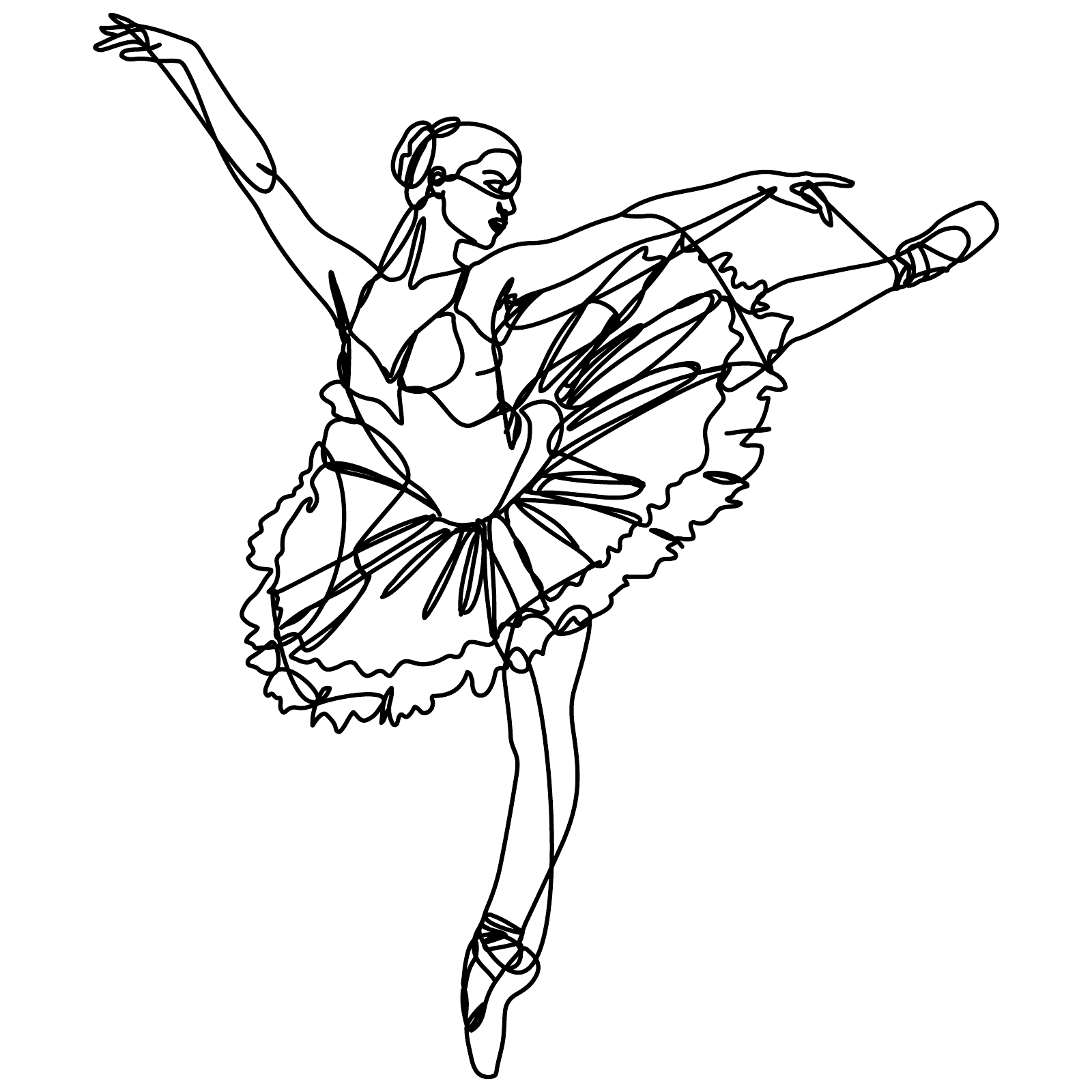}}%
        \end{subfigure}
        \\[\lrspace]
        \begin{subfigure}{\textwidth}
            \fbox{\includegraphics[width=\textwidth, height=\textwidth]{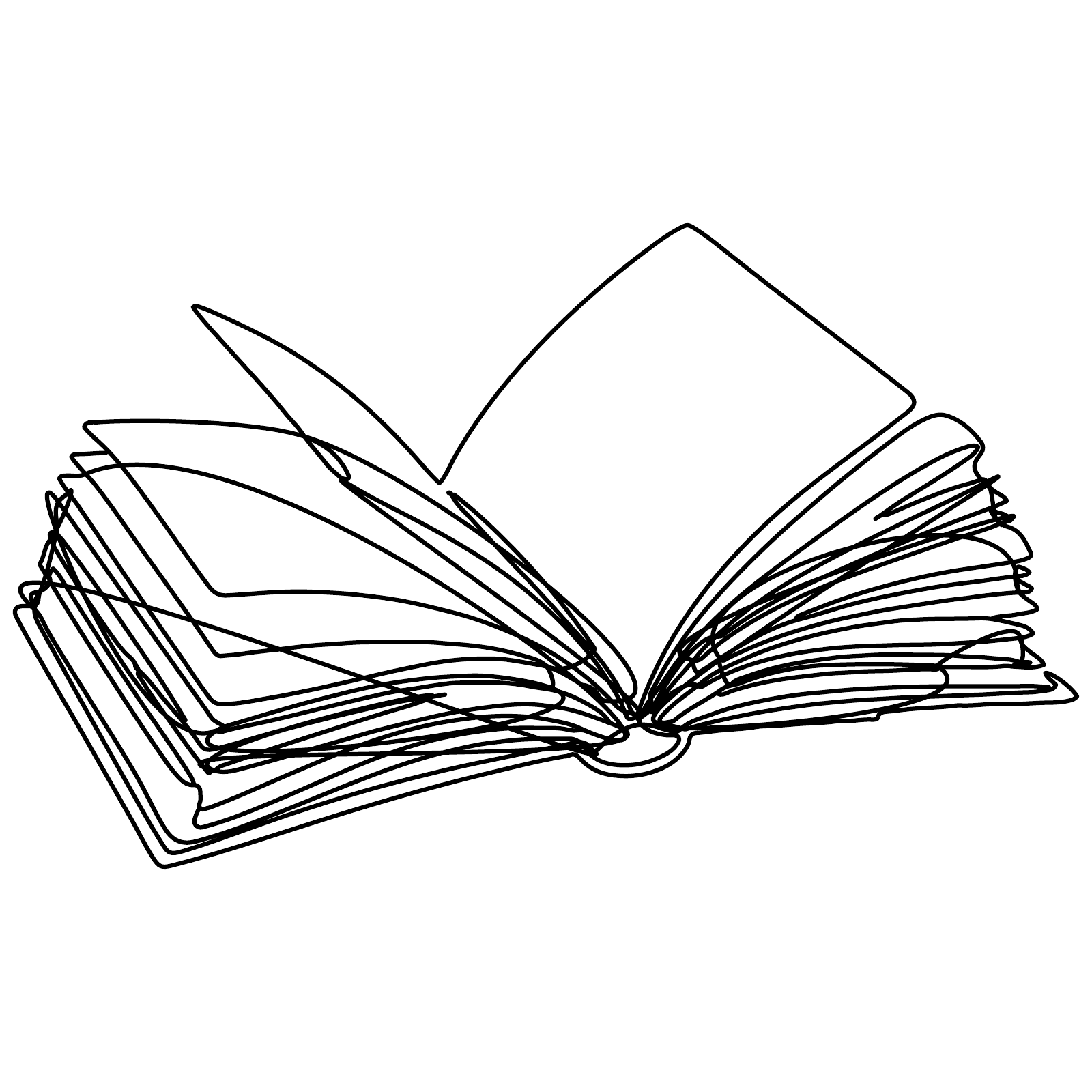}}%
        \end{subfigure}
        \\[\lrspace]
        \begin{subfigure}{\textwidth}
            \fbox{\includegraphics[width=\textwidth, height=\textwidth]{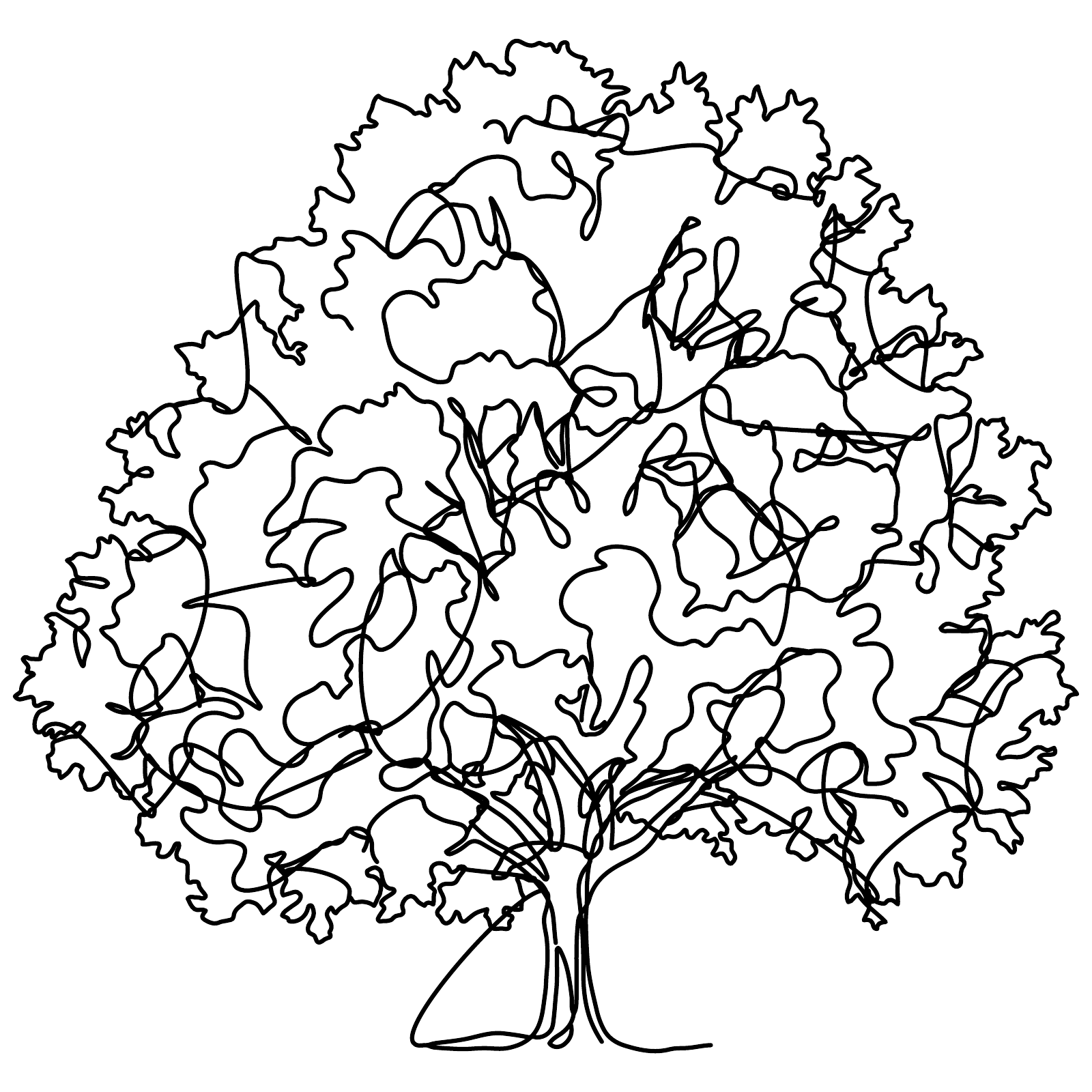}}%
        \end{subfigure}
        \\[\lrspace]
        \begin{subfigure}{\textwidth}
            \fbox{\includegraphics[width=\textwidth, height=\textwidth]{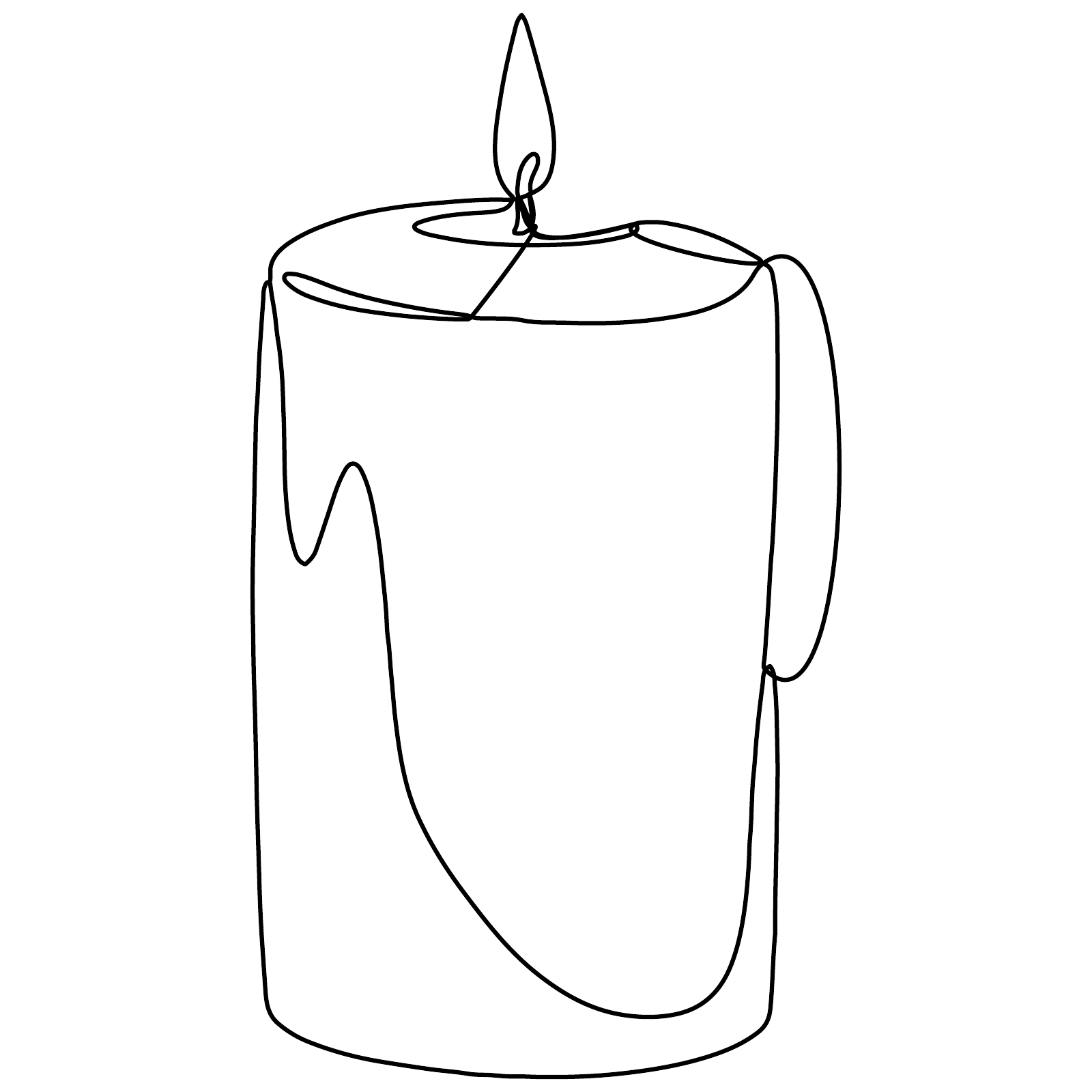}}%
        \end{subfigure}
        \\[\lrspace]
        \begin{subfigure}{\textwidth}
            \fbox{\includegraphics[width=\textwidth, height=\textwidth]{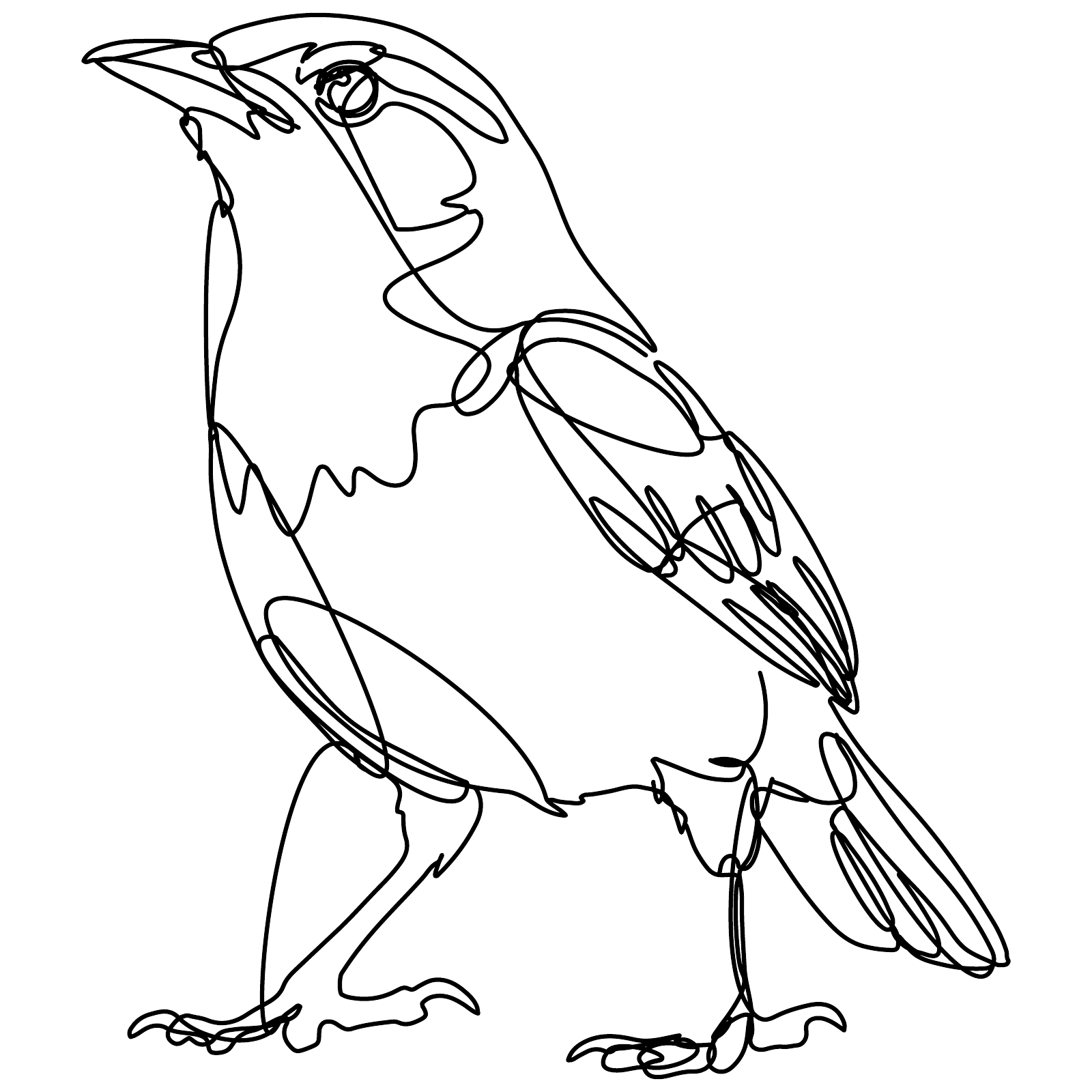}}%
        \end{subfigure}
        \\[\lrspace]
        \begin{subfigure}{\textwidth}
            \fbox{\includegraphics[width=\textwidth, height=\textwidth]{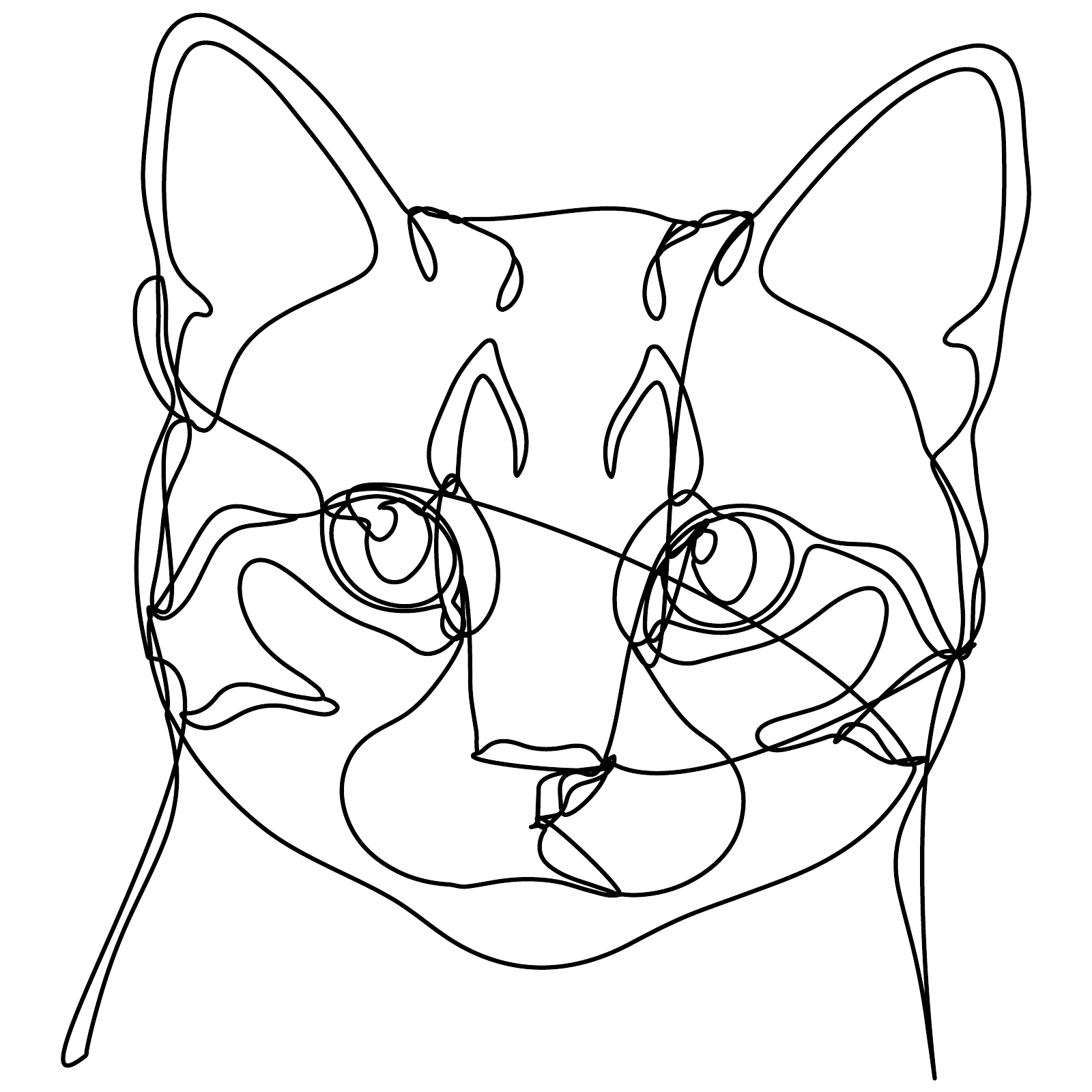}}%
        \end{subfigure}
        \\[\lrspace]
        \begin{subfigure}{\textwidth}
            \fbox{\includegraphics[width=\textwidth, height=\textwidth]{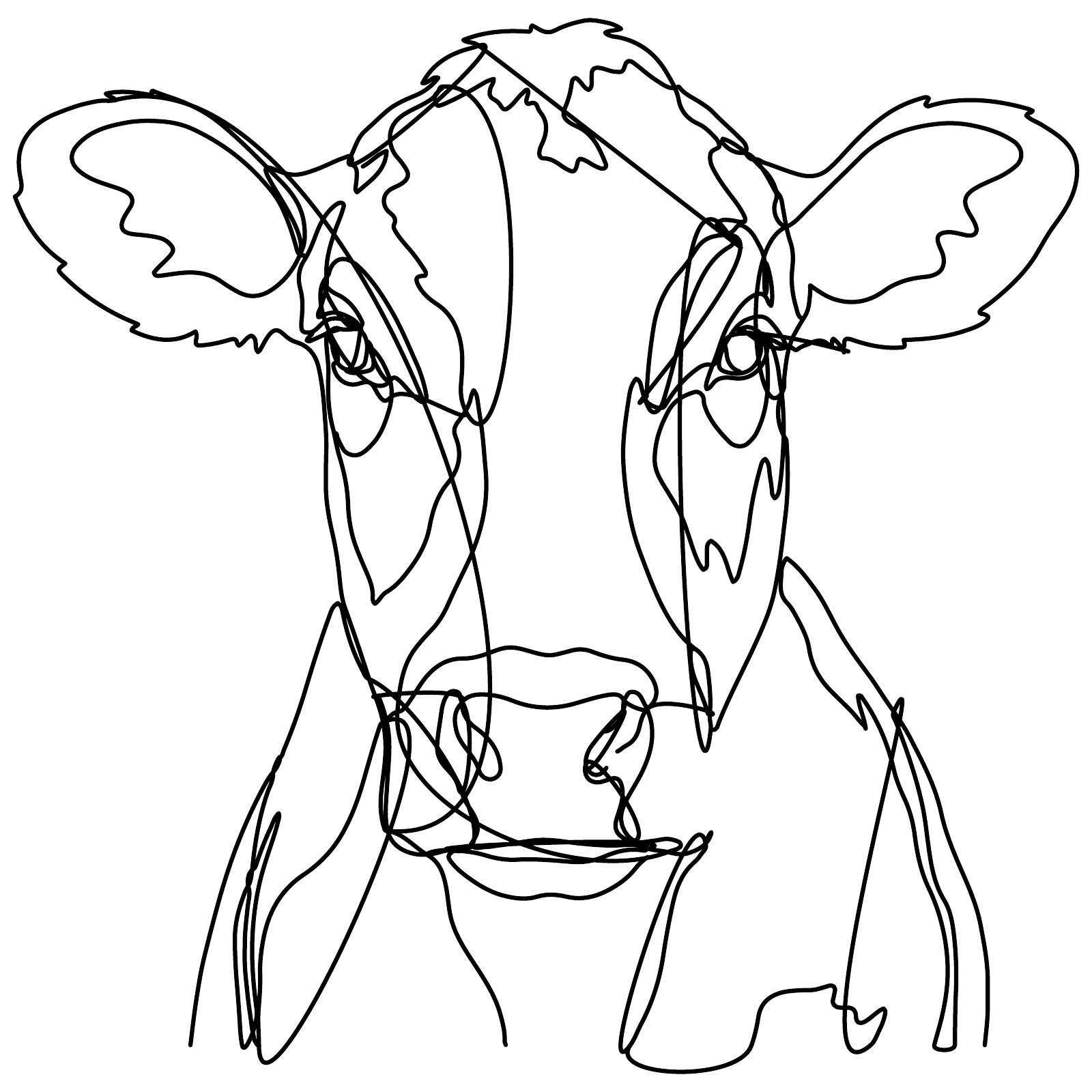}}%
        \end{subfigure}
        \vspace{-15pt}
        \caption*{\scriptsize Gemini (connected)}
    \end{subfigure}%
    & \begin{subfigure}[t]{\cwidth}
        \begin{subfigure}{\textwidth}
            \fbox{\includegraphics[width=\textwidth, height=\textwidth]{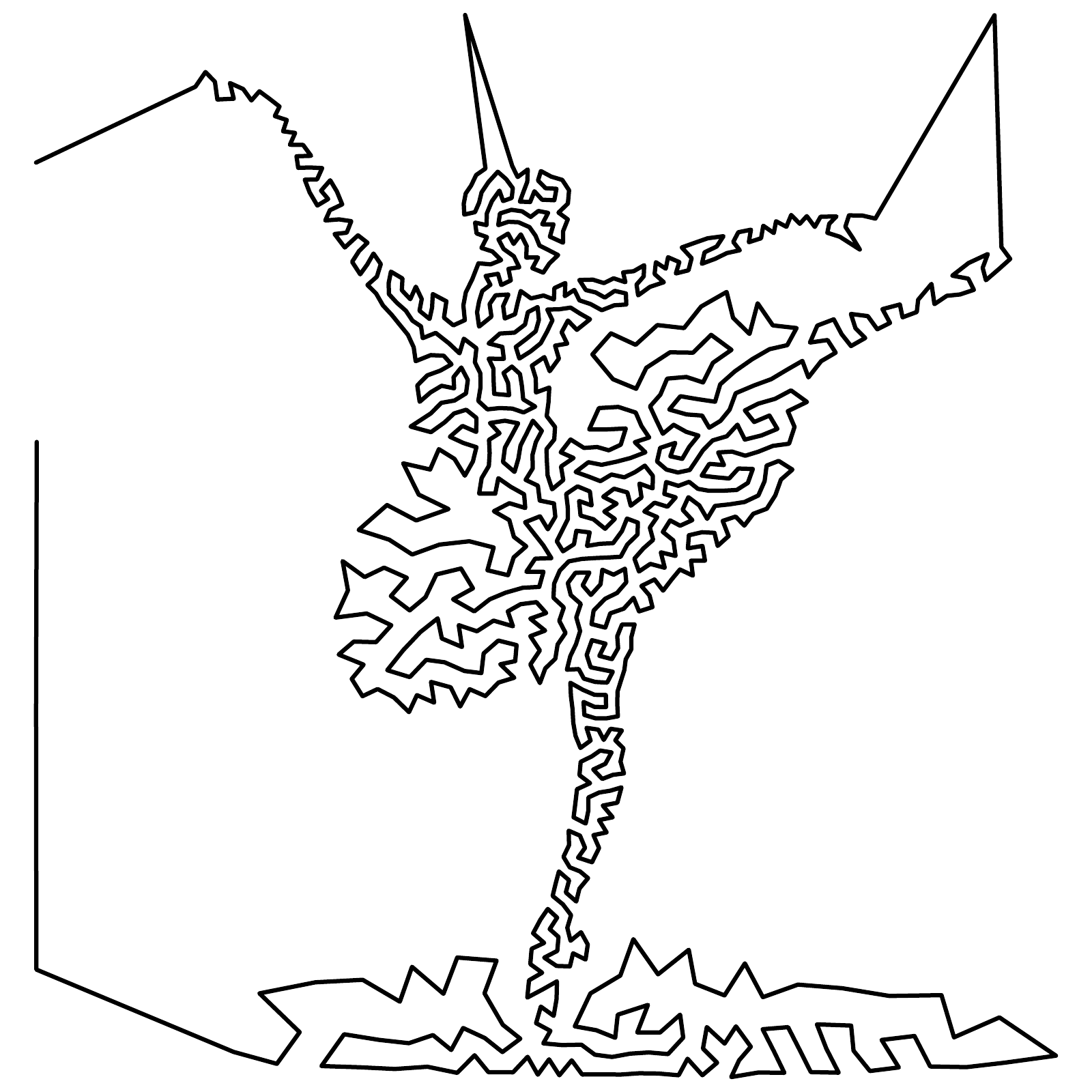}}%
        \end{subfigure}
        \\[\lrspace]
        \begin{subfigure}{\textwidth}
            \fbox{\includegraphics[width=\textwidth, height=\textwidth]{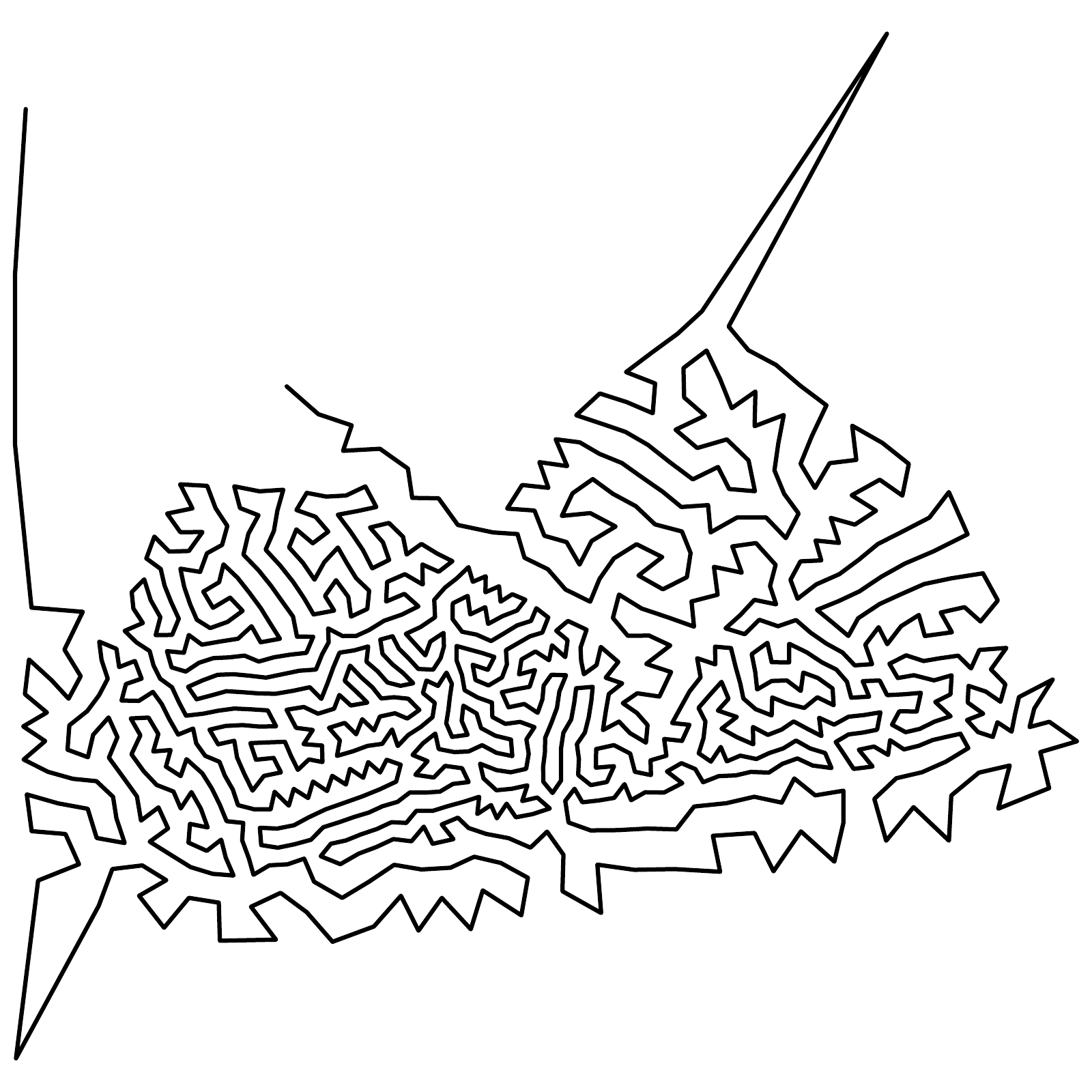}}%
        \end{subfigure}
        \\[\lrspace]
        \begin{subfigure}{\textwidth}
            \fbox{\includegraphics[width=\textwidth, height=\textwidth]{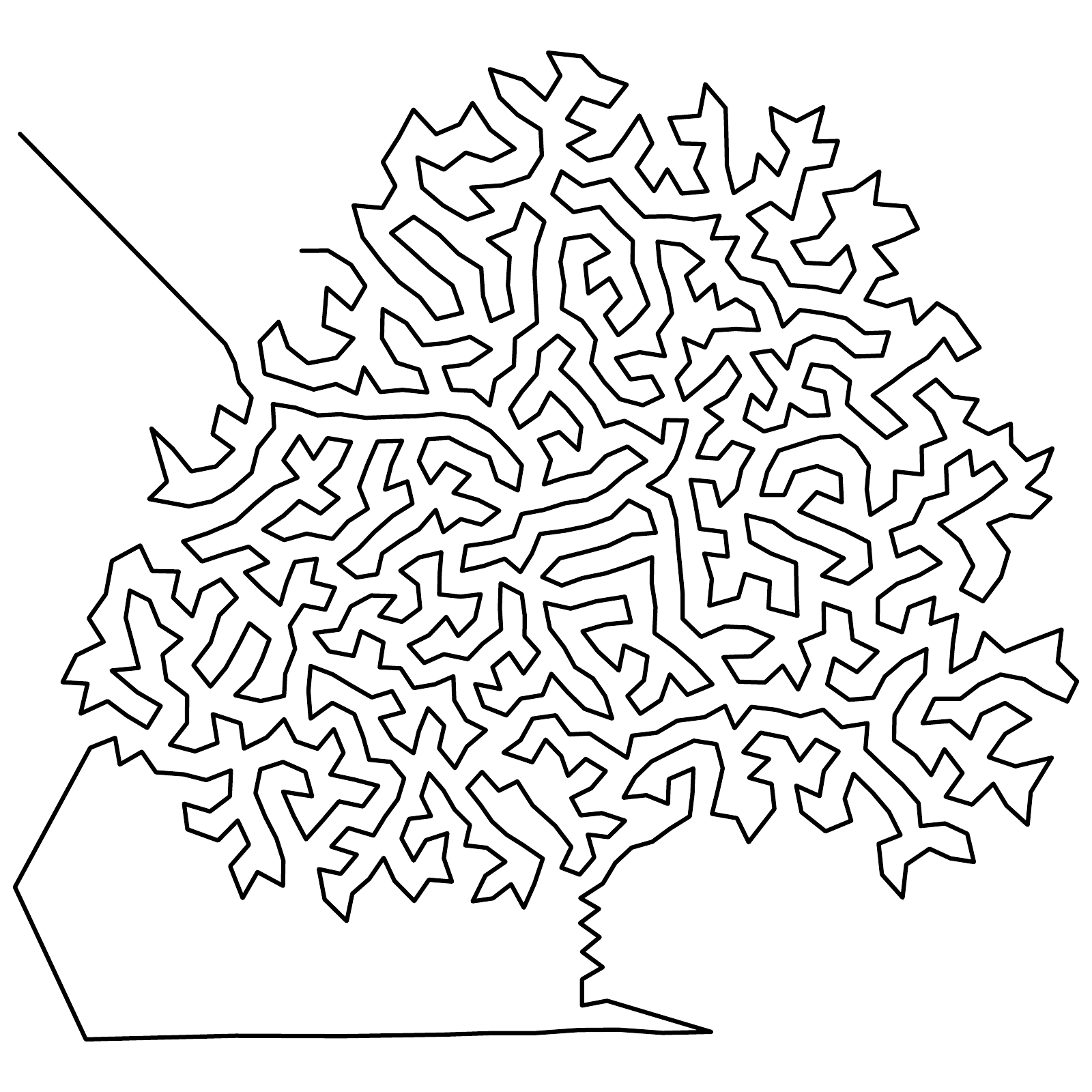}}%
        \end{subfigure}
        \\[\lrspace]
        \begin{subfigure}{\textwidth}
            \fbox{\includegraphics[width=\textwidth, height=\textwidth]{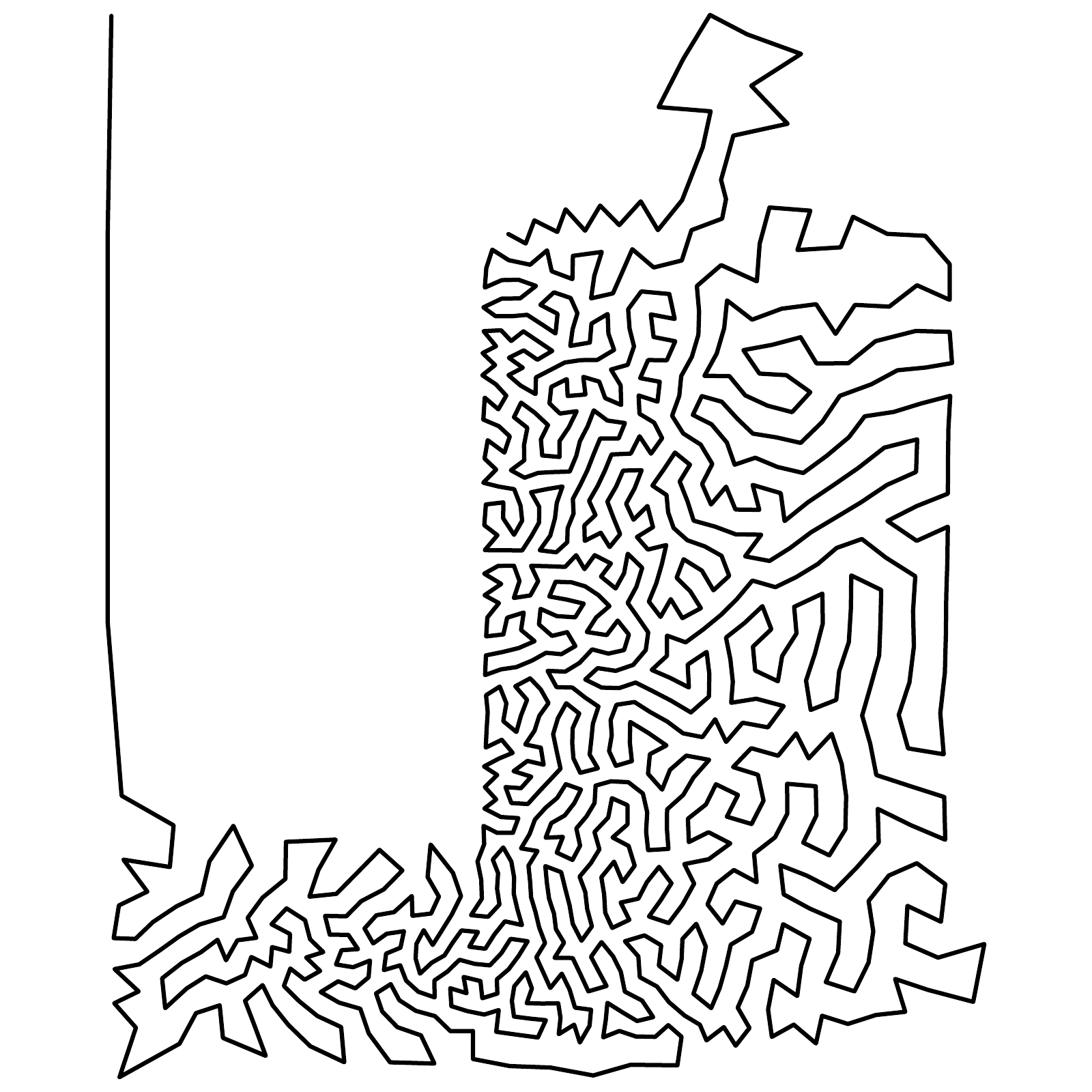}}%
        \end{subfigure}
        \\[\lrspace]
        \begin{subfigure}{\textwidth}
            \fbox{\includegraphics[width=\textwidth, height=\textwidth]{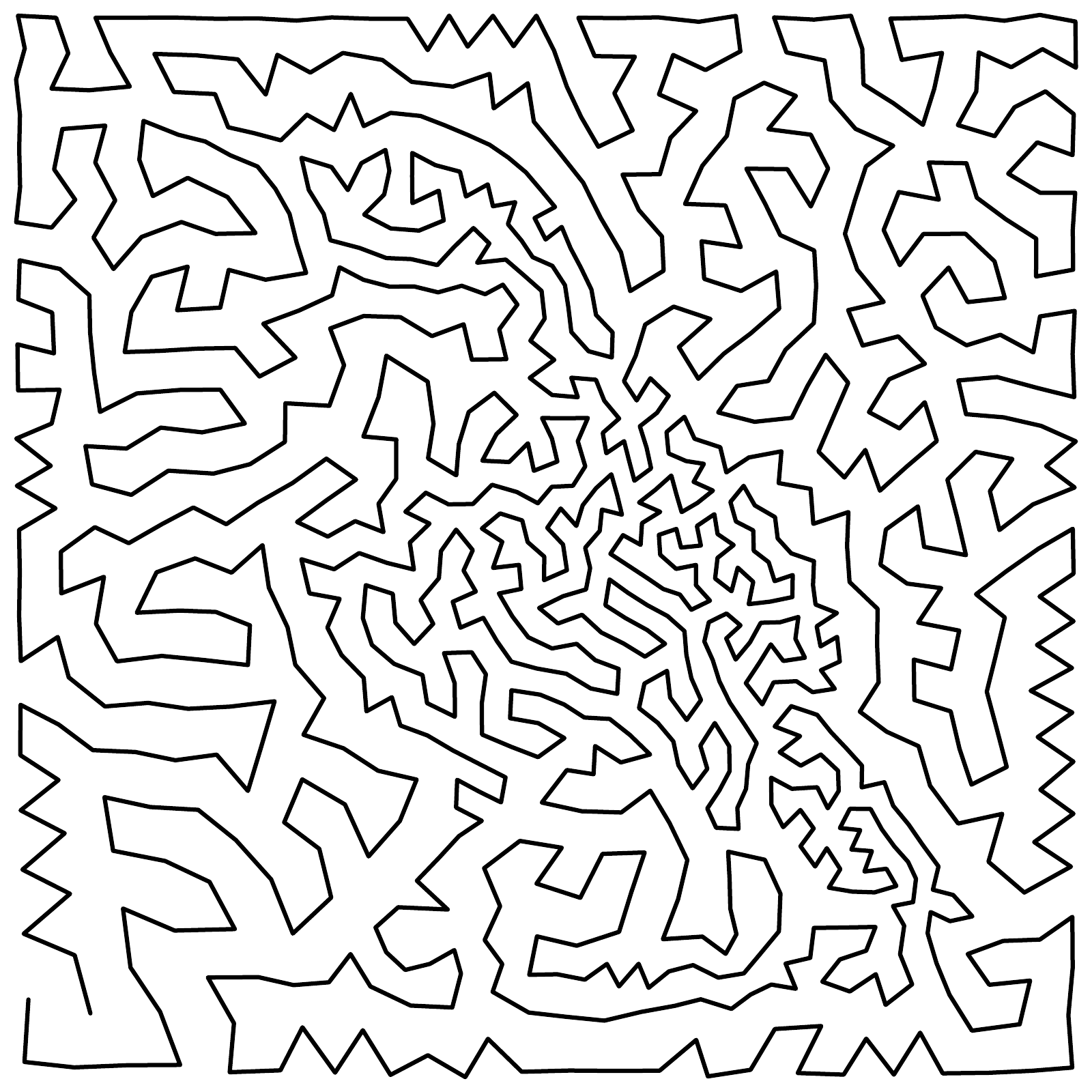}}%
        \end{subfigure}
        \\[\lrspace]
        \begin{subfigure}{\textwidth}
            \fbox{\includegraphics[width=\textwidth, height=\textwidth]{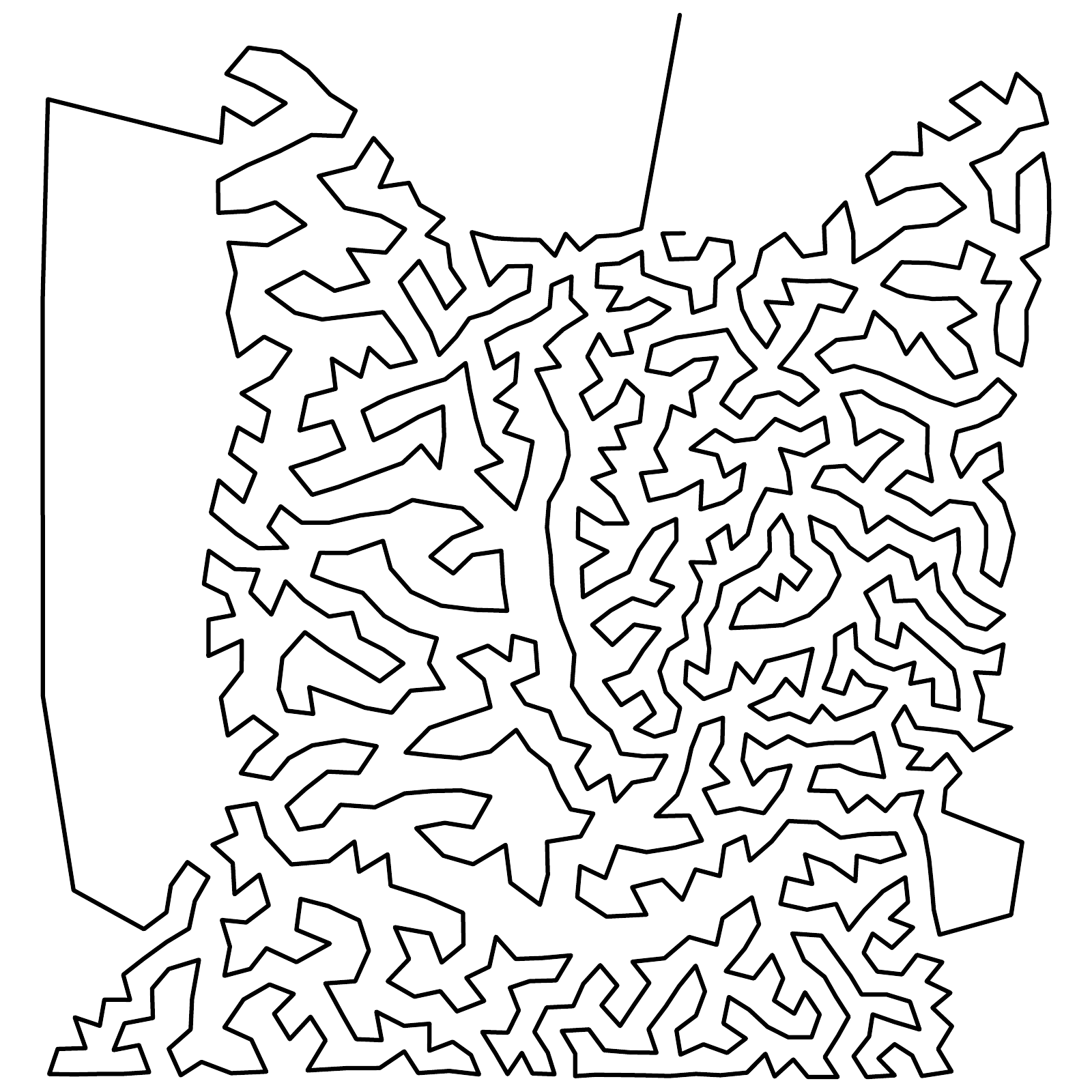}}%
        \end{subfigure}
        \\[\lrspace]
        \begin{subfigure}{\textwidth}
            \fbox{\includegraphics[width=\textwidth, height=\textwidth]{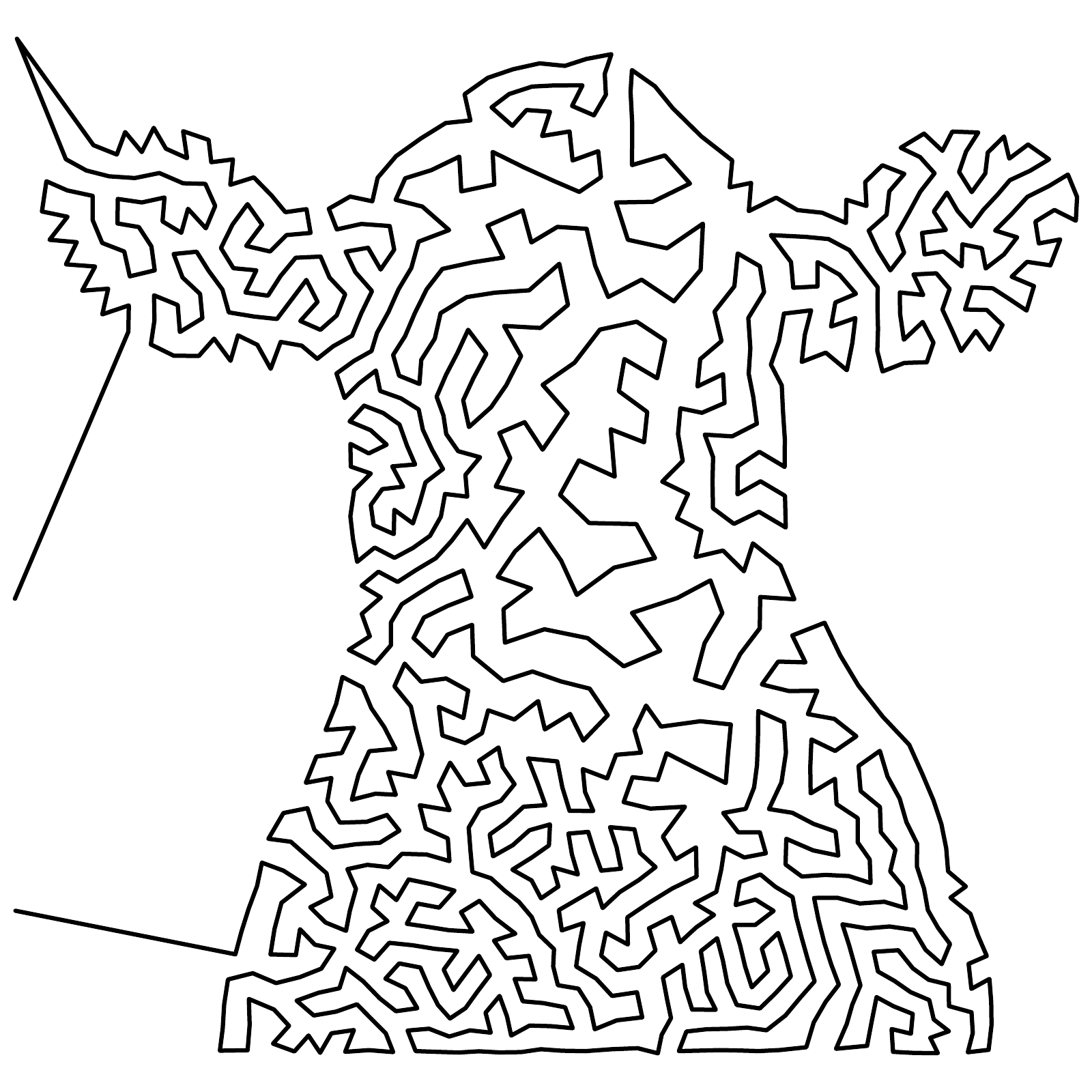}}%
        \end{subfigure}
        \vspace{-15pt}
        \caption*{\scriptsize TSP art \cite{kaplan.bosch2005}}
    \end{subfigure}%
    & \begin{subfigure}[t]{\cwidth}
        \begin{subfigure}{\textwidth}
            \fbox{\includegraphics[width=\textwidth]{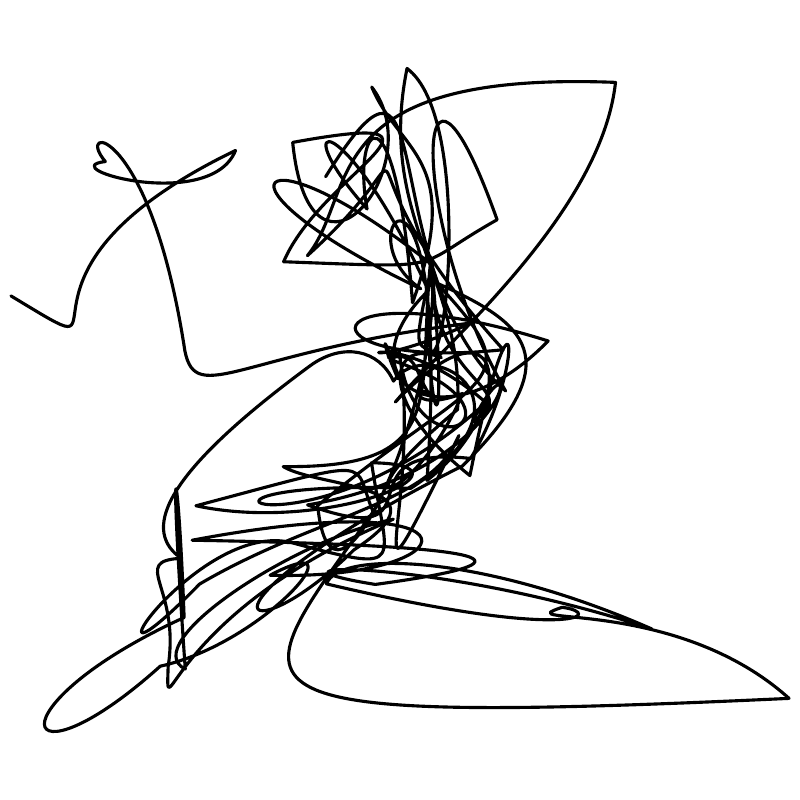}}%
        \end{subfigure}
        \\[\lrspace]
        \begin{subfigure}{\textwidth}
            \fbox{\includegraphics[width=\textwidth]{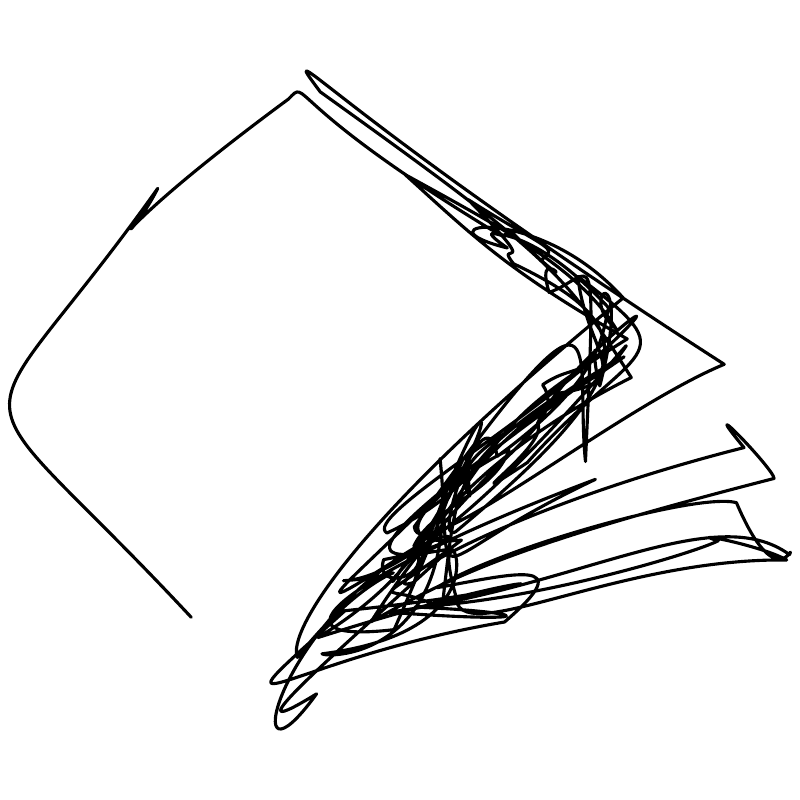}}%
        \end{subfigure}
        \\[\lrspace]
        \begin{subfigure}{\textwidth}
            \fbox{\includegraphics[width=\textwidth]{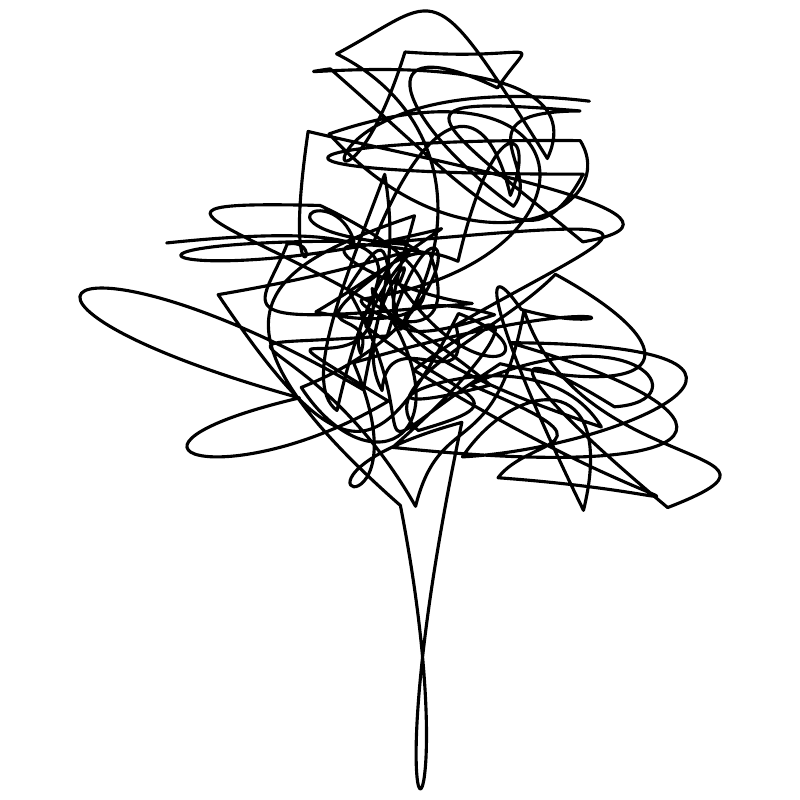}}%
        \end{subfigure}
        \\[\lrspace]
        \begin{subfigure}{\textwidth}
            \fbox{\includegraphics[width=\textwidth]{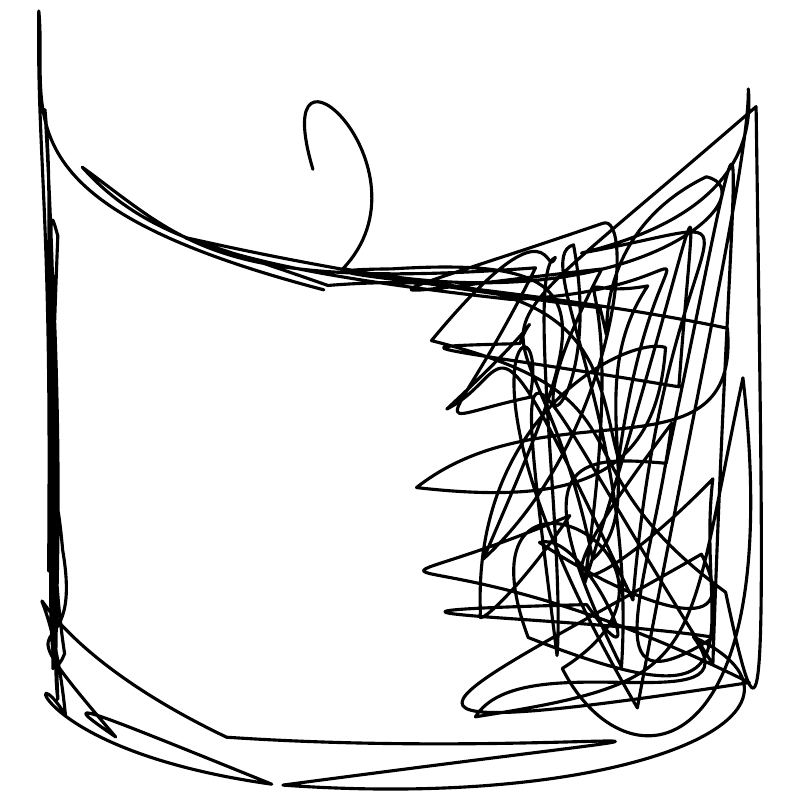}}%
        \end{subfigure}
        \\[\lrspace]
        \begin{subfigure}{\textwidth}
            \fbox{\includegraphics[width=\textwidth]{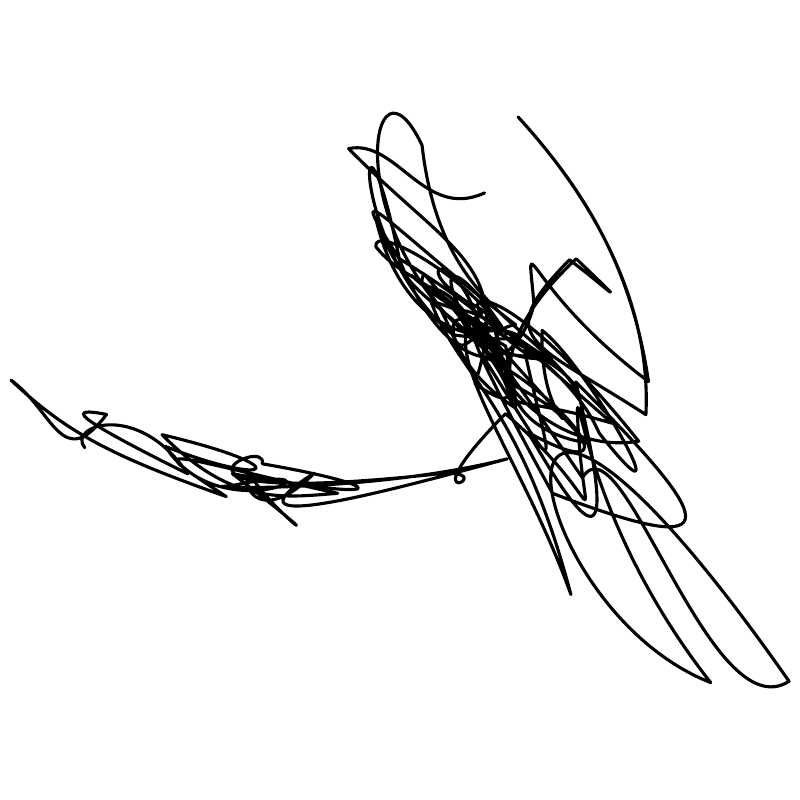}}%
        \end{subfigure}
        \\[\lrspace]
        \begin{subfigure}{\textwidth}
            \fbox{\includegraphics[width=\textwidth]{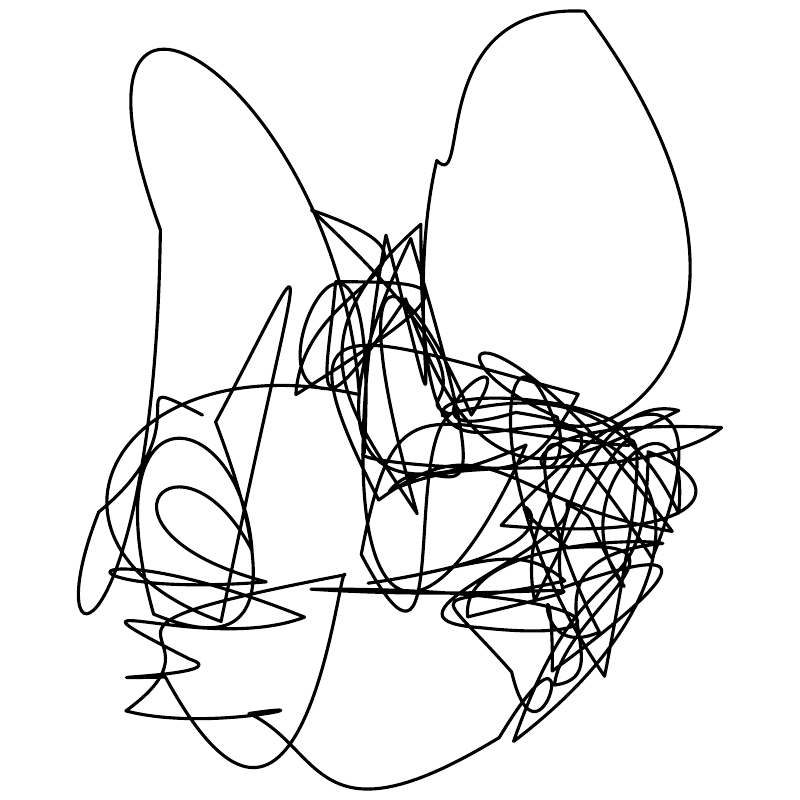}}%
        \end{subfigure}
        \\[\lrspace]
        \begin{subfigure}{\textwidth}
            \fbox{\includegraphics[width=\textwidth]{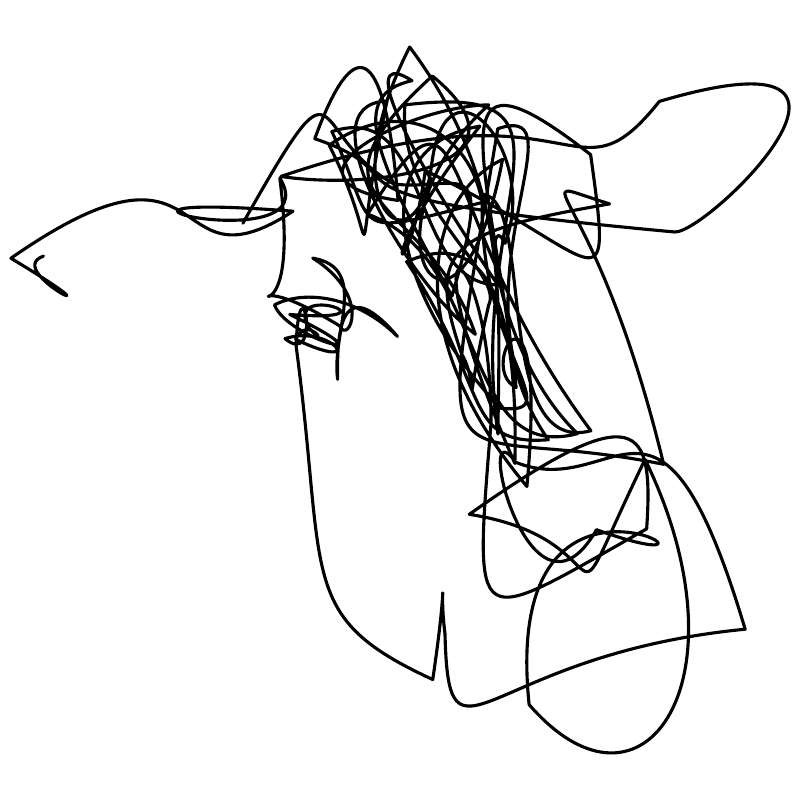}}%
        \end{subfigure}
        \vspace{-15pt}
        \caption*{\scriptsize ControlSketch \cite{arar.etal2025}}
    \end{subfigure}
    & \begin{subfigure}[t]{\cwidth}
        \begin{subfigure}{\textwidth}
            \fbox{\includegraphics[width=\textwidth]{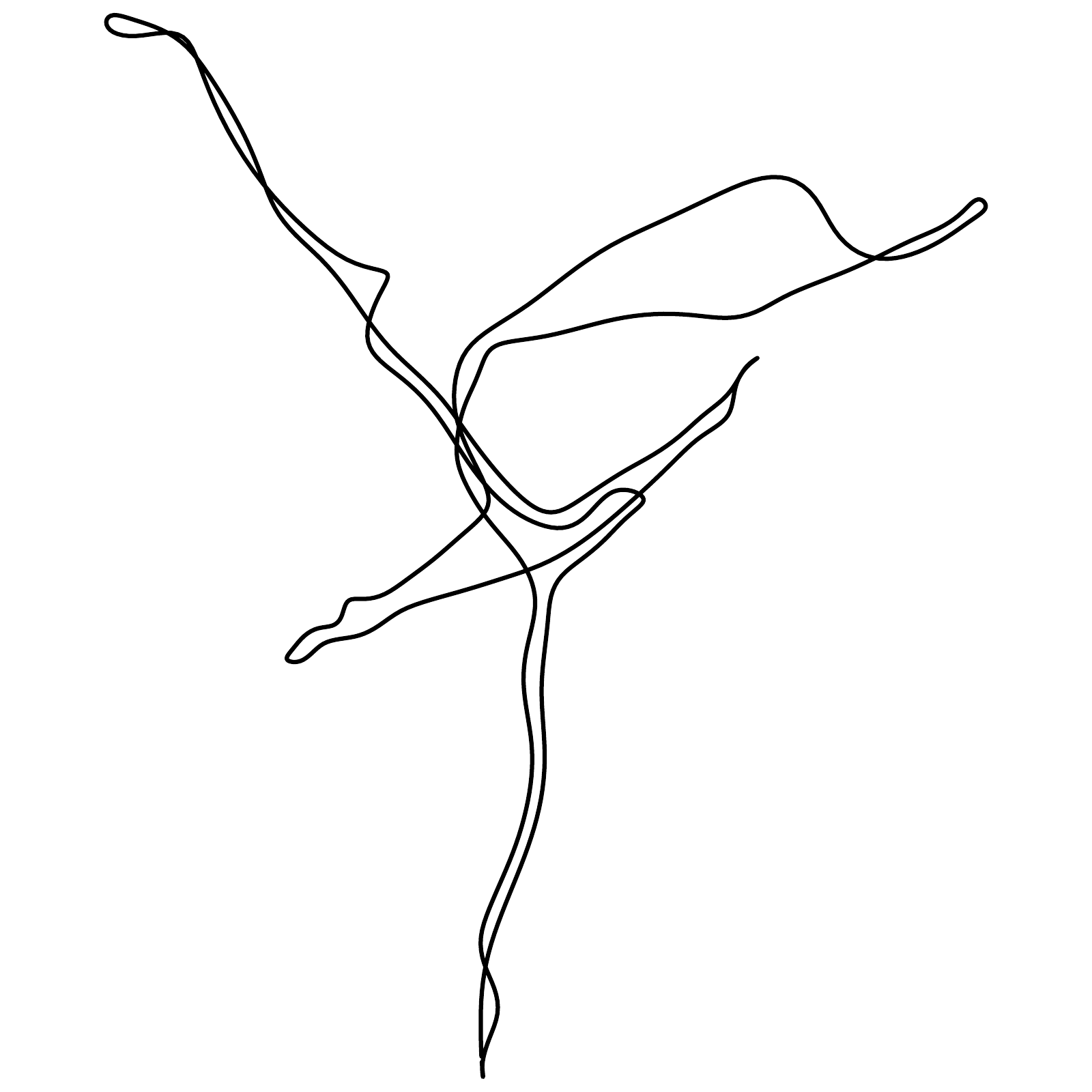}}%
        \end{subfigure}
        \\[\lrspace]
        \begin{subfigure}{\textwidth}
            \fbox{\includegraphics[width=\textwidth]{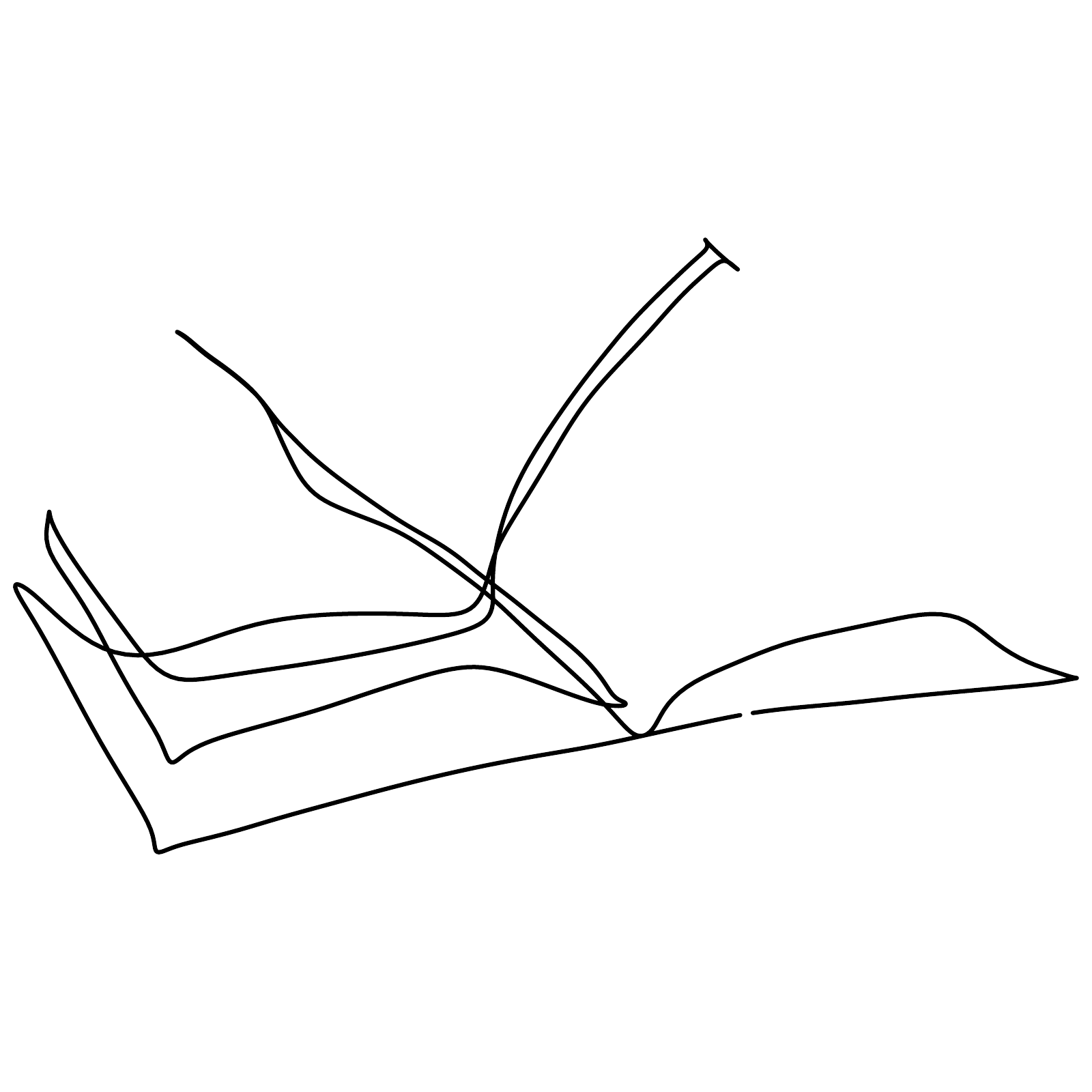}}%
        \end{subfigure}
        \\[\lrspace]
        \begin{subfigure}{\textwidth}
            \fbox{\includegraphics[width=\textwidth]{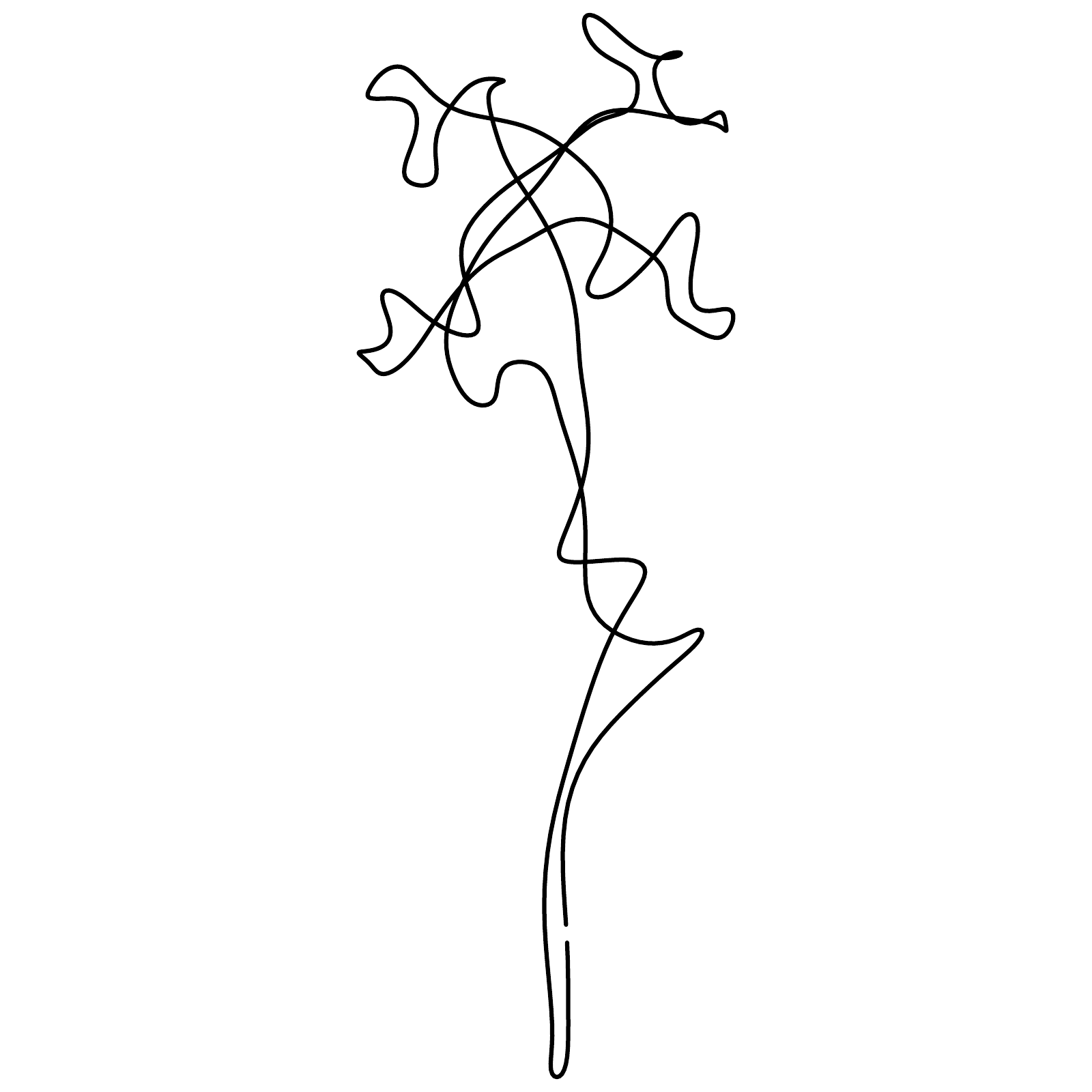}}%
        \end{subfigure}
        \\[\lrspace]
        \begin{subfigure}{\textwidth}
            \fbox{\includegraphics[width=\textwidth]{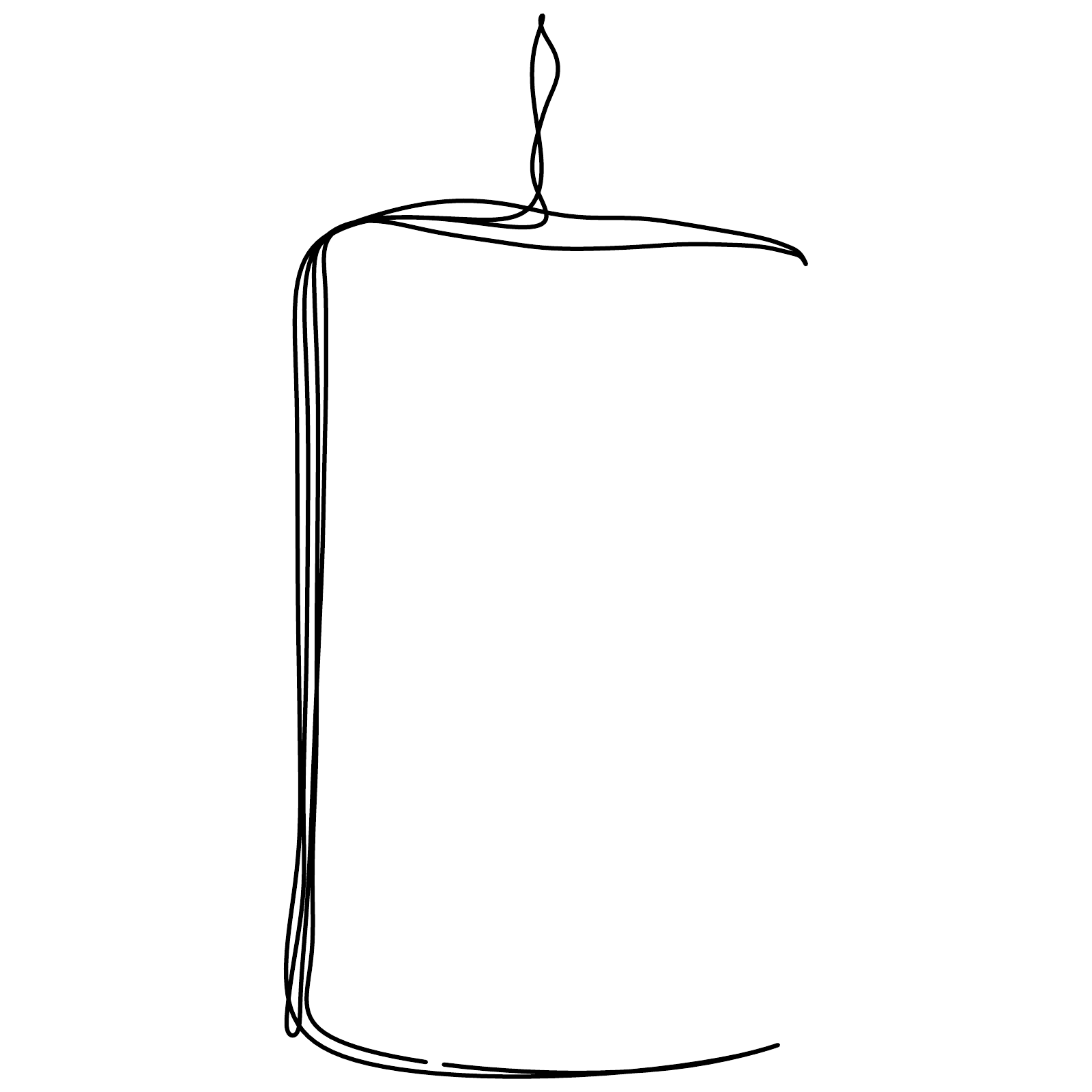}}%
        \end{subfigure}
        \\[\lrspace]
        \begin{subfigure}{\textwidth}
            \fbox{\includegraphics[width=\textwidth]{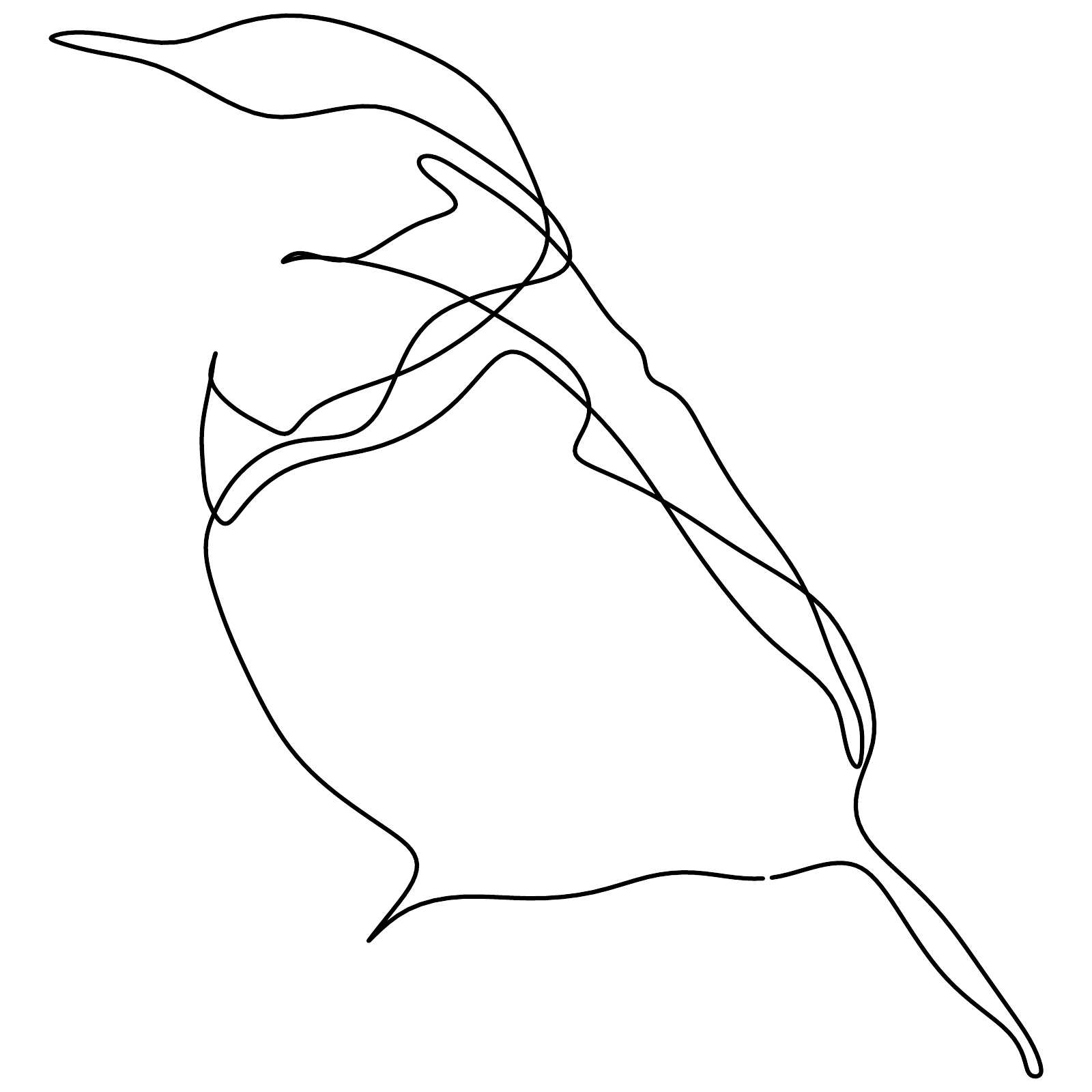}}%
        \end{subfigure}
        \\[\lrspace]
        \begin{subfigure}{\textwidth}
            \fbox{\includegraphics[width=\textwidth]{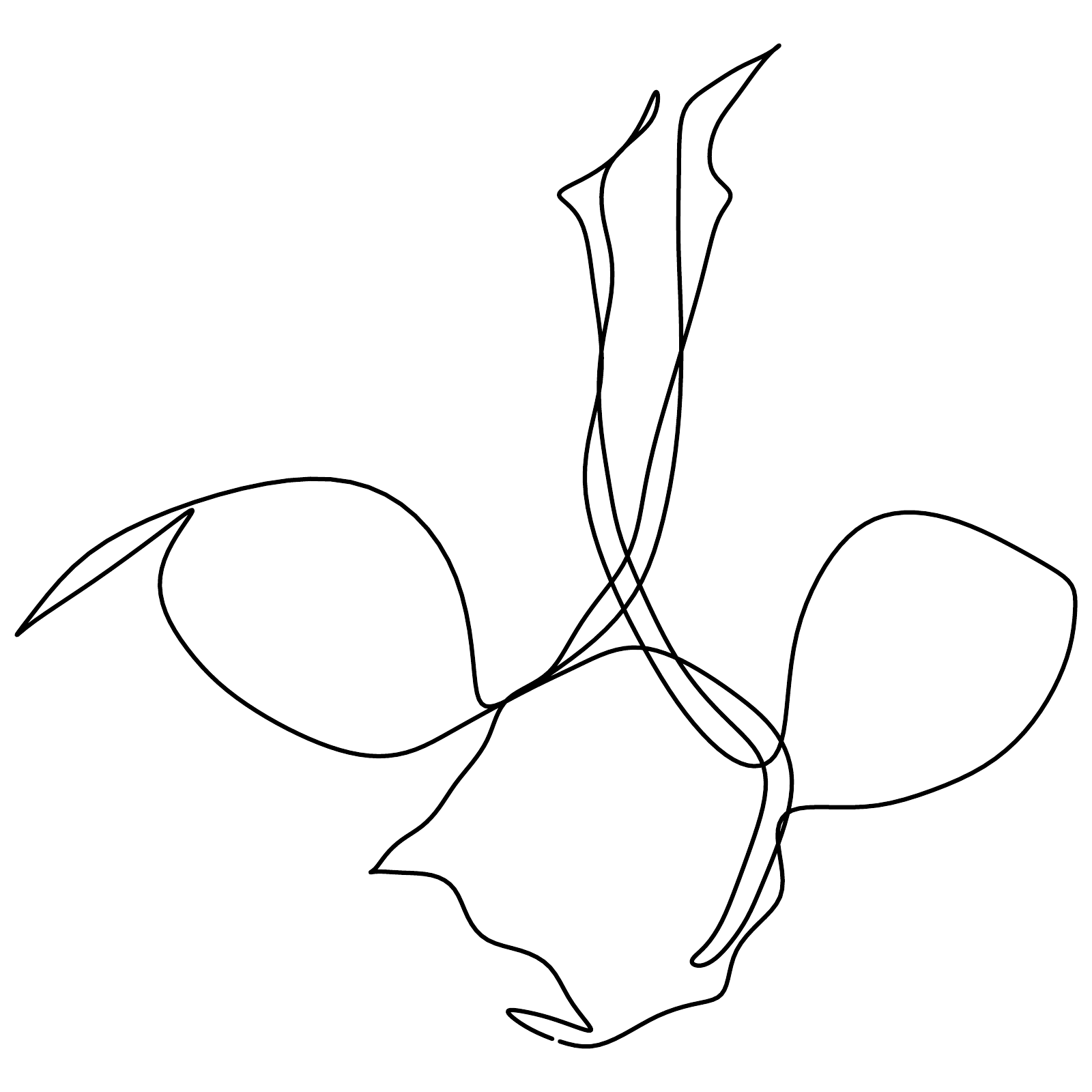}}%
        \end{subfigure}
        \\[\lrspace]
        \begin{subfigure}{\textwidth}
            \fbox{\includegraphics[width=\textwidth]{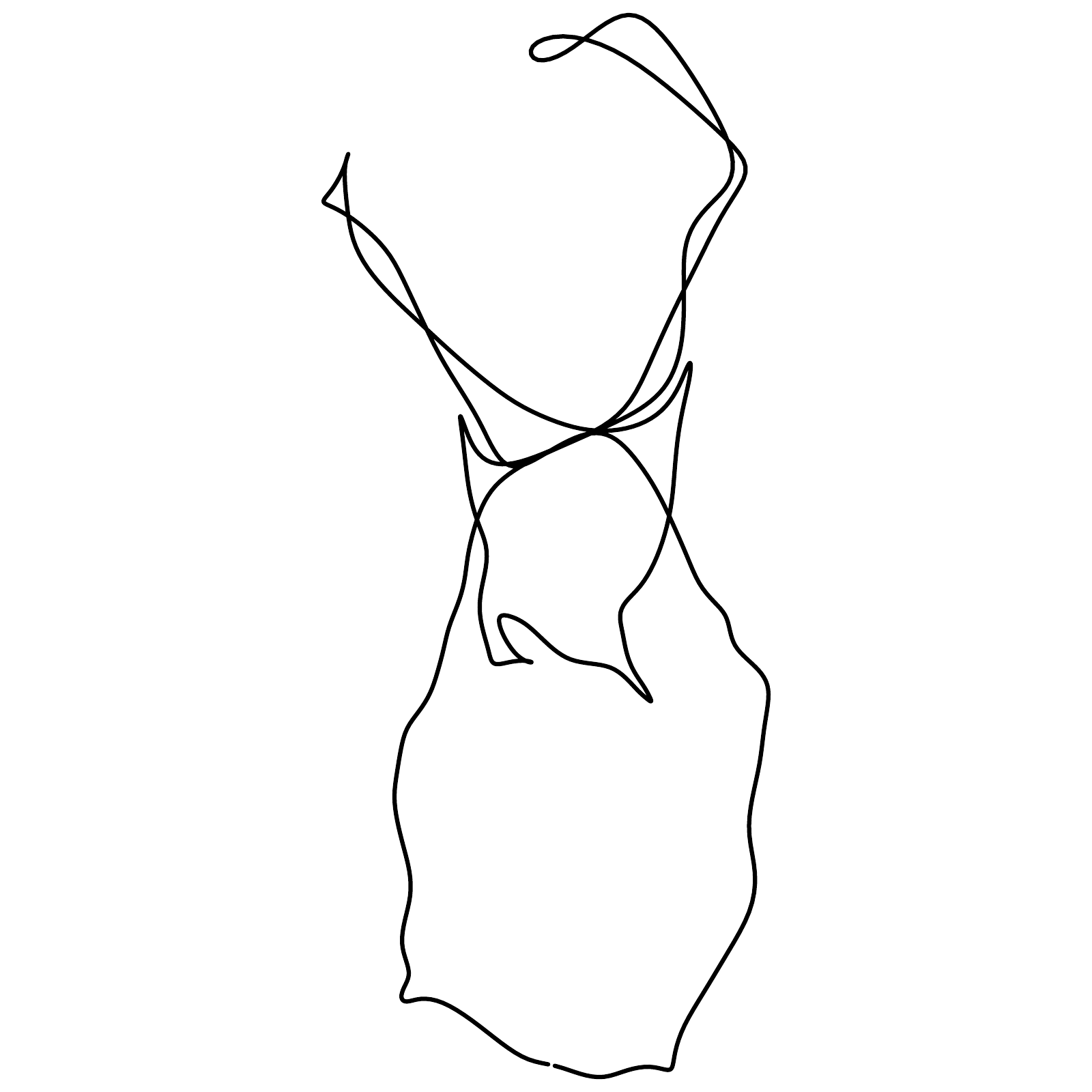}}%
        \end{subfigure}
        \vspace{-15pt}
        \caption*{\scriptsize 3D Wire Art \cite{tojo.etal2024}}
    \end{subfigure}
    & \begin{subfigure}[t]{\cwidth}
        \begin{subfigure}{\textwidth}
            \fbox{\includegraphics[width=\textwidth]{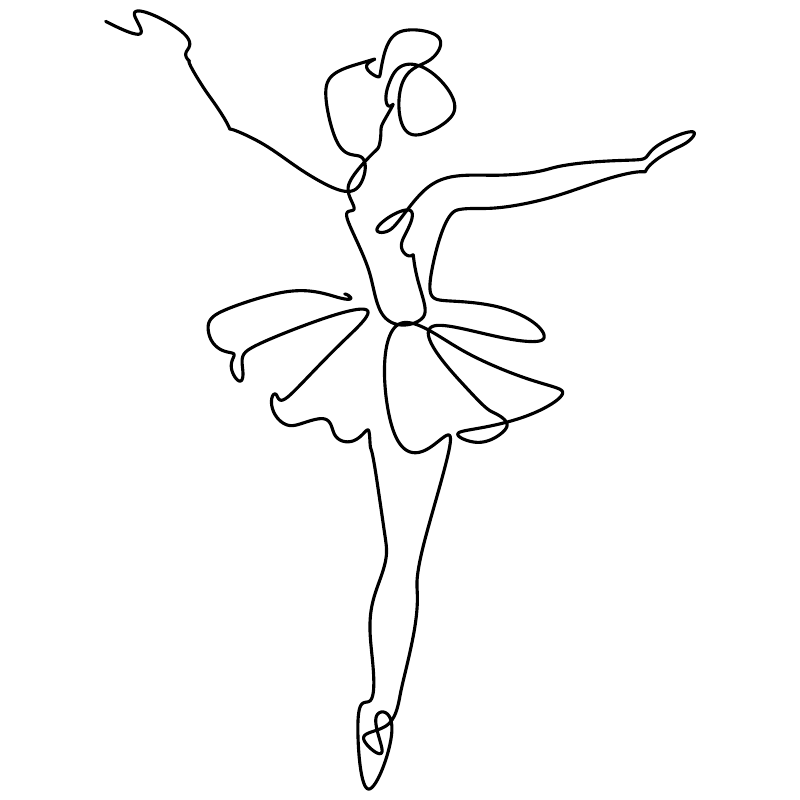}}%
        \end{subfigure}
        \\[\lrspace]
        \begin{subfigure}{\textwidth}
            \fbox{\includegraphics[width=\textwidth]{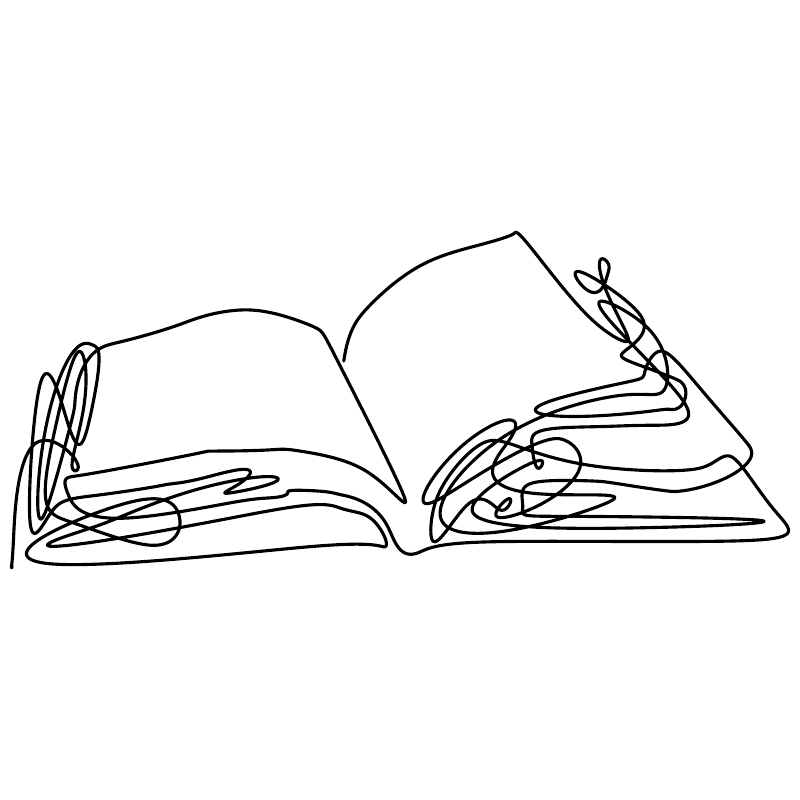}}%
        \end{subfigure}
        \\[\lrspace]
        \begin{subfigure}{\textwidth}
            \fbox{\includegraphics[width=\textwidth]{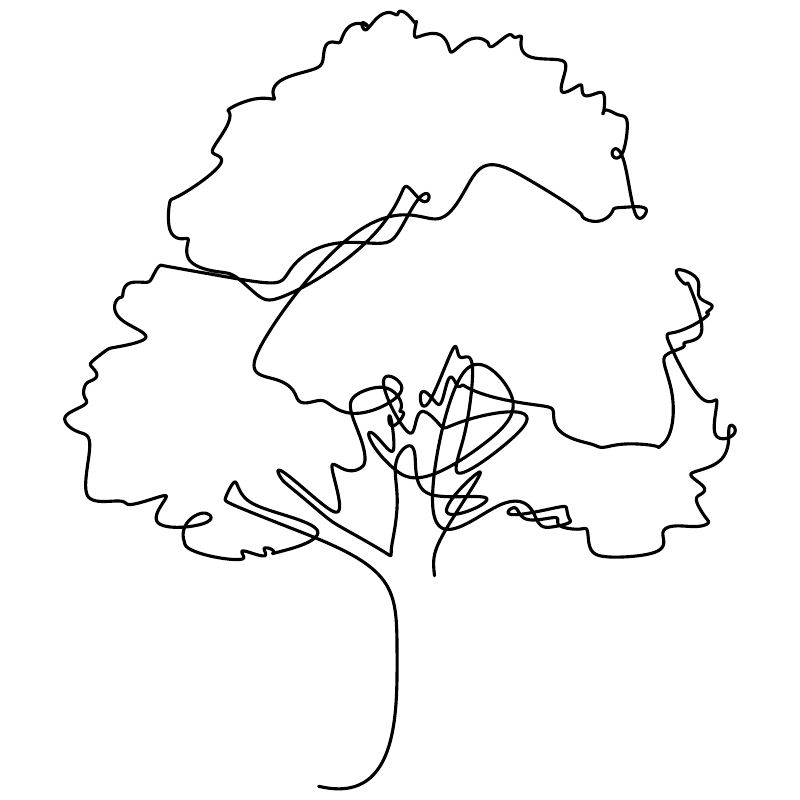}}%
        \end{subfigure}
        \\[\lrspace]
        \begin{subfigure}{\textwidth}
            \fbox{\includegraphics[width=\textwidth]{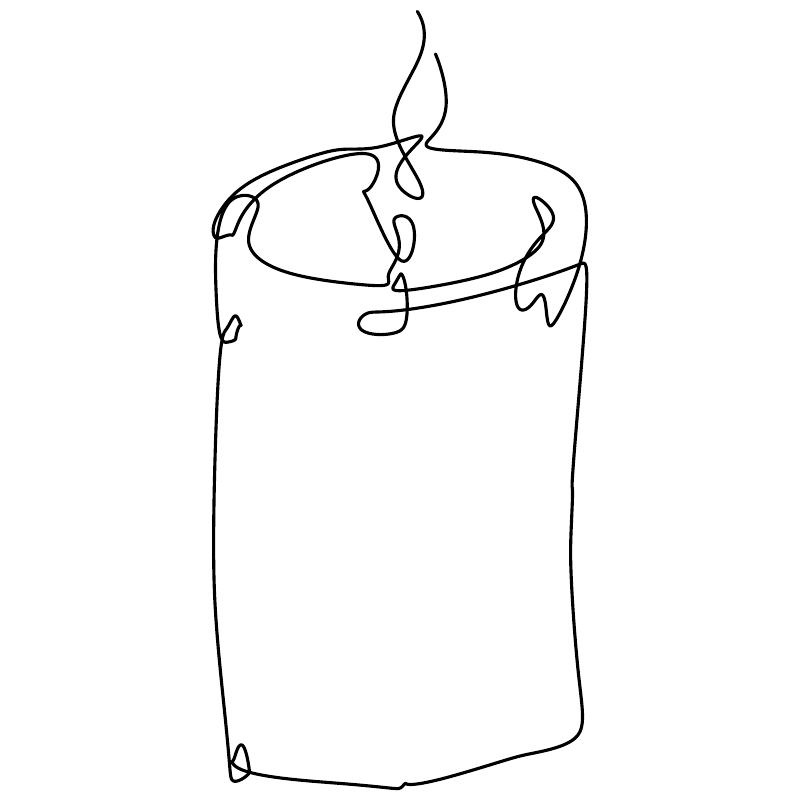}}%
        \end{subfigure}
        \\[\lrspace]
        \begin{subfigure}{\textwidth}
            \fbox{\includegraphics[width=\textwidth]{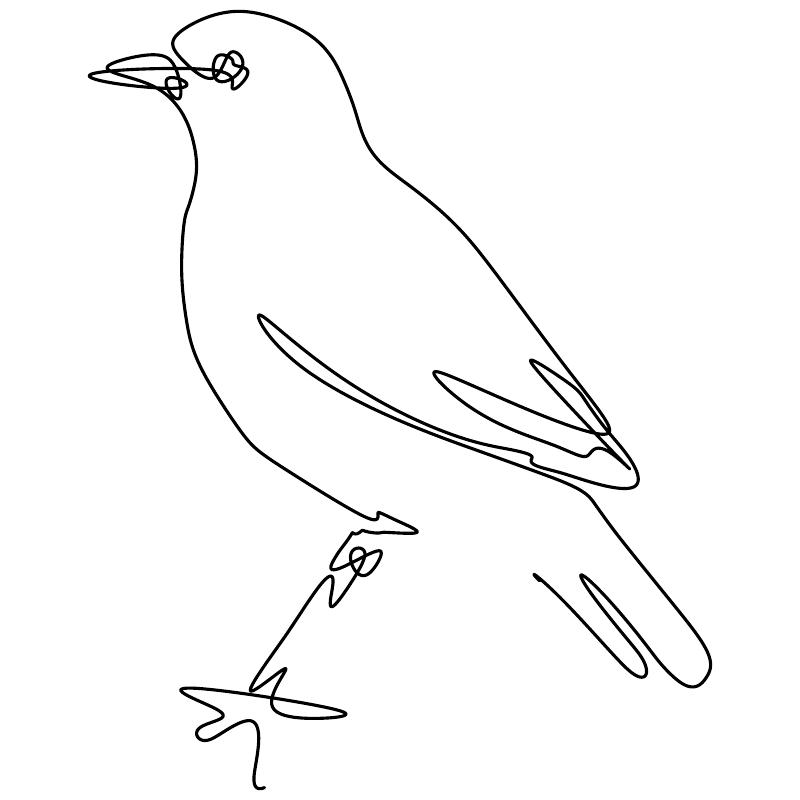}}%
        \end{subfigure}
        \\[\lrspace]
        \begin{subfigure}{\textwidth}
            \fbox{\includegraphics[width=\textwidth]{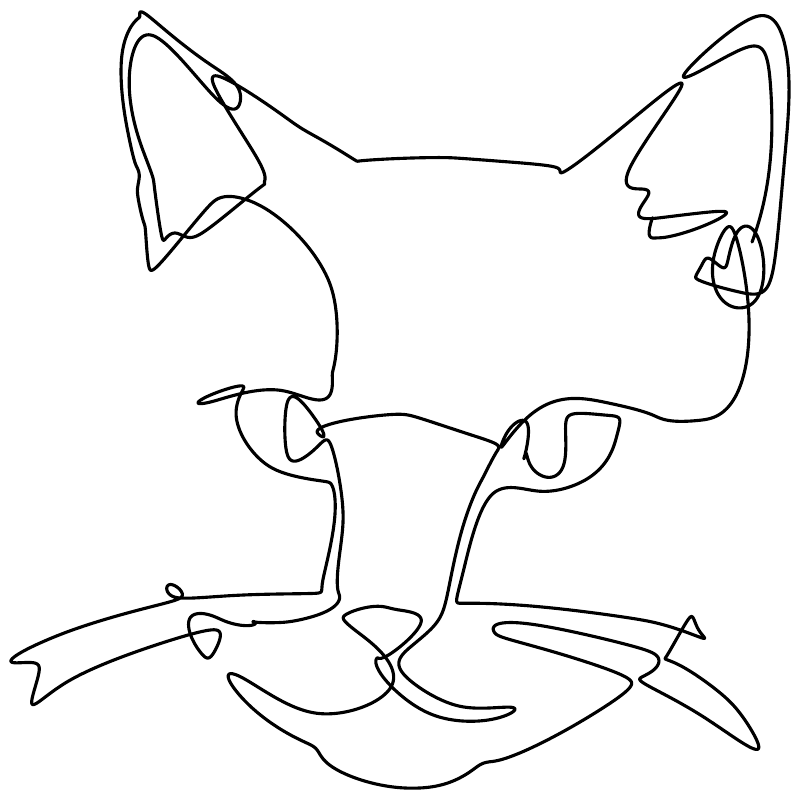}}%
        \end{subfigure}
        \\[\lrspace]
        \begin{subfigure}{\textwidth}
            \fbox{\includegraphics[width=\textwidth]{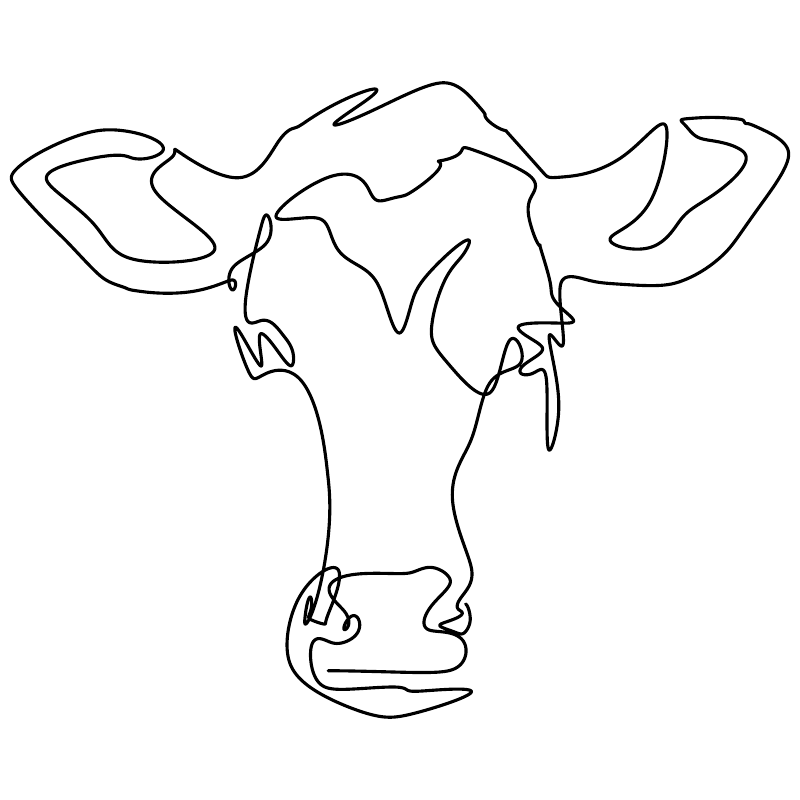}}%
        \end{subfigure}
        \vspace{-15pt}
        \caption*{\scriptsize Our method}
    \end{subfigure}%
    \end{tabular}
    \caption{Qualitative comparison of various line drawing generation methods. Text-to-image models (Gemini in this case) produce raster images and contain multiple lines. We present here the results obtained after applying the single-line conversion strategy (see \secref{sec:results}).}
    \label{fig:extra_comparison}
\end{figure*}

\clearpage
\bibliographystyle{eg-alpha-doi} 
\bibliography{reference}

\clearpage
\appendix

\section{Background}
\label{app:background}
We summarize the core ideas in diffusion models and score distillation sampling, which we use in our method to generate single-line drawings.

\subsection{Diffusion models}

Given a set of images from distribution $p_{\text{data}}$, generative models learn to sample new images from this distribution. Diffusion models, a particular class of generative models, are trained to reverse a gradual diffusion process \cite{sohldickstein2015deep,ho2020denoising}. For the forward diffusion process, a data sample $x_0 \sim p_{\text{data}}$ is perturbed over $T$ steps by incrementally adding noise, eventually becoming $x_T \sim \mathcal{N}(0, \sigma_T)$. Variance-preserving formulations \cite{song2021scorebased} define the noisy sample after $t$ steps as $x_t = \alpha_t x_0 + \epsilon \sigma_t$, where $\epsilon \sim \mathcal{N}(0, 1)$, and $(\alpha_t, \sigma_t)$ are parameters of the diffusion process \cite{kingma2023variational}. Given a timestep $t$ and a noisy sample $x_t$, the diffusion model, often implemented as a CNN \cite{CNNLeCun1989} or Transformer\cite{vaswani2017attentionneed}, produces an estimate $\epsilon_\phi(x_t; t)$ of the perturbation $\epsilon$. In other words, the model attempts to infer the noise that was combined with the clean sample $x_0$ to produce $x_t$.

Diffusion models can also be conditioned on auxiliary information $y$, for example, a text caption or another image. This modifies the target distribution so that generated samples align with the given condition. A widely used technique for this is classifier-free guidance (CFG) \cite{ho2022classifierfree}. CFG `amplifies' the conditioned prediction $\epsilon_\phi(x_t; y, t)$ relative to the unconditioned prediction $\epsilon_\phi(x_t; \emptyset, t)$ by applying a scaling factor $s \in \mathds{R}$ to the difference:
\begin{equation*}
\epsilon_{s, \phi}(x_t; y, t) = \epsilon_\phi(x_t; y, t) \ + \ s \,(\epsilon_\phi(x_t; y, t) - \epsilon_\phi(x_t; \emptyset, t)).
\end{equation*}

\subsection{Score distillation sampling}
\label{sec:SDS_background}

While state-of-the-art diffusion models benefit from internet-scale image datasets, more specialized data (e.g., vector images, 3D meshes) are often not available at such a scale and in formats exposing the internal representation. A common strategy to overcome this issue is to distill the ``knowledge'' of a pre-trained image diffusion model into the target representation~\cite{poole2023dreamfusion}.
The intuition is to use a pre-trained 2D diffusion model as a critic to optimize a parametric target representation. More concretely, the target representation is rendered to a 2D image via differentiable rasterization, and random noise is added. The pre-trained diffusion model then evaluates this noisy image to predict the added noise, conditioned on a given prompt or additional structural guidance, such as ControlNet \cite{zhang.etal2023}. The difference between the model's predicted noise and the artificially injected noise yields an update direction that points toward how the rendered pixels should change to better match the conditioning. Because this update direction aligns with the score function of the diffusion model, the process is called score distillation sampling (SDS). This gradient on the 2D image is then backpropagated through the differentiable renderer to optimize the underlying target parameters. Originally, DreamFusion~\cite{poole2023dreamfusion} introduced SDS to update the parameters of a neural radiance field (NeRF) \cite{mildenhall2020nerf}.

Formally, given a set of parameters $\theta$ and  a differentiable rasterizer $\mathcal{R}$, we denote $x = \mathcal{R}(\theta)$ and $x_t = \alpha_t x + \epsilon \sigma_t$ as above for a given time step $t$. The gradient of the SDS loss is then given by \begin{equation*} \label{eq:sds}
    \nabla_{\theta} \mathcal{L}_\text{SDS} = \mathbb{E}_{t, \epsilon}\left[\left( \epsilon_{s, \phi}(x_t; y, t) - \epsilon \right) \frac{\partial x}{\partial \theta}\right],
\end{equation*}
where $\mathbb{E}_{t, \epsilon}$ is the expectation.
During the optimization, $\nabla_{\theta} \mathcal{L}_\text{SDS}$ is used to update $\theta$ via standard backpropagation methods \cite{kingma.ba2015}. Variants of SDS have been successfully applied beyond NeRFs, enabling the use of pre-trained diffusion models for vector graphics \cite{jain.etal2023,xing.etal2024}, font design \cite{iluz2023wordasimage}, cross-stitch embroidery \cite{Binninger:SDpiXL:2024}, Escher tilings \cite{aigerman2023generative}, mesh textures \cite{youwang2024paintit} and full 3D structures such as 3D Gaussians \cite{tang2024dreamgaussian}. In our work, we employ DiffVG \cite{li.etal2020} as a differentiable rasterizer. 

\section{LoRA additional results}
\label{app:lora}

\begin{figure}[t]
    \centering
	\small
	\setlength{\tabcolsep}{1pt}
    \setlength{\fboxsep}{0pt}
    \begin{tabular}{ >{\centering\arraybackslash}p{0.33\linewidth} >{\centering\arraybackslash}p{0.33\linewidth} >{\centering\arraybackslash}p{0.33\linewidth} }    
    \adjincludegraphics[width=\linewidth,trim={{0.0\width} {0.17\height} {0.0\width} {0.17\height}},clip]{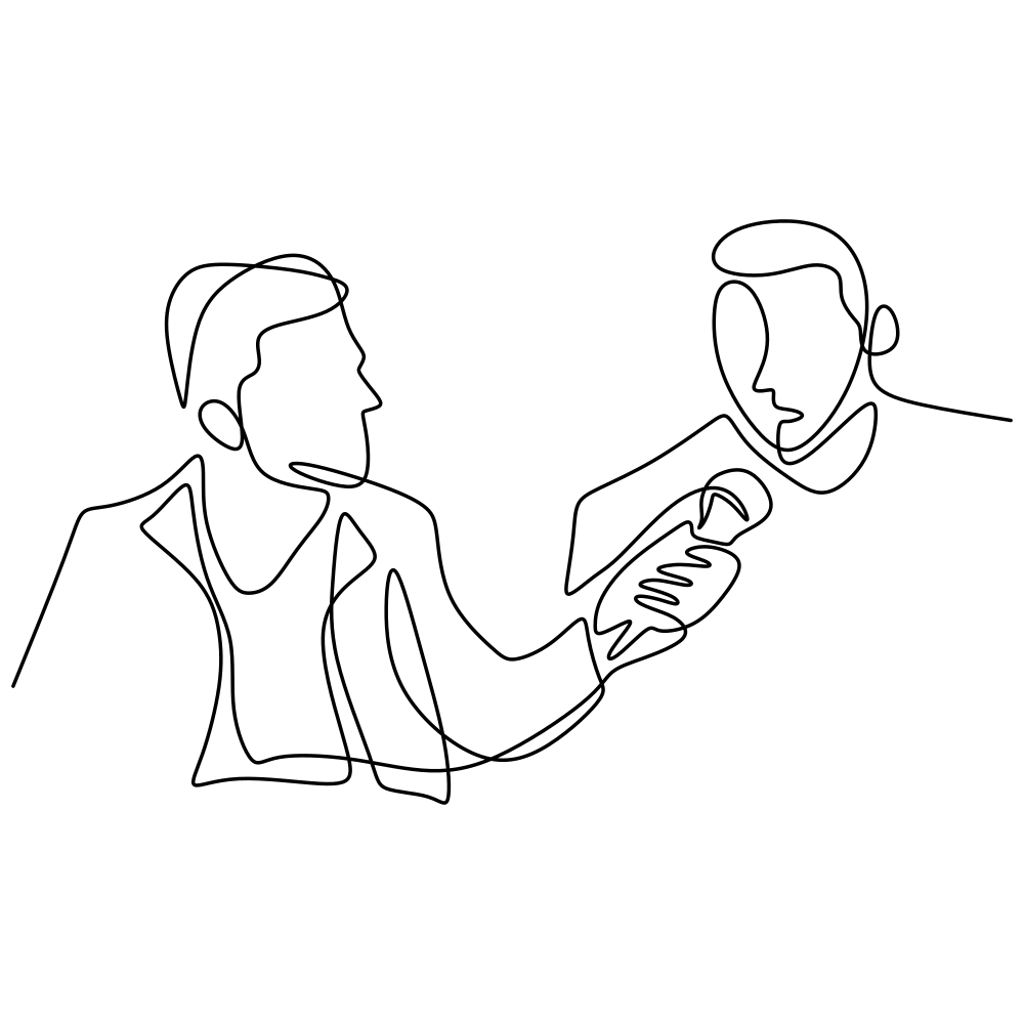}
    & \adjincludegraphics[width=\linewidth,trim={{0.0\width} {0.17\height} {0.0\width} {0.17\height}},clip]{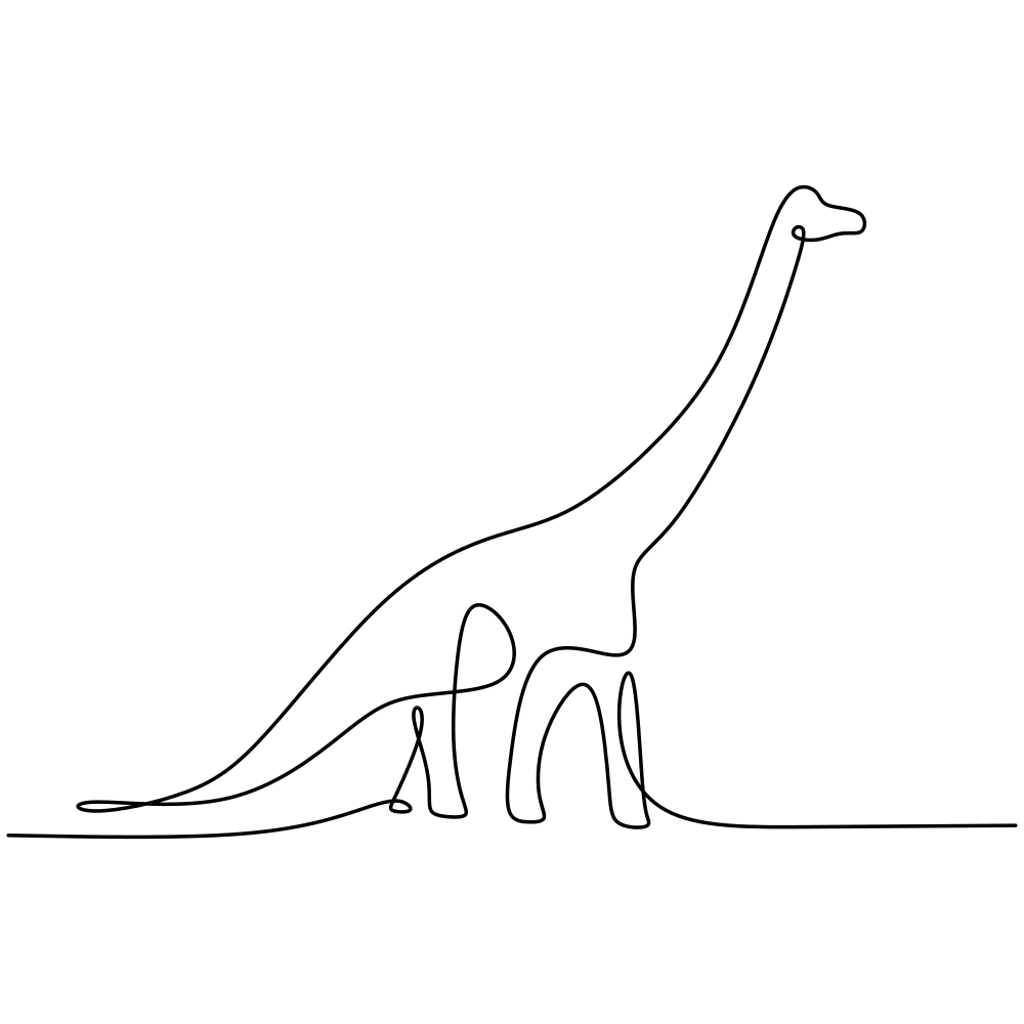}
    & \adjincludegraphics[width=\linewidth,trim={{0.0\width} {0.17\height} {0.0\width} {0.17\height}},clip]{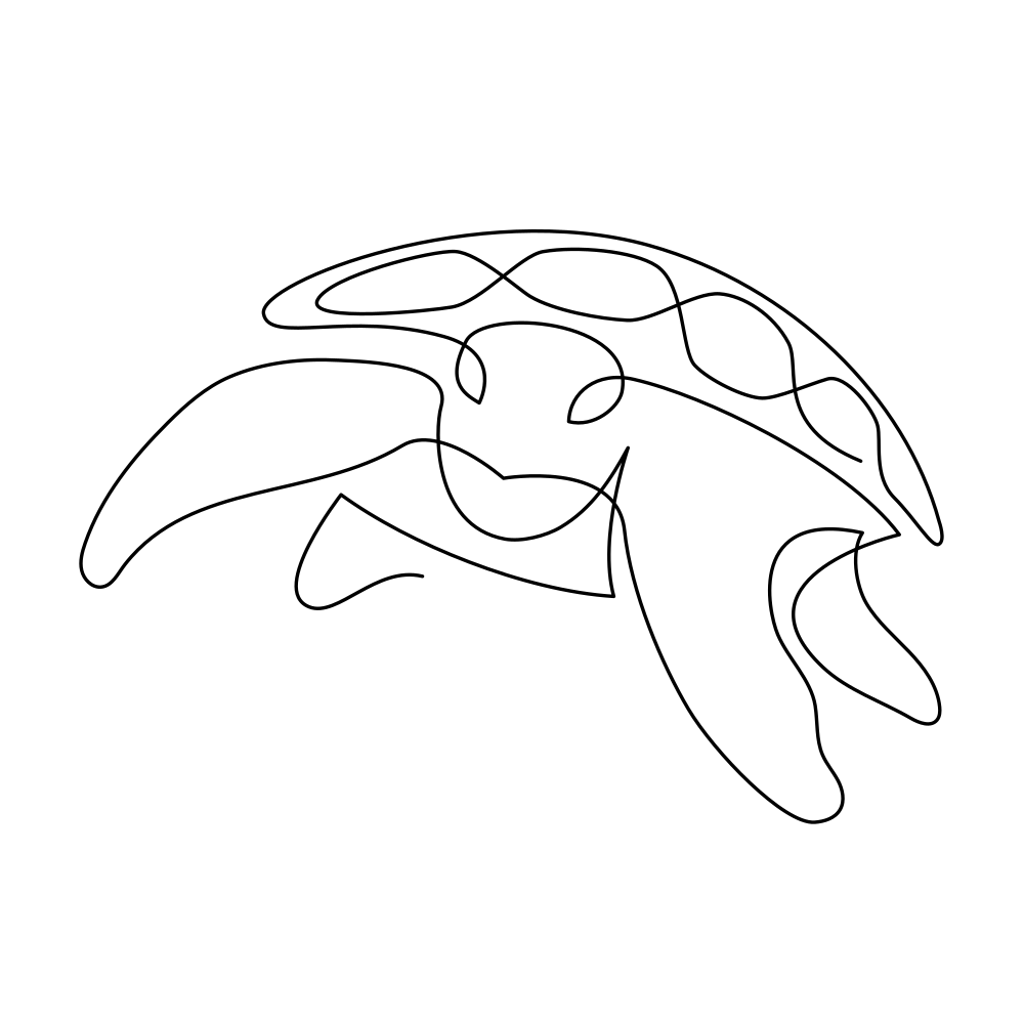}\\
    \parbox{0.96\linewidth}{\raggedright\scriptsize A single line drawing of two men, one interviewing the other with a microphone, rendered in black on a white background.}
    &\parbox{0.96\linewidth}{\raggedright\scriptsize A continuous line drawing of a dinosaur, rendered in black on a white background.}
    &\parbox{0.96\linewidth}{\raggedright\scriptsize A single line drawing of a turtle, created using a continuous line that flows and curves to form the features of the turtle.}
    \end{tabular}
    \caption{Sample from the training data of the LoRA model. Top: target image. Bottom: input prompt.}
    \label{fig:lora_training}
\end{figure}

Here, we present some additional information about the LoRA model we trained. First, some samples from the training dataset are presented in \figref{fig:lora_training}.

\begin{figure}[t]
    \setlength{\fboxsep}{0pt}
    \centering
    \adjincludegraphics[width=0.32\linewidth,trim={{0.0\width} {0.175\width} {0.0\width} {0.225\width}},clip]{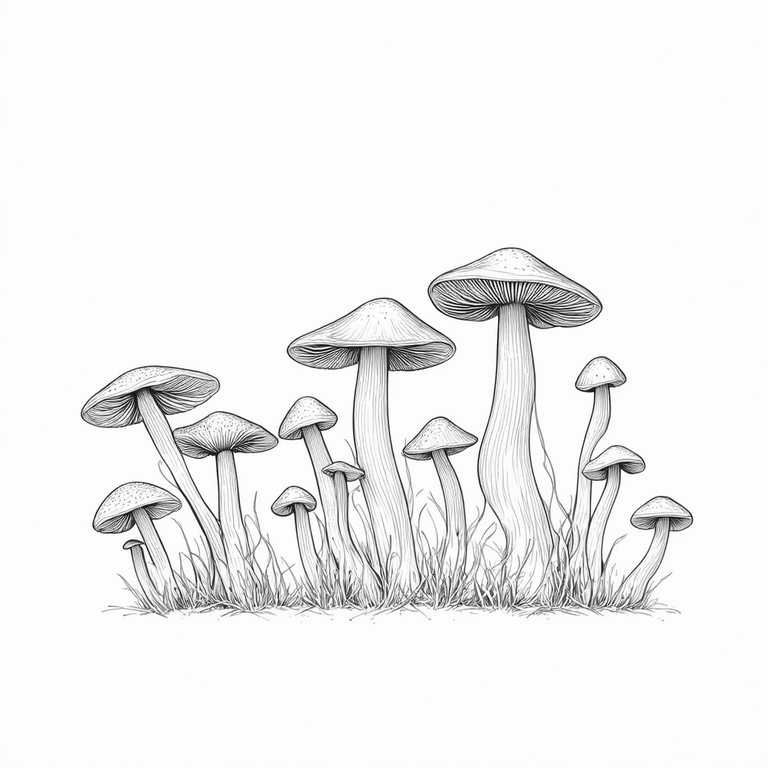}%
    \adjincludegraphics[width=0.32\linewidth,trim={{0.0\width} {0.2\width} {0.0\width} {0.2\width}},clip]{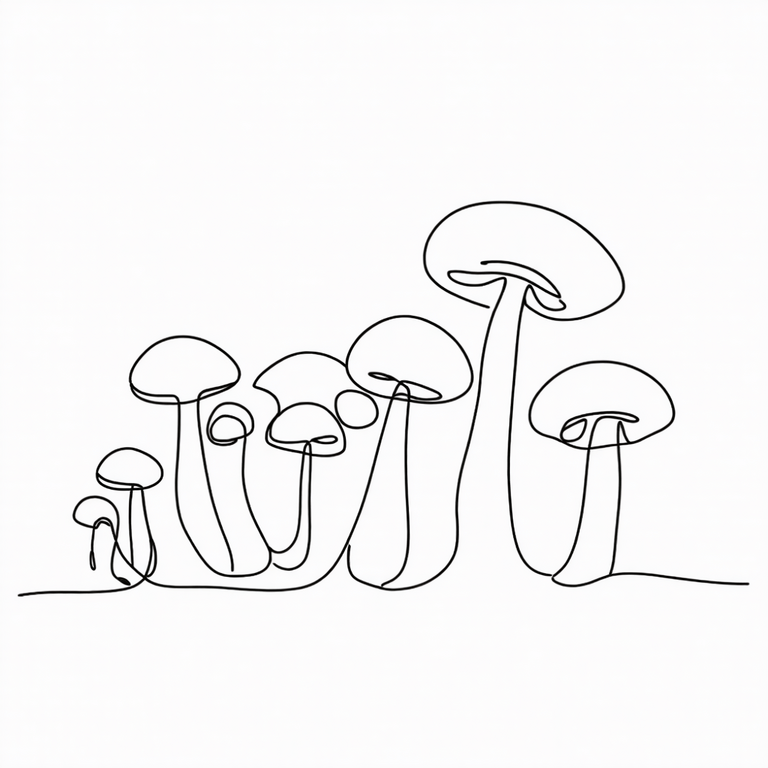}%
    \adjincludegraphics[width=0.32\linewidth,trim={{0.0\width} {0.2\width} {0.0\width} {0.2\width}},clip]{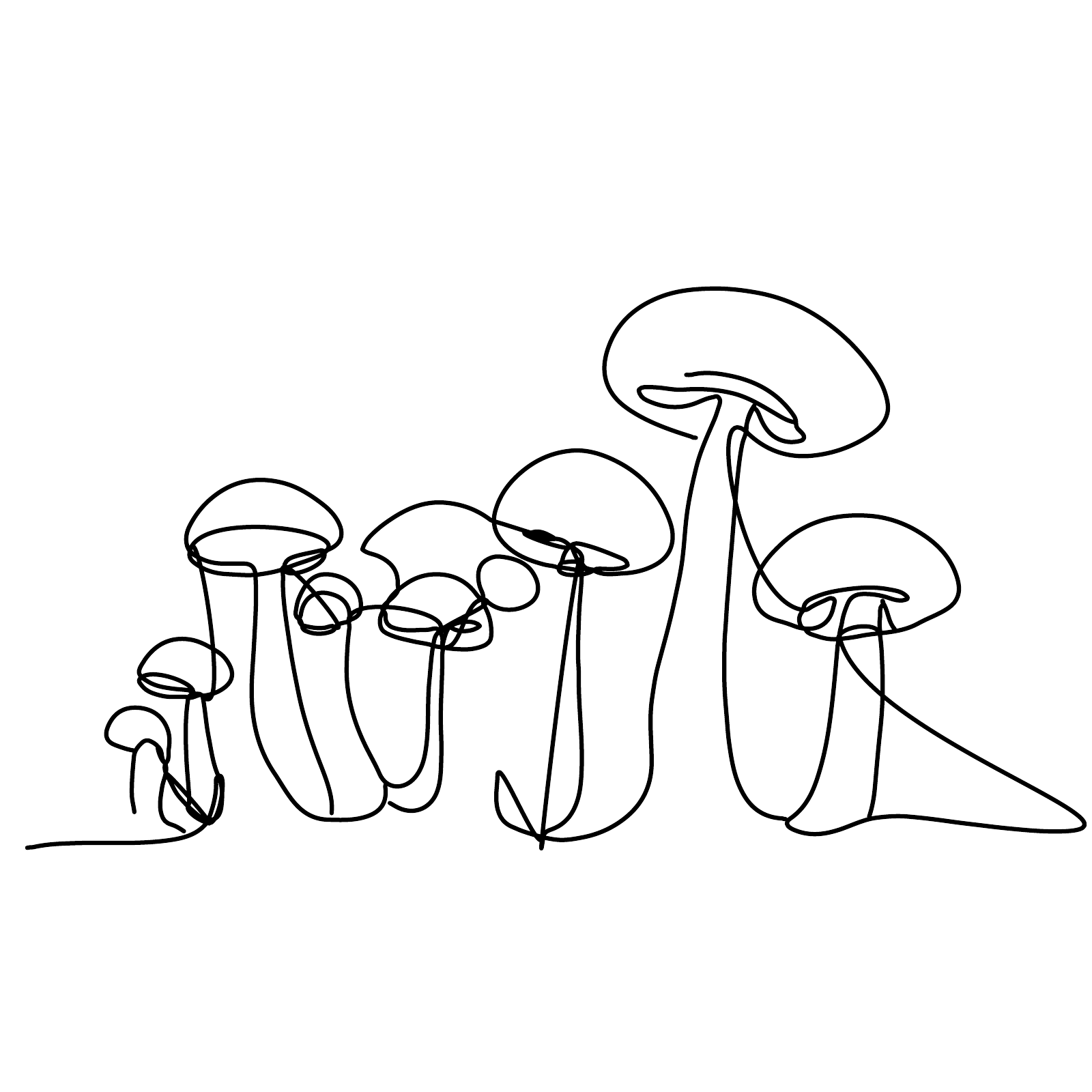}\\%
    \adjincludegraphics[width=0.32\linewidth,trim={{0.0\width} {0.2\width} {0.0\width} {0.225\width}},clip]{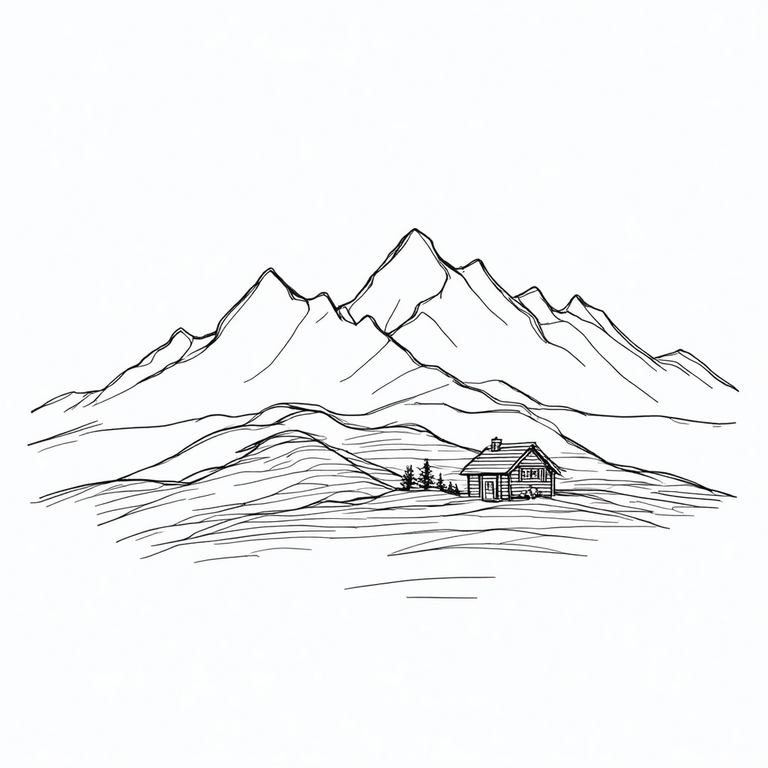}%
    \adjincludegraphics[width=0.32\linewidth,trim={{0.0\width} {0.2\width} {0.0\width} {0.225\width}},clip]{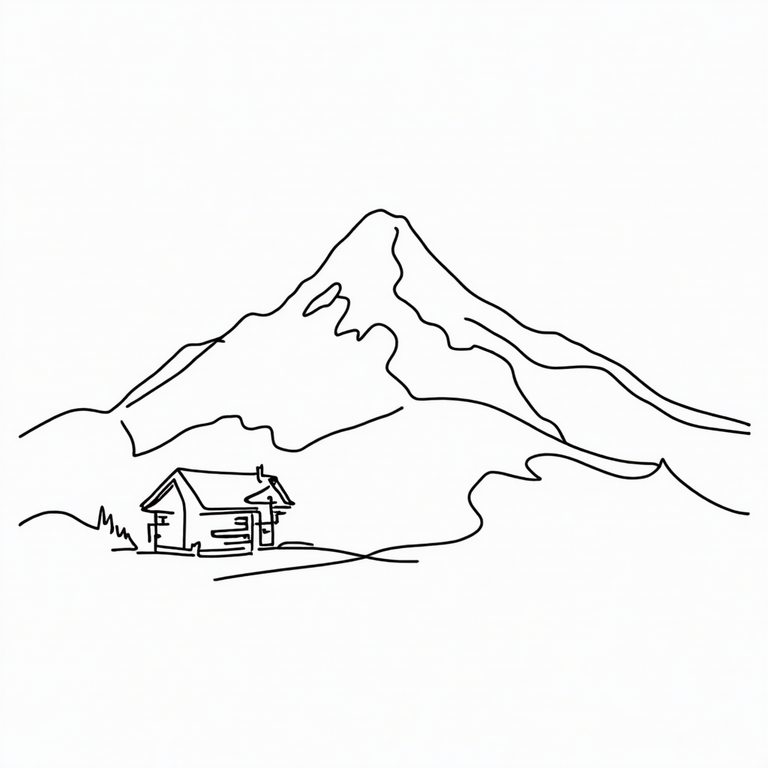}%
    \adjincludegraphics[width=0.32\linewidth,trim={{0.0\width} {0.2\width} {0.0\width} {0.225\width}},clip]{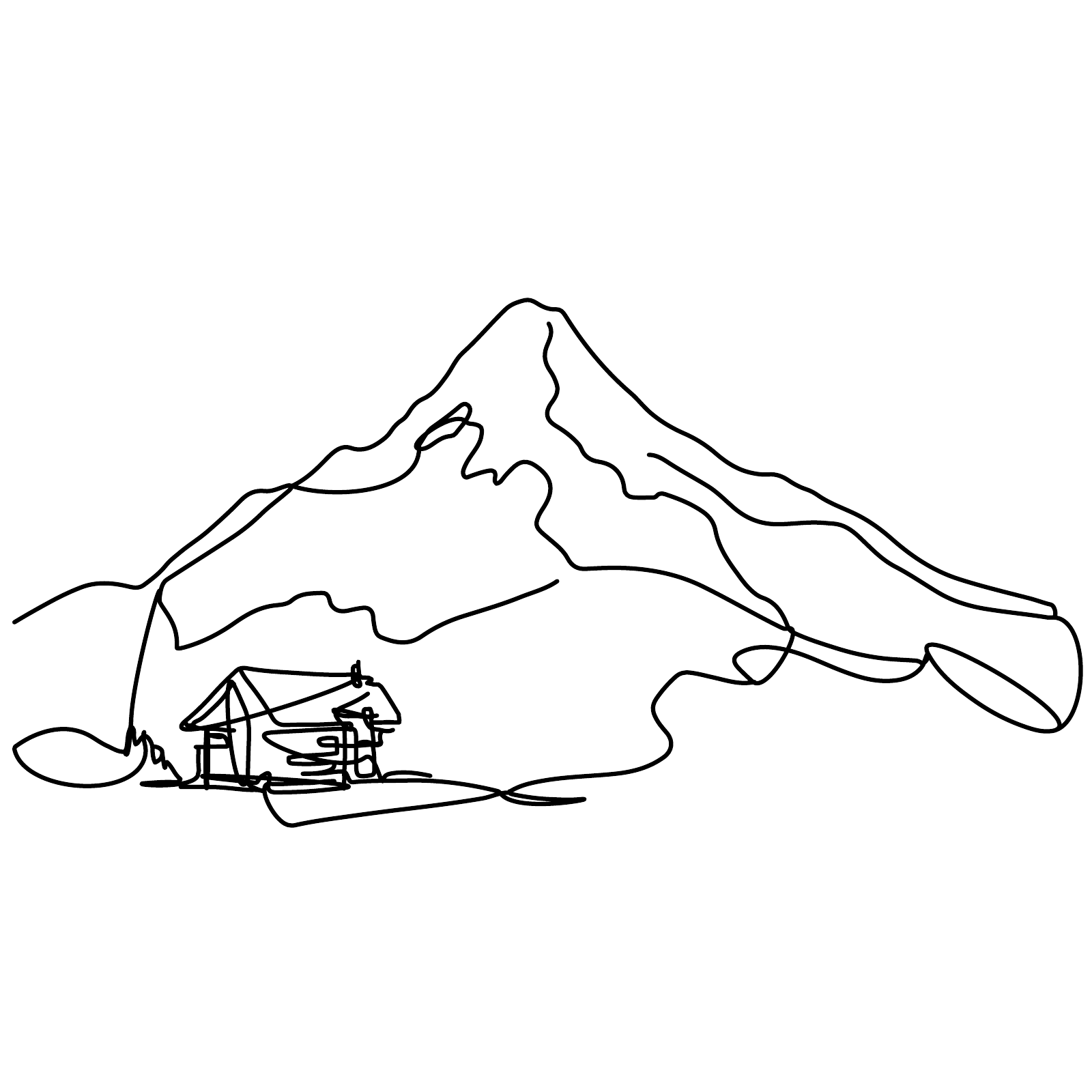}\\%
    \adjincludegraphics[width=0.32\linewidth,trim={{0.05\width} {0.2\width} {0.05\width} {0.225\width}},clip]{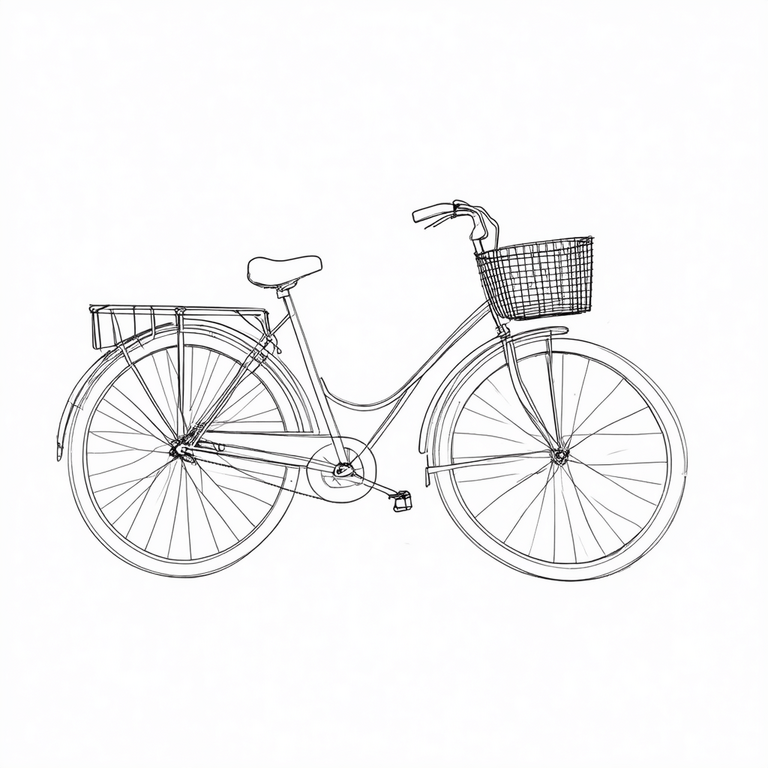}%
    \adjincludegraphics[width=0.32\linewidth,trim={{0.05\width} {0.2\width} {0.05\width} {0.225\width}},clip]{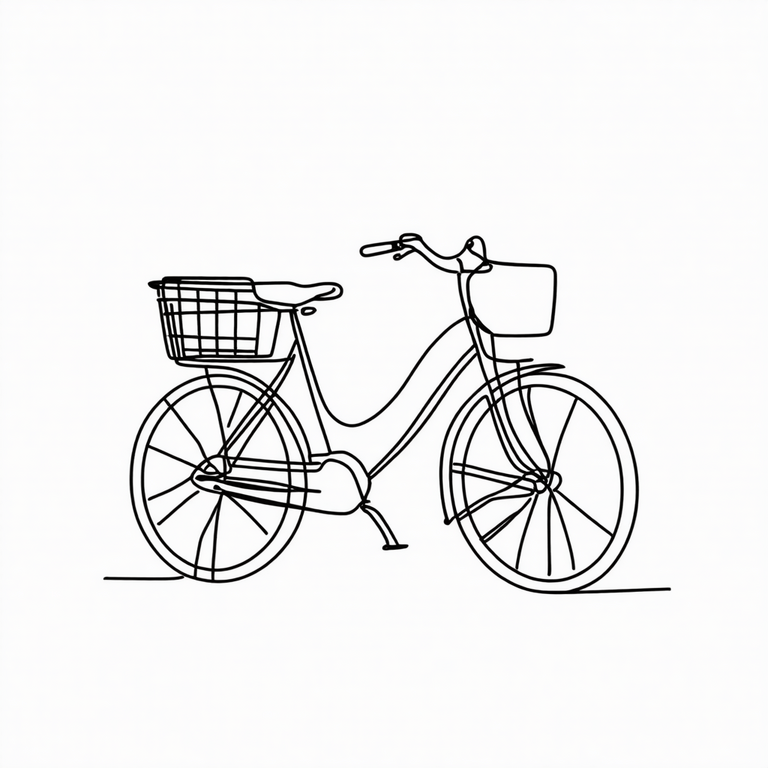}%
    \adjincludegraphics[width=0.32\linewidth,trim={{0.05\width} {0.2\width} {0.05\width} {0.225\width}},clip]{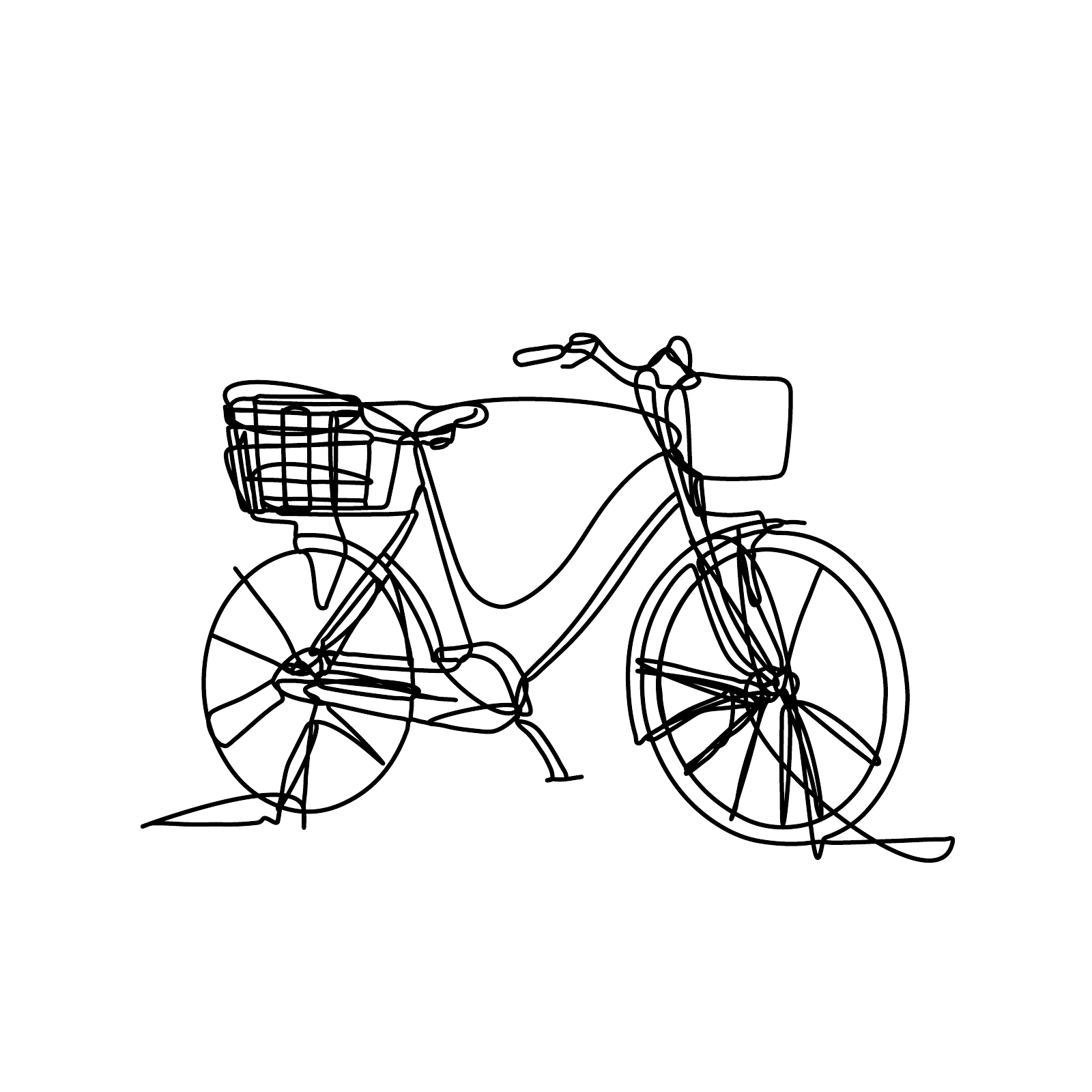}%
    \caption{Left: an image generated with the original weights of the diffusion model. Middle: an image generated using the LoRA weights trained on a dataset of single-line drawings. Right: the result of the LoRA model after applying line vectorization and the line connection strategy.}\label{fig:lora_example}
\end{figure}

We also present additional results created by our LoRA model in \figref{fig:lora_example}. Similar to \figref{fig:lora}, this figure shows examples of images generated using the original diffusion model and images generated using the diffusion model with the LoRA weights for the same prompt. While the result produced by the LoRA model is not a single-line drawing, it is much closer to the style of such drawings. The right column presents the results of the line connection strategy applied to the LoRA output. As can be observed for other methods where this strategy is applied, it leads to curves placed randomly and the overall impression that the drawing is a regular drawing with connected curves, but not a single-line drawing.

\section{Implementation details}
\label{app:implementation_details}

We implement our method in Python using PyTorch for automatic differentiation and DiffVG \cite{li.etal2020} for the differentiable rasterization, and based our implementation on ControlSketch~\cite{arar.etal2025}. We optimize the parameters of our curve using the Adam optimizer \cite{kingma.ba2015} for 4000 steps. The learning rate for the control points is set to 0.8. Because the weights for the URBS curve are in a different range (0--1) from the control points (0--512), we use a lower learning rate of 0.8/512. The parameters $\lambda_\text{rep}$ and $\lambda_\text{short}$ are set to 0.004 and 0.1, respectively;  $\lambda_\text{sparse}$ linearly increases from 0 to 2000 over the entire optimization. Initially, the number of control points is set to 385. We sample 5,000 points along the curve to rasterize it. On average, this corresponds to a distance of less than 1 pixel between consecutive sampled points, leading to maximal precision. The diffusion model used for computing the SDS loss is Stable Diffusion 3.5 (SD3.5) \cite{esser.etal2024}. On a machine with a 32\nobreakdashes-core Intel i9-13900K @ 5.5GHz CPU, 64GB RAM and an NVIDIA GeForce RTX 4090 GPU, the optimization runs in roughly 15 minutes. About 60\% of our approach is spent computing and back-propagating the SDS loss and 30\% on sampling and rasterizating the curve.

\section{Perceptual study details}
\label{app:user_study}

We present here details regarding the design and results of our perceptual study.

\begin{figure}[t]
    \centering
	\small
    \setlength{\fboxsep}{0pt}
    \includegraphics[width=\linewidth]{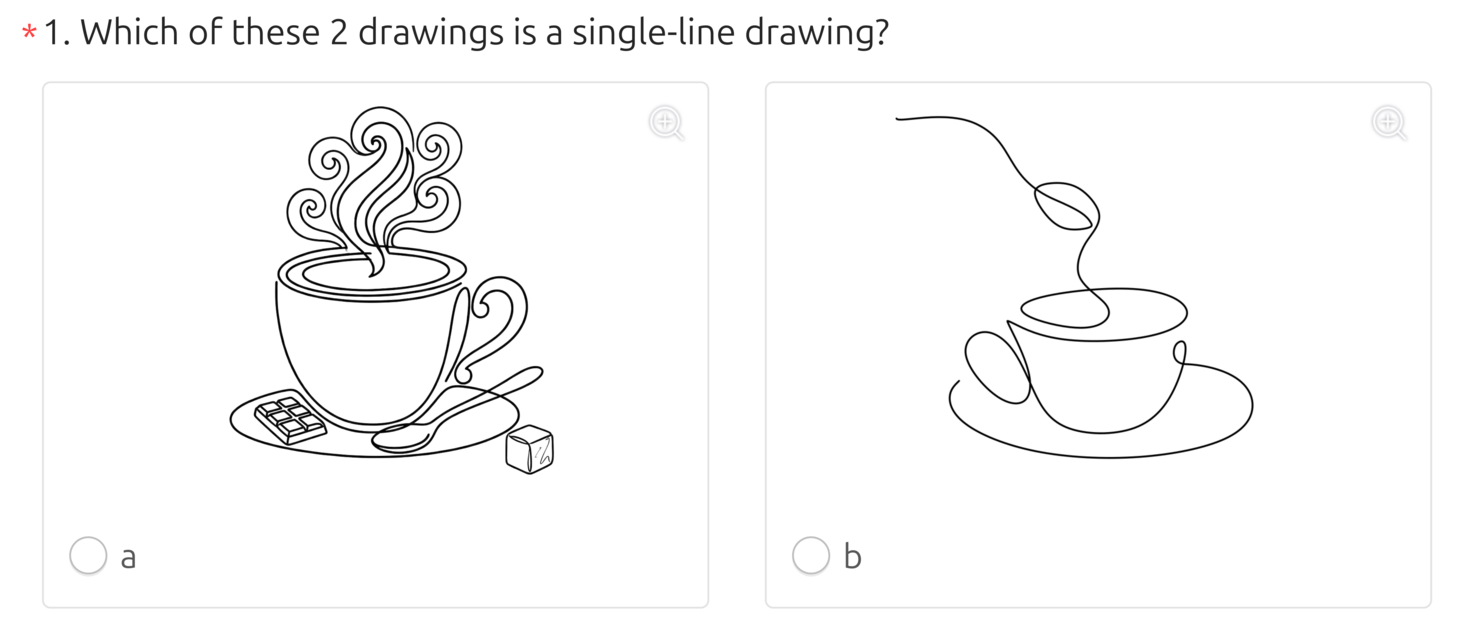}
    \includegraphics[width=\linewidth]{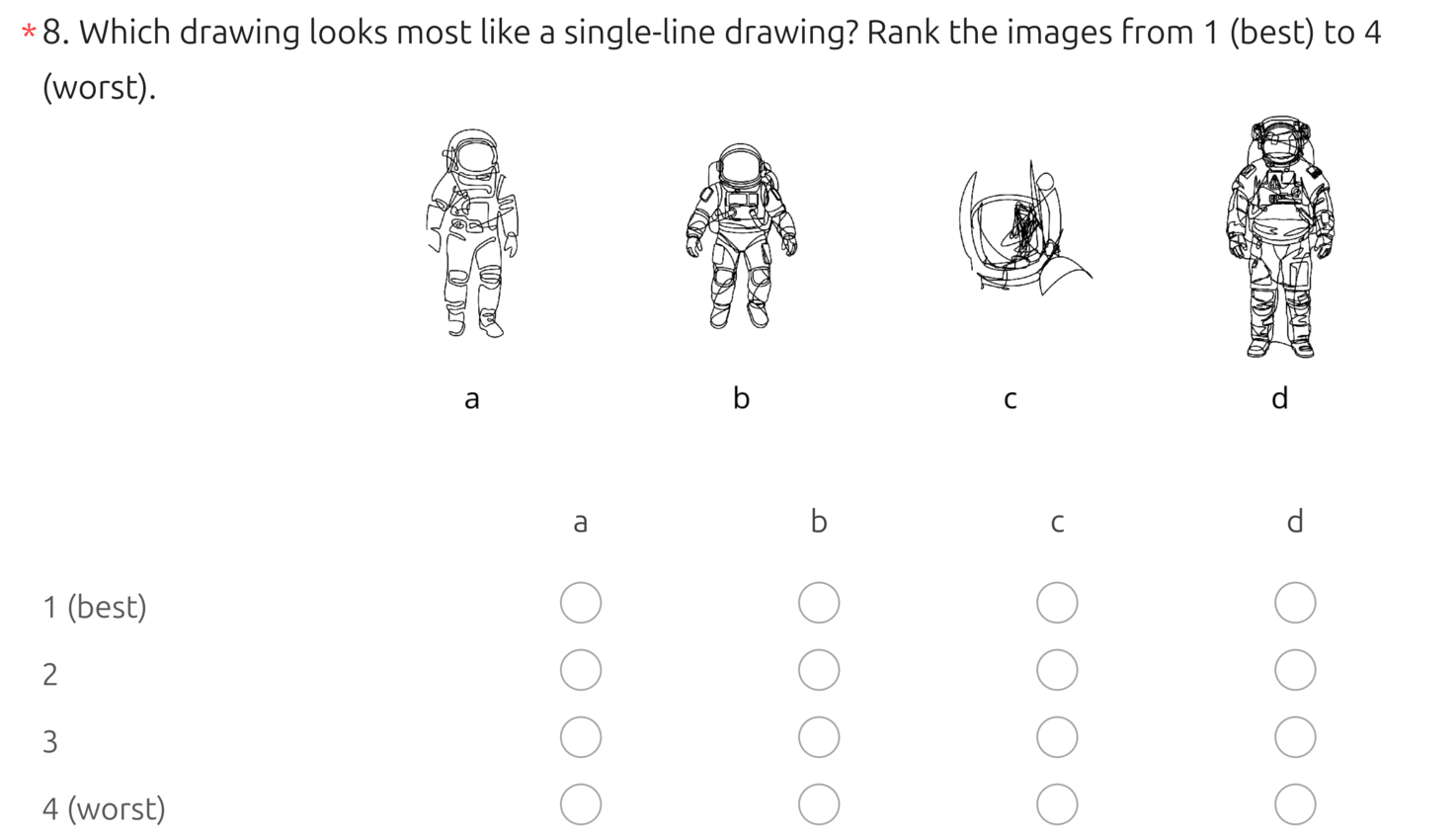}
    \caption{Questions used in the perceptual study. The top question is an example of the first 7 questions, while the bottom one is an example of the next 15 questions.}
    \label{fig:user_study}
\end{figure}

\textit{Design.\ } The study is divided into two parts. The first part consists of 7 questions where the user is presented with two drawings, one being a single-line drawing and the other not. In this part, the user has to distinguish which drawing is a single-line drawing (see \figref{fig:user_study}, top). The second part contains 15 questions and forms the core of the study. Here, participants are presented with 4 images, each created with a different method, as described in \secref{subsec:user_study}. All of these images are actual single-line drawings (in the sense that the vector representation contains a single stroke). Participants have to rank them based on their adherence to the single-line drawing aesthetic (see \figref{fig:user_study}, bottom).

\textit{Results.\ } The first part of the study assesses whether participants are capable of correctly distinguishing single-line drawings from regular drawings. Of the 49 respondents, only 2 made mistakes on these questions, one scoring 4/7 and the other 6/7. This demonstrates the users' ability to effectively distinguish these types of drawings. The second part of the study evaluates how effectively our method conveys the stylistic features that characterize a single-line drawing (simplicity, curve smoothness, abstraction). The results presented in \tabref{tab:userstudytable} show that our method is perceived to adhere most closely to the aesthetic of artist-created single-line drawings.

\end{document}